\documentclass[pra,reprint,notitlepage,superscriptaddress,nofootinbib,longbibliography]{revtex4-1}
\usepackage[margin=1.5cm,includehead,includefoot]{geometry}
\usepackage{amsmath,amssymb,bbm,mathrsfs,graphicx,txfonts}
\usepackage{comment,braket,framed}
\usepackage[bookmarks=true,colorlinks,linkcolor=blue,citecolor=blue,urlcolor=blue]{hyperref}
\usepackage[english]{babel}

\newcommand{\eq}[2]{\begin{eqnarray}\label{#1} #2 \end{eqnarray}}
\newcommand{\cre}[2]{#1^{\dagger}_{#2}}
\newcommand{\creo}[2]{#1^{\dagger}_{#2}}
\newcommand{\anno}[2]{#1^{\phantom{\dagger}}_{#2}}

\newcommand{\ann}[2]{#1_{#2}}

\newcommand{\R}{\mathbb{R}}
\newcommand{\Z}{\mathbb{Z}}
\newcommand{\cc}{\mathrm{c.c.}}
\newcommand{\hc}{\mathrm{H.c.}}
\newcommand{\id}{\mathbbm{1}}
\newcommand{\qnb}{\bar{\phi}_q = \bar{\phi}_q^{*} = 0}
\newcommand{\qn}{\phi_q = \phi_q^{*} = 0}
\newcommand{\ezr}{\eta_{\mathit{ZR}}}

\newcommand{\abs}[1]{\left\lvert #1 \right\rvert}
\newcommand{\tr}{\mathop{\mathrm{tr}}}
\newcommand{\Tr}{\mathop{\mathrm{Tr}}}

\renewcommand{\Im}{\mathop{\mathrm{Im}}}

\begin{document}

\title{Keldysh Field Theory  for Driven Open Quantum Systems}

% \author[1]{L. M. Sieberer,}
% \author[2]{M. Buchhold,}
% \author[2,3,4]{S. Diehl}

% \affil[1]{Department of Condensed Matter Physics, Weizmann Institute of Science,
%   Rehovot 7610001, Israel}

% \affil[2]{Institute of Theoretical Physics, TU Dresden, D-01062 Dresden,
%   Germany}
  
% \affil[3]{Kavli Institute for Theoretical Physics, University of California,
%   Santa Barbara, CA 93106, USA}
  
% \affil[4]{Institut f\"ur Theoretische Physik, Universit\"at zu K\"oln, D-50937
%   Cologne, Germany}

%   \maketitle

\author{L. M. Sieberer}
\affiliation{Department of Condensed Matter Physics, Weizmann Institute of Science,
  Rehovot 7610001, Israel}

\author{M. Buchhold}
\affiliation{Institute of Theoretical Physics, TU Dresden, D-01062 Dresden,
  Germany}

\author{S. Diehl}
\affiliation{Institute of Theoretical Physics, TU Dresden, D-01062 Dresden,
  Germany}
\affiliation{Kavli Institute for Theoretical Physics, University of California,
  Santa Barbara, CA 93106, USA}  
\affiliation{Institut f\"ur Theoretische Physik, Universit\"at zu K\"oln, D-50937
  Cologne, Germany}

\begin{abstract}
  Recent experimental developments in diverse areas --- ranging from cold atomic
  gases to light-driven semiconductors to microcavity arrays --- move systems
  into the focus which are located on the interface of quantum optics, many-body
  physics and statistical mechanics. They share in common that coherent and
  driven-dissipative quantum dynamics occur on an equal footing, creating
  genuine non-equilibrium scenarios without immediate counterpart in equilibrium
  condensed matter physics. This concerns both their non-thermal stationary states, as well as their many-body time evolution. It is a
  challenge to theory to identify novel instances of universal emergent
  macroscopic phenomena, which are tied unambiguously and in an observable way
  to the microscopic drive conditions. In this review, we discuss some recent
  results in this direction. Moreover, we provide a systematic introduction to
  the open system Keldysh functional integral approach, which is the proper
  technical tool to accomplish a merger of quantum optics and many-body physics,
  and leverages the power of modern quantum field theory to driven open quantum
  systems.
\end{abstract}

\maketitle

\tableofcontents

\section{Introduction}
\label{sec:intro}

Understanding the quantum many-particle problem is one of the grand challenges
of modern physics. In thermodynamic equilibrium, the combined effort of
experimental and theoretical research has made tremendous progress over the last
decades, revealing the key concepts of emergent phenomena and universality. This
refers to the observation, that the relevant degrees of freedom governing the
macrophysics may be vastly different from those of the microscopic physics, but
on the other hand are constrained by basic symmetries on the short distance
scale, restoring predictive power. Regarding the role and power of these
concepts in out-of-equilibrium situations, there is a large body of work in the
context of classical near equilibrium and non-equilibrium many-body physics and
statistical
mechanics~\cite{Hohenberg1977,Halpin-Healy1995,Jensen1998,Hinrichsen2000,Tauber2014a}. However,
analogous scenarios and theoretical tools for non-equilibrium quantum systems
are much less developed.

This review addresses recent theoretical progress in an important and uprising
class of dynamical non-equilibrium phases of quantum matter, which emerge in
driven open quantum systems, where a Hamiltonian is not the only resource of
dynamics. This concerns both non-equilibrium stationary states, but also the
dynamics of such ensembles. Strong motivation for the exploration of non-thermal
stationary states comes from a recent surge of experiments in diverse areas: in
cold atomic gases~\cite{Bloch2008,Lewenstein2007,Cirac2012}, hybrid light-matter
systems of Bose-Einstein condensates placed in optical cavities are
created~\cite{Baumann2010,Ritsch2013}, or driven Rydberg ensembles are
prepared~\cite{Schauss12}; light-driven semiconductor heterostructures realize
Bose-Einstein condensation of
exciton-polaritons~\cite{Kasprzak2006,Carusotto2013}; coupled microcavity arrays
are pushed to a regime demonstrating both strong coupling of light and
matter~\cite{Houck2012} and scalability~\cite{Underwood2012}; large ensembles of
trapped ions implement varieties of driven spin
models~\cite{Britton2012,Blatt2012}. All those systems are genuinely made of
quantum ingredients, and share in common that coherent and driven-dissipative
dynamics occur on equal footing. This creates close ties to typical setups in
quantum optics. But on the other hand, they exhibit a continuum of 
degrees of freedom, characteristic for many-body physics \footnote{This includes the cases of extended spatial continuum and lattice systems. In both cases, a continuum of momentum modes obtains, underlying the characteristics of many-body problems at long wavelength.}. Systems located at
this new interface are not guaranteed to thermalize, due to the absence of
energy conservation and the resulting breaking of detailed balance by the
external drive at the microscale. They rather \emph{converge to non-equilibrium stationary 
  states of matter}, creating scenarios without counterpart in condensed,
equilibrium matter. This rules out conventional theoretical equilibrium concepts
and techniques to be used, and calls for the development of new theoretical
tools. The physical framework sparks broader theoretical questions on the
existence of new phases of bosonic~\cite{Lechner2013,Altman2015} and
fermionic~\cite{Piazza2014,Keeling2014,Brennecke2015} matter, the nature of
phase transitions in such driven
systems~\cite{Raftery2014,Hartmann2008,Ritsch2013, DallaTorre2010,Sieberer2013},
and the observable consequences of quantum mechanics at the largest
scales~\cite{VanHorssen2015a,Marino2015}. Beyond stationary
states~\cite{Hsiang2015}, a fundamental challenge is set by the \emph{time
  evolution} of interacting quantum systems, which is currently explored
theoretically~\cite{Rey2004,Berges2004a,Calabrese2006,Mitra2011,Koghee2014,Carusotto2010,Larre2015}
and experimentally in cold
atomic~\cite{Gring2012,Trotzky2012,Meinert2013,Hung2013,Jaskula2012} and
photonic systems~\cite{Nardin2009}. A key goal is to identify universal
dynamical regimes that hold beyond specific realizations or precise initial
conditions. Combining idealized closed system evolution with the intrinsic open
system character of any real world experiment takes this setting to the next
stage, and exhibits emergent dynamics markedly different from closed systems
both for short~\cite{heating,Buchholdmethod} and long evolution
times~\cite{Poletti2013,Cai2013,Poletti2013,Schachenmayer2014,Daley2014,Olmos2012,MarcuzziDP,Lang2016}.

The interplay of coherent and driven-dissipative dynamics can be a natural
consequence of the driving necessary to maintain a certain many-body
state. Going one step further, it is possible to exploit and further develop the
toolbox of quantum optics for the driven-dissipative manipulation of many-body
systems. Recently, it has been recognized that the concept of dissipative state
preparation in quantum optics~\cite{Poyatos1996,Bose1999} can be developed into
a many-body context, both
theoretically~\cite{Diehl2008,Verstraete2009,Weimer2010,Muller2012} and
experimentally~\cite{Krauter2011,Barreiro2011,Schindler2013}. Suitably tailored
dissipation then does not necessarily act as an adversary to subtle quantum
mechanical correlations such as phase coherence or
entanglement~\cite{Eisert2010,Kastoryano2011,Horstmann2013,Hoening2012,Hafezi2014,Lang2015}. In
contrast, it can even \emph{create} these correlations, and dissipation then
represents the dominant resource of many-body dynamics. In particular, even
topologically ordered states in spin systems~\cite{Kitaev06} or of fermionic
matter~\cite{Hasan2010,Qi2011} can be induced dissipatively (\cite{Weimer2010}
and~\cite{Diehl2011,Bardyn2012,Budich2015a}, respectively). These developments
open up a new arena for many-body physics, where the quantum mechanical
microscopic origin is of key importance despite a dominantly dissipative
dynamics.

Summarizing, this is a fledging topical area, where first results underpin the
promise of these systems to exhibit genuinely new physics. In the remainder of
the introduction, we will discuss in some more detail the three major
challenges which emerge in these systems. The first one concerns the
\emph{identification of novel macroscopic many-body phenomena}, which witness the
microscopic driven open nature of such quantum systems. Second, we anticipate
the \emph{theoretical machinery}, which allows us to perform the transition from micro-
to macrophysics in a non-equilibrium context in practice. Third, we
describe some representative \emph{experimental platforms}, which motivate the
theoretical efforts, and in which the predictions can be further explored.

\subsection{New phenomena} 

As pointed out above, one of the key goals of the research reviewed here is the
identification of new macroscopic many-body phenomena, which can be uniquely
traced back to the microscopic driven open nature of such systems, and do not
have an immediate counterpart in equilibrium systems.

The driven nature common to the systems considered here can always be associated
to the fact that the underlying microscopic Hamiltonian is \emph{ time
  dependent}, with a time dependence relating to external driving fields such as
lasers. When such an ensemble (i) in addition has a natural partition into a
``system'' and a ``bath'' --- a continuum of modes well approximated by harmonic
oscillators with short memory ---, (ii) the system-bath coupling is weak
compared to a typical energy scale of the ``system'' Hamiltonian and (iii)
linear in the bath creation and annihilation operators (so that they can be
integrated out straightforwardly), then an effective (still microscopic)
description in terms of combined Hamiltonian and driven-dissipative Markovian
quantum dynamics of the ``system'' ensues. The ``system'' dynamics obtains by
tracing out the bath variables.

The resulting effective microdynamics is ``\emph{non-equilibrium},'' in a sense
sharpened in Sec.~\ref{sec:therm-equil-sym}. \emph{This not only concerns the
  time evolution, but also holds for the non-equilibrium stationary states. }
More precisely, the above situation implies an explicit breaking of detailed
balance, since the ``system'' energy is not conserved due to the explicitly
time-dependent drive.

What do we actually mean by \textit{``detailed balance''} and \textit{``thermal
  equilibrium?''} In an operational sense, the principle of detailed balance
states that there is a partition invariance for the temperature (or, more
generally, the noise level) present in the system: an arbitrary bipartition of
the system can be chosen, one part can be traced out, and the resulting
subsystem will be at the same temperature (noise level) as the total
system. This partition invariance is the condition for a globally well-defined
temperature characteristic for systems in thermal equilibrium. More formally,
thermal equilibrium can be detected by means of so-called
fluctuation-dissipation relations (FDRs). These connect the two fundamental
observables in physical systems --- correlation and response functions (see
Sec.~\ref{sec:examples}). In the case of thermal equilibrium, the connection is
dictated by the particle statistics alone. It is then given by the Bose- and
Fermi-distributions, respectively. Deviations from this universal form, which
has only two free parameters (temperature and chemical potential, relating to
the typical conserved quantities energy and particle number), provide a
necessary requirement for non-equilibrium conditions.

In non-equilibrium stationary states, no such general form exists. We will
encounter a concrete and simple example in the context of the \emph{driven Dicke model}
(a cavity photon coupled to a collective spin) in Sec.~\ref{sec:ising}, where
the form of the FDR depends on the observable we are choosing (e.g., the
position or momentum correlations and responses).

On the other hand, thermal FDRs can \emph{emerge}
at long wavelength, even though the microscopic dynamics manifestly breaks
detailed balance~\cite{Sieberer2013}. In particular, in three dimensions and close to the critical point of driven-dissipative Bose-condensation, a degeneracy of critical exponents indicates a \emph{universal
asymptotic thermalization}, in the sense of an \emph{emergent} thermal FDR. A similar
phenomenology is observed in a disordered multimode extension of the Dicke
model, see Sec.~\ref{sec:multi}. This underlines the strongly attractive nature
of the thermal equilibrium fixed point at low frequencies. Still in these
systems, non-equilibrium conditions leave their traces in the
dynamical response of the system, in terms of information that does not enter the FDR at leading order. For example, the critical behavior is
characterized by a \emph{fine structure in a new and independent critical exponent},
which measures decoherence, and \emph{whose value distinguishes equilibrium and
non-equilibrium dynamics}, see Sec.~\ref{sec:critical-dynamics-3d}. 

Instead of emergent thermal behavior indicating the fadeout
of non-equilibrium conditions upon coarse graining, also the opposite behavior
is possible. For example, \emph{low dimensional ($d\leq2$) bosonic systems} at low
noise level, such as exciton-polaritons well above threshold, are not attracted to the 
equilibrium fixed point as their three dimensional counterparts, but \emph{rather flow
to the non-equilibrium fixed point} of the
Kardar-Parisi-Zhang~\cite{Kardar1986} universality class, see
Sec.~\ref{sec:absence-algebr-order}. This can be interpreted as a universal and
indefinite increase of the non-equilibrium strength, which is triggered even if
the violation of detailed balance at the microscopic level is very small.

Universal non-equilibrium phenomena can also occur in the \emph{time evolution of
driven open systems}. For example, intriguing \emph{scaling laws} describing algebraic
decoherence \cite{Cai2013}, anomalous diffusion \cite{Poletti2012}, or glass-like behavior \cite{Poletti2013,Lesanovsky2013,Lesanovsky2014,Olmos2014a} have been
identified in the long time asymptotics of driven spin systems close to the
stationary states. Conversely, the short time behavior of driven open lattice
bosons shows universal scaling laws directly witnessing the non-equilibrium
drive, see Sec.~\ref{sec:heating-dynamics}. This scaling can be related to a
strongly pronounced non-equilibrium shape of the time-dependent distribution
function in the early stages of evolution, and be traced back to conservation
laws of the driven-dissipative generator
of dynamics.

The above discussion mainly focuses on the difference between equilibrium and
non-equilibrium systems on the macroscopic level of observation. Another
direction, still much less developed, concerns the \emph{distinction between classical
and quantum effects}. Again, although the quantum
mechanical description is necessary at a microscopic level, the
persistence of quantum effects at the macroscale is not guaranteed. This is
mainly due to the Markovian noise level inherent to such quantum
systems. Nevertheless, systems with suitably engineered driven-dissipative
dynamics show typical quantum mechanical phenomena such as phase coherence \cite{Diehl2008,Verstraete2009,Diehl2010D,Eisert2010,Horstmann2013,Hoening2012},
entanglement \cite{Krauter2011,Barreiro2011,Kastoryano2011,Schindler2013}, or topological order \cite{Weimer2010,Diehl2011,Bardyn2012,Budich2015a,Lang2015}. Especially fermionic systems, which
 do not possess a classical limit, are promising in this direction.

\subsection{Theoretical concepts and techniques} 

The development of theoretical tools needed to perform the 
transition from micro- to macro-physics in driven open quantum systems is still
in progress, as a topic of current research. A reason for the preliminary status
of the theory lies in the fact that two previously rather independent
disciplines --- quantum optics and many-body physics --- need to be unified on a
technical level.
 
\emph{Quantum optical systems} are well described microscopically in terms of
Markovian quantum master equations, which treat coherent Hamiltonian and
driven-dissipative dynamics on  equal footing. To solve such equations both
for their dynamics and their stationary states, powerful techniques have been
devised. This comprises efficient exact numerical techniques for small enough
systems, such as the quantum trajectories
approach~\cite{Dalibard1992,Plenio1998,Daley2014}. But it also includes analytical
approaches such as perturbation theory for quantum master equations, e.g., in the frame of the
Nakajima-Zwanzig projection operator technique~\cite{Gardiner2000,Breuer2002}, or
mappings to $P$, $W$, or $Q$ representations~\cite{Carusotto2005}, casting the
problem from a second quantized formulation into partial differential equations.

A characteristic feature of traditional quantum optical systems is the finite
spacing of the few energy levels which play a role. When considering systems
with a spatial continuum of degrees of freedom instead, the energy levels become
continuous. This does not mean that the microscopic modelling in terms of a
quantum master equation is inappropriate: for the driven-dissipative terms in
such an equation, the assumption of spatially independent dissipative processes
(such as atomic loss or spontaneous emission) is still valid as long as the
emitted wavelength of radiation is well below the spatial resolution at the
scale where the microscopic model is defined. Indeed, in this situation,
destructive interference of radiation justifies the description of driven
dissipation in terms of incoherent processes. However, under these circumstances
the smallness of a microscopic expansion parameter no longer guarantees the
smallness of the associated perturbative correction. Here the reason is that in
perturbation theory, one is summing over intermediate states with propagation
amplitudes down to the longest wavelengths. This can lead to infrared
divergences in naive perturbation theory --- a circumstance that found its
physical interpretation and technical remedy in equilibrium in terms of the
renormalization group~\cite{Goldenfeld1992,CardyBook}. We emphasize that it is
precisely this situation of long wavelength dominance which underlies much of
the universality, i.e., insensitivity to microscopic details, which is
encountered when moving from the microscale to macroscopic observables in
many-particle systems.

The modern framework to understand \emph{many-particle problems in thermodynamic
  equilibrium} is in terms of the functional integral formulation of quantum
field theory. The spectrum of its application covers a remarkable range of
energy scales, from ultracold atomic gases to condensed matter systems with
strong correlations to quantum chromodynamics and the quantum theory of
gravity. It provides us with a well-developed toolbox of techniques, such as
\emph{diagrammatic perturbation theory} including sophisticated resummation
schemes. But it also encompasses non-perturbative approaches, which often
capitalize on the flexibility of the functional integral when it comes to
picking the relevant degrees of freedom for a given problem. This is the
challenge of \emph{emergent phenomena}, whose solution typically is strongly
scale dependent~\cite{Zinn-Justin2002}. Familiar examples include an efficient
description of emergent Cooper pair or molecular degrees of freedom in
interacting fermion systems, or vortices which conveniently parameterize the
long-wavelength physics of interacting bosons in two dimensions. The description
of the change of physics with scale was given its mathematical foundation in
terms of the \emph{renormalization group} already mentioned above, yet another
tool developed and most clearly formulated in a functional language. Finally,
the functional integral based on a single scalar quantity --- the system's
action, which encodes all the dynamics on the microscopic scale --- is a
convenient framework when it comes to the classification of \emph{symmetries and
  associated conservation laws}, and their use in devising approximation schemes
respecting them.

To put it short, while the driven open many-body systems are well described by
microscopic master equations, the traditional techniques of quantum optics
cannot be used efficiently -- at least not in the case where the generic
complications of many-body systems start to play a role. Conversely, their
driven open character makes it impossible to approach these problems in the
framework of equilibrium many-body physics. This situation calls for a
\emph{merger of the disciplines of quantum optics and many-body physics} on a
technical level. On the numerical side, progress has been made in one spatial
dimension recently by combining the method of quantum trajectories with powerful
density matrix renormalization group
algorithms~\cite{Daley2010,Bernier2013,Bonnes14,Bonnes14arx},
see~\cite{Daley2014} for an excellent review on the topic. For more analytical
approaches, the \emph{Keldysh functional or path
  integral}~\cite{Schwinger1961,Schwinger1960,Keldysh1965,Mahanthappa1962,Bakshi1963,Bakshi1963a}
is ideally suited (but see~\cite{Koch14} for a recent systematic perturbative
approach to lattice Lindblad equations with extensions to sophisticated
resummation schemes~\cite{Koch15}, and~\cite{Degenfeld15,Weimer15,Finazzi15} for
advanced variational techniques). Conceptually, the latter captures the most
general situation in many-body physics --- the dynamics of a density matrix
under an arbitrary temporally local generator of dynamics. We refer
to~\cite{Kamenev2011,Calzetta2008} for an introduction to the Keldysh functional
integral. In our context, it can be derived by a direct functional integral
quantization of the Markovian quantum master equation. This procedure results in
a simple translation table from the master equation to the key player in the
associated Keldysh functional integral, the \emph{Markovian action}. At first
sight, the complexity of the non-equilibrium Keldysh functional integral is
increased by the characteristic ``doubling'' of degrees of freedom compared to
thermal equilibrium. However, it should be noted that it is precisely this
feature which relates the Keldysh functional integral much closer to real time
(von Neumann) evolution familiar from quantum mechanics. We will demonstrate
this in Sec.~\ref{sec:deriv-keldysh-acti}, making the Markovian action a rather
intuitive object to work with. Furthermore, when properly harnessing symmetries,
the complexity of calculations can often be made comparable to thermodynamic
equilibrium. Most importantly however, the powerful toolbox of quantum field
theory is opened in this way. It is thus possible to leverage the full power of
sophisticated techniques from equilibrium field theory to driven open many-body
systems.

\emph{Relation to classical dynamical field theories} --- The quantum mechanical
Keldysh formulation reduces to the so-called \emph{Martin-Siggia-Rose (MSR)
  functional integral} in the (semi-) classical limit, in turn equivalent to a
stochastic Langevin equation formulation~\cite{Kamenev2011}. This statement will
be made precise and discussed in Sec.~\ref{sec:semicl-limit-keldysh}. A large
amount of work has been dedicated to this limit in the past. On the one hand,
this concerns dynamical aspects of equilibrium statistical mechanics, and we
refer to the classic work by Hohenberg and Halperin~\cite{Hohenberg1977} for an
overview. This work shows that, while the static universal critical behavior is
determined by the symmetries and the dimensionality of the problem, the
dynamical critical behavior is sensitive to additional dynamical conservation
laws. This leads to a fine structure, defining dynamical universality classes
which are denoted by models A--J~\cite{Hohenberg1977}. These models also provide
a convenient framework to describe the statistics of work, summarized in
Jarzynski's work theorem~\cite{Jarzynski1997,Jarzynski1997a} and Crooks'
relation~\cite{Crooks1999}. On the other hand, non-equilibrium situations are
captured as well. Here, we highlight in particular genuine non-equilibrium
universality classes, which are not smoothly connected to the equilibrium
models. Among them is the problem of reaction-diffusion models~\cite{CardyRev}
including directed percolation~\cite{Hinrichsen2000}, which is relevant to
certain chemical processes (for an implementation of this universality class
with driven Rydberg gases, see~\cite{MarcuzziDP}). Another key example is
surface growth, described by the Kardar-Parisi-Zhang equation~\cite{Kardar1986},
giving rise to a non-equilibrium universality class which is at the heart of
driven phenomena such as the growth of bacterial colonies or the spreading of
fire.

In the same class of approaches ranges the
so-called \emph{Doi-Peliti functional integral}~\cite{Doi1976,Peliti1985}, which is a
functional representation of classical master equations, and may be viewed as an
MSR theory with a specific, highly non-linear appearance of the field variables. It
reduces to the conventional MSR form in a leading order Taylor expansion of
the field non-linearities. A comprehensive overview of models, methods, and
physical phenomena in the (semi-)classical limit is provided in~\cite{Tauber2014a}.

We also note that the usual mean field theory, where correlation functions are
factorized into products of field amplitudes and which is often used as an
approximation to the quantum master equation in the literature, corresponds to a
further formal simplification of the semi-classical limit. Here the effects of
noise are neglected completely. Conversely, the semi-classical limit represents
a systematic extension of mean field theory, which includes the Markovian noise
fluctuations. This level of approximation is referred to as \emph{optimal path
  approximation} in the literature on MSR functional
integrals~\cite{Kamenev2011,Tauber2014a}.

In many cases, even though the microscopic description is in terms of a quantum
master equation, at long distances the Keldysh field theory reduces to a
semi-classical MSR field theory. The reason is the finite
Markovian noise level that such systems exhibit generically, as explained in
Sec.~\ref{sec:semicl-limit-keldysh}. The prefix ``semi'' refers to the fact that
phase coherence may still persist in such circumstances --- the situation is
comparable to a Bose-Einstein condensate at finite temperature. Recently
however, situations have been identified where the drastic simplifications of
the semi-classical limit do not apply. In particular, this occurs in systems
with \emph{dark states} --- pure quantum states which are dynamical fixed points
of driven-dissipative evolution~\cite{Diehl2008,Diehl2010a}. In these cases,
classical dynamical field theories are inappropriate, which calls for
the development of \emph{quantum dynamical field
  theories}~\cite{Marino2015}. These developments are just in their beginnings.

In this review, we concentrate on systems composed of bosonic degrees of
freedom. However, it is also possible to address spin systems in terms of
functional integrals~\cite{Sachdev2011}, and simple models systems have been
analyzed in this way, see~\cite{DallaTorre2013} and Sec.~\ref{sec:ising}. More
sophisticated approaches to spin systems were elaborated in the context of
multimode optical cavities in~\cite{Buchhold2013}, and systematically for
various symmetries for lattice systems
in~\cite{Maghrebi2015}. Fermi statistics is also conveniently implemented in the
functional integral formulation. This is relevant, e.g., for driven open Fermi gases
in optical cavities~\cite{Keeling2014,Piazza2014a,Brennecke2015} or
lattices~\cite{Bernier2014}, or dissipatively stabilized topological fermion
matter~\cite{Diehl2011,Bardyn2012,Budich2015a}.

\subsection{Experimental platforms}
\label{sec:exper-platf}

The progress in controlling, manipulating, detecting and scaling up driven open
quantum systems to many-body scenarios has been impressive over the last
decade. Here we sketch the basic physics of three representative platforms, and
indicate the relevant microscopic theoretical models in the frame of the
Markovian quantum master equation. In later sections, we will translate this
physics into the language of the Keldysh functional integral. For each of these
platforms, excellent reviews exist, which we refer to at the end of
Sec.~\ref{sec:outline} together with further literature on open systems. The
purpose of this section is to give an overview only, and to put the respective
platforms into their overarching context as driven open quantum systems with
many degrees of freedom.

\subsubsection{Cold atoms in an optical cavity, and microcavity arrays: driven-dissipative spin-boson
  models}
\label{sec:cold-atoms-cavity}

Cavity quantum electrodynamics (cavity QED), with its focus on strong
light-matter interactions, is a growing field of research, which has experienced
several groundbreaking advances in the past few years. Historically, these
systems were developed as few or single atom experiments, detecting the
radiation properties of atoms, which are strongly coupled to a quantized light
field. The focus has recently been shifted towards loading more and more atoms
inside a cavity. Thereby, not only single particle dynamics in strong radiation
fields can be probed, but also collective, macroscopic phenomena, which are
driven by light-matter interactions. For an excellent review on this topic,
covering important experimental and theoretical developments,
see~\cite{Ritsch2013}. In the experiments, cold atoms are loaded inside an
optical or microwave cavity, for which the coherent interaction between the
atomic internal states and a single cavity mode dominates over dissipative
processes~\cite{Mabuchi2002}. The atoms absorb and emit cavity photons, thereby
changing their internal states. Due to this process, the spatial modulation of
the intra-cavity light field induces a coupling of the cavity photons to the
atomic internal state as well as their motional degree of
freedom~\cite{Hood2000}. In this way, cavity photons mediate an effective
atom-atom interaction, which leads to a back-action of a single atom on the
motion of other atoms inside the cavity. This cavity mediated interaction is
long-ranged in space and represents the source of collective effects in cold
atomic clouds within cavity QED experiments. One hallmark of collective dynamics
in cavity QED has been the observation of self-organization of a Bose-Einstein
condensate in an optical cavity. This is accompanied by a Dicke phase transition
via breaking of a discrete $Z_2$-symmetry of the underlying
model~\cite{Nagy2010,Baumann2010,Baumann2011,Maschler2005}.

Although the dominant dynamics in these systems is coherent, there are
dissipative effects which cannot be discarded. These are the loss of cavity
photons due to imperfections in the cavity mirrors, or spontaneous emission
processes of atoms, which emit photons into transverse modes. These effects
modify the dynamics of the system on the longest time scales. Therefore, they
become relevant for macroscopic phenomena, such as phase transitions and
collective dynamics. For instance, dissipative effects have been shown
theoretically~\cite{Oztop2012,Kulkarni13} and experimentally~\cite{Landig2015}
to modify the critical exponent of the Dicke transition compared to its zero
temperature value. This illustrates that for the analysis of collective
phenomena in cavity QED experiments, the dissipative nature of the system has to
be taken into account properly~\cite{DallaTorre2013,Brennecke2013}.

Typically, for a cavity field which has a very narrow spectrum, the atomic
internal degrees of freedom can be reduced to two internal states, whose
transitions are nearly resonant to the photon frequency. The operators acting on these two
internal states can be represented by Pauli
matrices, making them equivalent to a spin-$1/2$ degree of freedom. A
very important model in the framework of cavity QED is the Dicke model~\cite{Hepp1973,Wang1973,Emary2003}, which
describes $N$ atoms (i.e., two-level systems) coupled to a single quantized
photon mode. This is expressed by the Hamiltonian (here and in the following
we set $\hbar = 1$)
 \eq{Spin1a}{
  H_{\text{D}}=\omega_0
  a^{\dagger}a+\frac{\omega_z}{2}\sum_{i=1}^{N}\sigma^z_i+\frac{g}{\sqrt{N}}\sum_{i=1}^N\sigma^x_i\left(a^{\dagger}+a\right).}
Here, $\omega_0$ is the photon frequency,
$\sigma^z=|1\rangle\langle1|-|0\rangle\langle0|$ represents the splitting of the
two atomic levels with energy difference $\omega_z$. 
$\sigma^x=|0\rangle\langle1|+|1\rangle\langle0|$ describes the coherent
excitation and de-excitation of the atomic state proportional to the atom-photon
coupling strength $g$. The Dicke model features a discrete $Z_2$ Ising
symmetry: it is invariant under the transformation
$(a^{\dagger},a,\sigma^x_i)\mapsto(-a^{\dagger},-a,-\sigma^x_i)$. In
the thermodynamic limit, for $N\rightarrow\infty$, it features a phase
transition, which spontaneously breaks the Ising symmetry. Crossing the transition, the system  enters a
superradiant phase, characterized by finite expectation values
$ \langle a\rangle\neq0,\langle\sigma^x_i\rangle\neq0$. This describes
condensation of the cavity photons, i.e., the formation of a macroscopically occupied, coherent
intra-cavity field, and a ``ferromagnetic'' ordering of the atoms in the
$x$-direction.

Although the Dicke model is a standard model for cavity QED experiments in the ultra-strong coupling limit $\sqrt{N}g > \omega_z\omega_0$, it has been realized only very
recently in cold atom experiments, where an entire BEC was placed inside an
optical cavity~\cite{Baumann2010}. It has been shown that this setup maps to a Dicke
model, with a ``collective'' spin degree of freedom~\cite{Baumann2011}. Here, the detuning of the pump laser was chosen such that the atoms
effectively remain in the internal ground state, but acquire a characteristic
recoil momentum when scattering with a cavity photon. This scattering creates a
collective, motionally excited state, which replaces the role of an individual,
internally excited atom. The experimental realization of a superradiance
transition in the Dicke model is usually inhibited, since the required coupling
strength by far exceeds the available value of the atomic dipole
coupling. However, for the BEC in the cavity, the energy scales of the excited
modes are much lower than the optical scale of the atomic modes. In this way, the
superradiance transition indeed became experimentally accessible. This was
inspired by a theoretical proposal using two balanced Raman channels between
different internal atomic states  inside an optical cavity, which reduced
the effective level splitting of the internal states to much lower energy
scales~\cite{Dimer2007}.

In addition to the unitary dynamics represented by the Dicke model, the cavity
is subject to permanent photon loss due to imperfections in the cavity
mirrors. For high finesse cavities, the coupling of the intra cavity photons to
the surrounding vacuum radiation field is very weak, and the latter can be eliminated
in a Born-Markov approximation~\cite{Gardiner2000}. This results in a Markovian quantum
master equation for the system's density matrix \eq{Spin2a}{
  \partial_t\rho=-i[H_{\text{D}},\rho]+\mathcal{L}_d \rho, } where $\rho$ is the
density matrix for the intra cavity system and $\mathcal{L}_d$ adds dissipative
dynamics to the coherent evolution of the Dicke model. For a vacuum radiation
field, it is given by \eq{Spin3}{ \mathcal{L}_d \rho=\kappa\left(L\rho
    L^{\dagger}-\tfrac{1}{2}\{L^{\dagger}L,\rho\}\right). } The Lindblad operator $L$ acting on the density matrix describes pure
photon loss ($L = a$) with an effective loss rate $\kappa$; the latter depends on system
specific parameters, but is typically the smallest scale in the master
equation~\eqref{Spin2a}~\cite{Ritsch2013}.

In generic cavity experiments, there are also atomic spontaneous emission
processes. The atoms scatter a laser or cavity photon out of the cavity, and
this represents a source of decoherence. This process has been considered in
Ref.~\cite{DallaTorre2013}, and leads to an effective decay rate of the atomic
excited state and therefore to a dephasing of the atoms, described by additional
Lindblad operators $L_i = \sigma^z_i$ \cite{Murch2008}. However, these losses
are at least three orders of magnitude smaller than the cavity decay rate and
typically not considered~\cite{Baumann2010,Brennecke2013}. The basic model for
cavity QED with cold atoms is therefore represented by the Dicke model with
dissipation, formulated in terms of the master equation~\eqref{Spin2a}. Its
dynamics is discussed in Sec.~\ref{sec:neqspin}, including the Dicke
superradiance transition.

The Dicke model Hamiltonian takes the form $H_\text{D} = H_c + H_s + H_{cs}$,
where $H_{c,s,cs}$ represent the bosonic cavity, spin, and spin-boson sectors,
respectively. There are many directions to go beyond the Dicke model with single
collective spin, still keeping the basic feature of coupling spin to boson
(cavity photon) degrees of freedom.  One direction -- relevant to future cold
atom experiments -- is to consider multimode cavities instead of a single
one. In particular, in conjunction with quenched disorder, intriguing analogies
to the physics of quantum glasses can be established in this way
\cite{Strack2010G,Buchhold2013}. Here, the global coupling of all spins to a
single mode $g_i \equiv g$ is replaced by random couplings $g_{i,\ell}$, where
the index $\ell$ now refers to a collection of cavity modes. The many-body
physics of such an open system is discussed in Sec.~\ref{sec:neqspin}.

The basic building block of the Dicke model is the spin-boson term of the form
$H_{sc} \sim (a+a^\dag) \sigma^x$, i.e. a Rabi type non-linearity that preserves
the $Z_2$ symmetry of $H_{c,s}$.  In the context of circuit quantum
electrodynamics, a natural many-body generalization of Hamiltonians with a
spin-boson interaction is to consider entire arrays of microcavities (instead of
considering many modes within a single cavity). These cavities can be coupled to
each other by single photon tunnelling processes between adjacent cavities,
giving rise to Hubbard-type hopping terms $\sim J a_i^\dag a_j + \text{h. c.}$,
where $i,j$ now label the spatial index of the cavities. In cirquit QED, strong
non-linearities can be generated, e.g., by coupling to adjacent qubits made of
Cooper pair boxes~\cite{Angelakis2007,Hartmann06}. This gives rise to many-body
variants of the Rabi model~\cite{Schmidt2013}, whose phase diagrams have been
studied recently~\cite{Schiro13,Schiro2012b,Schiro2015}. Furthermore, for the
implementation of lattice Dicke models with large collective spins, the use of
hybrid quantum systems consisting of superconducting cavity arrays coupled to
solid-state spin ensembles have been proposed~\cite{Zou14}. Spontaneous
collective coherence in driven-dissipative cavity arrays has been studied
in~\cite{Ruiz-Rivas2014}.

In many physical situations (away from the ultra-strong coupling limit), a form
of the spin-cavity interaction alternative to the Rabi term is more appropriate,
with non-linear building block $H_{cs} \sim a\sigma^+ + a^\dag \sigma^-$. In
fact, this form results naturally from the weak coupling rotating wave
approximation of a driven spin-cavity problem. For a single cavity mode and
spin, the resulting model is the Jaynes-Cummings model, which in contrast to the
Dicke model possesses a continuous $U(1)$ phase rotation symmetry under
$a \to e^{i\theta}a, \sigma^- \to e^{i\theta}\sigma^-$.

Clearly, when such systems are driven coherently via a Hamiltonian $H_d = \Omega (a + a^\dag)$ or suitable multimode generalizations thereof, both the $U(1)$ and even the $Z_2$ symmetries of the above models are broken explicitly. Coherent drive is usually the simplest way to compensate for unavoidable losses due to cavity leakage, although incoherent pumping schemes are conceivable using multiple qubits~\cite{Marcos2012}. An advantage of such schemes is that the symmetries of the underlying dynamics are preserved or less severely corrupted in this way (cf. the discussion in Sec.~\ref{sec:symm-keldysh-acti}). In other platforms, such as exciton-polariton systems (cf. the subsequent section), incoherent pumping is more natural from the outset. 

All the systems discussed here represent genuine instances of driven open many-body systems. Besides the coherent drive, they undergo dissipative processes, which have to be taken into account for a proper understanding of their time evolution and stationary states. A generic feature of these processes is their locality. For the effective spin degrees of freedom, the typical processes are qubit decay (local Lindblad operators $L_i = \sigma^-_i$) and dephasing ($L_i = \sigma^z_i$). For the bosonic component, local single-photon loss is  dominant, $L_i = a_i$.

Under various circumstances, such as a low population of the excited spin states, the latter degrees of freedom can be integrated out. In this limit, their physical effect is to generate Kerr-type bosonic non-linearities, giving rise to driven open variants of the celebrated Bose-Hubbard model. These models can even be brought into the correlation dominated regime~\cite{Angelakis2007,Hartmann06,Hartmann2008,Tomadin2010}. Oftentimes, these approximations on the spin sector actually apply, and it is both useful and interesting to study these effective low frequency bosonic theories instead of the full many-body spin-boson problems, see Refs.~\cite{Kepesidis12,Boite13,Jin14,Boite14,Biella15} for recent work in this direction.

\subsubsection{Exciton-polariton systems: driven open interacting bosons}
\label{sec:excit-polar-syst}

Exciton-polaritons are an extremely versatile experimental platform, which is
documented by the richness of physical phenomena that have been studied in these
systems both in theory and experiment. For a comprehensive account of the
subject, we refer to a number of excellent review
articles~\cite{Carusotto2013,Deng2010,Byrnes2014}. A Keldysh functional integral
approach is discussed in Refs.~\cite{Keeling2010,Szymanska2012}, which provides both a microscopic derivation of an exciton-polariton model and a mean field analysis including Gaussian fluctuations. At this
point, we content ourselves with a short introduction, with the aim of showing
that in a suitable parameter regime, exciton-polaritons very naturally provide a
test-bed to study Bose condensation phenomena out of thermal
equilibrium. Similar physics can also arise in a variety of other systems,
including condensates of photons~\cite{Klaers2010},
magnons~\cite{Demokritov2006}, and potentially
excitons~\cite{Alloing14}. Remarkably, even cold atoms could be brought to
condense in a non-equilibrium regime, where continuous loading of atoms balances
three-body losses~\cite{Falkenau2011}, or in atom laser
setups~\cite{Mewes1997,Robins2008,Robins2013}.

\begin{figure}
  \centering
  \includegraphics[width=.8\linewidth]{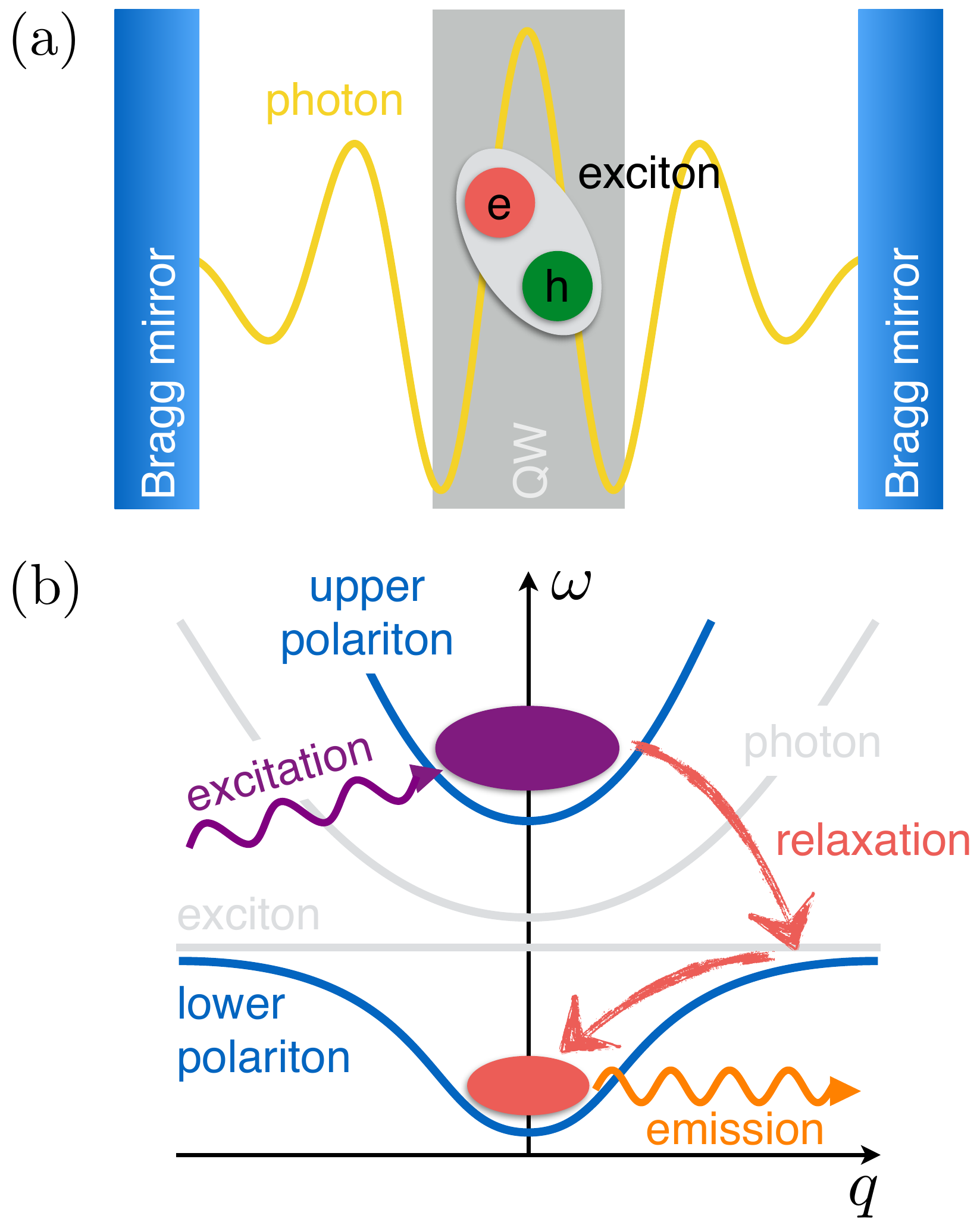}
  \caption{(a) Schematic of two Bragg mirrors forming a microcavity, in which a
    quantum well (QW) is embedded. In the regime of strong light-matter
    interaction, the cavity photon and the exciton hybridize and form new
    eigenmodes, which are called exciton-polaritons. (b) Energy dispersion of
    the upper and lower polariton branches as a function of in-plane momentum
    $q$. In the experimental scheme illustrated in this figure (cf.\
    Ref.~\cite{Kasprzak2006}), the incident laser is tuned to highly excited
    states of the quantum well. These undergo relaxation via emission of phonons
    and scattering from polaritons, and accumulate at the bottom of the lower
    polariton branch. In the course of the relaxation process, coherence is
    quickly lost, and the effective pumping of lower polaritons is
    incoherent.}
  \label{fig:XPs_schematic}
\end{figure}

A basic experimental setup for exciton-polaritons consists of a planar
semiconductor microcavity embedding a quantum well (see
Fig.~\ref{fig:XPs_schematic} (a)). This setting allows for a strong coupling of
cavity light and matter in the quantum well, as originally proposed
in~\cite{imamoglu96}. The free dynamics of the elementary excitations of this
system --- i.e., of cavity photons and Wannier-Mott excitons --- is described
by the quadratic Hamiltonian~\cite{Carusotto2013}
\begin{equation}
  \label{eq:95}
  H_0 = H_C + H_X + H_{\mathit{C-X}},
\end{equation}
where the parts of the Hamiltonian involving only photons and excitons,
respectively, take the same form, which is given by (here the index $\alpha$
labels cavity photons, $\alpha = C$, and excitons, $\alpha = X$,
respectively)\footnote{In Ref.~\cite{Keeling2010,Szymanska2012}, a different model for
  excitons is used: they are assumed to be localized by disorder, and
  interactions are included by imposing a hard-core constraint.}
\begin{equation}
  \label{eq:100}
  H_{\alpha} = \int \frac{d \mathbf{q}}{\left( 2 \pi \right)^2} \sum_{\sigma}
  \omega_{\alpha}(q) a_{\alpha, \sigma}^{\dagger}(\mathbf{q})
  a_{\alpha, \sigma}(\mathbf{q}).
\end{equation}
Field operators $a_{\alpha, \sigma}^{\dagger}(\mathbf{q})$ and
$a_{\alpha, \sigma}(\mathbf{q})$ create or destroy a photon or
exciton (note that both are bosonic excitations) with in-plane momentum
$\mathbf{q}$ and polarization $\sigma$ (there are two polarization states of the
exciton which are coupled to the cavity mode~\cite{Carusotto2013}). For
simplicity, we neglect polarization effects leading to an effective spin-orbit
coupling~\cite{Carusotto2013}. Due to the confinement in the transverse ($z$)
direction, i.e., along the cavity axis, the motion of photons in this direction
is quantized as $q_{z, n} = \pi n/l_z$, where $n$ is a positive integer, and
$l_z$ is the length of the cavity. In writing the Hamiltonian~\eqref{eq:100}, we
are assuming that only the lowest transverse mode is populated, which leads to a
quadratic dispersion as a function of the in-plane momentum
$q = \abs{\mathbf{q}} = \sqrt{q_x^2 + q_y^2}$:
\begin{equation}
  \label{eq:101}
  \omega_C(q) = c \sqrt{q_{z, 1}^2 + q^2} = \omega_C^0 +
  \frac{q^2}{2 m_C} + O(q^4).
\end{equation}
Here, $c$ is the speed of sound, $\omega_C^0 = c q_{z,1}$, and the effective
mass of the photon is given by $m_C = q_{z,1}/c$. Typically, the value of the
photon mass is orders of magnitude smaller than the mass of the exciton, so that
the dispersion of the latter appears to be flat on the scale of
Fig.~\ref{fig:XPs_schematic} (b).

Upon absorption of a photon by the semiconductor, an exciton is generated. This
process (and the reverse process of the emission of a photon upon radiative
decay of an exciton) is described by
\begin{equation}
  \label{eq:102}
  H_{\mathit{C-X}} = \Omega_R \int \frac{d \mathbf{q}}{\left( 2 \pi \right)^2}
  \sum_{\sigma} \left( a_{X, \sigma}^{\dagger}(\mathbf{q}) a_{C,
      \sigma}(\mathbf{q}) + \hc \right),
\end{equation}
where $\Omega_R$ is the rate of the coherent interconversion of photons into
excitons and vice versa. The quadratic Hamiltonian~\eqref{eq:95} can be
diagonalized by introducing new modes --- the lower and upper
exciton-polaritons, $\psi_{\mathrm{LP}, \sigma}(\mathbf{q})$ and
$\psi_{\mathrm{UP}, \sigma}(\mathbf{q})$ respectively, which are linear
combinations of photon and exciton modes. The dispersion of lower and upper
polaritons is depicted in Fig.~\ref{fig:XPs_schematic} (b). In the regime of
strong light-matter coupling, which is reached when $\Omega_R$ is larger than
both the rate at which photons are lost from the cavity due to mirror
imperfections and the non-radiative decay rate of excitons, it is appropriate to
think of exciton-polaritons as the elementary excitations of the system.

In experiments, it is often sufficient to consider only lower polaritons in a
specific spin state, and to approximate the dispersion as
parabolic~\cite{Carusotto2013}. Interactions between exciton-polaritons
originate from various physical mechanisms, with a dominant contribution
stemming from the screened Coulomb interactions between electrons and holes
forming the excitons. Again, in the low-energy scattering regime, this leads to
an effective contact interaction between lower polaritons. As a result, the
low-energy description of lower polaritons takes the form (in the following we
drop the subscript indices in
$\psi_{\mathrm{LP}, \sigma}$)~\cite{Carusotto2013}
\begin{equation}
  \label{eq:103}
  H_{\mathrm{LP}} = \int d\mathbf{x} \left[ \psi^{\dagger}(\mathbf{x})
    \left( \omega_{\mathrm{LP}}^0 - \frac{\nabla^2}{2 m_{\mathrm{LP}}} \right)
    \psi(\mathbf{x}) + u_c \psi^{\dagger}(\mathbf{x})^2 \psi(\mathbf{x})^2
  \right].
\end{equation}
While this Hamiltonian is quite generic and arises also, e.g., in cold bosonic
atoms in the absence of an external potential, the peculiarity of exciton-polaritons
is that they are excitations with relatively short lifetime. In turn, this necessitates
continuous replenishment of energy in the form of laser driving in order to
maintain a steady population. In Fig.~\ref{fig:XPs_schematic} (b), we consider
the case in which the excitation laser is tuned to energies well above the lower
polariton band. The thus created high-energy excitations are deprived of their
excess energy via phonon-polariton and stimulated polariton-polariton
scattering. Eventually, they accumulate at the bottom of the lower polariton
band. As a consequence of multiple scattering processes, the coherence of the
incident laser field is quickly lost, and the effective pumping of lower
polaritons is incoherent.

A phenomenological model for the dynamics of the lower polariton field, which
accounts for both the coherent dynamics generated by the
Hamiltonian~\eqref{eq:103} and the driven-dissipative one described above, was
introduced in Ref.~\cite{Wouters2007a}. It involves a dissipative
Gross-Pitaevskii equation for the lower polariton field that is coupled to a
rate equation for the reservoir of high-energy excitations. However, for the
study of universal long-wavelength behavior in Sec.~\ref{sec:bosons}, this
degree of microscopic modeling is actually not required: indeed, any (possibly
simpler) model that possesses the relevant symmetries (see the discussion at the
beginning of Sec.~\ref{sec:bosons}) will yield the same universal physics. Such
a model can be obtained by describing incoherent pumping and losses of lower
polaritons by means of a Markovian master equation:\footnote{While this approach
  captures the universal behavior, we note that non-Markovian effects can be of
  key importance for other properties~\cite{Wouters2010,Ciuti2006,Chiocchetta2014}.}
\begin{equation}
  \label{eq:104}
  \partial_t \rho = -i [H_{\mathrm{LP}}, \rho] + \mathcal{L}_d \rho,
\end{equation}
where $\mathcal{L}_d \rho$ encodes incoherent single-particle pumping and
losses, as well as two-body losses:
\begin{equation}
  \label{eq:6}
  \mathcal{L}_d \rho =\int d\mathbf{x} \left( \gamma_p \mathcal{D}[\psi(\mathbf{x})^{\dagger}] \rho + \gamma_l
  \mathcal{D}[\psi(\mathbf{x})] \rho + 2 u_d \mathcal{D}[\psi(\mathbf{x})^2]
  \rho \right),
\end{equation}
where
\begin{equation}
  \label{eq:92}
  \mathcal{D}[L] \rho = L \rho L^{\dagger} - \frac{1}{2} \{ L^{\dagger} L, \rho \}
\end{equation}
reflects the Lindblad form, and $\gamma_p$, $\gamma_l$, $2 u_d$ are the rates of
single-particle pumping, single-particle loss, and two-body loss,
respectively. The inclusion of the non-linear loss term ensures saturation of
the pumping. An analogous mechanism is implemented in the above-mentioned
Gross-Pitaevskii description. More precisely, in the spirit of universality, the
above quantum master equation ~\eqref{eq:104} and the above mentioned
phenomenological model reduce to precisely the same low frequency model for
bosonic degrees of freedom upon taking the semiclassical limit in the Keldysh
path integral associated to Eq.~\eqref{eq:104} (see
Sec.~\ref{sec:semicl-limit-keldysh} for its implementation), and integrating out
the upper polariton reservoir in the phenomenological model.

%\sd{ TO BE FLESHED OUT:} lasing as second-order phase transition~\cite{Graham1970,DeGiorgio1970} \lsc{what should we say about this  refs are included at beginning of Sec on FRG}

\subsubsection{Cold atoms in optical lattices: heating dynamics}
\label{sec:cold-atoms-optical-lattices}

In recent years, experiments with cold atoms in optical lattices have shown
remarkable progress in the simulation of many-body model systems both in and out
of equilibrium. A particular strength of cold atom experiments is the unprecedented
tuneability of model parameters, such as the local interaction
strength and the lattice hopping amplitude. This becomes possible by, e.g., manipulation of the
lattice laser and external magnetic fields. It comes along with a very weak coupling
of the system to the environment, such that the dynamics can often be seen as
 isolated on relevant time scales for typical measurements of static,
equilibrium correlations. However, more and more experiments start to
investigate the realm of non-equilibrium phenomena with cold atoms, e.g., by
letting systems prepared in a non-equilibrium initial condition relax in
time towards a steady
state~\cite{Hacker2010,Meinert2013,Meinert2014,Gring2012,Preiss2015}. With these
experiments, time scales are reached, for which the dissipative coupling to the
environment becomes visible in experimental observables. Such dissipation may even hinder the
system from relaxation towards a well defined steady state.

A relevant example of a dissipative coupling is decoherence of an atomic
cloud induced by spontaneous emission of atoms in the
lattice~\cite{Pichler2010,Pichler2013}. In this way, the
many-body system is heated up, and therefore
driven away from the low entropy state in which it was prepared initially. A
detailed discussion of the microscopic physics and its long time dynamics can
be found in~\cite{Schachenmayer2014,Poletti2013,Cai2013}, see also the review article~\cite{Daley2014}. For bosonic atoms in
optical lattices, the coherent dynamics is described by
the Bose-Hubbard Hamiltonian~\cite{Fisher1989,Jaksch1998} \eq{IntroI}{
  H_{\text{BH}}=-J\sum_{\langle l,m\rangle} b^{\dagger}_lb_m+\frac{U}{2}\sum_l
  n_l(n_l-1), } which models bosonic atoms in terms of the creation and
annihilation operators $[\anno{b}{l},\creo{b}{m}]=\delta_{lm}$ in the lowest
band of a lattice with site indices $l,m$. The atoms hop between neighboring
lattice sites with an amplitude $J$, and experience an on-site repulsion
$U$. The lattice potential $V(x)$, which leads to the second quantized
form of the Bose-Hubbard model, is created by the superposition of
counter-propagating laser beams in each spatial dimension. The coherent laser field couples two internal atomic states via stimulated absorption
and emission, which leads to the single particle Hamiltonian \eq{IntroII}{
  H_{\text{atom}}=\frac{\hat{\bf
      p}^2}{2m}-\frac{\Delta}{2}\sigma^z-\sigma^x\frac{\Omega(\hat{\bf x})}{2},
} where $\hat{\bf p}$ is the atomic momentum operator, $\Delta$ is the detuning
of the laser from the atomic transition frequency, and $\Omega(\hat{\bf x})$ is
the laser field at the atomic position. For large detuning, the excited state of the atom can be
traced out, which leads to the lattice Hamiltonian \eq{IntroIII}{
  H_{\text{atom}}^{\text{eff}}=\frac{\hat{\bf p}^2}{2m}+\frac{|\Omega(\hat{\bf
      x})|^2}{4\Delta}. } This describes a lattice potential for the ground state
atoms generated by the spatially modulated AC Stark shift. Adding an atomic
interaction potential and expanding both the single particle
Hamiltonian~\eqref{IntroIII} and the interaction in terms of Wannier states, the
leading order Hamiltonian is the Bose-Hubbard
model~\eqref{IntroI}~\cite{Jaksch1998}.

\begin{figure}[t]
\centering
\includegraphics[width=0.8\linewidth]{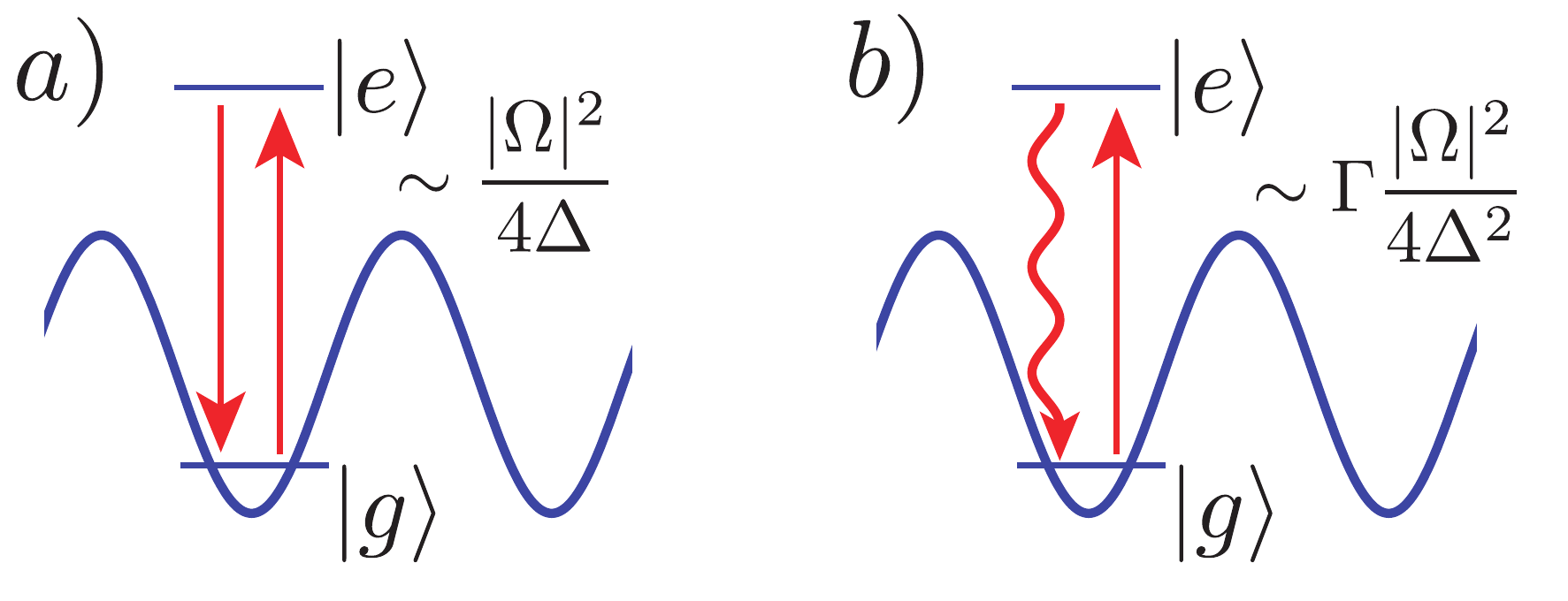}
    \caption
    {\label{fig:Deco} Illustration of the coherent (a) and the
      incoherent (b) contribution to the dynamics stemming from the coupling of
      the atoms to a laser with amplitude $|\Omega|$ and large detuning $\Delta$. (a)
      Via the AC Stark effect, stimulated absorption and emission lead to a
      coherent periodic potential with amplitude
      $\frac{|\Omega|^2}{4\Delta}$. (b) Stimulated absorption and subsequent
      spontaneous emission lead to effective decoherence of the atomic state
      with rate $\Gamma\frac{|\Omega|^2}{4\Delta}$, where $\Gamma$ is the
      microscopic spontaneous emission rate.}
\end{figure}

In the semi-classical treatment of the atom-laser interaction,
spontaneous emission events are neglected, as their probability is
typically very small (see below for a more precise statement). They can be taken into account on the basis of optical
Bloch equations~\cite{Pichler2010}, which leads, after elimination of the excited
atomic state, to an additional, driven-dissipative term in the atomic dynamics. It describes the
decoherence of the atomic state due to spontaneous emission, i.e., position
dependent random light scattering. The leading order contribution to the
dynamics in the basis of lowest band Wannier states is captured by the master
equation \eq{IntroIV}{
  \partial_t\rho=-i[H_{\text{BH}},\rho]-\gamma\sum_l
  [ n_l,[ n_l,\rho]], } where $n_l=\creo{b}{l}\anno{b}{l}$ is
the local atomic density, and $\rho$ is the many-body density matrix. For a red
detuned laser the rate $\gamma=\Gamma\frac{|\Omega|^2}{4\Delta^2}$ is proportional to the microscopic spontaneous emission rate $\Gamma$ and the
laser amplitude $|\Omega|$.  Note the suppression of the scale $\gamma$ by a factor $\Gamma/\Delta \ll 1$ for large detuning, compared to the strength of $H_{\text{atom}}^{\text{eff}}$. The coherent and incoherent contribution of the
atom-laser coupling to the dynamics is illustrated in Fig.~\ref{fig:Deco}.

The dissipative term in the master equation~\eqref{IntroIV}
leads to an energy increase $\langle H_{\text{BH}}\rangle(t)\sim t$ linear in
time, and thus to heating. Furthermore, it introduces decoherence in the number state basis, i.e., it
projects the local density matrix on its diagonal in Fock space and leads to a
decrease of the coherences in time~\cite{Daley2014,Schachenmayer2014,Poletti2013,Cai2013}. Starting from a 
low entropy state at $t=0$, the heating leads to a crossover from coherence
dominated dynamics at short and intermediate times~\cite{heating} to a
decoherence dominated dynamics at long
times~\cite{Schachenmayer2014,Poletti2013,Cai2013}. In one dimension, both
regimes have been analyzed extensively both numerically (with a focus on
decoherence dominated dynamics) as well as analytically, and display several
aspects of non-equilibrium universality, see Sec.~\ref{sec:heating-dynamics}. Therefore, heating in interacting
lattice systems represents a crucial example for universality in out-of-equilibrium
dynamics, which can be probed by cold atom experiments.

\subsection{Outline and scope of this review}
\label{sec:outline}

The remainder of this review is split into two parts.

Part~\ref{part:theor-backgr} develops the theoretical framework for the
efficient description of driven open many-body quantum systems. In
Sec.~\ref{sec:keldysh-path-integr}, we begin with a direct derivation of the
open system Keldysh functional integral from the many-body quantum master
equation (Sec.~\ref{sec:deriv-keldysh-acti}). We then discuss in
Sec.~\ref{sec:single-mode-cavity} in detail a simple example: the damped and
driven optical cavity. This allows the reader to familiarize with the functional
formalism. In particular, the key players in terms of observables ---
correlation and response functions --- are described. We also point out a number
of exact structural properties of Keldysh field theories, which hold beyond the
specific example.  Another example is introduced in
Sec.~\ref{sec:driv-diss-cond}: there we discuss the mean field theory of
condensation in a bosonic many-body system with particle losses and pumping. The
semi-classical limit of this model and its validity are the content of
Sec.~\ref{sec:semicl-limit-keldysh}. This is followed by a discussion of
symmetries and conservation laws in the Keldysh formalism in
Sec.~\ref{sec:symm-keldysh-acti}. In particular, we point out a symmetry that
allows one to distinguish equilibrium from non-equilibrium conditions. Finally,
an advanced field theoretical tool --- the open system functional
renormalization group --- is introduced in Sec.~\ref{sec:open-sys-FRG}.

Part~\ref{part:applications} harnesses this formalism to generate an
understanding of the physics in different experimental platforms. We begin in
Sec.~\ref{sec:neqspin} with simple but paradigmatic spin models with discrete
Ising $Z_2$ symmetry in driven non-equilibrium stationary states. In particular, we
discuss the physics of the driven open Dicke model in Sec.~\ref{sec:ising}. This
is followed by an extended variant of the latter in the presence of disorder and
a multimode cavity, which hosts an interesting spin and photon glass phase in
Sec.~\ref{sec:multi}. Sec.~\ref{sec:bosons} is devoted to the non-equilibrium
stationary states of bosons with a characteristic $U(1)$ phase rotation
symmetry: the driven-dissipative condensates introduced in
Sec.~\ref{sec:driv-diss-cond}. After some additional technical developments
relating to $U(1)$ symmetry in Sec.~\ref{sec:dens-phase-repr}, we discuss
critical behavior at the Bose condensation transition in three dimensions. In
particular, we show the decrease of a parameter quantifying non-equilibrium
strength in this case (Sec.~\ref{sec:critical-dynamics-3d}). The opposite
behavior is observed in two (Sec.~\ref{sec:absence-algebr-order}) and one
(Sec.~\ref{sec:kpz-scaling-1d}) dimensions. Finally, leaving the realm of
stationary states, in Sec.~\ref{sec:heating-dynamics} we discuss an application
of the Keldysh formalism to the time evolution of open bosonic systems in one
dimension, which undergo number conserving heating processes. We set up the
model in Sec.~\ref{sec:intll}, derive the kinetic equation for the distribution
function in Sec.~\ref{sec:kineq}, and discuss the relevant approximations and
physical results in Secs.~\ref{sec:scb} and~\ref{sec:heatison},
respectively. Conclusions are drawn in Sec.~\ref{sec:conclusions}. Finally,
brief introductions to functional differentiation and Gaussian functional
integration are given in Appendices~\ref{sec:funct-diff}
and~\ref{sec:gauss-funct-integr}.

Reflecting the bipartition of this review, the scope of it is twofold.  On the
one hand, it develops the Keldysh functional integral approach to driven open
quantum systems ``from scratch'', in a systematic and coherent way. It starts
from the Markovian quantum master equation representation of driven dissipative
quantum dynamics~\cite{Gardiner2000,Breuer2002,Barnett1997}, and introduces an
equivalent Keldysh functional integral representation. Direct contact is made to
the language and typical observables of quantum optics. It does not require
prior knowledge of quantum field theory, and we hope that it will find the
interest of --- and be useful for --- researchers working on quantum optical
systems with many degrees of freedom.  On the other hand, this review documents
some recent theoretical progresses made in this conceptual framework in a more
pedagogical way than the original literature. We believe that this not only
exposes some interesting physics, but also demonstrates the power and
flexibility of the Keldysh approach to open quantum systems.

This work is complemetary to excellent reviews putting more emphasis on the
specific experimental platforms partially mentioned above: The physics of driven
Bose-Einstein condensates in optical cavities is reviewed
in~\cite{Ritsch2013}. A general overview of driven ultracold atomic systems,
specifically in optical lattices, is provided in~\cite{Daley2014}, and systems
with engineered dissipation are described in~\cite{Muller2012}. Detailed
accounts for exciton-polariton systems are given
in~\cite{Carusotto2013,Deng2010,Byrnes2014}; specifically, we refer to the
review~\cite{Keeling2010} working in the Keldysh formalism. The physics of
microcavity arrays is discussed
in~\cite{Hartmann2008,Tomadin2010,Houck2012,Schmidt2013}, and trapped ions are
treated in~\cite{Leibfried2003,Blatt2012}. We also refer to recent reviews on
additional upcoming platforms of driven open quantum systems, such as Rydberg
atoms~\cite{Marcuzzi2014a} and opto-nanomechanical
settings~\cite{marquardt09}. For a recent exposition of the physics of quantum
master equations and to efficient numerical techniques for their solution,
see~\cite{Daley2014}.

%%% Local Variables:
%%% mode: latex
%%% TeX-master: "dds_review"
%%% End:

\part{Theoretical background}
\label{part:theor-backgr}

\section{Keldysh functional integral for driven open systems}
\label{sec:keldysh-path-integr}

In this part, we will be mainly concerned with a Keldysh field theoretical
reformulation of the stationary state of Markovian many-body quantum master
equations. As we also demonstrate, this opens up the powerful toolbox of modern
quantum field theory for the understanding of such systems.

The quantum master equation, examples of which we have already encountered in
Sec.~\ref{sec:exper-platf}, describes the time evolution of a reduced system
density matrix $\rho$ and reads~\cite{Gardiner2000,Breuer2002}
\begin{equation}
  \label{eq:109}
  \partial_t \rho = \mathcal{L} \rho = - i [H, \rho] + \sum_{\alpha}
  \gamma_{\alpha} \left( L_\alpha \rho L_\alpha^{\dagger} - \frac{1}{2} \{ L_\alpha^{\dagger} L_\alpha, \rho \} \right),
\end{equation}
where the operator $\mathcal{L}$ acts on the density matrix $\rho$ ``from both
sides'' and is often referred to as \emph{Liouville superoperator} or
\emph{Liouvillian} (sometimes this term is reserved for the second contribution
on the RHS
of Eq.~\eqref{eq:109} alone). There are two contributions to the
Liouvillian: first, the commutator term, which is familiar from the von Neumann
equation, describes the coherent dynamics generated by a system Hamiltonian $H$;
the second part, which we will refer to as the \emph{dissipator} $\mathcal{D}$
\footnote{We use the term ``dissipation'' here for all kinds of environmental
  influences on the system which can be captured in Lindblad form, including
  effects of decay and of dephasing/decoherence.}, describes the dissipative
dynamics resulting from the interaction of the system with an environment, or
``bath.''  It is defined in terms of a set of so-called \emph{Lindblad
  operators} (or \emph{quantum jump} operators) $L_{\alpha}$, which model the
coupling to that bath. The dissipator has a characteristic Lindblad
form~\cite{Lindblad1976,Kossakowski1972}: it contains an anticommutator term
which describes dissipation; in order to conserve the norm $\tr(\rho)$ of the
system density matrix, this term must be accompanied by fluctuations. The
corresponding term, where the Lindblad operators act from both sides onto the
density matrix, is referred to as \emph{recycling} or \emph{quantum jump}
term. Dissipation occurs at rates $\gamma_{\alpha}$ which are non-negative, so
that the density matrix evolution is completely positive, i.e., the eigenvalues
of $\rho$ remain positive under the combined dynamics generated by $H$ and
$\mathcal{D}$~\cite{Bacon2001}. If the index $\alpha$ is the site index in an
optical lattice or in a microcavity array, or even a continuous position label
(in which case the sum is replaced by an integral), in a translation invariant
situation there is just a single scale $\gamma_{\alpha} = \gamma$ for all
$\alpha$ associated to the dissipator.

The quantum master equation~\eqref{eq:109} provides an accurate description of a
system-bath setting with a strong separation of scales. This is generically the
case in quantum optical systems, which are strongly \emph{driven} by external
classical fields. More precisely, there must be a large energy scale in the bath
(as compared to the system-bath coupling), which justifies to integrate out the
bath in second-order time-dependent perturbation theory. If in addition the bath
has a broad bandwidth, the combined Born-Markov and rotating-wave approximations
are appropriate, resulting in Eq.~\eqref{eq:109}. A concrete example for such a
setting is provided by a laser-driven atom undergoing spontaneous
emission. Generic condensed matter systems do not display such a scale
separation, and a description in terms of a master equation of the
type~\eqref{eq:109} is not justified.\footnote{Davies'
  prescription~\cite{Davies74} allows one to also describe equilibrium systems
  in terms of operatorial master equations, however with collective Lindblad
  operators ensuring detailed balance conditions (cf.\ also~\cite{Breuer2002}).}
However, the systems discussed in the introduction, belong to the class of
systems which permit a description by Eq.~\eqref{eq:109}. We refer to them as
\emph{driven open many-body quantum systems}.

Due to the external drive these systems are out of thermodynamic
equilibrium. This statement will be made more precise in
Sec.~\ref{sec:therm-equil-sym} in terms of the absence of a dynamical symmetry
which characterizes any system evolving in thermodynamic equilibrium, and which
is manifestly violated in dynamics described by Eq.~\eqref{eq:109}. Its absence
reflects the lack of energy conservation and the epxlicit breaking of detailed
balance. As stated in the introduction, the main goal of this review is to point
out the macroscopic, observable consequences of this microscopic violation of
equilibrium conditions.

\subsection{From the quantum master equation to the Keldysh functional integral}
\label{sec:deriv-keldysh-acti}

In this section, starting from a many-body quantum master equation
Eq.~\eqref{eq:109} in the operator language of second quantization, we derive an
equivalent Keldysh functional integral. We focus on stationary states, and we
discuss how to extract dynamics from this framework in
Sec.~\ref{sec:heating-dynamics}. Our derivation of the Keldysh functional
integral applies to a theory of bosonic degrees of freedom. If spin systems are
to be considered, it is useful to first perform the typical approximations
mapping them to bosonic fields, and then proceed along the construction below
(see Sec.~\ref{sec:neqspin} and
Refs.~\cite{DallaTorre2013,Maghrebi2015}). Clearly, this amounts to an
approximate treatment of the spin degrees of freedom; for an exact (equilibrium)
functional integral representation for spin dynamics, taking into account the
full non-linear structure of their commutation relations, we refer to
Ref.~\cite{Altland2010}. For fermionic problems, the construction is analogous
to the bosonic case presented here. However, a few signs have to be adjusted to
account for the fermionic anticommutation
relations~\cite{Altland2010,Kamenev2011}.

\begin{figure}
  \centering  
  \includegraphics[width=\linewidth]{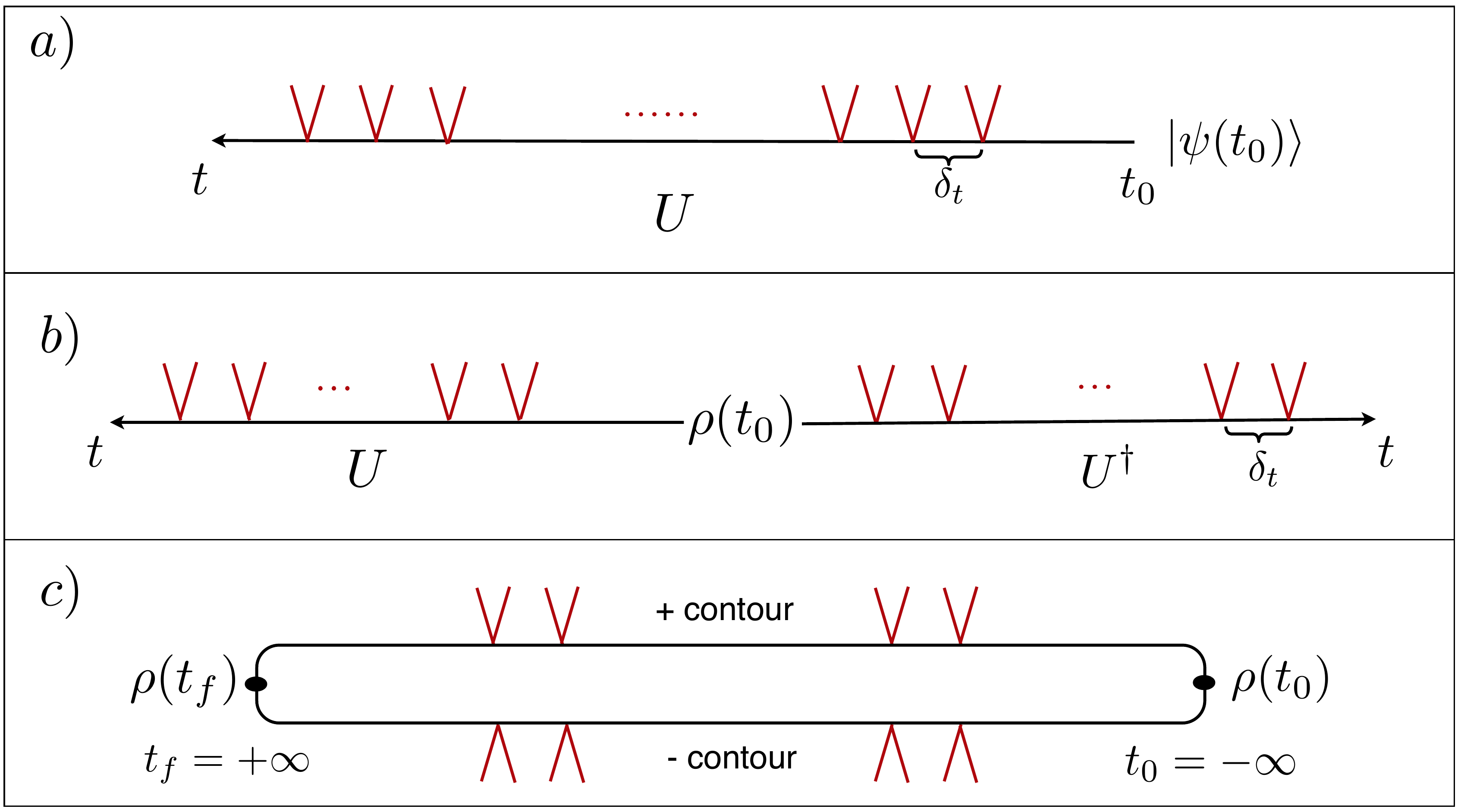}
  \caption{Idea of the Keldysh functional integral. a) According to the
    Schr\"odinger equation, the time evolution of a pure state \emph{vector} is
    described by the unitary operator
    $U(t,t_0) = e^{- i H \left( t-t_0 \right)}$.  In the Feynman functional
    integral construction, the time evolution is chopped into infinitesimal
    steps of length $\delta_t$, and completeness relations in terms of coherent
    states are inserted in between consecutive time steps. This insertion is
    signalled by the red arrows. b) In contrast, if the state is mixed, a
    density \emph{matrix} must be evolved, and thus two time branches are
    needed. As explained in the text, the dynamics need not necessarily be
    restricted to unitary evolution. The most general time-local (Markovian)
    dynamics is generated by a Liouville operator in Lindblad form. c) For the
    analysis of the stationary state, we are interested in the real time analog
    of a partition function $Z = \tr \rho(t_f)$, starting from $t_0 =-\infty$ and
    running until $t_f=+\infty$. The trace operation connects the two time
    branches, giving rise to the closed Keldysh contour.}
  \label{fig:Keldysh_idea}
\end{figure}

The basic idea of the Keldysh functional integral can be developed in simple
terms by considering the Schr\"odinger \textit{vs.} the von Neumann equation,
\begin{align}\label{eq:schhe}
  i \partial_t |\psi(t)\rangle & = H  |\psi(t) \rangle & \Rightarrow & & |\psi(t) \rangle
  & = U(t,t_0) |\psi(t_0) \rangle, \nonumber \\
  \partial_t \rho (t) & = - i [ H, \rho(t) ] & \Rightarrow & & \rho (t) & = U(t,t_0) \rho(t_0) U^\dag(t,t_0),
\end{align}
where $U(t,t_0) = e^{- i H \left( t-t_0 \right)}$ is the unitary time evolution
operator. In the first case, a real time path integral can be constructed along
the lines of Feynman's original path integral formulation of quantum
mechanics~\cite{Feynman1948}. To this end, a Trotter decomposition of the
evolution operator
\begin{equation}
  \label{eq:106}
  e^{-iH \left(t-t_0 \right)}= \lim_{N\rightarrow \infty} \left( \id - i \delta_t
    H\right)^N,
\end{equation}
with $\delta_t=\frac{t-t_0}{N}$, is performed. Subsequently, in between the
factors of the Trotter decomposition, completeness relations in terms of
coherent states are inserted in order to make the (normal ordered) Hamilton
operator a functional of classical field variables. This is illustrated in
Fig.~\ref{fig:Keldysh_idea} a), and we will perform these steps
explicitly and in more detail below in the context of open many-body
systems. Crucially, we only need one set of field variables representing
coherent Hamiltonian dynamics, which corresponds to the forward evolution of the
Schr\"odinger state vector. It is also clear that --- noting the formal analogy
of the operators $e^{-iH \left(t-t_0 \right)}$ and $e^{-\beta H}$ --- this
construction can be leveraged over to the case of thermal equilibrium, where the
``Trotterization'' is done in imaginary instead of real time.

In contrast to these special cases, the von Neumann equation for general mixed
state density matrices cannot be rewritten in terms of a state \emph{vector}
evolution, even in the case of purely coherent Hamiltonian dynamics.\footnote{
  Of course, the evolution of an $M\times M$ matrix, where $M$ is the dimension
  of the Hilbert space, can be formally recast into the evolution of a vector of
  length $M\times M$. While such a strategy is often pursued in numerical
  approaches~\cite{Daley2014}, it does in general not allow for a physical
  interpretation.} Instead, it is necessary to study the evolution of a state
\emph{matrix}, which transforms according to the integral form of the von
Neumann equation in the second line of Eq.~\eqref{eq:schhe}. Therefore, we have
to apply the Trotter formula and coherent state insertions on \emph{both} sides
of the density matrix. This leads to the \emph{doubling} of degrees of freedom,
characteristic of the Keldysh functional integral. Moreover, time evolution can
now be interpreted as occurring along two branches, which we denote as the
\emph{forward} and a \emph{backward} branches, respectively (cf.\
Fig.~\ref{fig:Keldysh_idea} b)). Indeed, this is an intuitive and natural
feature of evolving matrices instead of vectors.

So far, we have concentrated on closed systems which evolve according to purely
Hamiltonian dynamics. However, we can allow for a more general generator of
dynamics and still proceed along the two-branch strategy. The most general (time
local) evolution of a density matrix is given by the quantum master
equation~\eqref{eq:109}. Its formal solution reads
\begin{equation}
  \label{eq:22}
  \rho(t)=e^{\left( t -t_0 \right) \mathcal{L}} \rho(t_0) \equiv
  \lim_{N\rightarrow\infty}  \left( \id + \delta_t\mathcal{L}\right)^N\rho(t_0).
\end{equation}
The last equality gives a meaning to the formal solution in terms of the Trotter
decomposition: at each infinitesimal time step, the exponential can be expanded
to first order, such that the action of the Liouvillian superoperator is just
given by the RHS of the quantum master equation~\eqref{eq:109}; At finite times,
the evolved state is given by the concatenation of the infinitesimal Trotter
steps.

If we restrict ourselves to the stationary state of the system,\footnote{We
  assume that it exists. We thus exclude scenarios with dynamical limit cycles,
  for simplicity.} but want to evaluate correlations at arbitrary time
differences, we should extend the time branch from an initial time in the
distant past $t_0 \to -\infty$ to the distant future, $t_f \to +\infty$. In
analogy to thermodynamics, we are then interested in the so-called Keldysh
partition function $Z = \tr \rho(t) =1$. The trace operation contracts the
indices of the time evolution operator as depicted in
Fig.~\ref{fig:Keldysh_idea} c), giving rise to the closed time path or Keldysh
contour. Conservation of probability in the quantum mechanical system is
reflected in the time-independent normalization of the partition function. In
order to extract physical information, again in analogy to statistical
mechanics, below we introduce sources in the partition function. This allows us
to compute the correlation and response functions of the system by taking
suitable variational derivatives with respect to the sources.

After this qualitative discussion, let us now proceed with the explicit
construction of the Keldysh functional integral for open systems, starting from
the master equation~\eqref{eq:109}. As mentioned above, the Keldysh functional
integral is an unraveling of Liouvillian dynamics in the basis of coherent
states, and we first collect a few important properties of those. Coherent
states are defined as (for simplicity, in the present discussion we restrict
ourselves to a single bosonic mode $b$)
$\ket{\psi} = \exp( \psi b^{\dagger} ) \ket{\Omega}$, where $\ket{\Omega}$
represents the vacuum in Fock space. (Note that according to this definition,
which is usually adopted in the discussion of field integrals~\cite{Negele1998},
the state $\ket{\psi}$ is not normalized.) A key property of coherent states is
that they are eigenstates of the annihilation operator, i.e.,
$b\ket{\psi} = \psi\ket{\psi}$, with the complex eigenvalue $\psi$. Clearly,
this implies the conjugate relation
$\bra{\psi}b^\dag = \bra{\psi}\psi^{*}$.\footnote{Note that the creation
  operator cannot have eigenstates due to the fact that there is a minimal
  occupation number of a bosonic state. In particular, the coherent states are
  not eigenstates. We rather have the relations
  $b^\dag\ket{\psi} = \left( \partial/\partial \psi \right) \ket{\psi},
  \bra{\psi}b = \left( \partial/\partial \psi^{*} \right) \bra{\psi}$.}
The overlap of two non-normalized coherent states is given by
$\braket{\psi | \phi} = e^{\psi^{*} \phi}$, and the completeness relation reads
$\id = \int \frac{d \psi d \psi^{*}}{\pi} e^{- \psi^{*} \psi} \ket{\psi}
\bra{\psi}$.

The starting point of the derivation is Eq.~\eqref{eq:22}, and we focus first on
a single time step, as in the usual derivation of the coherent state functional
integral~\cite{Negele1998}. That is, we decompose the time evolution from $t_0$
to $t_f$ into a sequence of small steps of duration $\delta_t = (t_f - t_0)/N$,
and denote the density matrix after the $n$-th step, i.e., at the time
$t_n = t_0 + \delta_t n$, by $\rho_n = \rho(t_n)$. We then have
\begin{equation}
  \label{eq:25}
  \rho_{n + 1}= e^{\delta_t \mathcal{L}} \rho_n = \left( \id + \delta_t
    \mathcal{L} \right) \rho_n + O(\delta_t^2).
\end{equation}
As anticipated above, we proceed to represent the density matrix in the basis of
coherent states. For instance, $\rho_n$ at the time $t_n$ can be written as
\begin{multline}
  \label{eq:28}
  \rho_n = \int \frac{d \psi_{+, n} d\psi_{+, n}^{*}}{\pi} \frac{d \psi_{-, n} d \psi_{-,
    n}^{*}}{\pi} e^{- \psi_{+, n}^{*} \psi_{+, n} - \psi_{-, n}^{*} \psi_{-, n}} \\
  \times \braket{\psi_{+, n} | \rho_n | \psi_{-, n}} \ket{\psi_{+,
      n}}\bra{\psi_{-, n}}.
\end{multline}
As a next step, we would like to express the matrix element
$\braket{\psi_{+, n + 1} | \rho_{n + 1} | \psi_{-, n + 1}}$, which appears in
the coherent state representation of $\rho_{n + 1}$, in terms of the
corresponding matrix element at the previous time step $t_n$.  Inserting
Eq.~\eqref{eq:28} in Eq.~\eqref{eq:25}, we find that this requires us to
evaluate the ``supermatrixelement''

\begin{widetext}
  \begin{multline}
    \label{eq:29}
    \braket{\psi_{+, n + 1} | \mathcal{L}( \ket{\psi_{+, n}}\bra{\psi_{-, n}}) |
      \psi_{-, n + 1}} = -i \left( \braket{\psi_{+, n + 1} | H | \psi_{+, n}}
      \braket{\psi_{-, n} | \psi_{-, n + 1}} - \braket{\psi_{+, n + 1} |
        \psi_{+, n}} \braket{\psi_{-, n} | H | \psi_{-, n + 1}} \right) \\ +
    \sum_{\alpha} \gamma_{\alpha} \left[ \braket{\psi_{+, n + 1} | L_{\alpha} |
        \psi_{+, n}} \braket{\psi_{-, n} | L_{\alpha}^{\dagger} | \psi_{-, n +
          1}} - \frac{1}{2} \left( \braket{\psi_{+, n + 1} |
          L_{\alpha}^{\dagger} L_{\alpha} | \psi_{+, n}} \braket{\psi_{-, n} |
          \psi_{-, n + 1}} + \braket{\psi_{+, n + 1} | \psi_{+, n}}
        \braket{\psi_{-, n} | L_{\alpha}^{\dagger} L_{\alpha} | \psi_{-, n + 1}}
      \right) \right].
  \end{multline}
  Without loss of generality, we assume that the Hamiltonian is normal
  ordered. Then, a matrix element $\braket{\psi | H | \phi}$ of the Hamiltonian
  between coherent states can be obtained simply by replacing the creation
  operators by $\psi^{*}$ and the annihilation operators by $\phi$. The same is
  true for matrix elements of $L, L^{\dagger},$ and $L^{\dagger} L$, after
  performing the commutations which are necessary to bring these operators to
  the form of sums of normal ordered expressions (see Ref.~\cite{Sieberer2014}
  for a detailed discussion of subtleties related to normal ordering). Then we
  obtain by re-exponentiation
  \begin{equation}
    \label{eq:30}
    \braket{\psi_{+, n + 1} | \rho_{n + 1} | \psi_{-, n + 1}} = \int
    \frac{d \psi_{+, n} d \psi_{+, n}^{*}}{\pi} \frac{d \psi_{-, n} d \psi_{-, n}^{*}}{\pi} e^{i
      \delta_t \left( - \psi_{+, n} i \partial_t \psi_{+, n}^{*} - \psi_{-, n}^{*}
        i \partial_t \psi_{-, n} - i \mathcal{L}(\psi_{+, n + 1}^{*}, \psi_{+, n},
        \psi_{-, n + 1}^{*}, \psi_{-, n}) \right)} \braket{\psi_{+, n} | \rho_n
      | \psi_{-, n}} + O(\delta_t^2),
  \end{equation}
\end{widetext}
where we are using the shorthand suggestive notation
$\partial_t \psi_{\pm, n} = (\psi_{\pm, n + 1} - \psi_{\pm, n})/\delta_t$. The
time derivative terms emerge from the overlap of neighboring coherent states at
time steps $n$ and $n+1$, combined with the weight factor in the completeness
relation for step $n$; the quantity
$\mathcal{L}(\psi_{+, n + 1}^{*}, \psi_{+, n}, \psi_{-, n + 1}^{*}, \psi_{-,
  n})$
is the supermatrixelement in Eq.~\eqref{eq:29}, divided by the above-mentioned
overlaps:
\begin{equation}
  \label{eq:85}
  \mathcal{L}(\psi_{+, n + 1}^{*}, \psi_{+, n},
  \psi_{-, n + 1}^{*}, \psi_{-, n}) = \frac{\braket{\psi_{+, n + 1} |
      \mathcal{L}( \ket{\psi_{+, n}}\bra{\psi_{-, n}}) |
      \psi_{-, n + 1}}}{\braket{\psi_{+, n + 1} | \psi_{+, n}} \braket{\psi_{-, n} |
    \psi_{-, n + 1}}}.
\end{equation}
By iteration of Eq.~\eqref{eq:30}, the density matrix can be evolved from
$\rho(t_0)$ at $t_0$ to $\rho(t_f)$ at $t_f = t_N$. This leads in the limit
$N \to \infty$ (and hence $\delta_t \to 0$) to
\begin{equation}
  \label{eq:31}
  \begin{split}
    Z_{t_f,t_0} & = \tr \rho(t_f) = \tr e^{\left( t_f - t_0 \right) \mathcal{L}}
    \rho(t_0) \\ & = \int \mathscr{D}[\psi_{+}, \psi_{+}^{*}, \psi_{-},
    \psi_{-}^{*}] \, e^{i S} \braket{\psi_{+}(t_0) | \rho(t_0) | \psi_{-}(t_0)},
  \end{split}
\end{equation}
where the integration measure is given by
\begin{equation}
  \label{eq:84}
  \mathscr{D}[\psi_{+}, \psi_{+}^{*}, \psi_{-}, \psi_{-}^{*}] = \lim_{N \to
    \infty} \prod_{n = 0}^N \frac{d \psi_{+, n} d \psi_{+, n}^{*}}{\pi} \frac{d
    \psi_{-, n} d \psi_{-, n}^{*}}{\pi},
\end{equation}
and the Keldysh action reads
\begin{equation}
  \label{eq:21}
  S = \int_{t_0}^{t_f} dt \left( \psi_{+}^{*} i \partial_t \psi_{+} -
    \psi_{-}^{*} i \partial_t \psi_{-} - i \mathcal{L}(\psi_{+}^{*}, \psi_{+},
    \psi_{-}^{*}, \psi_{-}) \right).
\end{equation}
The coherent state representation of $\mathcal{L}$ in the exponent in
Eq.~\eqref{eq:30} comes with a prefactor $\delta_t$, so that to leading order
for $\delta_t \to 0$ it is consistent to ignore the difference stemming from the
bra vector at $n+1$ and the ket vector at $n$ in Eq.~\eqref{eq:85}. Assuming all
operators are normally ordered in the sense discussed above, we obtain
\begin{multline}
  \label{eq:63}
  \mathcal{L}(\psi_{+}^{*}, \psi_{+}, \psi_{-}^{*}, \psi_{-}) = - i \left(
    H_+ - H_- \right) \\ + \sum_{\alpha} \gamma_{\alpha} \left[ L_{\alpha,+}
    L^{*}_{\alpha,-} - \frac{1}{2} \left( L^{*}_{\alpha,+} L_{\alpha,+} +
      L^{*}_{\alpha,-} L_{\alpha,-} \right) \right],
\end{multline}
where $H_{\pm} = H(\psi_{\pm}^{*}, \psi_{\pm})$ contains fields on the $\pm$
contour only, and the same is true for $L_{\alpha, \pm}$. We clearly recognize
the Lindblad superoperator structure of Eq.~\eqref{eq:109}: operators acting on
the density matrix from the left (right) reside on the forward, + (backward, -)
contour. This gives a simple and direct translation table from the bosonic
quantum master equation to the Markovian Keldysh action~\eqref{eq:63}, with the
crucial caveat of normal ordering to be taken into account before performing the
translation.

\emph{Keldysh partition function for stationary states} --- When we are
interested in a stationary state, but would like to obtain information on
temporal correlation functions at arbitrarily long time differences, it is
useful to perform the limit $t_0 \to - \infty$, $t_f \to + \infty$ in
Eq.~\eqref{eq:31}. In an open system coupled to several external baths, it is
typically a useful assumption that the initial state in the infinite past does
not affect the stationary state --- in other words, there is a complete loss of
memory of the initial state. Under this physical assumption, we can ignore the
boundary term, i.e., the matrix element
$\braket{\psi_{+}(t_0) | \rho(t_0) | \psi_{-}(t_0)}$ of the initial density matrix
in Eq.~\eqref{eq:31}, and obtain for the final expression of the Keldysh
partition function
\begin{align}    
  \label{eq:3166666}
  Z & = \int \mathscr{D}[\psi_{+}, \psi_{+}^{*}, \psi_{-}, \psi_{-}^{*}] \, e^{i
      S} =1,
  \\ \label{eq:111}
  S & = \int_{-\infty}^{\infty} dt \left( \psi_{+}^{*} i \partial_t \psi_{+} -
      \psi_{-}^{*} i \partial_t \psi_{-} - i \mathcal{L}(\psi_{+}^{*}, \psi_{+},
      \psi_{-}^{*}, \psi_{-}) \right).
\end{align}

This setup allows us to study stationary states far away from thermodynamic
equilibrium as realized in the systems introduced in Sec.~\ref{sec:intro}, using
the advanced toolbox of quantum field theory. For the discussion of the time
evolution of the system's initial state, the typical strategy in practice is not
to start directly from Eq.~\eqref{eq:31} --- strictly speaking, this would
necessitate knowledge of the entire density matrix of the system, which in a
genuine many-body context is not available. Rather, the Keldysh functional
integral is used to derive equations of motion for a given set of correlation
functions. The initial values of the correlation functions have to be taken from
the physical situation under consideration. For interacting theories, the set of
correlations functions typically corresponds to an infinite hierarchy. The
possibility of truncating this hierarchy to a closed subset usually involves
approximations, which have to be justified from case to case.

The Keldysh partition function is normalized to 1 by construction. As
anticipated above, correlation functions can be obtained by introducing source
terms $J_{\pm} = \left( j_{\pm}, j_{\pm}^{*} \right)$ that couple to the fields
$\Psi_{\pm} = \left( \psi_{\pm}, \psi_{\pm}^{*} \right)$ (here and in the
following we denote spinors of a field and its complex conjugate by capital
letters),
\begin{equation}
  \label{eq:35}
  \begin{split}
    Z[J_+, J_-] & = \int \mathscr{D}[\Psi_+, \Psi_-] \, e^{i S + i \int_{t,
        \mathbf{x}} \left( J_{+}^{\dagger} \Psi_{+} - J_{-}^{\dagger} \Psi_{-}
      \right)} \\ & = \left\langle e^{i \int_{t, \mathbf{x}} \left(
          J_{+}^{\dagger} \Psi_{+} - J_{-}^{\dagger} \Psi_{-} \right)}
    \right\rangle,
  \end{split}
\end{equation}
where we abbreviate (switching now to a spatial continuum of fields)
$\int_{t, \mathbf{x}} = \int d t \int d^d \mathbf{x}$, the average is taken with
respect to the action $S$, and we have the normalization
$Z[J_+ = 0, J_- = 0] = 1$ in the absence of sources. Physically, sources can be
realized, e.g., by coherent external fields such as lasers; this will be made
more concrete in the following section. The source terms can be thought of as
shifts of the original Hamiltonian operator, justifying the von Neumann
structure indicated above.

\emph{Keldysh rotation} --- With these preparations, arbitrary correlation
functions can be computed by taking variational derivatives with respect to the
sources. However, while the representation in terms of fields residing on the
forward and backward branches allows for a direct contact to the second
quantized operator formalism, it is not ideally suited for practical
calculations. In fact, the above description contains a redundancy which is
related to the conservation of probability (this statement and the origin of the
redundancy is detailed below in Sec.~\ref{sec:single-mode-cavity}). This can be
avoided by performing the so-called Keldysh rotation, a unitary transformation
in the contour index or Keldysh space according to
\begin{equation}
  \label{eq:41}
  \phi_c = \frac{1}{\sqrt{2}} \left( \psi_{+} + \psi_{-} \right), \quad \phi_q
  = \frac{1}{\sqrt{2}} \left( \psi_{+} - \psi_{-} \right),
\end{equation}
and analogously for the source terms. The index $c$ ($q$) stands for
``classical'' (``quantum'') fields, respectively. This terminology signals that
the symmetric combination of fields can acquire a (classical) field expectation
value, while the antisymmetric one cannot. In terms of classical and quantum
fields, the Keldysh partition function takes the form
\begin{equation}
  \label{eq:14}
  \begin{split}
    Z[J_c, J_q] & = \int \mathscr{D}[\Phi_c, \Phi_q] e^{i S + i \int_{t,
        \mathbf{x}} \left( J_q^{\dagger} \Phi_c + J_c^{\dagger} \Phi_q \right)}
    \\ & = \Braket{e^{i \int_{t, \mathbf{x}} \left( J_q^{\dagger} \Phi_c +
          J_c^{\dagger} \Phi_q \right)}};
  \end{split}
\end{equation}
note in particular the coupling of the classical field
$\Phi_c = \left( \phi_c, \phi_c^{*} \right)$ to the quantum source
$J_q = \left( j_q, j_q^{*} \right)$, and vice versa. Apart from removing the
redundancy mentioned above, a further key advantage of this choice of basis is
that taking variational derivatives with respect to the sources produces the two
basic types of observables in many-body systems: correlation and response
functions. Of particular importance is the single particle Green's function,
which has the following matrix structure in Keldysh space (for a brief
introduction to functional differentiation see Appendix~\ref{sec:funct-diff};
note that here we are talking functional derivatives with respect to the
components of the spinors of sources
$J_{\nu} = \left( j_{\nu}, j_{\nu}^{*} \right)$ where $\nu = c, q$),
%\begin{widetext}
\begin{multline}\label{eq:greend}
  \begin{pmatrix}
    \langle \phi_c(t,\mathbf{x}) \phi^{*}_c(t',\mathbf{x}')\rangle_d &\langle
    \phi_c(t,\mathbf{x}) \phi^{*}_q(t',\mathbf{x}')\rangle_d  \\
    \langle \phi_q(t,\mathbf{x}) \phi^{*}_c(t',\mathbf{x}')\rangle_d &\langle
    \phi_q(t,\mathbf{x}) \phi^{*}_q(t',\mathbf{x}')\rangle_d
\end{pmatrix}
\\
\begin{aligned}
  & = - \left.
  \begin{pmatrix}
    \frac{\delta^2 Z}{\delta j^{*}_q(t,\mathbf{x})  \delta j_q(t',\mathbf{x}') }
    &\frac{\delta^2 Z}{\delta j^{*}_q(t,\mathbf{x})  \delta j_c(t',\mathbf{x}')
    }   \\
    \frac{\delta^2 Z}{\delta j^{*}_c(t,\mathbf{x}) \delta j_q(t',\mathbf{x}') }
    &\frac{\delta^2 Z}{\delta j^{*}_c(t,\mathbf{x}) \delta j_c(t',\mathbf{x}') }
\end{pmatrix}\right\rvert_{J_c = J_q = 0} \\ & = i \begin{pmatrix}
  G_d^K(t-t', \mathbf{x}- \mathbf{x}') & G_d^R(t-t', \mathbf{x}- \mathbf{x}')  \\
  G_d^A(t-t', \mathbf{x}- \mathbf{x}') & 0
\end{pmatrix}.
\end{aligned}
\end{multline}
% \end{widetext}
In the last equality, in addition to stationarity (time translation invariance)
we have assumed spatial translation invariance. $G^{R/A/K}$ are called retarded,
advanced, and Keldysh Green's function, and --- in the terminology of
statistical mechanics --- the index $d$ stands for \emph{disconnected} averages
obtained from differentiating the partition function $Z$; the zero component is
an exact property and reflects the elimination of redundant information (see
Sec.~\ref{sec:single-mode-cavity} below). We anticipate that the retarded and
advanced components describe responses, and the Keldysh component the
correlations. The physical meaning of the Green's function is discussed in the
subsequent subsection by means of a concrete example.

\emph{ Keldysh effective action } --- At this point we need one last technical
ingredient: the effective action, which is an alternative way of encoding the
correlation and response information of a non-equilibrium field
theory~\cite{Amit/Martin-Mayor,Calzetta1988a}. We first introduce the new
generating functional
\begin{equation}
  \label{eq:15}
  W[J_c, J_q] = - i \ln Z[J_c, J_q].
\end{equation}
Differentiation of the functional $W$ generates the hierarchy of so-called
\emph{connected} field averages, in which expectation values of lower order
averages are subtracted; for example,
$\langle \phi_c(t,\mathbf{x}) \phi^{*}_c(t',\mathbf{x}')\rangle = \langle
\phi_c(t,\mathbf{x}) \phi^{*}_c(t',\mathbf{x}')\rangle_d -\langle
\phi_c(t,\mathbf{x})\rangle_d \langle \phi^{*}_c(t',\mathbf{x}')\rangle_d$
describes the density of particles which are not condensed. In a compact
notation, where $W^{(2)}$ denotes the second variation as in
Eq.~\eqref{eq:greend},
\begin{multline}
  \label{eq:greenc}
  \left. W^{(2)}(t - t', \mathbf{x} - \mathbf{x}') \right\rvert_{J_c = J_q =0}
  \\ = -
  \begin{pmatrix}
    G^K(t-t', \mathbf{x}- \mathbf{x}') & G^R(t-t', \mathbf{x}- \mathbf{x}')  \\
    G^A(t-t', \mathbf{x}- \mathbf{x}') & 0
  \end{pmatrix}.
\end{multline}
The effective action $\Gamma$ is obtained from the generating functional $W$ by
a change of active variables. More precisely, the active variables of the
functional $W$, which are the external sources $J_{c, q}$, i.e.,
$W = W[J_c, J_q]$, are replaced by the field expectation values
$\bar{\Phi}_{\nu} = \left( \bar{\phi}_{\nu}, \phi_{\nu}^{*} \right)$ where
$\nu = c, q$, i.e., $\Gamma = \Gamma[\bar{\Phi}_{c}, \bar{\Phi}_{q}]$. (We
remind the reader that capital letters denote spinors of fields and their
complex conjugates.) The field expectation values are defined as (the notation
indicates that these expectation values are taken in the presence of the sources
taking non-zero values)
\begin{equation}
  \label{eq:wphi}
  \bar{\Phi}_{\nu} = \langle \Phi_{\nu} \rangle \rvert_{J_c,J_q} = \delta
  W/\delta J_{\nu'}^{*} = \left( \delta W/\delta j_{\nu'}^{*}, \delta W/\delta
    j_{\nu'} \right), 
\end{equation}
where $\nu' = q$ for $\nu = c$ and vice versa. Switching to these new variables
is accomplished by means of a Legendre transform familiar from classical mechanics,
\begin{equation}
  \label{eq:16}
  \Gamma[\bar{\Phi}_c,\bar{\Phi}_q] = W[J_c, J_q] - \int_{t, \mathbf{x}} \left( J_c^{\dagger}
    \bar{\Phi}_q + J_q^{\dagger} \bar{\Phi}_c \right).
\end{equation}
We now proceed to show that the difference between the action $S$ and the
effective action $\Gamma$ lies in the inclusion of both statistical and quantum
fluctuations in the latter. To this end, instead of working with
Eq.~\eqref{eq:16}, we represent $\Gamma$ as a functional
integral~\cite{Amit/Martin-Mayor}, with the result
\begin{equation}
  \label{eq:1700}
  e^{i \Gamma[\bar{\Phi}_c,\bar{\Phi}_q]} = \int \mathscr{D}[\delta
  {\Phi}_c,\delta {\Phi}_q] \, e^{i S[\bar{\Phi}_c + \delta
    {\Phi}_c,\bar{\Phi}_q + \delta {\Phi}_q] -i \tfrac{\delta
      \Gamma}{\delta \bar{\Phi}_c^T} \delta   {\Phi}_q -i \tfrac{\delta
      \Gamma}{\delta \bar{\Phi}_q^T} \delta  {\Phi}_c}.
\end{equation}
On the right hand side, we have used the property of the Legendre transformation
\begin{equation}
  \label{eq:var}
  \delta \Gamma/\delta \bar{\Phi}_c = -J_q^{*} , \quad \delta \Gamma/\delta \bar{\Phi}_q = -J_c^{*}.
\end{equation} 
Equation~\eqref{eq:1700} is obtained by exponentiating ($i$ times)
Eq.~\eqref{eq:16}, and using the explicit functional integral representation of
the Keldysh partition function in $W = - i\ln Z$, Eq.~\eqref{eq:14}. We have
introduced the notation $ {\Phi}_\nu = \bar{\Phi}_\nu + \delta {\Phi}_\nu$. This
implies $\langle \delta \Phi_{\nu} \rangle \rvert_{J_c,J_q} = 0$ by construction
(the average is taken in the presence of non-vanishing sources), and moreover
allows us to use
$\mathscr{D}[ {\Phi}_c, {\Phi}_q] = \mathscr{D}[\delta {\Phi}_c,\delta
{\Phi}_q]$
in the functional measure. Note the appearance of the field fluctuation
$\delta \Phi_{\nu}$ in the last term in the functional
integral~\eqref{eq:1700}. This is due to the explicit subtraction of the source
term in Eq.~\eqref{eq:16}.

Equation~\eqref{eq:1700} simplifies when we consider vanishing external sources
$J_c = J_q=0$. This case is particularly intuitive, as it shows that the
effective action $\Gamma[\bar{\Phi}_c,\bar{\Phi}_q]$ corresponds to
supplementing the action $S[\bar{\Phi}_c, \bar{\Phi}_q]$ by all possible
fluctuation contributions, expressed by the functional integration over
$\delta \bar{\Phi}_c$ and $\delta \bar{\Phi}_q$.

The variational condition on $\Gamma$, Eq.~\eqref{eq:var} (technically
reflecting the change of active variables via Eq.~\eqref{eq:16}), precisely takes
the form of the equation of motion derived from the principle of least action
(in the presence of an external source or force term) familiar form classical
mechanics. Here, however, it governs the dynamics of the full effective
action. In light of the above interpretation of the effective action,
Eq.~\eqref{eq:var} thus promotes the conventional classical action principle to
full quantum and statistical status.

Conversely, neglecting the quantum and statistical fluctuations in
Eq.~\eqref{eq:1700}, we directly arrive at
$\Gamma[\bar{\Phi}_c,\bar{\Phi}_q] = S[\bar{\Phi}_c, \bar{\Phi}_q]$. This
approximation is appropriate at intermediate distances\footnote{At very long
  wavelengths, gapless fluctuations of the Goldstone mode lead to infrared
  divergences in perturbation theory and invalidate this power counting
  argument~\cite{Pata73,Zwerger04}.} in the case where there is a
macroscopically occupied condensate: we may then count
$\bar\Phi_c = O(\sqrt{N})$, with $N$ the extensive number of particles in the
condensate, while fluctuations $\delta\Phi_c,\delta\Phi_q= O(1)$, as well as
$\bar\Phi_q= O(1)$ for the noise field, which cannot acquire a non-vanishing
expectation value. This crude approximation of the full effective action
reproduces the standard Gross-Pitaevski mean-field theory, if we consider a
generic Hamiltonian with kinetic energy and local two-body collisions, and drop
the dissipative contributions to the action.

The functions generated by the effective action functional (via taking
variational derivatives with respect to its variables
$ \bar{\Phi}_c, \bar{\Phi}_q$) are called the one-particle irreducible (1PI) or
amputated vertex functions~\cite{Amit/Martin-Mayor}. A crucial and useful
relation between the 1PI vertex functions and connected correlation functions
(generated by $W$) is
\begin{multline}
  \label{eq:33}
  - \int_{t'', \mathbf{x}''}
  \begin{pmatrix}
    0 & \frac{\delta^2 \Gamma}{\delta  \bar\phi_c^{*}(t, \mathbf{x}) \delta
       \bar\phi_q(t'', \mathbf{x}'')} \\ \frac{\delta^2 \Gamma}{\delta \bar\phi_q^{*}(t,
      \mathbf{x}) \delta  \bar\phi_c(t'', \mathbf{x}'')} & \frac{\delta^2
      \Gamma}{\delta  \bar\phi_q^{*}(t, \mathbf{x}) \delta  \bar\phi_q(t'', \mathbf{x}'')}
  \end{pmatrix} \\ \times
  \begin{pmatrix}
    \frac{\delta^2 W}{\delta j_q^{*}(t'', \mathbf{x}'') \delta j_q(t',
      \mathbf{x}')} & \frac{\delta^2 W}{\delta j_q^{*}(t'', \mathbf{x}'') \delta
    j_c(t', \mathbf{x}')} \\ \frac{\delta^2 W}{\delta j_c^{*}(t'', \mathbf{x}'')
  \delta j_q(t', \mathbf{x}')} & 0
  \end{pmatrix}
  = \delta(t - t') \delta(\mathbf{x} - \mathbf{x}') \id.
\end{multline}
This relation follows directly from Eqs.~\eqref{eq:wphi} and~\eqref{eq:var},
using the chain rule. Combined with Eq.~\eqref{eq:greenc}, it states that the
second variation of the effective action is precisely the full inverse Green's
function.
 
Typically, an exact evaluation of the effective action is not possible --- it
would constitute the full solution of the interacting non-equilibrium many-body
problem. However, powerful analytical tools have been developed for the analysis
of such problems, ranging from systematic diagrammatic perturbation theory over
the efficient introduction of emergent degrees of freedom to genuine
non-perturbative approaches such as the functional renormalization group. The
latter technique is discussed in Sec.~\ref{sec:open-sys-FRG}. Examples in which
such strategies were put into practice are discussed in
Part~\ref{part:applications} of this review.

%%% Local Variables:
%%% mode: latex
%%% TeX-master: "dds_review"
%%% End:

\subsection{Examples}
\label{sec:examples}

In this section, we bring the rather formal considerations of the previous one
to life by considering explicit examples for driven-dissipative systems. To be
specific, in Sec.~\ref{sec:single-mode-cavity} we consider the decay of a
single-mode cavity. This is arguably one of the simplest examples that combines
coherent and dissipative dynamics: the system itself consists of a single
bosonic mode (the cavity photon), has a quadratic Hamiltonian and linear
Lindblad operators. Hence, the Keldysh action is quadratic and the functional
integral can be solved exactly. Hence, we are able to obtain an explicit
expression for the generating functional defined in Eq.~\eqref{eq:14}, which
allows us to conveniently study several properties of the Keldysh formalism:
what is known as the causality structure, the analyticity properties of the
Green's functions, and the intuitive and transparent way in which spectral and
statistical properties are encoded in the formalism. Moreover, we compare these
findings with the case of a bosonic mode in equilibrium. The presence of the
properties discussed in this section is not restricted to the non-interacting
case. Much rather, they prevail also in the presence of interactions, and are
thus exact properties of non-equilibrium field theories. We point this out
alongside the examples.

In Sec.~\ref{sec:driv-diss-cond}, we consider an example for an interacting
bosonic many-body system, which contains non-linearities not only due to
particle-particle interactions, but also in the dissipative contribution to the
dynamics. These features are realized experimentally in exciton-polariton
systems. In the present section, we restrict ourselves to a mean-field analysis
of this system and defer a discussion of the role of fluctuations to
Sec.~\ref{sec:bosons}.

\subsubsection{Single-mode cavity, and some exact properties}
\label{sec:single-mode-cavity}

The master equation describing the decay of photons in a single-mode cavity
takes the general form of Eq.~\eqref{eq:109}, with $H = \omega_0 a^{\dagger} a$,
where $a^{\dagger}$ and $a$ are creation and annihilation operators of photons,
and $\omega_0$ is the frequency of the cavity mode. Assuming the external
electromagnetic field to be in the vacuum state, there is only a single term in
the sum over $\alpha$ with $L = a$, describing the decay of the cavity field at
a rate $2 \kappa$ (the factor of 2 is chosen for convenience). The corresponding
Keldysh action is given by~\cite{DallaTorre2013}
\begin{multline}
  \label{eq:4045612}
  S = \int_t \left\{ a_{+}^{*} \left( i \partial_t - \omega_0 \right) a_{+} -
    a_{-}^{*} \left( i \partial_t - \omega_0 \right) a_{-} \right. \\ \left. - i
    \kappa \left[ 2 a_{+} a_{-}^{*} - \left( a_{+}^{*} a_{+} + a_{-}^{*} a_{-}
      \right) \right] \right\},
\end{multline}
where $a_{\pm}, a^*_{\pm}$ represent the complex photon field.
Performing the basis rotation to classical and quantum fields as in
Eq.~\eqref{eq:41} in Sec.~\ref{sec:deriv-keldysh-acti}, and going to Fourier
space, the action becomes
\begin{equation}
  \label{eq:39}
  S = \int_{\omega} \left( a_c^{*}(\omega), a_q^{*}(\omega) \right)
  \begin{pmatrix}
    0 & P^A(\omega) \\ P^R(\omega) & P^K
  \end{pmatrix}
  \begin{pmatrix}
    a_c(\omega) \\ a_q(\omega)
  \end{pmatrix},
\end{equation}
where we used the shorthand
$\int_{\omega} \equiv \int_{-\infty}^{\infty} \frac{d \omega}{2 \pi}$. Furthermore, we have 
\begin{equation}
  \label{eq:26}
  P^R(\omega) = P^A(\omega)^{*} = \omega - \omega_0 + i \kappa,\quad P^K = 2 i \kappa.
\end{equation}
$P^{R/A}$ are the inverse retarded and advanced Green's functions, and $P^K$ is
the Keldysh component of the inverse Green's function. To see this, we evaluate
the generating functional~\eqref{eq:14} by Gaussian integration (cf.\
Appendix~\ref{sec:gauss-funct-integr}). Then, the generating
functional~\eqref{eq:15} for connected correlation functions is given by
\begin{equation}
  \label{eq:45}
  W[J_c, J_q] = - \int_{\omega} \left( j_q^{*}(\omega), j_c^{*}(\omega) \right)
  \begin{pmatrix}
    G^K(\omega) & G^R(\omega) \\ G^A(\omega) & 0
  \end{pmatrix}
  \begin{pmatrix}
    j_q(\omega) \\ j_c(\omega)
  \end{pmatrix}.
\end{equation}
According to Eq.~\eqref{eq:greenc}, the second variation (i.e., the matrix in
the above equation, see Appendix~\ref{sec:funct-diff}) represents the Green's
function. It is obtained by inversion of the matrix in the action in
Eq.~\eqref{eq:39},
\begin{align}
  \label{eq:42}  
  G^R(\omega) & = G^A(\omega)^{*} = \frac{1}{P^R(\omega)} = \frac{1}{\omega - \omega_0 +
                i \kappa}, \\ \label{eq:54}
  G^K(\omega) & = - G^R(\omega) P^K G^A(\omega) = - \frac{i 2 \kappa}{\left(
                \omega - \omega_0 \right)^2 + \kappa^2},
\end{align}
i.e., the matrix in the action indeed represents the inverse Green's
function. We now summarize a few key structural properties that can be gleaned
from this explicit discussion. Indeed, as we argue below, these properties are
valid in general.

\emph{Conservation of probability} --- There is a zero matrix entry in
Eq.~\eqref{eq:39}, or, equivalently, in Eq.~\eqref{eq:45}.  Technically, as
anticipated above, this property reflects a redundancy in the $\pm$ basis and
eliminates it. This simplifies practical calculations in the Keldysh basis.
Physically, this property ensures the normalization of the partition function
($Z = \tr \rho (t) =1$), and can thus be interpreted as manifestation of the
conservation of probability ($\partial_t\tr \rho(t) =0$), which is an exact
property of physical problems. This can be seen as follows: consider the more
general property, which implies the vanishing matrix element in the quadratic
sector, $S[a_c, a_c^{*}, a_q = 0, a_q^{*} = 0] = 0$. Any Keldysh action
associated to the Liouville operator Eq.~\eqref{eq:109}, has this property, as
can be seen by setting $\psi_+ = \psi_-$ (i.e., $\phi_q =0$) in
Eq.~\eqref{eq:63}. Indeed, this operation on the Keldysh action may be
interpreted as taking the trace in the operator based master equation
Eq.~\eqref{eq:109}: in this way, using the cyclic property of the trace allows
us to shift all operators to one side of the density matrix, leading to the
cancellation of terms such that $\partial_t\tr \rho(t) =0$. 

We still need to
argue that the above property of the classical action also holds for the full
theory, i.e., the effective action,
\begin{equation}
  \label{eq:43}
  \Gamma[\bar a_c, \bar a_c^{*}, \bar a_q = 0, \bar a_q^{*} = 0] = 0,
\end{equation}
or, more schematically in the notation of Sec.~\ref{sec:deriv-keldysh-acti}, $\Gamma [\bar \Phi_c, \bar \Phi_q =0] =0$. To this end, we note that $W[J_c,J_q=0 ]=0$ ($Z[J_c,J_q=0 ] =1$) holds actually for \emph{arbitratry} classical sources $J_c$ (whereas in Sec.~\ref{sec:deriv-keldysh-acti} we worked additionally with $J_c=0$ for conceptual clarity): any term $\sim J_c$ can be absorbed into the underlying Hamiltonian, describing nothing but a Hamiltonian contribution in the presence of a classical external potential, and thus it cannot affect the normalization property of the theory. The above properties are equivalent: 
\begin{eqnarray}
\Gamma [\bar \Phi_c, \bar \Phi_q =0] =0 \Leftrightarrow W[J_c,J_q=0 ]=0.
\end{eqnarray}
This is seen using the definition of the Legendre transform Eq.~\eqref{eq:16} and the definition of the quantum field in terms of Eq.~\eqref{eq:wphi} for one direction of the mutual implication. The other direction results from the involutory property of the Legendre transform \footnote{We thank F. Tonielli for pointing out this compact argument.}.

%This can be done using an RG continuity argument. The functional renormalization group equation (cf.\ Eq.~\eqref{eq:FRG}) smoothly connects the classical action $S$ to the full effective action $\Gamma$. At the level of $S$, the property holds. Performing the first RG step, there is no way of generating a nonvanishing matrix element, due to the zero in the Green's function associated to the classical action. Iterating this argument down to the full effective action establishes the property also for the latter object. 

Sometimes, the property of conservation of probability -- expressed in the effective action formalism as $\Gamma [\bar \Phi_c, \bar \Phi_q =0] =0$ --  is referred to as ``causality structure'' in the literature.

\emph{Hermeticity properties of the Green's functions} --- As can be read
off from Eqs.~\eqref{eq:42} and~\eqref{eq:54}, $G^R(\omega)$ and $G^A(\omega)$
are Hermitian conjugates, and $G^K(\omega)$ is anti-Hermitian. These properties are exact, as can be read off from the definition of the Green's function in terms of functional derivatives in Eq.~\eqref{eq:greend}. Equivalently,
these properties hold for the corresponding components of the inverse Green's
function (cf. Eq.~\eqref{eq:33}).

\emph{Analytic structure} --- The poles of $G^R(\omega)$ are located at
$\omega = \omega_0 - i \kappa$, i.e., in the lower half of the complex plane
(accordingly, the poles of $G^A(\omega)$ are in the upper half). In the real
time domain, this implies that $G^R$ describes the \emph{retarded} response (and
accordingly, $G^A$ the \emph{advanced}): indeed, taking the inverse Fourier
transform, we obtain
\begin{equation}
  \label{eq:44}
  G^R(t) = - i \theta(t) e^{- \left( i \omega_0 + \kappa \right) t},
\end{equation}
where $\theta(t)$ is the Heaviside step function. Hence, the response of the
system, if it is perturbed at $t = 0$, is retarded (non-zero only for $t > 0$)
and decays. 

The analytic structure of retarded and advanced Green's functions is a general
property of Keldysh actions, too; one may then think of these Green's functions
as renormalized, full single particle Green's functions connected to the bare,
microscopic ones via renormalization, and thus preserving the analyticity
properties. We note, however, that the pole of the retarded bosonic Green's
function can approach the real axis from below via tuning of microscopic
parameters. The touching point typically signals a physical instability: beyond
that point, the description of the system must be modified qualitatively. Such a
scenario is discussed in the next subsection.

\emph{Connection to the operator formalism} --- It is sometimes useful to
restore the precise relation between the operator formalism and the functional
integral description at the level of the Green's functions. At the single
particle level, they read
\begin{align}
    \label{eq:48}
  G^R(t, t') & = - i \theta(t - t') \langle [a(t), a^{\dagger}(t')] \rangle,\\
  \label{eq:51}
  G^K(t, t') & = - i \langle \{ a(t), a^{\dagger}(t') \} \rangle,
\end{align}
and we note again that in stationary state
$G^{R/A/K}(t, t') = G^{R/A/K}(t - t')$. These relations are exact and can be
obtained from going back to the $\pm$ basis: the retarded Green's function is
\begin{equation}
  \begin{split}
    G^R(t,t') & = -i\langle a_{c}(t)a^*_q(t')\rangle \\ & =-\frac{i}{2}\langle
    \left(a_{+}(t)+a_-(t)\right)(a^*_+(t')-a_-^*(t'))\rangle \\
    & = -\frac{i}{2}\left(\langle T a(t)a^{\dagger}(t')\rangle+\langle
      [a(t),a^{\dagger}(t')]\rangle-\langle\tilde{T}a(t)a^{\dagger}(t')\rangle
    \right) \\
    & = -i\theta(t-t')\langle[a(t),a^{\dagger}(t')]\rangle,
  \end{split}
\end{equation}
where $T, \tilde{T}$ are the time-ordering, anti-time-ordering operators, which
lead to a cancellation of the commutator for $t'>t$ (see Ref.~\cite{Kamenev2011}
for a more detailed discussion of time ordering on the Keldysh contour). The
Keldysh Green's function in the operator formalism is obtained the same way,
\begin{equation}
  \begin{split}
    G^K(t,t') & = -i\langle a_{c}(t)a^*_c(t')\rangle \\ & =-\frac{i}{2}\langle
    \left(a_{+}(t)+a_-(t)\right)(a^*_+(t')+a_-^*(t'))\rangle \\
    & = -\frac{i}{2}\left(\langle T a(t)a^{\dagger}(t')\rangle+\langle
      \{a(t),a^{\dagger}(t')\}\rangle+\langle\tilde{T}a(t)a^{\dagger}(t')\rangle
    \right) \\
    & = -i\langle\{a(t),a^{\dagger}(t')\}\rangle.
  \end{split}
\end{equation}

\emph{Response vs. correlation functions} --- In order to generate response and
correlation functions directly from the partition function, it is convenient to
introduce source fields $j_+, j_- $ and express the partition function as the
average~(cf.\ Eq.~\eqref{eq:35})
\begin{equation}
  Z[j_+,j_-] = \langle e^{-iS_j} \rangle=\int\mathcal{D}[a^*_+,a^*_-,a_+,a_-] e^{iS-iS_j},
\end{equation}
where the source action is defined as 
\begin{equation}
  \label{eq:S_h}
  S_j = \int_t \left(j^*_+ a_+-j^*_-a_- + \cc \right)=\int_t \left(j^*_c
    a_q+j^*_qa_c+ \cc \right).
\end{equation}
In the second step, we have performed a Keldysh rotation. Due to the
normalization of the Keldysh path integral $Z[j_c=0, j_q=0]=1$, and as a
consequence, expectation values of $n$-point functions of the fields $a^*, a$
can be expressed via $n$-th order functional derivatives of the partition
function with respect to the source fields (see
Appendix~\ref{sec:funct-diff}). For example
\begin{equation}
  \begin{split}
    \langle a_c(t)\rangle & = \left. i\frac{\delta Z(j_c, j_q)}{\delta
        j_q^*(t)}\right|_{j_c=j_q=0},\\ 
    G^K(t,t') & = -i\langle a_c(t)a^*_c(t')\rangle=\left. i\frac{\delta^2 Z(j_c,
        j_q)}{\delta j_q^*(t)\delta j_q(t')}\right|_{j_c=j_q=0},\\
    G^R(t,t') & = -i\langle a_c(t)a^*_q(t')\rangle=\left. i\frac{\delta^2 Z(j_c,
        j_q)}{\delta j_q^*(t)\delta j_c(t')}\right|_{j_c=j_q=0}.
  \end{split}
\end{equation}
Due to causality, the field $j_q$ has to be zero in any physical setup and the
introduction of this field only serves as a technical tool to compute
expectation values via derivatives. On the other hand, the field $j_c$ can, in
principle, be different from zero and we will see in the following what the
physical meaning of this classical source field is, and in which respect it
generates the response function.

\begin{figure}
  \centering
  \includegraphics[width=\linewidth]{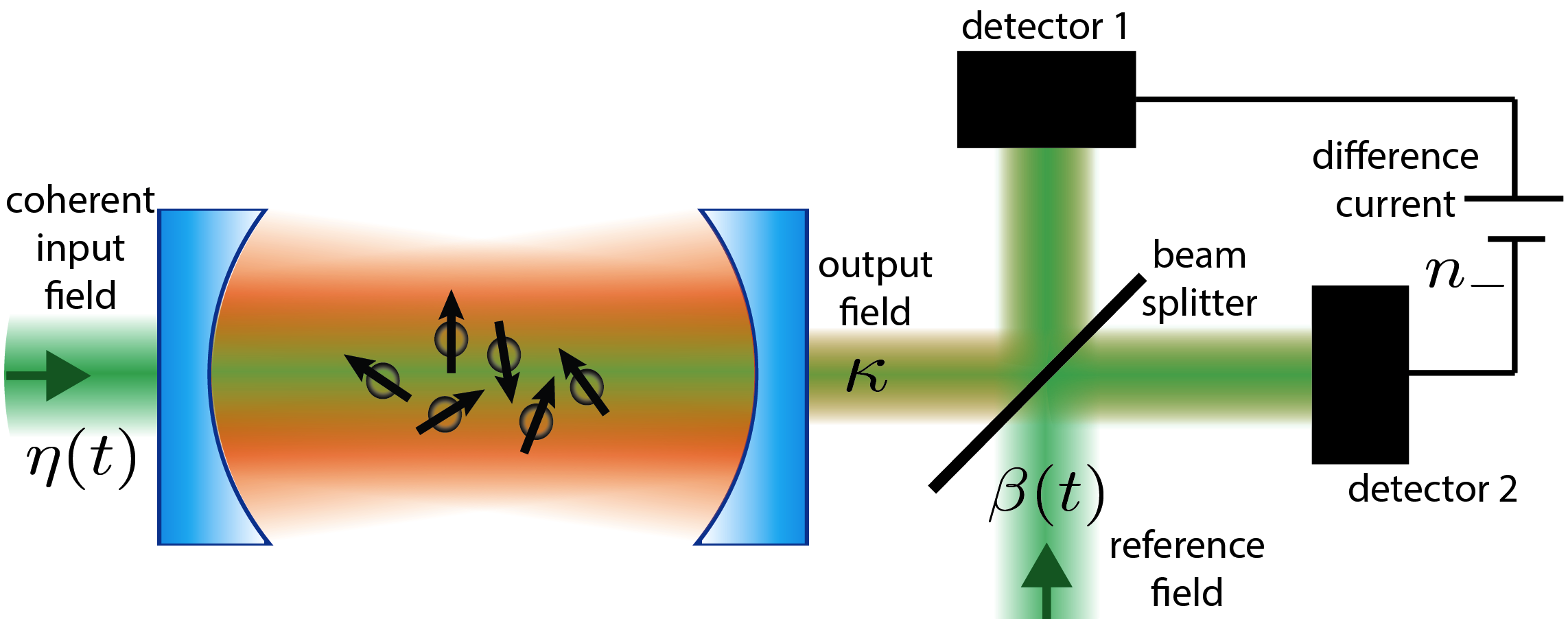}
  \caption{Illustration of a homodyne detection measurement which determines the
    response function $G^R(t,t')$ of the cavity photons, see
    Ref.~\cite{Buchhold2013}. The system of atoms and photons inside the cavity
    is perturbed by the laser field $\eta(t)$, entering the cavity through the
    left mirror. The response of the system is encoded in the light field which
    is leaking out of the cavity on the right mirror with a rate $\kappa$. It
    can be measured by a standard homodyne detection measurement in which the
    reference laser $\beta(t)$ and a beam splitter are used in order to obtain
    information on the system's coherence $\langle a_c(t)\rangle$. Figure copied
    from Ref.~\cite{Buchhold2013}. (Copyright (2013) by The American Physical
    Society.)}
\label{fig:HomDyn}
\end{figure}

\emph{Responses:} The retarded Green's function, often called synonymously
response function, describes the linear response of a system which is perturbed
by a weak external source field. As an illustrative example, consider a driven
cavity system consisting of atoms and photons, which is considered to be in a
stationary state. Due to imperfections in the cavity mirrors, there is a finite
rate with which a photon is escaping the cavity, or an external photon is
entering the cavity. This process is expressed by the Hamiltonian
\begin{equation}
  \label{eq:107}
  H_p=\sqrt{2\kappa}\left(b^{\dagger}a+a^{\dagger}b\right),
\end{equation}
where the operators $b,b^{\dagger}$ represent photons outside the
cavity. Shining a laser through the cavity mirror, the operators $b,b^{\dagger}$
can be replaced by the coherent laser field $j(t),j^*(t)$, which oscillates with
the laser frequency $\omega_j$. The corresponding Hamiltonian,
describing the complete system is (for the present purposes we can leave the
Hamiltonian $H$ of photons and atoms in the cavity unspecified)
\begin{equation}
  \label{eq:36}
  H_j  = H +\sqrt{2 \kappa} \left( j^{*}(t) a + j(t) a^{\dagger} \right).
\end{equation}
Since the fields $j,j^*$ are classical external fields, they are equal on the plus and
minus contour and the Keldysh action in the presence of the laser is
\begin{equation}
  S_j  = S - \int_t \left( j^{*}(t) a_q(t) + j(t) a_q^{*}(t) \right).\label{Ac38}
\end{equation}
This action is very similar to the action in Eq.~\eqref{eq:S_h}, with a linear
source term, which is however $j_+=j_-=j$. In the present example, the meaning
of the source term is physically very transparent: it is nothing but the
coherent laser field that is coupled to the cavity photons. For a weak source
field, one is interested in the first order correction of observables induced by
the coupling to the source. In the present case, this is the coherent light
field inside the cavity $\langle a_c(t)\rangle$. Up to first order in $j(t)$ it
is given by
\begin{equation}
    \label{eq:46}
    \begin{split}
      \langle a_c(t) \rangle_j & = \langle a_c(t) \rangle_{j=0} - i
      \sqrt{\kappa}
      \int_{t'} \langle a_c(t) a^*_q(t') \rangle_{j=0} j(t')\nonumber\\
      & = \langle a_c(t) \rangle_{j=0}+ \sqrt{\kappa} \int_{t'}G^R(t-t')j(t').
    \end{split}
\end{equation}
The retarded Green's function is therefore a measure of the system's response to
an external perturbation. An experimental setup, with which one can measure the
coherent light field and therefore the response function of the cavity via
so-called homodyne detection, is illustrated in Fig.~\ref{fig:HomDyn}.
A closely related function of interest is the \emph{spectral function}  
\begin{equation}
  \label{eq:49}
  A(\omega) = - 2 \Im G^R(\omega).
\end{equation}
It is the distribution of excitation levels of the system, i.e., when adding a
single photon with frequency $\omega$ to the system, $A(\omega)$ is the
probability to hit the system at resonance. Indeed, it can be
shown~\cite{Altland2010} that the spectral function is positive,
$A(\omega) > 0$, and fulfills the sum rule
\begin{equation}
  \label{eq:108}
  \int_{\omega} A(\omega) = \langle[a,a^{\dagger}]\rangle=1. 
\end{equation}
For our example of a single-mode cavity, the spectral function is given by
\begin{equation}
  \label{eq:50}
  A(\omega) = \frac{2 \kappa}{\left( \omega - \omega_0 \right)^2 + \kappa^2},
\end{equation}
i.e., it is a Lorentzian, which is centered at the cavity frequency $\omega_0$
and has a half-width at half-maximum given by $\kappa$. Note that for
$\kappa \to 0$, the photon number states become exact eigenstates and the
spectral density reduces to a $\delta$-function peaked at $\omega_0$,
$A(\omega) = 2 \pi \delta(\omega - \omega_0).$ As these considerations
illustrate, the retarded Green's function contains essential information on the
system's response towards an external perturbation, and on the spectral
properties.

\emph{Correlations:} The Keldysh Green's function contains elementary
information on the system's correlations and the occupation of the individual
quantum mechanical modes. A prominent example of a correlation function in
quantum optics is the photonic $g^{(2)}$ correlation function. It is defined as
the four-point correlator
\begin{equation}\label{g2}
g^{(2)}(t,\tau)=\frac{\langle a^{\dagger}(t)a^{\dagger}(t+\tau)a(t+\tau)a(t) \rangle}{|\langle a^{\dagger}(t)a(t)\rangle|^2},
\end{equation}
and it is proportional to the intensity fluctuations of the intracavity
radiation field $g^{(2)}(t,\tau)\propto\langle I(t)I(t+\tau)\rangle$. In the
limit of $\tau\rightarrow0$, the $g^{(2)}$ correlation function reveals the
statistics of the cavity photons, i.e., it demonstrates super-Poissonian
($g^{(2)}(0) > 1$) or sub-Poissonian statistics ($g^{(2)}(0) < 1$) as an effect
of the light-matter interactions. In the absence of interactions (as in our
example), it is straightforward to show that
\begin{equation}
  \label{eq:110}
  g^{(2)}(t,\tau)=1+\frac{|G^K(t,\tau+t)+G^R(t,\tau+t)-G^A(t,\tau+t)|^2}{| iG^K(t,t)-1|^2}.
\end{equation}
For equal times, the retarded and advanced Green's functions satisfy
$G^R(t,t)-G^A(t,t)=-i$, indicating the expected photon-bunching at
$\tau\rightarrow0$. On the other hand, in the presence of interactions,
Eq.~\eqref{g2} is modified and contains additional higher order terms, as well
as off-diagonal contributions. However, it remains a measure of the photonic
statistics in the cavity due to the physical relation to the intensity
fluctuations.

A particularly relevant and instructive limit of the two-time Keldysh Green's
function Eq.~\eqref{eq:51} concerns equal times $t = t'$, which in general
describes static correlation functions, or covariances in the quantum optics
language. In the cavity example, it yields the \emph{mode occupation} number,
\begin{equation}
  \label{eq:53}
  i G^K(t, t) = 2 \langle a^{\dagger} a \rangle + 1.
\end{equation}
The appearance of the combination $2 \langle a^{\dagger} a \rangle +1 $ is
rather intuitive when taking into account the relation to the operatorial
formalism. Indeed, in operator language,
$2 a^\dag a + 1 = \{a,a^\dag\}= \{a^\dag,a\}$ is invariant under permutation of
the operators $a,a^\dag$, and therefore lends itself to a direct functional
integral representation, which in fact carries no information on operator
ordering.\footnote{The same structural argument based on operator ordering
  insensitivity of the functional representation demonstrates why an ordinary
  Euclidean functional integral yields the time ordered Green's functions, cf.,
  e.g., Ref.~\cite{Negele1998}.} At the same time, the anticommutator carries
all the physical information on state occupation, while the commutator carries
none, $[a,a^\dag ] =1$ (note
$a^\dag a = \tfrac{1}{2} (\{a,a^\dag\} - [a,a^\dag])$).

Then, the explicit form of the Keldysh Green's function stated in
Eq.~\eqref{eq:54} leads to the result
\begin{equation}
  \label{eq:55}
  \langle a^{\dagger} a \rangle = \frac{1}{2} \left( i \int_{\omega}
    G^K(\omega) - 1 \right) = 0,
\end{equation}
showing that the cavity is empty in the steady state---which is not surprising in
the absence of external pumping.
  
We note that in the master equation formalism, information on the static or
spatial correlations is most easily accessible --- only the state (density
matrix) has to be known, but not the dynamics acting on it. Temporal correlation
functions can be extracted using the quantum regression
theorem~\cite{Walls2008}. In the Keldysh formalism, spatial and temporal
correlation functions are treated on an equal footing.
  
We can concisely summarize the above discussion of the response and Keldysh
Green's functions of the system, at the risk of oversimplifying:
\begin{framed}
  \noindent Responses $G^{R/A} \leftrightarrow$ spectral information on
  excitations: which excitations are there? \\
  Correlations $G^{K} \leftrightarrow$ statistical information on excitations:
  how are the excitations occupied?
\end{framed}

\emph{Relation to thermodynamic equilibrium} --- Before we move on, let us
briefly discuss how we have to modify the formalism in order to describe a
cavity in thermodynamic equilibrium. In general, a system is in thermodynamic
equilibrium at a temperature $T = 1/\beta$, if it is in a Gibbs state
$\rho = e^{- \beta H}/Z$ where $Z = \tr e^{- \beta H}$, and its dynamics is
coherent and generated by the Hamiltonian $H$. (In particular, dissipative
dynamics described by a term in the Lindblad form in Eq.~\eqref{eq:109} is not
compatible with equilibrium conditions: see
Refs.~\cite{Talkner1986,Ford1996,Sieberer2015}, and the discussion in
Sec.~\ref{sec:therm-equil-sym}.) Specifying the thermal density matrix at $t_0$
in the Keldysh functional integral in Eq.~\eqref{eq:31} explicitly in terms of
its matrix elements is rather inconvenient, especially if one is interested in
steady state properties and wants to take the limit $t_0 \to - \infty.$ An
alternative is suggested by the observation that thermodynamic equilibrium can
be established in the system if it is weakly coupled to a thermal bath. Then,
the finite decay rate $\kappa$ in the retarded Green's function in
Eq.~\eqref{eq:42} should be replaced by an infinitesimally weak system-bath
coupling $\delta \to 0$.  Additionally, the Keldysh Green's function has to be
modified in such a way that the integral in Eq.~\eqref{eq:55} yields the Bose
distribution function $n(\omega_0)$ implying thermal occupations of the cavity
mode,
\begin{equation}
  \label{eq:57}
  \langle a^{\dagger} a \rangle = n(\omega_0) = \frac{1}{e^{\beta \omega_0} - 1}.
\end{equation}
This can be achieved by replacing the Keldysh component of the inverse Green's
function in Eq.~\eqref{eq:39} by
$P^K(\omega) = i 2 \delta \left( 2 n(\omega) + 1 \right)$. We emphasize the key
structural difference of the thermal Keldysh component, which is strongly
frequency dependent, to the Markovian case discussed previously, where this
entry is frequency independent --- this gives a strong hint that Markovian
systems can behave quite differently from systems in thermal equilibrium. With
$P^R(\omega) = P^A(\omega)^{*} = \omega - \omega_0 - i \delta$ we obtain the
equilibrium Green's functions
\begin{align}
  \label{eq:58}
  G^R(\omega) & = G^A(\omega)^{*} = \frac{1}{\omega - \omega_0 + i \delta}, \\ 
  G^K(\omega) & = - i 2 \pi \delta(\omega - \omega_0) \left( 2 n(\omega) + 1 \right).
\end{align}
Note that the Green's functions obey a \emph{thermal fluctuation-dissipation
  relation (FDR)}, which for the present example of a single bosonic mode reads
\begin{equation}
  \label{eq:56}
  G^K(\omega) = \left( 2 n(\omega) + 1 \right) \left( G^R(\omega) - G^A(\omega)
  \right).
\end{equation}
This is discussed in more detail in Sec.~\ref{sec:therm-equil-sym}. For the
time being, let us mention that this construction to describe thermodynamic
equilibrium by adding infinitesimal dissipative terms does not only work in the present
case of a quadratic action but can also be applied in the interacting
case. Then, the construction ensures that the free Green's functions (i.e., the
ones obtained by ignoring the interactions) obey an FDR. If the non-linear terms
in the action obey the equilibrium symmetry discussed in
Sec.~\ref{sec:therm-equil-sym} (which is the case for generic interaction
terms), this property is shared by the full Green's functions of the non-linear
system.

\emph{General Fluctuation-Dissipation Relations} ---
While Eq.~\eqref{eq:56} is valid only in thermodynamic equilibrium, it is always
possible to parameterize the (anti-Hermitian) Keldysh Green's function in terms
of the retarded and advanced Green's functions (which, as discussed above, are
Hermitian conjugates of each other) and a Hermitian matrix $F = F^{\dagger}$ in
the form~\cite{Kamenev2011,Altland2010}
\begin{equation}
  \label{eq:FDR-general}
  G^K = G^R \circ F - F \circ G^A,
\end{equation}
where $\circ$ denotes convolution. In this parametrization, $F$ is the
distribution function, which describes the distribution of (quasi-) particles
over the modes of the system. For a non-equilibrium steady state, $F$ is
time-translational invariant and its Fourier transform in frequency space
$F(\omega)$ represents the energy resolved occupation of (quasi-) particle
modes.  On the other hand, as we discuss in detail in Sec.~\ref{sec:kineq}, for
the case of a time-evolving system, for which time translation invariance is
absent, the Wigner transform of $F$ corresponds to the instantaneous local
distribution function.  For the important case of thermodynamic equilibrium (of
a bosonic system), it is $F(\omega) = \coth(\beta \omega/2) = 2 n(\omega) + 1$,
and Eq.~\eqref{eq:FDR-general} reduces to~\eqref{eq:56}. For the case where the
bosonic Green's function has a matrix structure in Nambu space, a subtlety
concering the preservation of the symplectic structure of that space arises,
cf.\ Sec.~\ref{sec:kineq}. There, also a time-dependent variant of the
non-equilibrium fluctuation-dissipation relations is discussed.

\subsubsection{Driven-dissipative condensate}
\label{sec:driv-diss-cond}

In the previous section, we discussed the simple case of a quadratic Keldysh
action, which allowed us to perform the Keldysh functional integral
explicitly. An additional simplification resulted from the fact that we were
considering only a single bosonic mode. Let us now consider a genuine many-body
problem, which is non-linear and in which the system consists of a continuum of
modes. To be specific, in this section we discuss the model introduced in
Sec.~\ref{sec:excit-polar-syst} for a bosonic many-body system with interactions
and non-linear loss processes (i.e., with a loss rate that is proportional to
the density) in addition to the linear dissipative terms which were already
present in the example of the single-mode cavity. Then, for a specific value of
the mean-field density $\rho_0$, non-linear loss and linear pump exactly balance
each other and the system reaches a stationary state. If $\rho_0$ is different
from zero, this signals the presence of a condensate, which is accompanied by
the breaking of a specific phase rotation symmetry as will be discussed in
detail in Sec.~\ref{sec:symm-keldysh-acti}, and the establishment of long-range
order. In Sec.~\ref{sec:bosons}, we give a detailed account of the influence of
fluctuations on this driven-dissipative condensation transition. Here, we
content ourselves with a mean-field analysis, which serves to illustrate some of
the field theoretical concepts introduced in Sec.~\ref{sec:deriv-keldysh-acti}
--- in particular, the effective action and field equations --- in a simple
setting.

In the basis of classical and quantum fields, the Keldysh action associated with
the quantum master equation~\eqref{eq:104} reads
\begin{multline}
  \label{eq:112}  
  S = \int_{t, \mathbf{x}} \left\{ \phi_q^{*} \left( i \partial_t + K_c \nabla^2 - r_c + i r_d
    \right) \phi_c + \cc \right. \\ - \left[ \left( u_c - i u_d \right) \left(
      \phi_q^{*} \phi_c^{*} \phi_c^2 + \phi_q^{*} \phi_c^{*} \phi_q^2 \right) +
    \cc \right] \\ \left. + i 2 \left( \gamma + 2 u_d \phi_c^{*} \phi_c \right)
    \phi_q^{*} \phi_q \right\},
\end{multline}
where $K_c = 1/(2 m_{\mathrm{LP}})$ and $r_c = \omega_{\mathrm{LP}}^0$; as
additional parameters, we introduced the noise level
$\gamma = ( \gamma_l + \gamma_p)/2$ and the spectral mass or gap
$r_d = ( \gamma_l - \gamma_p )/2$. Hence, the rates of losses and pumping add up
to the total noise level; in contrast, the difference of these rates enters in
the spectral gap $r_d$, which becomes negative when the rate of incoherent
pumping exceeds the single-particle loss rate, signaling the physical
instability against condensation. In a mean-field analysis of the condensation
transition, we perform a saddle-point approximation of the functional integral
in Eq.~\eqref{eq:1700}. To leading order, fluctuations around the field
expectation values are completely neglected. The expectation values are then
obtained as spatially homogenous and stationary solutions to the classical field
equations (here, ``classical'' refers to the fact that these field equations are
derived from the classical (or: bare, microscopic) action $S$ discarding
fluctuations, in contrast to the field equations in Eq.~\eqref{eq:var}, which
involve the effective action)
\begin{equation}
\label{eq:59}
  \frac{\delta S}{\delta \phi_c^{*}} = 0, \quad \frac{\delta S}{\delta
    \phi_q^{*}} = 0.
\end{equation}
As already mentioned above in Sec.~\ref{sec:single-mode-cavity}, there are no
terms in the action Eq.~\eqref{eq:16} with zero power of both $\phi_q^{*}$ and
$\phi_q$, and the same is clearly true for $\delta S/\delta \phi_c^{*}$.
Therefore, the first equation in~\eqref{eq:59} is solved by $\phi_q =
0$. Inserting this condition into the second equation, we have
\begin{equation}
\label{eq:60}
  \left[ - r_c + i r_d - \left( u_c - i u_d \right) \abs{\phi_0}^2 \right] \phi_0 = 0.
\end{equation}
The solution $\phi_c = \phi_0$ is determined by the imaginary part of
Eq.~\eqref{eq:60}: for $r_d \geq 0$, in the so-called symmetric phase, the
classical field expectation value is zero, $\rho_0 = \abs{\phi_0}^2 = 0$,
whereas for $r_d < 0$ we have a finite condensate density $\rho_0 = - r_d/u_d$.
Taking the real part of Eq.~\eqref{eq:60}, we obtain for the parameter $r_c$ the
relation $r_c = u_c r_d/u_d$. This condition can always be satisfied by proper
choice of a rotating frame, i.e., by performing a gauge transformation
$\phi_c \mapsto \phi_c e^{-i \omega t}$ such that $r_c \mapsto r_c - \omega$. In
the original (laboratory) frame this simply means that the condensate amplitude
oscillates at a finite frequency.

In a first step beyond mean field theory, quadratic fluctuations around the
mean-field order parameter can be investigated within a Bogoliubov or tree-level
expansion: we set $\phi_c = \phi_0 + \delta \phi_c, \phi_q = \delta \phi_q$ in
the action Eq.~\eqref{eq:112} and expand the resulting expression to second
order in the fluctuations $\delta \phi_{c, q}$. The inverse retarded, advanced
and Keldysh Green's functions now become $2 \times 2$ matrices in the space of
Nambu spinors
$\delta \Phi_{\nu} = \left( \delta \phi_{\nu},\delta \phi_{\nu}^{*} \right)$. In
particular, we have in the frequency and momentum domain
\begin{equation}
  \label{eq:40456}
  \begin{split}
    P^R(\omega,\mathbf{q}) & =
    \begin{pmatrix}
      \omega - K_c q^2 - \left( u_c - i u_d \right) \rho_0 & - \left( u_c - i
        u_d \right) \rho_0  \\
      - \left( u_c + i u_d \right) \rho_0 & - \omega - K_c q^2 - \left( u_c + i
        u_d \right)
    \end{pmatrix}, \\
    P^A(\omega,\mathbf{q})) & = P^R(\omega, \mathbf{q})^{\dagger}, \\ P^K & = i
    \gamma \id.
  \end{split}
\end{equation}

The excitation spectrum is obtained from the condition
$\det P^R(\omega,\mathbf{q}) = 0$. Indeed, this is the condition for the
field equation of the fluctuations
$ P^R(\omega,\mathbf{q}) \delta \Phi_{c}(\omega,\mathbf{q}) =0$ to have
nontrivial solutions. This yields~\cite{Wouters2007a}
\begin{equation}
  \label{eq:61}
  \omega_{1,2}^R = - i u_d \rho_0 \pm \sqrt{K_c q^2 \left( K_c q^2 + 2 u_c \rho_0 \right) - \left(
      u_d \rho_0 \right)^2}.
\end{equation}
We note that due to the tree-level shifts $\propto \rho_0$ the above described
instability for $r_d < 0$ is lifted: both poles are consistently located in the
lower complex half-plane, indicating a physically stable situation with decaying
single-particle excitations. For $u_d = 0$, Eq.~\eqref{eq:61} reduces to the
standard Bogoliubov result~\cite{Lifshitz1980}, where for $q \to 0$ the
dispersion is phononic, $\omega_{1,2}^R = \pm c q$
($c = \sqrt{2 K_c u_c \rho_0}$ the speed of sound), whereas particle-like
behavior $\omega_{1,2}^R \sim K_c q^2$ is obtained at high momenta. Here, due to
the presence of two-body loss $u_d \neq 0$, the dispersion is qualitatively
modified: while at high momenta the dominant behavior is still
$\omega_{1,2}^R \sim K_c q^2$, at low momenta we find purely incoherent
diffusive, non-propagating modes $\omega_1^R \sim - i \frac{K_c u_c}{u_d} q^2$
and $\omega_2^R \sim - i 2 u_d \rho_0$. In particular, for $q = 0$ we have
$\omega_1^R = 0$: this is a dissipative Goldstone
mode~\cite{Wouters2006,Wouters2007a,Szymanska2006}, associated with the
spontaneous breaking of the global $U(1)$ symmetry in the ordered phase. The
existence of such a mode is not bound to the mean-field approximation, but
rather is an exact property guaranteed by the $U(1)$ invariance of the effective
action, even in the present case of a driven-dissipative condensate. We will
come back to this point in Sec.~\ref{sec:goldstone}.

%%% Local Variables:
%%% mode: latex
%%% TeX-master: "dds_review"
%%% End:

\subsection{Semiclassical limit of the Keldysh action}
\label{sec:semicl-limit-keldysh}

The theoretical description of a many-body system depends crucially on the scale
(this could be a length, time, or energy scale --- in practice these scales can
be expressed in terms of each other) on which it is observed and, in particular,
on the relation between this observation scale and the intrinsic scales of the
system. For example, at finite temperature $T$ above a quantum critical point,
non-trivial quantum critical behavior can be observed at moderate energy scales
which are larger than $T$ but smaller than the Ginzburg scale where fluctuations
start to dominate over the mean field effects~\cite{Sachdev2011}. On the other
hand, classical thermal critical behavior, which is perfectly described by
taking the semiclassical limit of the underlying quantum theory, sets in at
energy scales below $T$. Out of thermodynamic equilibrium, the analogue of a
finite temperature is Markovian noise. In the Keldysh action, this corresponds
to a constant term in the Keldysh sector of the inverse Green's function, i.e.,
a constant noise vertex. Such a term is present in the action given in
Eq.~\eqref{eq:112} and, therefore, we expect that in the long-wavelength limit
this action can actually be simplified by taking the semiclassical
limit~\cite{Kamenev2011,Altland2010}. We refer to it as the semiclassical limit,
as, e.g., effects of quantum mechanical phase coherence are not necessarily
suppressed in this limit, as we will see. A useful equilibrium analogy is the
physics of Bose-Einstein condensates at finite but low temperatures, where phase
coherence still persists.

\begin{figure*}
  \centering  
  \includegraphics[width=.8\textwidth]{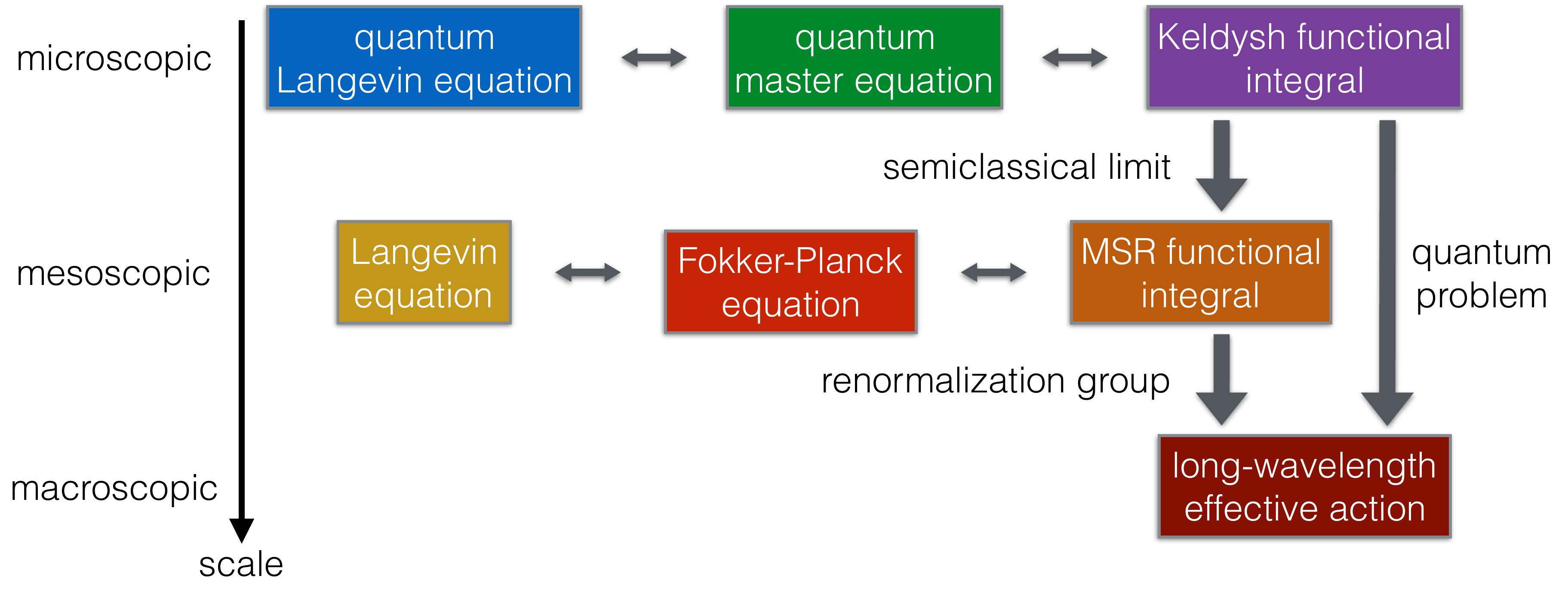}
  \caption{Equivalent descriptions on varying length scales. The quantum and
    classical Langevin equations are stochastic equations of motions for the
    field operators, or for classical field variables. In contrast, the
    descriptions in the middle column are deterministic equations of motion,
    where either a density operator or a probability distribution (diagonals of
    a density matrix) are evolved. In the functional integral formulation, the
    basic object in both quantum and classical cases is an action, which is
    averaged over all possible realizations of field configurations. The
    semiclassical limit is valid at mesoscopic scales as dicsussed in
    Sec.~\ref{sec:semicl-limit-keldysh}. An effective description at macroscopic
    scales can be obtained by means of renormalization group methods (see
    Sec.~\ref{sec:open-sys-FRG}). Generically in Markovian systems, in order to
    reduce the complexity of the problem it is useful to first perform the
    semiclassical limit before doing a renormalization group
    computation. However, in driven open quantum systems there are also
    circumstances where this is inappropriate, and the full quantum problem has
    to be analyzed, cf.~\cite{Marino2015}.}
  \label{fig:quantum_to_classical}
\end{figure*}

Formally, the suitability of the semiclassical limit can be understood in terms
of ideas that lie at the basis of the renormalization group (RG). The latter
provides a recipe for finding an effective description that is valid on large
length scales, starting from a microscopic theory. Various RG schemes exist,
which allow to systematically eliminate fluctuations on short scales and infer
their influence on the effective description of the system on large scales; cf.\
Sec.~\ref{sec:open-sys-FRG}. The most basic but still informative RG approach
however consists in simply \emph{ignoring} the effect of short-scale
fluctuations: one subdivides momentum space into a slow and a fast region, with
$q_s \in [0, \Lambda/b]$ and $q_f \in [\Lambda/b, \Lambda]$ where $\Lambda$ is
the UV cutoff (given by the inverse of some microscopic length scale below which
the theoretical description is not valid any more, e.g., a lattice spacing) and
$b > 1$, and omits all contributions to the action which involve fluctuations
with fast momenta. In a second step, one rescales all momenta with $b$ to
restore the original range of momenta $q \in [0, \Lambda]$, and the thus
obtained effective long-wavelength description can be compared to the original
one. As a result, due to this simple RG transformation couplings in the action
are rescaled as $g \mapsto g b^{d_g}$, where $d_g$ is known as the canonical
scaling dimension of $g$. Evidently, for $d_g > 0$, $g$ grows under
renormalization and hence $g$ is a relevant coupling, while it shrinks under
renormalization and, hence, is irrelevant at long wavelength for $d_g < 0$.  For
the case of a marginal coupling with $d_g = 0$, this level of approximation is
not conclusive as to whether the coupling becomes larger or smaller under
renormalization. Classifying the coupling parameters of an action according to
this scheme is known as canonical scaling analysis, or power counting. A useful
approximation consists in neglecting all irrelevant couplings. On the other
hand, relevant couplings which are compatible with the symmetries of the model
-- even those, which are not present in the microscopic description --- should
be included in the long-wavelength effective action.

Let us perform such an analysis for the Keldysh action in Eq.~\eqref{eq:112}. We
can anticipate the canonical scaling dimensions by noting that they are just the
\emph{physical} dimensions, measured in powers of the momentum. We first focus
on the vicinity of the threshold for condensation, i.e., when
$r_d \sim \gamma_l - \gamma_p \to 0$. Then the retarded and advanced inverse
Green's functions scale as $P^{R/A}\sim q^2$ (note that $\omega \sim q^2$ for
low-momentum excitations). The canonical dimension is thus positive,
$d_{P^{R/A}} =2$. On the other hand, as anticipated above, Markovian noise
analogous to a finite temperature corresponds to a momentum-independent noise
vertex, i.e., the term $\propto \phi_q^{*} \phi_q$ in the Keldysh action. In
other words, the Keldysh component of the inverse Green's function has vanishing
canonical scaling dimension and classifies as marginal,
$P^K = i 2 \gamma \sim q^0$. We furthermore use the natural scaling
$d^d \mathbf{x} \sim q^{-d}$ and $dt \sim 1/\omega \sim q^{-2}$, and the
condition of scale invariance of the action, $S \sim q^0$ --- this is a
requirement stemming from the fact that the action appears in the exponent of
the functional integral~\eqref{eq:31}, and thus must be dimensionless. We then
find the scaling dimensions of the fields from the Gaussian, quadratic part of
the action to be
\begin{equation}\label{eq:powerc}
\phi_c \sim q^{(d - 2)/2}, \quad \phi_q \sim q^{(d + 2)/2}.
\end{equation}
This in turn allows us to derive the scaling dimensions of the quartic terms,
and to classify their degree of relevance as pointed out above. In particular,
we find that in three spatial dimensions, any quartic term that includes more
than a single quantum field is irrelevant; the only non-irrelevant term of
higher order in the noise field is the quadratic noise vertex discussed
above. Therefore, omitting irrelevant terms amounts to keeping only the
classical vertex in the action Eq.~\eqref{eq:112}, i.e., to taking the
semiclassical limit~\cite{Kamenev2011,Altland2010}. Then, the Keldysh action
takes the form
\begin{multline}
  \label{eq:140}
  S = \int_{t, \mathbf{x}} \left\{ \phi_q^{*} \left[ i \partial_t + \left( K_c -
        i K_d \right) \nabla^2 - r_c + i r_d \right] \phi_c + \cc \right. \\
  \left. - \left[ \left( u_c - i u_d \right) \phi_q^{*} \phi_c^{*} \phi_c^2 +
      \cc \right] + i 2 \gamma \phi_q^{*} \phi_q \right\},
\end{multline}
where in addition to omitting irrelevant contributions we have added an
effective diffusion term $\propto i K_d \nabla^2$. Such a term can and will be
generated upon integrating out short-scale
fluctuations~\cite{Sieberer2013,Sieberer2014}. Moreover, a complex prefactor of
the term involving the derivative with respect to time, which also emerges upon
renormalization, can be absorbed into a redefinition of the fields.

Strictly speaking, the above analysis is valid in the vicinity of the critical
point only, where the inverse retarded and advanced Green's function show
scaling. However, as long as one is close enough to threshold
$\lvert \gamma_l - \gamma_p \rvert/(\gamma_l + \gamma_p)\ll1$, the canonical
power counting is expected to give a useful orientation in the problem.

Apart from providing a significant simplification, taking the semiclassical
limit also allows us to establish the connection~\cite{Zhou1980,Chou1985}
between the Keldysh functional integral formalism and the more traditional
formulation of dynamics close to a continuous phase transition in terms of
Langevin equations~\cite{Hohenberg1977,Tauber2014a}. In fact, the action in
Eq.~\eqref{eq:140} is fully equivalent to the following Langevin equation:
\begin{equation}
  \label{eq:17}
  i \partial_t \phi_c = \left[ - \left( K_c - i K_d \right) \nabla^2 + r_c - i
    r_d + \left( u_c - i u_d \right) \abs{\phi_c}^2 \right] \phi_c + \xi,
\end{equation}
where $\xi$ is a Markovian Gaussian noise source with zero mean,
$\langle \xi(t, \mathbf{x}) \rangle = 0,$ and second moment
$\langle \xi(t, \mathbf{x}) \xi(t', \mathbf{x}') \rangle = 2 \gamma \delta(t -
t') \delta(\mathbf{x} - \mathbf{x}')$.
This equivalence can be established by means of a Hubbard-Stratonovich
transformation of the noise vertex~\cite{Kamenev2011,Altland2010}, i.e., the
term $i 2 \gamma \phi_q^{*} \phi_q$. To wit, in the Keldysh functional integral
\begin{equation}
  \label{eq:39123}
  Z = \int \mathscr{D}[\phi_c, \phi_c^{*}, \phi_q, \phi_q^{*}] \, e^{i S[\phi_c,
    \phi_c^{*}, \phi_q, \phi_q^{*}]},
\end{equation}
where $S$ is the semiclassical action in Eq.~\eqref{eq:140}, we write the noise
vertex as a Gaussian integral over an auxiliary variable $\xi$ (cf.\
Appendix~\ref{sec:gauss-funct-integr}),
\begin{equation}
  \label{eq:94}
  e^{- 2 \gamma \int_{t, \mathbf{x}} \phi_q^{*}
    \phi_q} = \int \mathscr{D}[\xi, \xi^{*}] \, e^{- \frac{1}{2 \gamma} \int_{t,
      \mathbf{x}} \xi^{*} \xi - i \int_{t, \mathbf{x}} \left( \phi_q^{*} \xi -
      \xi^{*} \phi_q \right)}.  
\end{equation}
As a result, the exponent in the functional integral~\eqref{eq:39123} becomes
linear in the quantum field, and the corresponding integration can be performed,
resulting in a $\delta$-functional,
\begin{multline}
  \label{eq:40}
  Z = \int \mathscr{D}[\phi_c, \phi_c^{*},\xi, \xi^{*}] \, e^{- \frac{1}{2
      \gamma} \int_{t, \mathbf{x}} \xi^{*} \xi} \\ \times \delta \! \left(
    \left[ i \partial_t + \left( K_c - i K_d \right) \nabla^2 - r_c + i r_d -
      \left( u_c - i u_d \right) \abs{\phi_c}^2 \right] \phi_c - \xi \right) \\
  \times \delta \!  \left( \left[ - i \partial_t + \left( K_c + i K_d \right)
      \nabla^2 - r_c - i r_d - \left( u_c + i u_d \right) \abs{\phi_c}^2 \right]
    \phi_c^{*} - \xi^{*} \right).
\end{multline}
This expression can be interpreted as follows: for a given realization of the
noise field $\xi$, the $\delta$-functional restricts the functional integral
over $\phi_c$ to the manifold of solutions to the Langevin
equation~\eqref{eq:17}. The statistics of the noise field is determined by the
Gaussian weight factor in Eq.~\eqref{eq:40}. Correlation functions of classical
fields can then be calculated by picking a random realization of $\xi$ and
calculating the corresponding $\phi_c$ that solves the Langevin equation;
evaluating the correlation function for this solution and finally averaging the
result according to the Gaussian distribution of the noise
field. Equation~\eqref{eq:17} is, therefore, equivalent to the functional
integral~\eqref{eq:40} in that it allows for the evaluation of arbitrary
correlation functions.

Originally, Langevin equations like Eq.~\eqref{eq:17} have been introduced as a
phenomenological description of the coarse-grained dynamics of the order
parameter, and only later a functional integral approach --- known as the MSR
approach --- has been
developed~\cite{Martin1973,DeDominicis1976,Janssen1976,Bausch1976}. The action
derived in these references is formally equivalent to the one in
Eq.~\eqref{eq:140}, with $\phi_c$ taking the role of the order parameter field,
while the field that corresponds to $\phi_q$ is known as the response field.

In addition to the descriptions of semiclassical dynamical models in terms of a
semiclassical Keldysh (or MSR) functional integral or a Langevin equation, there
exists yet another equivalent approach, in which the stochastic Langevin
equation for the classical field or order parameter field is replaced by a
deterministic evolution equation for the probability distribution of the latter,
known as the Fokker-Planck equation. The derivation of the Fokker-Planck
equation can be found, e.g., in
Refs.~\cite{Altland2010,Kamenev2011,Tauber2014a,Chaikin1995}.
Figure~\ref{fig:quantum_to_classical} illustrates the equivalence of these
approaches, which are valid on a (coarse-grained) mesoscopic scale, and the
relations to microscopic and macroscopic descriptions.

%%% Local Variables:
%%% mode: latex
%%% TeX-master: "dds_review"
%%% End:

\subsection{Symmetries of the Keldysh action}
\label{sec:symm-keldysh-acti}

Symmetries take center stage in field theories. Translating the physics of the
quantum master equation to the Keldysh functional integral allows us to leverage
the power of symmetry considerations over to the context of open systems. In
this section, we discuss three different aspects of symmetries.

The presence of a first, discrete symmetry, considered in
Sec.~\ref{sec:therm-equil-sym}, allows us to conclude whether or not a quantum
or classical system is situated in the realm of thermodynamic equilibrium. This
symmetry is equivalent to the validity of thermal fluctuation-dissipation
relations (cf.\ Sec.~\ref{sec:single-mode-cavity}) for correlation functions of
any order and can be connected to energy conservation. Thermal equilibrium can
thus be diagnosed by means a of simple symmetry test on the Keldysh action. In
particular, we argue that Markovian quantum master equations explicitly violate
this symmetry, indicating non-equilibrium conditions. In the semiclassical
limit, we obtain a simple and intuitive geometric interpretation of the symmetry
and its absence under non-equilibrium drive in terms of the location of the
coupling constants of the effective action in the complex plane.

A second fundamental consequence of the presence of symmetries --- now,
continous global symmetries --- is the Noether theorem, stating that such
symmetries imply conserved charges. In Sec.~\ref{sec:noether-theorem-keldysh},
we discuss the Noether theorem in the context of the Keldysh formalism. Working
on the Keldysh contour, we have to distinguish between two symmetry
transformations, for each the forward and backward branches of the closed time
path. In the basis of classical and quantum fields, we can identify
``classical'' symmetry transformations, which act in the same way on fields on
the forward and backward branches, and ``quantum'' transformations, for which
the transformation of the fields on the backward branch is the inverse of the
transformation on the forward branch. Non-trivial conservation laws follow only
from the symmetry of the Keldysh action under \emph{quantum} transformations. We
illustrate this point with the example of the symmetry of the action of a closed
system with respect to space-time translations and phase rotations in
Sec.~\ref{sec:energy-momentum-particle-numb-cons}, which entails conservation of
energy and momentum and the number of particles, respectively. On the other
hand, in open systems, in which the number of particles is not conserved, only
\emph{classical} phase rotations are a symmetry of the action, and the
continuity equation that is implied by particle number conservation has to be
extended as detailed in Sec.~\ref{sec:extend-cont-equat-open-sys}.

Finally, another consequence of a global continuous symmetry is the existence of
gapless modes in case that this symmetry is broken spontaneously, as in a Bose
condensation transition. In a driven-dissipative condensate in which the number
of particles is not conserved (cf.\ Sec.~\ref{sec:driv-diss-cond}), we show that
the symmetry which is broken spontaneously at the condensation transition is the
classical phase rotation symmetry, and we work out the Goldstone theorem, which
guarantees the existence of a massless mode, for this case in
Sec.~\ref{sec:goldstone}.

\subsubsection{Thermodynamic equilibrium as a symmetry of the Keldysh action}
\label{sec:therm-equil-sym}

A system is in thermodynamic equilibrium at a temperature $T = 1/\beta$, if (i)
the density matrix is given by $\rho = e^{- \beta H}/Z$ where
$Z = \tr e^{-\beta H}$, and (ii) the very same Hamiltonian operator $H$
appearing in $\rho$ generates the unitary time evolution of the system,
$U = e^{-i H t}$. Condition (ii) implies that static correlations are in general
not sufficient to prove that a system is in thermal equilibrium; however,
dynamical correlations are: this can be inferred from the fact that the static
correlations of a physical system can always be encoded in a density matrix
$\rho$, and the latter can always be parameterized formally as an equilibrium
density matrix, $\rho = e^{-\beta H'}/Z'$, with a Hermitian operator $H'$. (In
other words, \emph{any} state can be thought of as a thermal state with respect
to \emph{some} Hamiltonian.) On the other hand, static (i.e., purely momentum-
or space-dependent) properties do not allow us to discriminate whether the
generator of dynamics coincides with $H'$. In sharp contrast, any dynamical
(i.e., frequency- or time-dependent) observable is manifestly governed by this
generator, as is easily seen in the Heisenberg picture (or a suitable
generalization to open systems). Response functions at finite frequency are such
dynamical observables, and the equilibrium conditions formulated above are
reflected in the fact that in thermodynamic equilibrium the response of the
system to an external perturbation at a frequency $\omega$ is related to thermal
fluctuations within the system at the same frequency by what is known as a
fluctuation-dissipation relation~\cite{Kubo1966} (FDR; for a discussion in the
context of non-equilibrium Bose-Einstein condensation see
Ref.~\cite{Chiocchetta2015}). Such a relation, which is equivalent to the
combination of the Kubo-Martin-Schwinger (KMS)
condition~\cite{Kubo1957,Martin1959} with time reversal~\cite{Jakobs2010}, is
valid for any pair of operators. In particular, for the case that these are the
basic field operators of a single-component Bose system at the temperature
$T = 1/\beta$, the FDR reads (note that this is a generalization of
Eq.~\eqref{eq:56} to the case of a spatial continuum of degrees of freedom)
\begin{equation}
  \label{eq:20}  
  G^K(\omega, \mathbf{q}) = \coth \! \left( \frac{\omega}{2 T} \right)
 ( G^R(\omega, \mathbf{q}) - G^A(\omega, \mathbf{q}))
\end{equation}
(in the presence of a chemical potential $\mu$, in the argument of the
trigonometric function we have to shift $\omega \to \omega -\mu$). In quantum
field theory, relations among Green's functions often follow as consequences of
a symmetry of the action. Then, they are known as Ward-Takahashi identities
associated with the
symmetry~\cite{Zinn-Justin2002,Peskin1995}.\footnote{Usually, the term
  ``Ward-Takahashi identity'' is reserved for relations that follow from a
  continuous symmetry. Here, we use it also in the context of discrete symmetry
  transformations.} This raises the question, whether FDRs and hence the
presence of thermodynamic equilibrium conditions are also connected to a
symmetry of the Keldysh action.

To make the point clear, let us rephrase the question: above we made the
observation, that the defining property of (canonical) thermal equilibrium is
the validity of FDRs. Importantly, these FDRs hold for correlation functions of
arbitrary order, i.e., not just the two-point Keldysh Green's function in
Eq.~\eqref{eq:20} can be expressed through response functions, but also any
higher order correlation function is determined by corresponding higher order
response functions. Then, the question --- which we answer in the affirmative
below --- is, whether this infinite hierarchy of FDRs expressing thermal
equilibrium is related to a \emph{structural} property of the theory. Indeed,
this structural property is a symmetry of the Keldysh action, i.e., a
transformation of the fields $\Psi \mapsto \mathcal{T}_{\beta} \Psi$ such that
\begin{equation}
  \label{eq:126}
  S[\Psi] = \tilde{S}[\mathcal{T}_{\beta} \Psi].
\end{equation}
Here, we denote $\Psi = \left( \psi_+, \psi_+^{*}, \psi_-, \psi_-^{*} \right)$,
and we specify the precise form of $\mathcal{T}_{\beta}$ below in
Eq.~\eqref{eq:18}. The tilde on the RHS of the equation indicates that all
external fields appearing, $S$ have to be replaced by their corresponding
time-reversed values. To name an example, the signs of magnetic fields have to
be inverted~\cite{Sieberer2015}. Evidently, discussing a single symmetry instead
of an infinite hierarchy of equations is much more elegant and practical:
checking whether a given Keldysh action obeys Eq.~\eqref{eq:126} is
straightforward and can be accomplished in finite time, in contrast to verifying
the full set of FDRs.

It is easily seen, how these FDRs can be deduced \emph{as a consequence}
of the symmetry of the Keldysh action~\eqref{eq:126}. To this end, we note by a
Keldysh rotation~\eqref{eq:41} and after Fourier transformation, the Green's
functions appearing in Eq.~\eqref{eq:20} can be expressed as a sum of averages
of the form
$\langle \psi_{\sigma}(t, \mathbf{x}) \psi_{\sigma'}^{*}(t', \mathbf{x}')
\rangle$.
Writing these explicitly as field integrals, and anticipating that a change of
integration variables $\Psi \to \mathcal{T}_{\beta} \Psi$ leaves the functional
measure $\mathscr{D}[\Psi]$ invariant~\cite{Sieberer2015}, we find
\begin{equation}
  \label{eq:128}
  \begin{split}
    \langle \psi_{\sigma}(t, \mathbf{x}) \psi_{\sigma'}^{*}(t', \mathbf{x}')
    \rangle & = \int \mathscr{D}[\Psi] \, \psi_{\sigma}(t, \mathbf{x})
    \psi_{\sigma'}^{*}(t', \mathbf{x}') e^{i S[\Psi]} \\ & = \int
    \mathscr{D}[\mathcal{T}_{\beta} \Psi] \, \mathcal{T}_{\beta}
    \psi_{\sigma}(t, \mathbf{x}) \mathcal{T}_{\beta} \psi_{\sigma'}^{*}(t',
    \mathbf{x}') e^{i S[\mathcal{T}_{\beta} \Psi]} \\ & = \langle
    \mathcal{T}_{\beta} \psi_{\sigma}(t, \mathbf{x}) \mathcal{T}_{\beta}
    \psi_{\sigma'}^{*}(t', \mathbf{x}') \rangle.
  \end{split}
\end{equation}
In the last equality, we used the symmetry of the Keldysh action~\eqref{eq:126}
(assuming for simplicity that there are no external fields). Inserting here the
explicit form of the symmetry transformation specified in Eq.~\eqref{eq:18} below
where $\beta = 1/T$ is the inverse temperature, it can be seen that
Eq.~\eqref{eq:128} is in fact equivalent to the
FDR~\eqref{eq:20}~\cite{Sieberer2015}.

By generalizing this argument to arbitrary field averages, one can establish the
\emph{full equivalence} between the infinite hierarchy of FDRs and the symmetry
property of the Keldysh action~\eqref{eq:126}~\cite{Sieberer2015}. Thus, the
symmetry of the Keldysh action under this transformation is a direct proof of
the presence of thermodynamic equilibrium conditions. The existence of a
symmetry that is related to thermal equilibrium has first been realized in the
context of classical stochastic
models~\cite{Janssen1976,Bausch1976,Janssen1979,Aron2010,Aron2014}, and these
considerations have been extended to the realm of quantum systems in
Refs.~\cite{Altland2010a,Sieberer2015}.

What is the explicit form of the transformation $\mathcal{T}_{\beta}$? We can
guess its essential ingredients by reminding ourselves that, as stated above
Eq.~\eqref{eq:20}, FDRs can be obtained from the KMS
condition~\cite{Kubo1957,Martin1959}. The latter reads, for operators
$A(t) = e^{i H t} A(0) e^{-i H t}$ and $B(t')$ in the Heisenberg representation,
\begin{equation}
  \label{eq:127}
  \langle A(t) B(t') \rangle = \langle B(t') A(t + i \beta) \rangle.
\end{equation}
Comparing this with Eq.~\eqref{eq:128} indicates that $\mathcal{T}_{\beta}$
involves a translation of $t$ into the complex plane by an amount proportional
to the inverse temperature $\beta$. Moreover, the order of operators on the RHS
of the KMS condition is \emph{reversed} as compared to the LHS. The original
time order can be restored by means of a time reversal
transformation~\cite{Messiah:II} --- this step is necessary in order to obtain a
time-ordered expression which can be expressed as a Keldysh field integral (by
construction, field integrals yield time-ordered averages~\cite{Negele1998}). A
careful analysis~\cite{Sieberer2015} leads to the precise form of the thermal
symmetry transformation ($\sigma = + (-)$ for fields on the forward (backward)
branch):
\begin{equation}
  \label{eq:18}
  \begin{split}
    \mathcal{T}_{\beta} \psi_{\sigma}(t, \mathbf{x}) & = 
    \psi_{\sigma}^{*}(-t + i \sigma \beta/2, \mathbf{x}), \\ \mathcal{T}_{\beta}
    \psi_{\sigma}^{*}(t, \mathbf{x}) & = 
    \psi_{\sigma}(-t + i \sigma \beta/2, \mathbf{x})
  \end{split}
\end{equation}
(in the presence of a chemical potential $\mu$, we have to multiply the RHS of
the first (second) line by $e^{\sigma \beta \mu/2} (e^{- \sigma \beta \mu/2})$).
The transformation $\mathcal{T}_{\beta}$ is a composition of complex conjugation
of the fields $\psi_{\sigma}$ and inversion of the sign of the time $t$ --- both
originating from the time reversal transformation ---, and translation of the
value of $t$ by an amount $i \sigma \beta/2$. The latter is induced by the KMS
condition Eq.~\eqref{eq:127}. We note that the translation of $t$ in
Eq.~\eqref{eq:18} takes opposite signs depending on whether a field on the
forward or on the backward branch is being transformed. As we show in
Sec.~\ref{sec:energy-momentum-particle-numb-cons} below, a similar form of time
translations is connected to conservation of energy: indeed, if the Keldysh
action is invariant under time translations
$\psi_{\sigma}(t, \mathbf{x}) \mapsto \psi_{\sigma}(t + \sigma s, \mathbf{x})$
with $s \in \R$, the total energy in the system is conserved. A crucial
difference from the time translation that is part of $\mathcal{T}_{\beta}$ is
that energy conservation requires invariance under shifts by an \emph{arbitrary}
real value $s$, whereas $\mathcal{T}_{\beta}$ involves a shift by the purely
imaginary value $i \beta/2,$ where $\beta = 1/T$ is \emph{fixed} and determined
by the temperature.

Under which conditions does the Keldysh action~\eqref{eq:111} with $\mathcal{L}$
defined in Eq.~\eqref{eq:63} have the thermal symmetry, i.e., under which
conditions does it describe a system in thermal equilibrium? Let us first
consider the parts of the action corresponding to unitary time evolution, i.e.,
the first two terms in Eq.~\eqref{eq:111} and the first line in
Eq.~\eqref{eq:63}. It is straightforward to check~\cite{Sieberer2015} that these
terms are symmetric if the Hamiltonian densities $H_{\pm}$ do not explicitly
depend on time. On the other hand, adding an external classical driving field
such as a laser, and thus breaking time translational invariance by making
$H_{\pm}$ time dependent, also the thermal symmetry is broken \footnote{The more precise statement is that there exists no rotating frame in which the explicit time dependence fully disappears from the problem.}. The violation of
the thermal symmetry by classical driving fields on the level of a microscopic
Hamiltonian description indicates that quantum master equations correspond to
genuine non-equilibrium
conditions~\cite{Sieberer2015,Talkner1986,Ford1996}. Physically, this is due to
the fact that a system for which such a description is appropriate is
necessarily driven.

In an effective description of a driven-dissipative system in terms of a
Markovian master equation in a rotating frame, there is often no explicit time
dependence --- indeed, our derivation of the dissipative Keldysh action in
Sec.~\ref{sec:deriv-keldysh-acti} started from Eq.~\eqref{eq:109} with
time-independent Hamiltonian. Even then, the thermal symmetry can be used to
diagnose non-equilibrium conditions~\cite{Sieberer2015}. Again, it is sufficient
to study only the time-translation part of $\mathcal{T}_{\beta}$: for
time-independent $H_{\pm}$ and $L_{\alpha, \pm}$ in Eq.~\eqref{eq:63}, the
Keldysh action is still invariant under time translations of the form
$\psi_{\sigma}(t, \mathbf{x}) \mapsto \psi_{\sigma}(t + s, \mathbf{x})$.
Importantly, this differs from the time translation that occurs in
$\mathcal{T}_{\beta}$ by the absence of a factor of $\sigma$, meaning that $t$
is shifted by the same amount on the forward and backward branches. Then, by
means of a simple shift of the integration variable $t$ in Eq.~\eqref{eq:111}
the original form of the action can be restored. This strategy fails for the
dissipative contributions in Eq.~\eqref{eq:63} that couple forward and backward
branches, when the transformation involves the contour index
$\sigma$.\footnote{We note that also dissipative contributions corresponding to
  a system in thermal equilibrium, as described below Eq.~\eqref{eq:57} in
  Sec.~\ref{sec:single-mode-cavity}, couple the forward and backward
  branches. However, the special form of these terms conspire with the fact that
  the temporal shift in $\mathcal{T}_{\beta}$ is determined by the inverse
  temperature $\beta = 1/T$ to make them invariant under $\mathcal{T}_{\beta}$.}
Again we reach the conclusion that quantum master equations describe genuine
non-equilibrium conditions, even though by a slightly different argument than in
the driven but purely Hamiltonian setting.

In some cases, the Markov and rotating-wave approximations leading to a
description in terms of a quantum master equation or equivalent Keldysh action
are applied in the absence of external driving fields. Then, these
approximations might still be justified to study the behavior of specific
observables, even though they explicitly break the
symmetry~\cite{Sieberer2015}. To give an example, if one attempts to study
thermalization of a system due to the coupling with a heat bath, and one
integrates out the bath using the above-mentioned approximations, the resulting
dynamics of the system will still lead to a thermal stationary
state. Correspondingly, all \emph{static} properties of the system will appear
thermal. However, as discussed at the beginning of the present section, to
unambiguously prove the presence of thermal equilibrium conditions, one has to
consider \emph{dynamical} signatures such as FDRs. Then, as a consequence of the
explicit violation of the thermal symmetry by the Markov and rotating-wave
approximations, the description of the system dynamics in terms of a quantum
master equation will lead to the wrong prediction that fluctuations in the
system do not obey an FDR.

A simple example that illustrates the above discussion is given by a single
bosonic mode $a$ with energy $\omega_0$, driven coherently at a frequency
$\omega_l$, and coupled to a bath of harmonic oscillators $b_{\mu}$ in thermal
equilibrium. The associated Keldysh action can be decomposed as
$S = S_s + S_{sb} + S_b$, where the action for the system consists of two parts,
$S_s = S_0 + S_l$ which read
\begin{equation}
  \label{eq:5}
  S_0 = \sum_{\sigma} \sigma \int_{\omega} a_{\sigma}^{*}(\omega) \left(
  \omega - \omega_0 \right) a_{\sigma}(\omega),
\end{equation}
and
\begin{equation}
  \label{eq:117}
  \begin{split}
    S_l & = \Omega \sum_{\sigma} \sigma \int dt \left( a_{\sigma}(t) e^{i
        \omega_l t} + a_{\sigma}^{*}(t) e^{-i \omega_l t} \right) \\
    & = \Omega \sum_{\sigma} \sigma \left( a_{\sigma}(\omega_l) +
      a_{\sigma}^{*}(\omega_l) \right).
  \end{split}
\end{equation}
$\Omega$ is the amplitude of the driving field, and due to the harmonic time
dependence of the drive it affects only the component $a(\omega_l)$ in frequency
space. The action for the bath is expressed most conveniently in the basis of
classical and quantum fields (cf.\ Eq.~\eqref{eq:41}). In addition to the
coherent part stemming from the oscillator frequencies, it involves
infinitesimal dissipative regularization terms specifying the thermal
equilibrium state of the bath at inverse temperature $\beta = 1/T$ as discussed
in Sec.~\ref{sec:single-mode-cavity},
\begin{multline}
  \label{eq:9}
  S_b = \sum_{\mu} \int_{\omega}
  \begin{pmatrix}
    b_{\mu, c}^{*}(\omega), b_{\mu, q}^{*}(\omega)
  \end{pmatrix}
  \\ \times
  \begin{pmatrix}
    0 & \omega - \omega_{\mu} - i \delta \\
    \omega - \omega_{\mu} + i \delta & i 2 \delta \coth(\beta \omega/2)
  \end{pmatrix}
  \begin{pmatrix}
    b_{\mu, c}(\omega) \\ b_{\mu, q}(\omega)
  \end{pmatrix}.
\end{multline}
Finally, the system-bath interaction with coupling strength $\lambda$
corresponds to the following contribution to the action:
\begin{equation}
  \label{eq:87}
  S_{sb} = \lambda \sum_{\sigma} \sigma \sum_{\mu} \int_{\omega} \left(
    a_{\sigma}^{*}(\omega) b_{\mu, \sigma}(\omega) + a_{\sigma}(\omega)
    b_{\mu, \sigma}^{*}(\omega) \right). 
\end{equation}
To check whether the transformation~\eqref{eq:18} is a symmetry of the action
even in the presence of the driving term in Eq.~\eqref{eq:117}, it is most
convenient to rewrite the transformation in frequency space. Moreover, for
future reference, we give the form of the transformation including a chemical
potential:
\begin{equation}
  \label{eq:116}
  \begin{split}
    \mathcal{T}_{\beta, \mu} a_{\sigma}(\omega) & = e^{- \sigma \beta \left( \omega -
      \mu\right)/2} a_{\sigma}^{*}(\omega), \\ \mathcal{T}_{\beta, \mu}
  a_{\sigma}^{*}(\omega) & = e^{\sigma \beta \left( \omega - \mu \right)/2}
  a_{\sigma}(\omega).
  \end{split}
\end{equation}
It is straightforward to check that both $S_b$ and $S_{sb}$ are invariant under
this transformation with $\mu = 0$~\cite{Sieberer2015}; the same holds true for
the contribution $S_0$ to the action of the system. However, the driving
part~\eqref{eq:117} becomes after the transformation
\begin{equation}
  \label{eq:118}
  S_l = \Omega \sum_{\sigma} \sigma \left( a_{\sigma}(\omega_l) e^{\sigma \beta
      \omega_l/2} + a_{\sigma}^{*}(\omega_l) e^{- \sigma \beta \omega_l/2} \right),
\end{equation}
where the appearance of the exponentials shows that the symmetry is
violated. One might wonder whether it is possible to restore the symmetry in a
rotating frame in which the explicit time dependence of the action is
eliminated, i.e., by introducing new variables $\tilde{a}_{\sigma}(t)$ and
$\tilde{b}_{\sigma}(t)$ rotating at the frequency of the driving field,
$\tilde{a}_{\sigma}(t) = a_{\sigma}(t) e^{i \omega_l t}$ and analogously for
$\tilde{b}_{\sigma}(t)$. In terms of these variables, the driving part of the
action becomes
\begin{equation}
  \label{eq:119}
  \begin{split}
    S_l & = \Omega \sum_{\sigma} \sigma \int dt \left( \tilde{a}_{\sigma}(t) +
      \tilde{a}_{\sigma}^{*}(t) \right) \\ & = \Omega \sum_{\sigma} \sigma
    \left( \tilde{a}_{\sigma}(0) + \tilde{a}_{\sigma}^{*}(0) \right),
  \end{split}
\end{equation}
i.e., the drive couples to the zero-frequency component of
$\tilde{a}_{\sigma}(\omega)$. Clearly, applying again the same transformation
Eq.~\eqref{eq:116} with $\mu = 0$ to the new variables, the driving term, as
well as $S_0$ in Eq.~\eqref{eq:5} and the system-bath coupling $S_{sb}$ in
Eq.~\eqref{eq:87} are invariant. However, in the action for the
bath~\eqref{eq:9}, as a consequence of the transformation to the rotating frame
the distribution function acquires an effective chemical potential and becomes
$\coth(\beta (\omega - \omega_l)/2)$. Hence, to leave this part of the action
invariant, the transformation Eq.~\eqref{eq:116} has to be applied with
$\mu = \omega_l$, and again the full action is not invariant under the
equilibrium transformation. No frame exists in which the reference to the driving scale $\omega_l$ were eliminated.

While this simple example demonstrates explicitly that generically external
driving takes a system out of thermal equilibrium, and how this is manifest in
the symmetry properties of the Keldysh action, it is interesting to note that
there are surprising exceptions to this rule, emerging as limiting cases. One of them has been identified in
Ref.~\cite{Hafezi2014a}. Along the lines of this reference, we discard the
driving term $S_l$ and instead consider a parametric system-bath coupling of the
form\footnote{In realistic systems the system-bath coupling usually also
  contains terms of the form $a_{\sigma}(t) b_{\sigma}(t)$ that are neglected in
  the rotating-wave approximation~\cite{Hafezi2014a}.}
\begin{equation}
  \label{eq:121}
  \begin{split}
    S_{sb} & = \lambda \sum_{\sigma} \sigma \sum_{\mu} \int dt \left(
      a_{\sigma}^{*}(t) b_{\mu, \sigma}(t) e^{-i \omega_p t} + a_{\sigma}(t)
      b_{\mu, \sigma}^{*}(t) e^{i \omega_p t} \right) \\ & = \lambda \sum_{\mu}
    \sum_{\sigma} \sigma \int_{\omega} \left( a_{\sigma}^{*}(\omega + \omega_p)
      b_{\mu, \sigma}(\omega) + a_{\sigma}(\omega + \omega_p) b_{\mu,
        \sigma}^{*}(\omega) \right),
  \end{split}
\end{equation}
in the limit $\lambda \to 0$. Then, the full action $S = S_0 + S_{sb} + S_b$,
where $S_0$ and $S_b$ are as above (Eqs.~\eqref{eq:5} and~\eqref{eq:9}), is
invariant if the system fields are transformed with
$\mathcal{T}_{\beta, \mu = \omega_p}$ and the bath oscillators with
$\mathcal{T}_{\beta, \mu = 0}$. In other words, the parametric coupling acts to
thermalize the system at the temperature of the bath while at the same time
shifting the chemical potential by $\omega_p$ with respect to the bath chemical
potential. Concomitantly, the FDR for the system variables takes the form of
Eq.~\eqref{eq:20} with $\omega \to \omega - \omega_p$, while for the bath it is
Eq.~\eqref{eq:20} without modification, and only FDRs for cross-correlations
between system and bath observables would reveal that there is no true
equilibrium in the sense of \emph{a single global temperature and chemical
  potential.} Such cross-correlations, however, are suppressed in the limit
$\lambda \to 0$. In this limit,  both subsystems decouple, and each of them exhibits thermal behavior.

\subsubsection{Semiclassical limit of the thermal symmetry}
\label{sec:semicl-limit-therm}

For many applications, as discussed in Sec.~\ref{sec:semicl-limit-keldysh}, the
Keldysh action in the semiclassical limit~\eqref{eq:140} is
appropriate. Correspondingly we should consider the semiclassical limit of the
transformation~\eqref{eq:18}. In the limit of high temperatures
$\beta = 1/T \to 0$, we can perform an expansion of the transformed fields, with
their arguments shifted by $i \sigma \beta/2$, in terms of derivatives. To leading
order,\footnote{Note that a contribution $\sim \partial_t \Phi_q$ in the first
  line is suppressed according to canonical power counting,
  cf.\ Sec.~\ref{sec:semicl-limit-keldysh}, Eq.~\eqref{eq:powerc}.} this yields
\begin{equation}
  \label{eq:19}
  \begin{split}
    \mathcal{T}_{\beta} \Phi_c(t,\mathbf{x}) & = \sigma_x \Phi_c(-t,\mathbf{x}), \\
    \mathcal{T}_{\beta} \Phi_q(t,\mathbf{x}) & = \sigma_x \left(
      \Phi_q(-t,\mathbf{x}) + \frac{i}{2 T} \partial_t \Phi_c(-t,\mathbf{x})
    \right),
  \end{split}
\end{equation}
where $\sigma_x$ is the usual Pauli matrix. ``High temperatures'' thus means
that the typical frequency scale of the field is much smaller than temperature;
indeed this recovers the intuition of a semiclassical limit. If we replace the
quantum field $\Phi_q$ by the response field
$\tilde{\Phi} = - i \sigma_z \Phi_q$, Eq.~\eqref{eq:19} takes the form of the
classical symmetry introduced in Ref.~\cite{Aron2010}. The FDR in the
semiclassical limit, which may be derived as a Ward-Takahashi of the symmetry
Eq.~\eqref{eq:19}, simplifies to the Raleigh-Jeans form
\begin{equation}
  \label{eq:20class}  
  G^K(\omega, \mathbf{q}) =  \frac{2T}{\omega}
  \left( G^R(\omega, \mathbf{q}) - G^A(\omega, \mathbf{q}) \right).
\end{equation}

\begin{figure}[htpb]
  \centering
  \includegraphics[width=.9\linewidth]{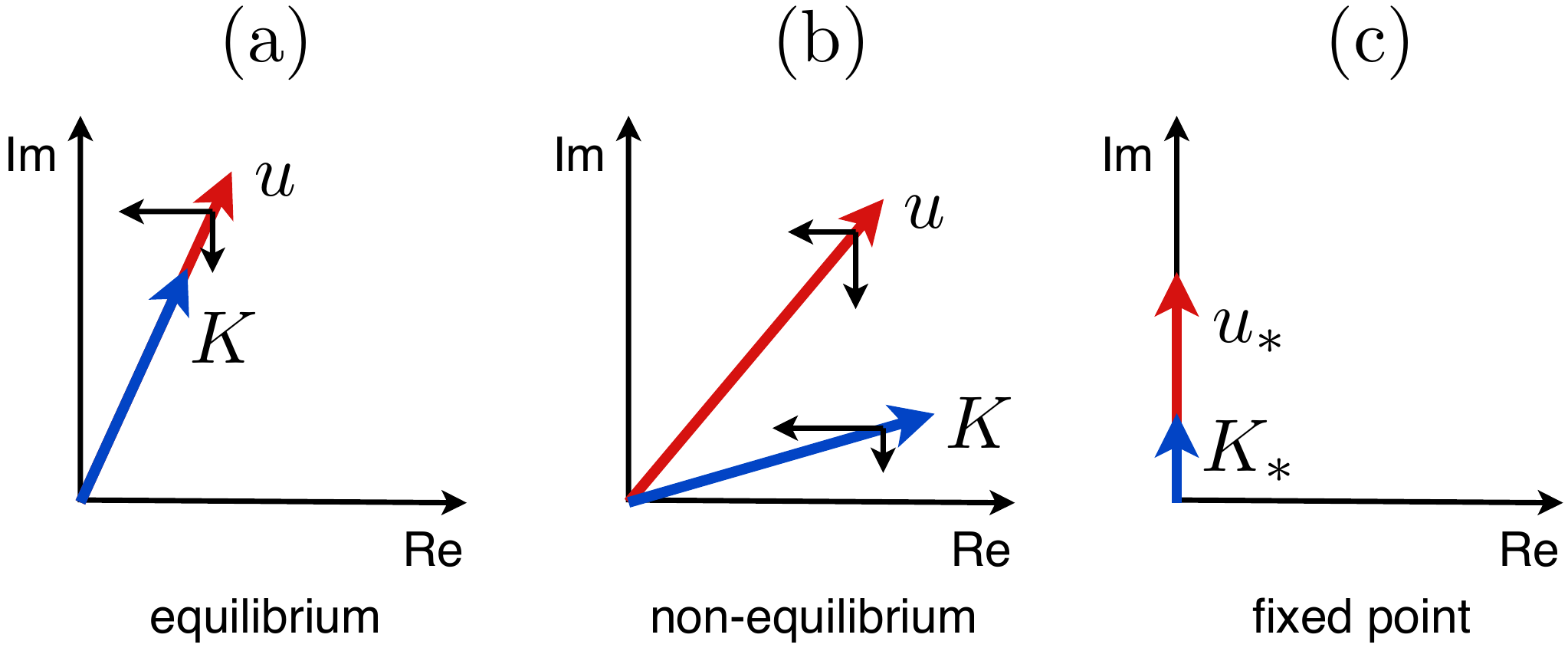}
  \caption{Location of the couplings in the Langevin equation~\eqref{eq:12} in
    the complex plane. Real (imaginary) parts describe reversible (irreversible)
    contributions to the dynamics. (a) An equilibrium system is characterized by
    the location of couplings on a single ray, reflecting detailed balance. The
    irreversible dynamics is not independent of the underlying reversible
    Hamiltonian dynamics, but rather generated by it. (b) In contrast, in a
    driven non-equilibrium system, generically there is a spread in the location
    of couplings, because the reversible and dissipative dynamics have different
    physical origins. (c) Near a critical point in three dimensions, the
    couplings flow strongly with scale and approach the imaginary axis
    (decoherence). The RG fixed point is purely dissipative. (Figure adapted
    from~\cite{Sieberer2013}.)}
  \label{fig:noneq}
\end{figure}

The thermal FDR is just one consequence of the presence of the symmetry
Eq.~\eqref{eq:19}. A second one concerns the possible values of the coupling
constants defining the action in the semiclassical limit. This allows us to
state precisely in which sense the driven-dissipative systems represent a
genuine non-equilibrium situation. To this end, it is most convenient to discuss
the equivalent Langevin equation Eq.~\eqref{eq:17}. We rewrite it by splitting
the deterministic parts on the RHS into reversible (coherent) and irreversible
(dissipative) contributions according to (here we replace $\phi_c = \psi$)
\begin{equation}
  \label{eq:12}
  i  \partial_t \psi = \frac{\delta H_c}{\delta \psi^{*}} - i  \frac{\delta
    H_d}{\delta \psi^{*}}  + \xi
\end{equation}
with effective coherent and dissipative Hamiltonians ($\alpha = c,d$)
\begin{equation}
  \label{eq:11}
  H_\alpha = \int_{t,\mathbf{x}} \left( K_\alpha \abs{\nabla \psi}^2 + r_\alpha
    \abs{\psi}^2 + \frac{u_\alpha}{2} \abs{\psi}^4 \right).
\end{equation}
It can be shown~\cite{Graham1990,Sieberer2014} that the presence of the
symmetry~\eqref{eq:19}, or, in physical terms, relaxation of a system to
thermodynamic equilibrium (a state with global detailed balance, where arbitrary
subparts are in equilibrium with each other) requires the condition
\begin{equation}
  \label{eq:168067}
  H_c = r H_d \quad \Leftrightarrow \quad  r = \frac{K_c}{K_d} = \frac{u_c}{u_d}.
\end{equation}
(Note that there is no condition on the ratio $r_c/r_d$ since the effective
chemical potential $r_c$ can always be adjusted by a gauge transformation
$\psi \mapsto \psi e^{-i \omega t}$ such that $r_c \mapsto r_c - \omega$ without
changing the physics.) In equilibrium dynamics, the ratio of real (reversible)
and imaginary (dissipative) parts is thus locked to one common value for all
couplings. This is illustrated in Fig.~\ref{fig:noneq} (a). In the complex plane
spanned by real and imaginary parts of the couplings $K = K_c + i K_d$ and
$u = u_c + i u_d$, they lie on one single ray. The intuition behind this seeming
fine-tuning is the following: a microscopically reversible dynamics starts from
a Hamiltonian functional $H_c$ alone, i.e., all couplings are located on the
real axis. Coarse graining the system from the microscopic to the macroscopic
scales introduces irreversible dynamics in the form of finite imaginary parts,
however preserving their location on a single ray: the ray just rotates under
coarse graining, but does not spread out. The geometric constraint is thus not
due to fine tuning, but results from the microscopic ``initial condition'' for
the RG flow, in combination with the presence of a symmetry. In stark contrast,
in a driven non-equilibrium system, the microscopic origins of reversible and
irreversible dynamics are independent, as illustrated in Fig.~\ref{fig:noneq}
(b). For example, in the microscopic description of Eq.~\eqref{eq:112} the rates
can be tuned fully independently from the Hamiltonian parameters --- they have
completely different physical origins.

Finally, we note that, while an explicit violation of the symmetry is present on
the microscopic scales on which the quantum master equation~\eqref{eq:109} or
the corresponding Keldysh action~\eqref{eq:21} provide a suitable description of
the system, in several cases it has been found that driven-dissipative systems
appear as approximately thermal at low
frequencies~\cite{Mitra2006,Wouters2006,Diehl2008,Diehl2010a,DallaTorre2010,Mitra2011,Mitra2012,Oztop2012,DallaTorre2013}. In
driven-dissipative condensates in three spatial dimensions, this thermalization
at low frequencies is particularly sharply reflected via the \emph{emergence} of
the thermal symmetry in the RG flow in this
regime~\cite{Sieberer2013,Sieberer2014}, cf.\ Fig.~\ref{fig:noneq} (c); this is
discussed in more detail in Sec.~\ref{sec:bosons}. On the other hand, there are
also cases where the opposite behavior occurs, and the non-equilibrium character
becomes more pronounced as one coarse grains to the macroscale. This occurs in
driven two dimensional systems, as explained in Sec.~\ref{sec:bosons}.

\subsubsection{The Noether theorem in the Keldysh formalism}
\label{sec:noether-theorem-keldysh}

A global continuous symmetry of the Keldysh action is a  transformation
$T_{\alpha}$ of the fields
$\Psi = \left( \psi_{+}, \psi_{+}^{*}, \psi_{-}, \psi_{-}^{*} \right)^T$ that
leaves the value of the action invariant, i.e.,
\begin{equation}
  \label{eq:69}
  S[T_{\alpha} \Psi] = S[\Psi],
\end{equation}
where $\alpha$ is a real time- and space independent parameter, and for $\alpha = 0$ the
transformation is the identity, $T_0 = \id$. The Noether theorem states that any
such global continuous symmetry entails the existence of a current $j$ with
components $j^{\mu},$ which obeys a continuity equation on average, i.e.,
$\langle \partial_{\mu} j^{\mu} \rangle = 0$ (here and in the following
summation over repeated indicies is implied), where $\partial_0 = \partial_t$,
and $\partial_{1,2, \dotsc d}$ are derivatives with respect to spatial
coordinates. Then, the integral over space $\mathscr{Q} = \int_{\mathbf{x}} j^0$
--- the Noether charge --- is an integral of motion, and we have
$\langle d \mathscr{Q}/dt \rangle = 0$. In the following, we prove this
relation, which states that $\mathscr{Q}$ is conserved \emph{on average}, in the
framework of the Keldysh formalism. We note however, that a global continuous
symmetry implies the even stronger statement $d \mathscr{Q}/d t = 0$ of
conservation of $\mathscr{Q}$ on the \emph{operator
  level}~\cite{Zinn-Justin2002}.

In order to prove the Noether theorem, it is sufficient to consider infinitesimal
transformations. Then, for $\alpha \ll 1$ we expand the transformation as
\begin{equation}
  \label{eq:687321}
  T_{\alpha} \Psi = \Psi + \alpha \mathfrak{G} \Psi + O(\alpha^2),
\end{equation}
where $\mathfrak{G}$ is called the generator of the transformation. In general,
$\mathfrak{G}$ is a $4 \times 4$ matrix with derivative operators as entries. We
consider specific examples below in
Sec.~\ref{sec:energy-momentum-particle-numb-cons}. In the following we assume
that as in Eq.~\eqref{eq:111} the action $S$ can be written in terms of a
Lagrangian density $\mathscr{L}$ as $S = \int_{t, \mathbf{x}} \mathscr{L}$, and that the
Lagrangian density is a local function of the fields and their first derivatives
with respect to time and space. This assumption is appropriate for most
practical purposes.

In an expansion of the LHS of Eq.~\eqref{eq:69} in powers of $\alpha$, each
coefficient has to vanish individually. The first-order contribution yields the
relation
\begin{equation}
  \label{eq:70}
  \int_{t, \mathbf{x}} \left( \frac{\partial \mathscr{L}}{\partial \Psi^T} \mathfrak{G} \Psi +
  \frac{\partial \mathscr{L}}{\partial \partial_{\mu} \Psi^T} \partial_{\mu}
  \mathfrak{G} \Psi \right) = 0,
\end{equation}
where
$\partial \mathscr{L}/\partial \Psi \equiv \left( \partial \mathscr{L}/\partial
  \psi_{+}, \partial \mathscr{L}/\partial \psi_{+}^{*}, \partial
  \mathscr{L}/\partial \psi_{-}, \partial \mathscr{L}/\partial \psi_{-}^{*}
\right)^T.$
In the cases we consider, Eq.~\eqref{eq:70} holds true because the integrand can
be written as the divergence of a vector field $f^{\mu}$, i.e., we have
\begin{equation}
  \label{eq:71}
  \partial_{\mu} f^{\mu} = \frac{\partial \mathscr{L}}{\partial \Psi^T} \mathfrak{G} \Psi +
  \frac{\partial \mathscr{L}}{\partial \partial_{\mu} \Psi^T} \partial_{\mu}
  \mathfrak{G} \Psi.
\end{equation}

To proceed we consider \emph{local} transformations, i.e., we consider
$T_{\alpha}$ with $\alpha = \alpha(t, \mathbf{x})$. We perform a change of
integration variables $\Psi \to T_{\alpha} \Psi$ in the partition
function. Then, assuming that the functional measure is invariant with respect
to the local transformation, we have
% (this is not the case in presence of what is known as anomalies)
% where this is not valid and which we leave aside in the present discussion.
\begin{equation}
  \label{eq:76}  
  Z = \int \mathscr{D}[\Psi] e^{i S[\Psi]} = \int \mathscr{D}[\Psi] e^{i S[T_{\alpha} \Psi]}.
\end{equation}
As before, we expand the RHS of this equality in a power series in
$\alpha$. Since by assumption the latter is a function of
$\left( t, \mathbf{x} \right)$, Eq.~\eqref{eq:69} does not hold true any
more. Instead, we find
\begin{equation}
  \label{eq:72}
  S[T_{\alpha} \Psi] = S[\Psi] + \int_{t, \mathbf{x}} \alpha \partial_{\mu} \left(
    f^{\mu} - \frac{\partial \mathscr{L}}{\partial \partial_{\mu} \Psi^T}
    \mathfrak{G} \Psi \right) + O(\alpha^2),
\end{equation}
where we used Eq.~\eqref{eq:71} and integration by parts to write the RHS in a
form that does not contain derivatives of $\alpha$ explicitly. Inserting
Eq.~\eqref{eq:72} in the exponential on the RHS of Eq.~\eqref{eq:76}, and
expanding the latter to first order in $\alpha$, we obtain the condition
\begin{equation}
  \label{eq:77}
  \int_{t, \mathbf{x}} \alpha \partial_{\mu} \left\langle f^{\mu} - \frac{\partial \mathscr{L}}{\partial \partial_{\mu}
      \Psi^T} \mathfrak{G} \Psi \right\rangle = 0.
\end{equation}
Since there are no restrictions on the choice of $\alpha$, this equation implies
that the divergence of the expectation value vanishes. The latter is just the
quantum Noether current,
\begin{equation}
  \label{eq:78}
  j^{\mu} = \frac{\partial \mathscr{L}}{\partial \partial_{\mu}
    \Psi^T} \mathfrak{G} \Psi - f^{\mu}.
\end{equation}
Note that $j^{\mu}$ is not unique: for constant $a, b^{\mu}$, the combination
$a j^{\mu} + b^{\mu}$ is also a conserved current. The associated Noether charge
$\mathscr{Q}$ corresponds to the integral over space of the zeroth component of the
Noether current,
\begin{equation}
  \label{eq:75}
  \mathscr{Q} = \int_{\mathbf{x}} j^0 = \int_{\mathbf{x}} \left( \frac{\partial \mathscr{L}}{\partial \partial_t
    \Psi^T} \mathfrak{G} \Psi - f^0 \right).
\end{equation}
While this derivation is quite general, a more concrete expression for the
Noether current can be obtained by restricting the form of the generator
$\mathfrak{G}$ in Eq.~\eqref{eq:687321}. In particular, in the next section we
consider the specific cases of classical and quantum transformations.

\subsubsection{Closed systems: energy, momentum and particle number
  conservation}
\label{sec:energy-momentum-particle-numb-cons}

\emph{Energy and momentum conservation} -- This property is expected for systems,
whose classical action is determined by a translationally invariant Hamiltonian
alone. Indeed, the construction of the Keldysh action for a Markovian master
equation in Sec.~\ref{sec:deriv-keldysh-acti} shows that terms which couple the
forward and backward branches are only contained in the dissipative parts of the
action~\eqref{eq:111}. In other words, they arise upon coupling the system to a
bath and integrating out the latter. On the other hand, the Keldysh action for a
closed system with unitary dynamics does not contain such terms and can be
written as
\begin{equation}
 \label{eq:80}
 S = \sum_{\sigma} \sigma \int_{t, \mathbf{x}} \mathscr{L}_{\sigma},
\end{equation}
where $\mathscr{L}_{+}$ and $\mathscr{L}_{-}$ are the Lagrangian densities
evaluated with fields on the forward and backward branches, respectively. % (In
% this case infinitesimal dissipative regularization terms have to be
% included~\cite{Kamenev2011,Altland2010}. These terms, however, do not affect
% the present argument: They should be interpreted as formal bookkeeping devices
% which indicate how the quadratic part of the action has to be inverted to
% obtain bare Green's functions corresponding to an equilibrium state, and not
% as proper contributions to the action.)
Given this structure of the action, let us consider a transformation
$T_{\alpha}$ which does not mix fields on the forward and backward
branches. Then, the generator $\mathfrak{G}$ has the block-diagonal structure
\begin{equation}
  \label{eq:2}
  \mathfrak{G} =
  \begin{pmatrix}
    \mathfrak{g}_{+} & 0 \\ 0 & \mathfrak{g}_{-}
  \end{pmatrix}.
\end{equation}
The cases of physical interest are the classical and quantum transformations
mentioned above, corresponding to the choices
$\mathfrak{g}_{+} = \mathfrak{g}_{-}$ and
$\mathfrak{g}_{+} = - \mathfrak{g}_{-}$, respectively. In both cases, the
Noether current can be represented as superposition of currents on the forward
and backward branches, which we define in terms of
$\mathfrak{g} \equiv \mathfrak{g}_{+}$ as
\begin{equation}
  \label{eq:1}
  j_{\sigma}^{\mu} = \frac{\partial \mathscr{L}_{\sigma}}{\partial \partial_{\mu}
    \Psi_{\sigma}^T} \mathfrak{g} \Psi_{\sigma} - f_{\sigma}^{\mu},
\end{equation}
where
\begin{equation}
  \label{eq:81}
  \partial_{\mu} f_{\sigma}^{\mu} = \frac{\partial \mathscr{L}_{\sigma}}{\partial
    \Psi_{\sigma}^T} \mathfrak{g} \Psi_{\sigma} +
  \frac{\partial \mathscr{L}_{\sigma}}{\partial \partial_{\mu} \Psi_{\sigma}^T} \partial_{\mu}
  \mathfrak{g} \Psi_{\sigma}.
\end{equation}
With this definition, $j_{-}$ can be obtained from $j_{+}$ simply by replacing
all instances of $\Psi_{+}$ appearing in $j_{+}$ by $\Psi_{-}$.  Then, the
symmetry of the action under a \emph{quantum} transformation yields a
\emph{classical} Noether current and \textit{vice versa}. These currents are
given by (as noted above we are free to choose a convenient multiplicative
normalization of the currents)
\begin{equation}
  \label{eq:3}
  j_c = \frac{1}{2} \left( j_{+} + j_{-} \right), \quad j_q = j_{+} - j_{-}.
\end{equation}
As pointed out in Sec.~\ref{sec:driv-diss-cond}, due to causality only the
classical component of the field can acquire a finite expectation value. The
same is true for the currents defined in Eq.~\eqref{eq:3}: in the presence of an
external field that induces a current, we have
$\langle j_{+} \rangle = \langle j_{-} \rangle = \langle j_c \rangle \neq 0,$
whereas $\langle j_q \rangle = 0.$ Hence, as anticipated above, the continuity
equation $\langle \partial_{\mu} j^{\mu}_q \rangle = 0$, which follows from the
symmetry under a classical transformation, does not entail a non-trivial Noether
charge. In the next section, we show that such a symmetry nevertheless does have
physical consequences, as it can be broken spontaneously in a condensation
transition.

Before, let us derive the Noether charge associated with the symmetry of a
Hamiltonian Keldysh action under contour-dependent --- i.e., quantum --
translations in space-time. To be specific, we consider the transformation
$\Psi_{\sigma}(X) \mapsto \Psi_{\sigma}(X + \sigma \alpha e_{\nu})$, where
$X = \left( t, \mathbf{x} \right)$, and $e_{\nu}$ is the basis vector in
direction of the $\nu$th coordinate of $d + 1$-dimensional space-time. This is a
symmetry of the action~\eqref{eq:80}: the shift of the coordinates can be
absorbed by performing a change of integration variables
$X \to X - \sigma \alpha e_{\nu}$ in the integral over the Lagrangian density
$\mathscr{L}_{\sigma}$. Clearly, the symmetry is violated in the presence of
dissipative terms that couple the forward and backward branches.

For space-time translations, the generators $\mathfrak{g}_{\pm}$ in
Eq.~\eqref{eq:2} acquire an additional index $\nu$ and are given by
$\mathfrak{g}_{\pm, \nu} = \pm \partial_{\nu}$.
% quantum transformation:
% \begin{equation}
%   \label{eq:4}
%   T_{\nu, \alpha} \Psi_{\sigma}(X) = \Psi_{\sigma}(X + \sigma \alpha e_{\nu}) = \Psi_{\sigma}(X) + \sigma
%   \alpha \partial_{\nu} \Psi_{\sigma}(X) + O(\alpha^2).
% \end{equation}
It follows that the vector field $f_{\sigma, \nu}^{\mu}$ in Eq.~\eqref{eq:81}
takes the form
$f_{\sigma, \nu}^{\mu} = \delta^{\mu}_{\nu} \mathscr{L}_{\sigma}$, and the
classical Noether current, which we denote as $T_{c, \nu}$, is given by
\begin{equation}
  \label{eq:62}
 T_{c, \nu}^{\mu} = \frac{1}{2} \sum_{\sigma} T^{\mu}_{\sigma, \nu}, \quad
 T^{\mu}_{\sigma, \nu} = \frac{\partial
   \mathscr{L}_{\sigma}}{\partial \partial_{\mu} \Psi_{\sigma}^T} \partial_{\nu}
 \Psi_{\sigma} - \delta^{\mu}_{\nu} \mathscr{L}_{\sigma},
\end{equation}
where $T_{\pm, \nu}^{\mu}$ are the components of the energy-momentum
tensor~\cite{Zinn-Justin2002,Peskin1995} on the forward and backward
branches. In particular, the component $T_{\pm, 0}^0$ is the energy density,
whereas $T_{\pm, i}^0$ is the density of momentum in the spatial direction
$i = 1, 2, \dotsc, d$. The spatial integrals over these densities yield the
associated Noether charges, e.g., if we express the Lagrangian density
$\mathscr{L}_{\sigma}$ in terms of a Hamiltonian density (as in
Eq.~\eqref{eq:111}, where $\mathcal{L}$ is given by Eq.~\eqref{eq:63} with
$\gamma_{\alpha} = 0$) as
\begin{equation}
  \label{eq:13}
  \mathscr{L}_{\sigma} = \frac{1}{2} \Psi_{\sigma}^{\dagger} i \sigma_z \partial_t
  \Psi_{\sigma} - H_{\sigma},
\end{equation}
we obtain the conserved energy density $\mathscr{E}$, which is as expected given
by the sum of the Hamiltonian densities on the forward and backward branches,
\begin{equation}
  \label{eq:16909090}
  \mathscr{E} = \frac{1}{2} \sum_{\sigma} \int_{\mathbf{x}} T_{\sigma, 0}^0 = \frac{1}{2}
  \int_{\mathbf{x}} \left( H_{+} + H_{-} \right).
\end{equation}
Along the same lines, conservation of angular momentum follows from the quantum symmetry
of the Keldysh action with respect to rotations of the spatial coordinates.

\emph{Number conservation} -- As another example, we consider phase rotations and their relation to particle
number conservation. In this case the classical transformation reads
$\psi_{\pm} \mapsto U_c \psi_{\pm} = e^{i \alpha} \psi_{\pm}$ and the quantum
transformation is
$\psi_{\pm} \mapsto U_q \psi_{\pm} = e^{\pm i \alpha} \psi_{\pm}$. Classical
phase rotations are a symmetry of the Keldysh action, if the fields appear only
in the combinations $\psi_{\sigma}^{*} \psi_{\sigma'}$. The quantum
transformation is more restrictive: in this case only products with
$\sigma = \sigma'$ are allowed. Hence, the symmetry with regard to quantum phase
rotations implies the classical symmetry. The generators of $U_q$ are given by
$\mathfrak{g}_{\pm} = \pm i \sigma_z$.  Then, with Eq.~\eqref{eq:6} and
inserting the representation of the Lagrangian density in Eq.~\eqref{eq:13} in
the expression for the Noether current~\eqref{eq:1}, we find that the latter can
be written as
$j_{\sigma} = \left( \rho_{\sigma}, \mathbf{j}_{\sigma} \right)^T,$ where
$\rho_{\sigma} = \abs{\psi_{\sigma}}^2$ is the density on the contour with index
$\sigma$ and the spatial components of the current are given by
\begin{equation}
  \label{eq:6345}
  \mathbf{j}_{\sigma} = \frac{1}{i 2 m} \left( \psi_{\sigma}^{*} \nabla \psi_{\sigma} -
    \psi_{\sigma} \nabla \psi_{\sigma}^{*} \right),
\end{equation}
which is just the ordinary quantum mechanical current for particles with
quadratic dispersion relation, evaluated on the closed time path. What is the
physical meaning of the classical and quantum components, which are defined in
Eq.~\eqref{eq:3}, of the current given in Eq.~\eqref{eq:6345}? If we introduce
in the Hamiltonian an externally imposed gauge field, in the Keldysh action it
would couple to the \emph{quantum} current, in analogy to the source terms in
the generating functional Eq.~\eqref{eq:14}. However, the observable effect of
such a gauge field is that it can induce a \emph{classical} current
$\langle j_c \rangle \neq 0$.

We close this section with a comment on the status of the $U_q(1)$ symmetry. It
will be present on the microscopic level for an action corresponding to a
Hamiltonian which commutes with the number operator. Since the symmetry
considerations are properly applied to the \emph{microscopic} action (and the
functional measure), the conservation law ensues. However, this does not imply
that the effective action must manifestly preserve a $U_q(1)$ symmetry. In fact,
classical models of number conserving dynamics~\cite{Hohenberg1977} do not
exhibit this symmetry. This leads to the picture that this symmetry is broken
spontaneously under RG. The precise workings of such a mechanism are, to the
best of our knowledge, not settled so far.

\subsubsection{Extended continuity equation in open systems}
\label{sec:extend-cont-equat-open-sys}

The conservation of particle number in a closed system follows from the
continuity equation for the classical Noether current that is associated with
the symmetry under quantum phase rotations. In an open system, this symmetry is
absent, and the continuity equation has to be extended to account for the
exchange of particles with the environment, as we discuss in the following.

To be specific, we consider the Keldysh action~\eqref{eq:112} for a system with
single-particle pump as well as single-particle and two-body loss. In order to
derive the extended continuity equation, as in Eq.~\eqref{eq:76} we perform a
change of integration variables $\Psi \to U_q \Psi$, where $U_q$ is a
\emph{local} quantum phase rotation, in the Keldysh partition function. Since
the partition function is invariant under this transformation, the coefficients
in an expansion of the partition function in powers of the phase shift must
vanish. In linear order, we find the condition
% \begin{multline}
%   \label{eq:7}
%   S[U_q \Psi] = S[\Psi] + \int_{t, \mathbf{x}} \alpha \sum_{\sigma = \pm} \left[ \partial_t
%     \rho_{\sigma} + \nabla \cdot \mathbf{j}_{\sigma} \vphantom{\left(
%         \psi_{-}^{*} \psi_{+} \right)^2} \right. \\ \left. -
%     \gamma_p \psi_{+}^{*} \psi_{-} + \gamma_l \psi_{-}^{*} \psi_{+} + 2 u_d
%     \left( \psi_{-}^{*} \psi_{+} \right)^2 \right] + O(\alpha^2).
% \end{multline}
\begin{equation}
  \label{eq:120}
  \langle \partial_{\mu} j_c^{\mu} \rangle -
  \gamma_p \langle \psi_{+}^{*} \psi_{-} \rangle + \gamma_l \langle \psi_{-}^{*}
  \psi_{+} \rangle + 4 u_d
  \langle \left( \psi_{-}^{*} \psi_{+} \right)^2 \rangle = 0.
\end{equation}
This should be compared to Eq.~\eqref{eq:77}: if $U_q$ were a symmetry of the
action, we would have found an ordinary continuity equation. This would have
been the case in the absence of the pumping and loss terms proportional to
$\gamma_p, \gamma_l,$ and $u_d$. The interpretation of these terms is as
follows: in a system that is perfectly isolated from its environment, the
temporal change of the density at a given point is due to the motion of
particles towards to or away from this point, i.e., the rate of change of
density is a source or sink for the mass flow as measured by the divergence of
the current $\langle \nabla \cdot \mathbf{j}_c \rangle$. In an open system, on
the other hand, there is in addition a \emph{dissipative
  current}~\cite{Gebauer2004,Avron2012} due to the exchange of particles between
the system and its environment. In particular, in Eq.~\eqref{eq:120} this
dissipative current has contributions from the pumping and loss terms.

It is interesting to note that Eq.~\eqref{eq:120} can also be obtained from the
equation of motion of the local density
$n(\mathbf{x}) = \psi^{\dagger}(\mathbf{x}) \psi(\mathbf{x})$ in the operator
formalism. From the master equation $\partial_t \rho = \mathcal{L} \rho$ with
Liouvillian $\mathcal{L}$ in Lindblad form given in Eq.~\eqref{eq:104}, it
follows that the time evolution of $n(t, \mathbf{x})$ in the Heisenberg
representation is given by
$\partial_t n(t, \mathbf{x}) = \mathcal{L}^{*} n(t, \mathbf{x})$ with the
adjoint Liouvillian defined by
$\tr( A \mathcal{L} B) = \tr(B \mathcal{L}^{*} A)$. We find
\begin{multline}
  \label{eq:105}
  \partial_t n(t, \mathbf{x}) = - \nabla \cdot \mathbf{j}(t,\mathbf{x}) +
  \gamma_p \psi(t,\mathbf{x}) \psi^{\dagger}(t, \mathbf{x}) \\ - \gamma_l n(t,
  \mathbf{x}) - 4 u_d \psi^{\dagger}(t, \mathbf{x})^2 \psi(t, \mathbf{x})^2,
\end{multline}
where first term encodes coherent dynamics and corresponds to the Heisenberg
commutator $i [H_{\mathrm{LP}}, n(t, \mathbf{x})]$, whereas the remaining
contributions incorporate the dissipative parts. Taking the average of this
relation and expressing the expectation values as Keldysh functional integrals
yields Eq.~\eqref{eq:120}.

Finally, we can obtain some intuition on the dissipative correction to the
current expectation value in mean field theory, where the correlators in
Eq.~\eqref{eq:120} factorize into products of single field expectation values
$\left\langle \psi_+ \right\rangle = \left\langle \psi_- \right\rangle =
\psi_0$.
We then recognize in the non-derivative terms ($\psi_0^*$ times) the LHS of
Eq.~\eqref{eq:60} (note that due to the factor of $1/\sqrt{2}$ in the Keldysh
rotation~\eqref{eq:41} the expectation values are related as
$\left\langle \phi_c \right\rangle = \phi_0 = \sqrt{2} \left\langle \psi_{\pm}
\right\rangle$,
and that $r_d = (\gamma_l - \gamma_p)/2$), which equals zero in a homogeneous
situation within mean field theory. In this way, we see that there is no
particle number current on average in a homogeneous driven-dissipative system in
stationary state, as expected.

\subsubsection{Spontaneous symmetry breaking and the Goldstone theorem}
\label{sec:goldstone}

Even though the pumping and loss terms in the Keldysh action in
Eq.~\eqref{eq:112} break the symmetry with respect to quantum phase rotations
$U_q$, the action is still symmetric under $U_c$ transformations. As a result,
even in the absence of particle number conservation there is a possibility of
spontaneous symmetry breaking. In particular, a finite average value
$\langle \phi_c \rangle \neq 0$ breaks the classical phase rotation symmetry in
a non-equilibrium condensation transition. Here we show that this spontaneous
symmetry breaking is accompanied by the appearance of a massless mode, i.e., a
mode with vanishing frequency and decay rate for zero momentum, which is known
as the Goldstone boson~\cite{Goldstone2008,Goldstone1962,Nambu1961}. We obtain
this result by carrying over the corresponding derivation from the equilibrium
formalism (see, e.g., Ref.~\cite{Amit/Martin-Mayor}) to the Keldysh framework.

The single-particle excitation spectrum is encoded in the poles of retarded and
advanced Green's functions, and such a pole at $\omega = q = 0$ is dubbed a
massless mode. Poles of the Green's functions correspond to roots of the
determinant of inverse propagator (see
Sec.~\ref{sec:deriv-keldysh-acti}). Hence, the presence of a massless mode can
be detected by checking whether the determinant of the mass matrix $M$
vanishes. The latter is determined by the inverse propagator
\begin{equation}
  \label{Propa1}
  P(\omega, \mathbf{q}) =
  \begin{pmatrix}
    0 & P^A(\omega, \mathbf{q}) \\
    P^R(\omega, \mathbf{q}) & P^K(\omega, \mathbf{q})
  \end{pmatrix}
  =G^{-1}(\omega, \mathbf{q}), 
\end{equation}
given exemplarily in Eq.~\eqref{eq:40456}, at $\omega = \mathbf{q} = 0$, i.e.,
$M = - P(0,0)$. Therefore, it is sufficient to consider frequency- and
momentum-independent field configurations or, in other words, homogeneous field
configurations, so that instead of the full effective action Eq.~\eqref{eq:1700}
we only have to deal with the effective potential
$U = - \left. \Gamma \right\rvert_{\mathrm{hom.}}/\Omega$, where
$\Omega$ is the quantization volume in space-time. Since the effective potential
$U$ inherits the symmetries of the effective action, it is a function of
precisely these combinations of fields which are invariant under classical or
quantum phase rotations. In the basis of classical and quantum fields, these
$U_c$-symmetric combinations are $\rho_{\nu \nu'} = \phi_{\nu}^{*} \phi_{\nu'}$,
where the indices $\nu$ and $\nu'$ can take the values $c$ and $q$. In the
following, we show that $U_c$ invariance is sufficient to guarantee the
existence of a massless mode.

For convenience we switch to a basis of real fields
$\chi = (\chi_{c,1}, \chi_{c,2}, \chi_{q,1}, \chi_{q,2})$ which correspond to
the real and imaginary parts of the complex classical and quantum fields, i.e.,
$\phi_{\nu} = \frac{1}{\sqrt{2}} \left( \chi_{\nu, 1} + i \chi_{\nu, 2} \right)$
for $\nu = c, q$. In this basis, the mass matrix reads
\begin{equation}
  \label{eq:8}
  M_{ij} = \left. \frac{\partial^2 U }{\partial \chi_i \partial \chi_j }
  \right\rvert_{\mathrm{ss}} = \sum_a \left[ \frac{\partial^2
      \rho_a}{\partial\chi_i \partial \chi_j} \frac{\partial
      U}{\partial \rho_a} \right]_{\mathrm{ss}}
    + \sum_{a, b} \left[ \frac{\partial \rho_a}{\partial \chi_i} \frac{\partial
      \rho_b}{\partial \chi_i}\frac{\partial^2 U}{\partial \rho_a \partial
      \rho_b} \right]_{\mathrm{ss}},
\end{equation}
where the subscript $\mathrm{ss}$ indicates that the fields should be set to
their average values in the stationary state. The indices $i$ and $j$ label the
components of the four-vector $\chi$ defined above (i.e., $\chi_1 = \chi_{c,1}$,
$\chi_2 = \chi_{c,2}$ etc.), and $a$ and $b$ are double indices, taking the
values $cc, cq, qc, qq$. Let us consider the first term on the RHS of
Eq.~\eqref{eq:8}: in the ordered phase, the classical field has a finite
expectation value in the stationary state. Without loss of generality we assume
that this value is real. Then, the field equations
\begin{equation}
  \label{eq:1511111}
  \left. \frac{\partial U}{\partial \chi_i} \right\rvert_{\mathrm{ss}} = \sum_a \left[ \frac{\partial \rho_a}{\partial \chi_i}
    \frac{\partial U}{\partial \rho_a} \right]_{\mathrm{ss}} = 0,
\end{equation}
which actually determine the average value of the fields $\chi_i$ in stationary
state, have the solution
$\left. \chi_i \right\rvert_{\mathrm{ss}} = \sqrt{2 \rho_0} \delta_{i, 1}$.
Performing the derivatives $\partial \rho_a/\partial \chi_i$ in
Eq.~\eqref{eq:1511111} explicitly and inserting
$\left. \chi_i \right\rvert_{\mathrm{ss}}$, we obtain the following conditions:
\begin{equation}
  \label{eq:10}
  \left. \frac{\partial U}{\partial \rho_{cc}} \right\rvert_{\mathrm{ss}} =
  \left. \frac{\partial U}{\partial \rho_{cq}} \right\rvert_{\mathrm{ss}} =
  \left. \frac{\partial U}{\partial \rho_{qc}} \right\rvert_{\mathrm{ss}} = 0.
\end{equation}
Therefore, only the derivative
$u_{qq} = [ \partial U/\partial \rho_{qq} ]_{\mathrm{ss}}$ contributes to the
first term on the RHS of Eq.~\eqref{eq:8}. Denoting mixed second derivatives of
the effective potential as
$u_{ab} = u_{ba} = [ \partial^2 U/\partial \rho_a\partial \rho_b
]_{\mathrm{ss}}$, we find that the mass matrix can be written as

\begin{widetext}
  \begin{equation}
    \label{eq:112456}
    M =
    \begin{pmatrix}
      0 & 0 & 0 & 0 \\
      0 & 0 & 0 & 0 \\
      0 & 0 & u_{qq} & 0 \\
      0 & 0 & 0 & u_{qq}
    \end{pmatrix}
    + \rho_0
    \begin{pmatrix}
      0 & 0 & u_{cc,cq} + u_{cc,qc}  &  i \left( u_{cc,cq} - u_{cc,qc} \right) \\
      0 & 0 & 0  & 0 \\
      u_{cc,cq} + u_{cc,qc} & 0 & \tfrac{1}{2} \left( u_{cq,cq} + 2 u_{cq,qc} +
        u_{qc,qc} \right)  & \tfrac{i}{2} \left( u_{cq,cq} - u_{qc,qc} \right)  \\
      i \left( u_{cc,cq} - u_{cc,qc} \right) & 0 & \tfrac{i}{2} \left( u_{cq,cq}
        - u_{qc,qc} \right) & - \tfrac{1}{2} \left( u_{cq,cq} - 2 u_{cq,qc} +
        u_{qc,qc} \right)  \\
    \end{pmatrix}.
\end{equation}
\end{widetext}
\noindent An entry $u_{cc,cc}$ is not ruled out by symmetry, however, it must
vanish due to conservation of probability, cf.\
Sec.~\ref{sec:single-mode-cavity}. Crucially, the retarded and advanced sectors
feature one zero eigenvalue. This proves the existence of a massless mode as a
consequence of spontaneous symmetry breaking for a theory with $U_c$ invariance.

While this analysis guarantees the existence of a zero mode (i.e., the complex
dispersion relation has the property $\omega(q = 0)= 0$), it does not provide a
statement on the functional dependence $\omega(q)$. The latter could be inferred
along the lines of
Refs.~\cite{Minami2015,Watanabe2012,Hidaka2013,Watanabe2014,Hayata2015}. As we
found in Sec.~\ref{sec:driv-diss-cond} in Bogoliubov approximation, in an open
system without (i.e., broken) $U_q$ symmetry and spontaneously broken
$U_c$ symmetry, the leading behavior at low momenta is diffusive,
$\omega (q) \sim - i D q^2$, with a real diffusion coefficient $D$. This has to
be contrasted to closed systems, in which microscopically both $U_q$ and $U_c$
symmetry are present. There, the leading behavior in a phase with spontaneously
broken $U_c$ symmetry is that of coherent sound waves,
$\omega (\mathbf{q}) \sim c q$, with real speed of sound $c$. A phenomenological
justification of this behavior is given in Ref.~\cite{Hohenberg1977}, where this
can be understood as a consequence of a coupling to additional slow modes
relating to particle number conservation. In these phenomenological models,
$U_q$ symmetry is absent. This suggests a scenario of an additional spontaneous
breakdown of $U_q$ symmetry upon coarse graining, but this issue seems not to be
settled to date.

%%% Local Variables:
%%% mode: latex
%%% TeX-master: "dds_review"
%%% End:

\subsection{Open system functional renormalization group}
\label{sec:open-sys-FRG}

In Sec.~\ref{sec:critical-dynamics-3d}, we investigate dynamical criticality of
the Bose condensation transition in driven open systems, motivated by many-body
ensembles such as exciton-polariton
condensates~\cite{Kasprzak2006,SnokeBook,Wouters2007a,Szymanska2006,Carusotto2013}. There
again, coherent dynamics naturally competes with dissipation in the form of
incoherent particle losses and pumping. The situation parallels a laser
threshold~\cite{Graham1970,DeGiorgio1970}, however with a continuum of spatial
degrees of freedom. This ingredient, however, causes the characteristic
long-wavelength divergences of many-body problems in their symmetry broken
phase, or at a critical point. It implies that perturbation theory necessarily
breaks down, even when the interaction constants are small, due to a continuum
of modes without a gap, which form the intermediate states and are summed over
in many-body perturbation theory. This calls for the development of efficient
functional integral techniques able to cope with these problems. Our method of
choice is the functional renormalization group based on the Wetterich
equation~\cite{Wetterich1993a}, which we briefly introduce here in its Keldysh
formulation. This approach offers the particularly attractive feature of not
being restricted to the critical point.

The functional renormalization group equation (for reviews
see~\cite{Berges2002,Delamotte2012,Blaizot2006,Pawlowski2007,Boettcher2012})
constitutes an exact reformulation of the functional \emph{integral}
representation of a quantum many-body problem in terms of a functional
\emph{differential} equation. In this, it is strongly distinct from, e.g.,
perturbative field theoretical renormalization group approaches, which
concentrate exclusively on the critical surface of a given problem. Instead, it
may be viewed as an alternative and potentially more tractable tool for the
analysis of the complete many-body problem, also on length scales well below the
correlation length near criticality. Indeed, it has proven a very versatile tool
in many different physical context, ranging from quantum
dots~\cite{Jakobs2007,Jakobs2010a,Karrasch2010}, ultracold
atoms~\cite{Boettcher2012}, strongly correlated electrons~\cite{Metzner2012},
classical stochastic models~\cite{Canet2007,Mesterhazy2013}, quantum
chromodynamics~\cite{Pawlowski2007,Rosten2012}, to quantum
gravity~\cite{Reuter1998}. Here we give a brief overview of the general concept
adapted to non-equilibrium
systems~\cite{Gezzi2007,Karrasch2010,Jakobs2007,Jakobs2010a,Berges2008,Berges2009,Berges2012,Gasenzer2008,Sieberer2013,Sieberer2014,Canet2004a,Canet2005,Canet2006,Canet2004,Canet2011a,Canet2010,Canet2011,Canet2012}. It
is used for the discussion of critical behavior in driven open quantum systems
in Sec.~\ref{sec:bosons}, with applications to a broader non-equilibrium
many-body context left for future work.

The transition from the action $S$ to the effective action $\Gamma$ consists in
the inclusion of both statistical and quantum fluctuations into the latter (cf.\
Eq.~\eqref{eq:1700}). In the functional renormalization group approach based on
the Wetterich equation~\cite{Wetterich1993a}, the functional integral over
fluctuations is carried out stepwise. To this end, an infrared regulator is
introduced, which suppresses the fluctuations with momenta less than an infrared
cutoff scale $\Lambda$. This is achieved by adding to the action
in~\eqref{eq:14} a term
\begin{equation}
  \label{eq:500}
  \Delta S_{\Lambda} = \int_{t, \mathbf{x}} \left( \phi_c^{*},\phi_q^{*} \right)  
  \begin{pmatrix}
    0 & R_{\Lambda} \\
    R_{\Lambda}^{*} & 0
  \end{pmatrix}  
  \begin{pmatrix}
    \phi_c \\ \phi_q
  \end{pmatrix},
\end{equation}
with a \emph{cutoff} or \emph{regulator function} $R_{\Lambda}$. Some key
structural properties are indicated below, but apart from these properties the
choice of the cutoff is flexible and problem-specific. The resulting
cutoff-dependent generating functional and its logarithm (cf.\
Eq.~\eqref{eq:15}) are denoted by, respectively, $Z_{\Lambda}$ and
$W_{\Lambda}$. Then, the scaled-dependent effective action $\Gamma_{\Lambda}$ is
defined by modifying the Legendre transform in Eq.~\eqref{eq:16} according to
\begin{multline}
  \label{eq:91}
  \Gamma_{\Lambda}[\bar{\Phi}_c,\bar{\Phi}_q] = W_{\Lambda}[J_c,J_q] \\ -
  \int_{t, \mathbf{x}} \left( J_c^{\dagger} \bar{\Phi}_q + J_q^{\dagger}
    \bar{\Phi}_c \right) - \Delta S_{\Lambda}[\bar{\Phi}_c,\bar{\Phi}_q].
\end{multline}
The subtraction of the $\Delta S_{\Lambda}$ on the RHS guarantees that the only
difference between the functional integral representations for $\Gamma$ and
$\Gamma_{\Lambda}$ is the inclusion of the cutoff term in the latter,
\begin{multline}
  \label{eq:24}
  e^{i \Gamma_{\Lambda}[\bar{\Phi}_c,\bar{\Phi}_q]} = \int \mathscr{D}[\delta
  {\Phi}_c,\delta {\Phi}_q] \, e^{i S[\bar{\Phi}_c + \delta
    {\Phi}_c,\bar{\Phi}_q + \delta {\Phi}_q]} \\ \times e^{-i \tfrac{\delta
      \Gamma}{\delta \bar{\Phi}_c^T} \delta{\Phi}_q -i \tfrac{\delta
      \Gamma}{\delta {\Phi}_q^T} \delta {\Phi}_c + i \Delta
    S_{\Lambda}[ \delta {\Phi}_c,\delta {\Phi}_q]}.
\end{multline}
Physically, $\Gamma_{\Lambda}$ can thus be viewed as the effective action for
averages of fields over a coarse-graining volume with a size
$\sim \Lambda^{- d}$, where $d$ is the spatial dimension.

Note that we chose the form of the cutoff action $\Delta S_{\Lambda}$ such that
it modifies the inverse retarded and advanced propagators in Eq.~\eqref{eq:500}
only. This is sufficient to regularize possible infrared divergences, which
result from poles of the retarded and advanced propagators being located at the
origin of the complex frequency plane. A typical choice in practical
calculations is
\begin{equation}
  \label{eq:83}
    R_{\Lambda}(q^2) \sim \Lambda^2, \quad q/\Lambda \to 0,
\end{equation}
giving the inverse propagators a mass $\sim \Lambda^2$. In this way,
fluctuations with wavelength $\gtrsim \Lambda^{-1}$ are effectively cut
off. Therefore, for any finite $\Lambda$, the technical problem of infrared
divergences is under control.

The main usefulness of the so-modified effective action, however, lies in the
fact that it smoothly interpolates between the action $S$ for
$\Lambda \to \Lambda_0$, where $\Lambda_0$ is the ultraviolet cutoff of the
problem (e.g., the inverse lattice spacing), and the full effective action $\Gamma$ for
$\Lambda \to 0$. This is ensured by the following requirements on the
cutoff~\cite{Berges2009}:
\begin{equation}
  \label{eq:2666666}
  \begin{aligned}
    R_{\Lambda}(q^2) & \sim \Lambda_0^2, & \Lambda & \to \Lambda_0, \\
    R_{\Lambda}(q^2) & \to 0, & \Lambda & \to 0.
  \end{aligned}
\end{equation}
Under the condition that $\Lambda_0$ exceeds all energy scales in the action,
for $\Lambda \to \Lambda_0$ we may evaluate the functional
integral~\eqref{eq:24} in the stationary phase approximation. Then, we find to
leading order $\Gamma_{\Lambda_0} \sim S$ --- in the absence of fluctuations
(suppressed by the cutoff mass gap $\sim \Lambda_0^2$), the effective action
approaches the classical, or microscopic one. The evolution of
$\Gamma_{\Lambda}$ from this starting point in the ultraviolet to the full
effective action in the infrared for $\Lambda \to 0$ is described by an exact
flow equation --- the Wetterich equation~\cite{Wetterich1993a} --- which was
adapted to the Keldysh framework in~\cite{Jakobs2007,Jakobs2010a,Gasenzer2008,Berges2009}. It reads
\begin{equation}
  \label{eq:FRG}
  \partial_{\Lambda} \Gamma_{\Lambda} = \frac{i}{2} \Tr \left[ \left(
      \Gamma^{(2)}_{\Lambda} +
      R_{\Lambda} \right)^{-1} \partial_{\Lambda} R_{\Lambda} \right],
\end{equation}
where $\Gamma_{\Lambda}^{(2)}$ and $R_{\Lambda}$ denote, respectively the second
variations of the effective action and the cutoff action $\Delta S_{\Lambda}$.
As anticipated above, the flow equation provides us with an alternative but
fully equivalent formulation of the functional integral~\eqref{eq:24} as a
functional differential equation. Like the functional integral, the flow
equation can not be solved exactly for most interesting problems. It is,
however, amenable to numerous systematic approximation strategies. For example,
in the vicinity of a critical point it is possible to perform an expansion of
the effective action $\Gamma_{\Lambda}$ in terms canonical scaling dimensions,
keeping only those couplings which are --- in the sense of the renormalization
group --- relevant or marginal at the phase transition, cf. the discussion in
Sec.~\ref{sec:semicl-limit-keldysh}, and see Sec.~\ref{sec:critical-dynamics-3d}
for applications.

%%% Local Variables:
%%% mode: latex
%%% TeX-master: "dds_review"
%%% End:

\part{Applications}
\label{part:applications}

\section{Non-equilibrium stationary states: spin models}
\label{sec:neqspin}

In this section, we discuss the steady state properties of many-body systems
consisting of atoms, which are coupled to the radiation field of a cavity, in
turn subject to dissipation in the form of permanent photon loss. The
corresponding low frequency field theory, a $0+1$-dimensional path integral for
real valued, Ising type fields corresponds to the simplest, non-trivial field
theoretic models and is therefore particular useful to get used to applications
of the Keldysh formalism for relevant physical setups. The basic model
describing the dynamics of atoms in a cavity is the Dicke model~\eqref{Spin1a}
with dissipation, as described in Sec.~\ref{sec:cold-atoms-cavity}, as well as
its extension to multiple cavity photon modes. Despite the simplicity of the
underlying field theory, it is a non-trivial task to solve it for its rich
many-body dynamics. This includes the critical behavior of a single mode cavity
at the superradiance transition as well as universal dynamics in the formation
of spin glasses in multi-mode cavities. We discuss these and further features of
the Dicke model here by putting a focus on the theoretical framework of solving
the corresponding Ising model on the Keldysh contour.

\subsection{Ising spins in a single-mode cavity}\label{sec:ising}

As for the equilibrium path integral, the Keldysh field theory for spin models
which obey the standard Ising $Z_2$ symmetry, is formulated in terms of real
fields, fluctuating in time and space. These models have become rather important
in the field of quantum optics, where the typical situation consists of a set of
atoms, modeled as two-level systems, coupled to the radiation field of a high
finesse cavity.  One important model in this context is the Dicke model, which
has been introduced in Eq.~\eqref{Spin1a} as the generic model for cavity QED
experiments with strong light-matter coupling. Its Hamiltonian is \eq{Spin1}{
  H=\omega_p
  a^{\dagger}a+\frac{\omega_z}{2}\sum_{i=1}^{N}\sigma^z_i+\frac{g}{\sqrt{N}}\sum_{i=1}^N\sigma^x_i\left(a^{\dagger}+a\right).}
The Dicke Hamiltonian represents an effective model which describes the
dynamics of a set of two-level atoms, labeled by the index $i$, inside a cavity,
which is pumped by a transverse laser. In this scenario, the coupling constant
$g$ describes the scattering of laser photons into the cavity, as well as the
reverse process, and therefore rotates in time with the laser frequency
$\omega_l$. In a rotating frame, this time dependence is gauged away in the Hamiltonian, resulting
in an effective shift $\omega_p=\omega_c-\omega_l$ of the bare cavity frequency
$\omega_c$. In addition to this external drive, the cavity is subject to permanent photon loss, due to
imperfections in the cavity mirrors. The photon loss is described by a weak
coupling of the intra-cavity photons to the environment, i.e., the vacuum
radiation field outside the cavity. For typical experimental parameters, this
coupling, which represents the rate with which the intra-cavity field and the
environment exchange photons, is much smaller than the typical relaxation time
of the environment and the latter can be traced out under the Born and Markov
approximation. This results in a Markovian master equation for the
system's density matrix, which reads (cf. Eqs.~(\ref{Spin2a},\ref{eq:109}) repeated for convenience) \eq{Spin2}{
  \partial_t\rho=-i[H,\rho]+\mathcal{L}\rho.  } In this equation, $\rho$ is the
density matrix, $H$ is the Dicke Hamiltonian~\eqref{Spin1} and $\mathcal{L}$ is
the dissipative Liouvillian, acting as on the density matrix as \eq{Spin3a}{
  \mathcal{L}\rho=\kappa\left(a\rho
    a^{\dagger}-\frac{1}{2}\{a^{\dagger}a,\rho\}\right) }
    with loss rate $\kappa$ for the cavity photons.
     
Given the Markovian master equation~\eqref{Spin2}, there are two common ways to
derive a corresponding path integral description for the dissipative Dicke
model, which are based on different representations of the atomic degree of
freedom in terms of field variables: the representation of the atomic degrees of
freedom as a collective spin and subsequent bosonization in terms of a
Holstein-Primakoff transformation~\cite{Emary2003,Oztop2012,Kulkarni13}, and the representation of the atomic
variables in terms of individual Ising fields. We discuss both approaches
in the following, putting a focus on the more general Ising representation,
which can also be applied in the case of multiple cavity modes and individual
atomic loss processes, where a representation of the atoms in terms of a single,
collective spin is no longer possible. In the regimes for which both approaches
can be applied, they are completely equivalent, as we demonstrate.

\subsubsection{Large-spin Holstein-Primakoff representation}

The single-mode Dicke Hamiltonian~\eqref{Spin1} has the property that all atomic
variables couple to the cavity photon mode via the same coupling constant
$g$. Consequently, the coupling of the photons to the sum of the individual
Pauli matrices can be replaced by the coupling of the cavity photon mode to a
large spin, \eq{Spin4}{ H=\omega_pa^{\dagger}a+\omega_z S^z
  +\frac{2g}{\sqrt{N}}S^x\left(a^{\dagger}+a\right),} where $S^z, S^x$ are spin
operators in a spin-$N/2$ Hilbert space. In order to find a path integral
representation for the Hamiltonian~\eqref{Spin4}, these spin operators can be
transformed into bosonic operators via the common Schwinger-boson or
Holstein-Primakoff transformations. For weak coupling $g$, the system remains
strongly polarized $|\langle S_z\rangle|\gg 1$ and the Holstein-Primakoff
transformation around the non-interacting ground state, which has the eigenvalue
$S_z=-N/2$, is the most natural choice. It reads \eq{Spin5}{
  S^z=b^{\dagger}b-\frac{N}{2}, \ \ \
  S^x=\left(\sqrt{N-b^{\dagger}b}\right)b+b^{\dagger}\left(\sqrt{N-b^{\dagger}b}\right),}
where $b,b^{\dagger}$ are bosonic operators. In the thermodynamic limit
$N\rightarrow\infty$, the square root in the $S^x$ operator is expanded in 
 powers of $1/N$, and the Hamiltonian is subsequently formulated on the Keldysh
contour. To zeroth order in $1/N$, the corresponding Keldysh action is
\begin{equation}
  \label{Spin6}
  S=\int_{\omega}\Phi^{\dagger}(\omega)
  \left(\begin{array}{cc}0
          &\left( G^A_{4\times4}\right)^{-1}\\
          \left(
          G^R_{4\times4}\right)^{-1}&\Sigma^K_{4\times4}\end{array}\right)\Phi(\omega),
\end{equation}
with the combined Nambu-Keldysh spinor
\begin{widetext}
  \eq{Spin6a}{ \Phi(\omega)=\left( a_c(\omega), a^*_c(-\omega), b_c(\omega),
      b^*_c(-\omega), a_q(\omega), a^*_q(-\omega), b_q(\omega),
      b^*_q(-\omega)\right)^T,}  the inverse retarded Green's function 
  in Nambu space
\begin{equation}
  \label{Spin7}
  \left(G^R_{4\times4}\right)^{-1}=\left(\begin{array}{cccc}\omega-\omega_p+i\kappa & 0& -g & -g\\ 0 & -\omega-\omega_p-i\kappa& -g& -g\\ -g & -g& \omega-\omega_z & 0\\
-g & -g & 0& -\omega-\omega_z \end{array}\right)
\end{equation}                           
\end{widetext}
and the Keldysh self-energy \eq{Spin8}{ \Sigma^K=2i\kappa \mbox{
    diag}(1,1,0,0).}  Due to the photon decay terms $\sim \kappa$ and the
atom-photon coupling $g$, the above theory is regularized even without
infinitesimal imaginary contributions in the atomic sector. When integrating out
the photons, the latter infinitesimal contributions are overwritten in any case
by the finite imaginary part of the photon Green's function, and it is therefore
reasonable to leave them out from the start.

\begin{figure}[t!]
  \centering
\includegraphics[width=\linewidth]{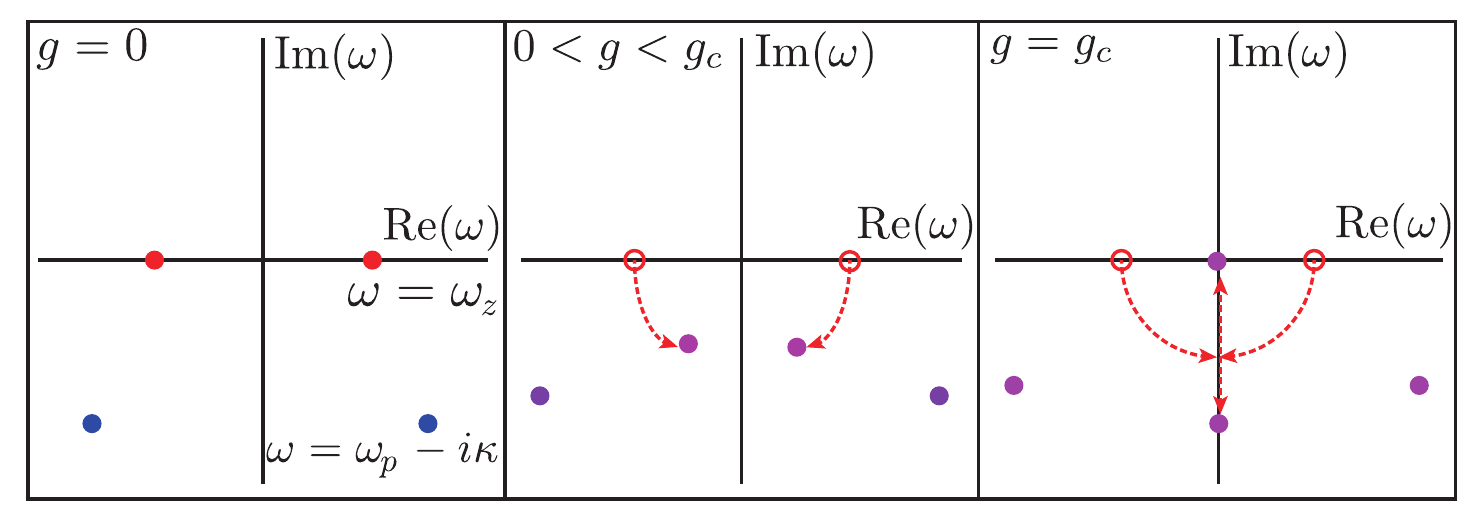}
    \caption
    {\label{fig:Poles} Illustration of the pole structure of the system's eigenmodes. The poles represent the location of the mode frequencies $\omega$ in the complex plane, the real part $\text{Re}(\omega)$ represents the energy of the mode, while the imaginary part $\text{Im}(\omega)$ is the mode's decay rate.
Without interactions, $g=0$, the system is described by the bare atomic, $\omega=\pm\omega_z$, and photonic modes $\omega=\pm\omega_p-i\kappa$. For finite interactions, the atoms and photons hybridize and form polaritonic modes, two of which move closer to the imaginary axis. At the critical point $g=g_c$, one single mode becomes critical at $\omega=0$ and the system undergoes a phase transition from a disordered phase to the $Z_2$ symmetry breaking superradiant phase. The classical nature of the transition in the presence of dissipation is expressed by the fact that the two zero energy modes $\text{Re}(\omega)=0$ become purely dissipative for couplings $g<g_c$ already (indicated with the red arrows).
}
\end{figure}

The excitation spectrum of the atom-photon system is encoded in the retarded
Green's function, and the poles, marking the excitation energies, fulfill the
requirement
\begin{equation}
  \label{Spin9}
  0\overset{!}{=}\det
  \left(G^R_{4\times4}\right)^{-1}\hspace{-0.2cm}=(\omega^2-\omega_z^2)[(\omega+i\kappa)^2-\omega_p^2]
  -4g^2\omega_p\omega_z.
\end{equation}
In the absence of atom-photon coupling, i.e., for
$g=0$, these are the non-interacting poles $\omega_{1,2}=\pm \omega_z$,
corresponding to the atomic transition frequencies and
$\omega_{3,4}=\pm \omega_p+i\kappa$, corresponding to the energy of a photon
$\omega_p$ and its decay rate $\kappa$. For $g>0$, the modes begin to hybridize
and the excitation energies are slightly modified compared to their
non-interacting values, see Fig.~\ref{fig:Poles}. For small coupling strength, the elementary excitations
can still be seen as weakly dressed atoms and photons, with excitation energies
close to the non-interacting values, while they strongly hybridized, inseparable degrees of
freedom for strong coupling strengths.

Above a critical coupling strength $g_c$, the ground state of the system breaks
the $Z_2$ symmetry and the atoms form a macroscopic spin aligned in the
$x$-direction, $\langle S_x\rangle= O(N)$, which is accompanied with a coherent
macroscopic population of the intra-cavity mode $\langle a\rangle \neq 0$. This symmetry
broken phase is called the superradiant phase of the cavity system. In the case
of a macroscopic expectation of $S_x$, the orthogonal $S_z$ component can no
longer be macroscopically large, which renders any expansion of the
Holstein-Primakoff operators~\eqref{Spin5} in $1/N$ invalid. This is expressed
by an instability of the quadratic theory at the superradiance transition,
revealed by the presence of  a zero energy mode, i.e., a pole at $\omega=0$ in
the excitation spectrum. According to Eq.~\eqref{Spin9}, this happens at
\begin{equation}
  \label{eq:82}
  g_c=\sqrt{\frac{(\kappa^2+\omega_p^2)\omega_z}{4\omega_p}}.
\end{equation}
The mode structure in the vicinity of the transition has been analyzed in
Ref.~\cite{DallaTorre2013}, where it has been found that in the presence of
photon decay, the critical mode becomes purely imaginary before the transition
happens, and the corresponding critical dynamics corresponds to a classical
finite temperature transition, see also Fig.~\ref{fig:Poles}. This is mirrored by the fact that the photons
effectively thermalize in the low energy regime and their correlations can be
described by a low energy effective temperature $T_{\mbox{\tiny eff}}$. The
effective temperature can be obtained via a fluctuation dissipation relation, as discussed in Eq.~\eqref{eq:FDR-general} but promoted to Nambu space (see Sec.~\ref{sec:kineq} for a discussion of the FDR in Nambu space). Solving this equation yields the photon distribution function in Nambu representation
\eq{Spin10}{ F_{\mbox{\tiny ph}}=\sigma^z+\sigma^x
  \frac{2g^2}{\omega_z\omega}\frac{\omega_z^2}{\omega^2-\omega_z^2}.  } For
large frequencies $\omega\gg \frac{g^2}{\omega_z}$, this corresponds to the zero
temperature distribution of the non-interacting system $F= \sigma^z$, which is
diagonal in the photon modes. The approach to this limit is, however, algebraic $\sim \omega^{-3}$, in contrast to an exponential approach according to large frequency behavior of the Bose-distribution function in equilibrium. On the other hand, for small frequencies
$\omega\ll \frac{g^2}{\omega_z}$, the occupation becomes essentially thermal and
purely off-diagonal $F=\frac{2T_{\mbox{\tiny eff}}}{\omega}\sigma^x$, with low
energy effective temperature $T_{\mbox{\tiny eff}}=\frac{g^2}{\omega_z}$, cf. Eq.~\eqref{eq:20}. In this low frequency limit, the elementary excitations are strongly hybridized polariton modes, which are diagonal in the photon quadratures $x$ and $p$. The absence of a global temperature scale in the present system reflects the fact that due to the Markovian dissipation, this interacting system with a discrete set of degrees of freedom is not able to achieve detailed balance between the particular modes, i.e. the $x$ and $p$ quadratures, for all energy scales.

One should note that $T_{\mbox{\tiny eff}}$ is not proportional to the decay rate $\kappa$
and consequently the zero temperature equilibrium limit is not simply obtained by taking the
limit $\kappa\rightarrow0$. The reason is that in the present setting, the
atom-photon coupling $g$ leads to a competition of unitary and dissipative
dynamics, which is reflected in the fact that for $g=0$, the eigenstate of the
Hamiltonian is a steady state of the dynamics while this is no longer the case
for finite $g$. As a consequence, the low energy effective temperature depends
on $g$ and vanishes in the limit $g\rightarrow0$.

A further consequence of the dissipative nature of the transition is that the
critical exponents correspond to the classical finite temperature equilibrium
transition. One such critical exponent of the superradiant transition is the
so-called photon flux exponent $\nu_x$, which describes the divergence of the
cavity photon number at the transition, i.e., $n\sim |g-g_c|^{-\nu_x}$. It is
obtained by frequency integration over the photonic correlation function
\eq{Spin11}{ 2n=\left(\int_{\omega}iG^K_{\mbox{\tiny ph}}(\omega)\right)-1\sim
  |g-g_c|^{-1}.}  The exponent $\nu_x=1$ found for the present non-equilibrium
transition coincides with the classical, finite temperature exponent for the
equilibrium Dicke model in line with the discussion above. On the other hand, at zero temperature, the exponent of the
corresponding quantum phase transition is found to be $\nu_x=1/2$.

The scaling behavior of the critical mode at the transition is expressed in terms of the dynamical critical exponent $\nu_t$. In the presence of dissipation, the excitations in the system decay exponentially in time, which results in an asymptotic exponential decay of real-time correlation functions
\eq{Spin12}{
\langle \{a(t),a^{\dagger}(0)\}\rangle=G^K_{\mbox{\tiny ph}}(t)\sim e^{-t/\xi_t}.
}
The characteristic decay time $\xi_t$ is determined by the slowest decaying mode in the system, i.e., by the mode with the smallest imaginary part in the frequency spectrum.  In the vicinity of the phase transition, this is the critical mode , which shows scaling
\eq{Spin13}{
\xi_t\sim \frac{1}{|g-g_c|},
}
i.e., a dynamical critical exponent $\nu_t=1$. This behavior is in stark contrast to a zero temperature quantum phase transition, where correlation functions do not decay over time but oscillate with characteristic frequencies $\omega_c$, the smallest of which indicates the distance to the phase transition and encodes the critical scaling behavior. The fact that in the present setting the critical mode becomes purely imaginary is a generic feature of dissipative phase transitions, which is discussed further in Sec.~\ref{sec:bosons}.

All of the above discussed results are encoded in the quadratic Keldysh action,
described by Eqs.~(\ref{Spin6}-\ref{Spin8}). The resulting critical behavior is then that of a non-interacting system, and is associated accordingly to a Gaussian fixed point of the renormalization group flow. This is in contrast to critical behavior in interacting systems, which are described by an interacting Wilson-Fisher fixed point, not smoothly connected to the previous one (cf. the discussion in Sec.~\ref{sec:critical-dynamics-3d}). This action describes the problem
up to $1/N$ corrections resulting from the expansion of the Holstein-Primakoff
bosons in $1/N$. It contains correctly the physics in the thermodynamic limit
$N\rightarrow\infty$ and captures the essential dynamics of the superradiance
transition. For finite systems, the higher order terms in $1/N$ can not be neglected and
have to be taken into account properly. Analytical approaches, relying on an
expansion of the Holstein-Primakoff operators up to first order in the ratio
$1/N$, have shown that the first order correction term leads to a quartic
contribution in the action, whose self-consistently determined one-loop
corrections are already in very good agreement with numerical
results~\cite{DallaTorre2013}. The latter have been obtained from a Monte-Carlo
wave function (MCWF) simulation of the master equation and agreed with the
analytical results even on the level of $N=10$ atoms.

\subsubsection{Effective Ising spin action for the single-mode cavity}

The Dicke model defined by the master equation~\eqref{Spin2} obeys the common
Ising $Z_2$ symmetry, which is spontaneously broken by the ground state of the
superradiant phase. Another approach to express the Dicke model in a Keldysh
path integral approach is therefore to represent the spin operators in terms of
real Ising field variables, $\sigma_x(t)\rightarrow\phi(t)$. In contrast to the
Holstein-Primakoff transformation applied in the previous section, the Ising
representation does not require a single, large spin to have a well defined
field theory representation, and is therefore also applicable in situations in
which the atom-photon coupling is spatially dependent $g\rightarrow g(x)$ and a
large spin representation becomes impossible. Furthermore, we use this approach
to describe the symmetry broken phase of the problem.

The atomic sector of the Hamiltonian~\eqref{Spin1} describes Ising spins in a transverse field and the corresponding universal low energy description for frequencies below the level spacing $\omega_z$ is obtained by transforming the quantum spin operators to classical real fields. According to the quantum to classical mapping~\cite{KogutRMP} this is possible in the scaling regime of the corresponding, i.e., for the present case in the vicinity of the $2$nd order superradiance and glass transitions. The present model represents a generalization of the zero-dimensional Ising model in a transverse field, for which the quantum to classical mapping is known~\cite{KogutRMP} and consists of the replacements
\eq{Spin14}{
\sigma^x_i(t)\rightarrow\phi_i(t)}
and
\eq{Spin15}{
\sigma^z_i(t)\rightarrow 1-\tfrac{2}{\omega_z^2}\left(\partial_t\phi_i(t)\right)^2.
}
The dynamic constraint $(\sigma^x_i(t))^2=1$ is implemented via the non-linear constraint
\eq{Spin16a}{
\delta(\phi_i^2(t)-1)=\int\mathcal{D}\lambda_l(t)\ e^{i\lambda_l(t)(\phi_i^2(t)-1)},
}
which introduces dynamical Lagrange parameters $\lambda_i(t)$ for each spin variable.
In this framework, the purely atomic part of the action on the $(\pm)$-contour is
\eq{Spin16}{
S_{\mbox{\tiny at}}=\tfrac{1}{\omega_z}\int_{\omega} \sum_{\alpha=\pm\atop i=1,..,N}\alpha\left[\lambda_{\alpha i}\left(\phi_{\alpha i}^2-1\right)-(\partial_t\phi_{\alpha i})^2\right].
}
The action for the atom-photon coupling and the bare photonic part is
straightforwardly derived according to the previous sections, and the
corresponding Keldysh action is
\begin{multline}
  \label{Spin17}
  S = \frac{1}{\omega_z}\int_{t}\sum_{i}(\phi_{ci},\phi_{qi})
  \left(\hspace{-0.15cm}\begin{array}{cc}\lambda_{qi}(t)&
                                                          \lambda_{ci}(t)+\partial_t^2\\ \lambda_{ci}(t)+\partial_t^2&\lambda_{qi}(t)\end{array}\hspace{-0.15cm}\right)\left(\hspace{-0.15cm}\begin{array}{c}\phi_{ci}\\ \phi_{qi}\end{array}\hspace{-0.15cm}\right) \\
+\frac{g}{\sqrt{N}}\int_t \sum_i\left(\phi_{ci}(a^*_q+a_q)+\phi_{qi}(a^*_c+a_c)\right)\\
+\int_t (a^*_c, a^*_q)\left(\hspace{-0.15cm}\begin{array}{cc}0 & i\partial_t-\omega_p-i\kappa\\ i\partial_t-\omega_p+i\kappa & 2i\kappa\end{array}\hspace{-0.15cm}\right)\left(\hspace{-0.15cm}\begin{array}{c}a_c\\ a_q\end{array}\hspace{-0.15cm}\right).
\end{multline}
It is quadratic in the fields $\phi, a, a^*$ as it was the case for the
Holstein-Primakoff action in the large-$N$ limit. However, the non-linear
constraint, i.e., the fact that $\lambda_{q,c}$ is a dynamic variable,
introduces higher order terms in the Ising fields. The  approach corresponding to
the large-$N$ expansion of the Holstein-Primakoff fields in the present
formalism (in the sense that the action becomes quadratic) is to treat the
Lagrange parameters as static mean field variables, i.e.,
$\lambda_{ci}(t)\rightarrow\lambda_c$ and
$\lambda_{qi}(t)\rightarrow\lambda_q$. Due to causality, a static mean quantum field
must be zero, $\lambda_q=0$. On the other hand, the value of
$\lambda_c$ has to be chosen such that the non-linear constraint is preserved on
average \eq{Spin18}{ \langle (\sigma^x_i(t))^2\rangle=\langle
  \phi_c^2(t)\rangle=\int_{\omega}iG^K_{\mbox{\tiny
      at}}(\omega)\overset{!}{=}1.}  This is equivalent to a vanishing variation of
the action with respect to the Lagrange parameters $\lambda_{c,q}$ and is known
as the ``soft-spin'' approach to spin models (i.e. the constraint is treated on the mean field level). It becomes exact in the
thermodynamic limit $N\rightarrow\infty$~\cite{Read1993,Read1995}.

Integrating out the $N$ individual atoms leads to the effective photon action
\eq{Spin18a}{
S=\frac{1}{2}\int_{\omega} A^{\dagger}(\omega)\left(\begin{array}{cc}0 & \left(G^{A}_{2\times2}\right)^{-1}\\ \left(G^{R}_{2\times2}\right)^{-1}& 2i\kappa\ \mbox{diag}(1,1)\end{array}\right)A(\omega),
}
with the Nambu spinor 
\eq{Spin19}{
A(\omega)=\left(\begin{array}{c}a_c(\omega)\\ a^*_c(-\omega)\\ a_q(\omega)\\ a^*_q(-\omega)\end{array}\right)
}
and the inverse retarded Green's function
\begin{equation}
  \label{Spin20}
  \left(G^R_{2\times2}\right)^{-1}=\left(\begin{array}{cc}\omega-\omega_p+i\kappa-\frac{2g^2\omega_z}{\omega^2-\lambda}& -\frac{2g^2\omega_z}{\omega^2-\lambda}\\-\frac{2g^2\omega_z}{\omega^2-\lambda} & \omega-\omega_p-i\kappa-\frac{2g^2\omega_z}{\omega^2-\lambda}
\end{array}\right).
\end{equation}
The poles of the Green's function are again determined by the zeros of the
determinant of the inverse Green's function, i.e., by the roots of the equation
\eq{Spin21}{
  0\overset{!}{=}(\lambda_c-\omega^2)\left((\omega+i\kappa)^2-\omega_p^2\right)+4\omega_p\omega_zg^2.}
For the non-interacting theory, i.e., for $g=0$, this determines the poles of
the atomic sector to be $\omega=\pm\sqrt{\lambda_c}$ and in turn fixes
$\lambda_c=\omega_z^2$ to be consistent with the microscopic theory. As long as
the system is not superradiant, the atom-photon interaction does not modify the
integral~\eqref{Spin18}, and for the entire parameter range of the normal phase,
we find $\lambda_c=\omega_z^2$ as for the non-interacting case. This again
yields the critical value of the coupling strength~\eqref{eq:82} we obtained
above from the Holstein-Primakoff approach. Consequently, the results from the
previous section are recovered in the Ising representation of the Keldysh path
integral.

We now turn to the description of the ordered phase. Directly at the superradiant transition, the system becomes unstable, which is indicated by a critical mode with frequency $\omega=0$. In order to stabilize the system, for coupling strengths $g>g_c$ the steady state breaks the $Z_2$ symmetry of the action, which is expressed in terms of a finite, symmetry breaking order parameter $\phi_c(\omega)\rightarrow\psi\delta(\omega)+\phi_c(\omega)$, and equivalently in the photon basis $a_c(\omega)\rightarrow a\delta(\omega)+a_c(\omega)$. These order parameters correspond to finite expectation values $\langle\sigma^x_i(t)\rangle=\psi$ and $\langle a(t)\rangle=a$ in the system's ground state. They modify the soft-spin condition~\eqref{Spin18} according to
\eq{Spin22}{
\langle (\sigma^x_i(t))^2\rangle=\psi^2+\int_{\omega}iG^K_{at}(\omega)\overset{!}{=}1,}
where $G^K_{at}$ is the Keldysh Green's function of the fluctuating variable $\phi$. As a consequence, in order to fulfill Eq.~\eqref{Spin22}, the Lagrange multiplier $\lambda_c$ becomes a continuous function of the order parameter, and therefore an implicit function of the coupling $g$ in the superradiant phase. The dependence of $\lambda_c$ can be determined from Eq.~\eqref{Spin21}. At the superradiant transition at $g=g_c$, this equation has one critical solution, i.e., the propagator has a pole at $\omega=0$, see Fig.~\ref{fig:Poles}. For larger coupling strengths, this pole crosses the origin and obtains a positive imaginary part, thereby rendering the system unstable. In order to compensate for this mechanism, $\lambda_c$ is modified such that the pole does not cross the real axis, i.e., remains at its value $\omega=0$ in the superradiant phase. Consequently $\lambda_c=\frac{4g^2\omega_p\omega_z}{\kappa^2+\omega_p^2}$ for $g>g_c$. Via Eq.~\eqref{Spin22}, this reveals the scaling of $\psi\sim\sqrt{|g-g_c|}$ when the transition is approached from inside the superradiant phase, which is the same critical behavior of the order parameter as for the equilibrium transition.

The results on the critical properties of the Ising field theory at the
superradiance transition conclude the discussion of the single mode Dicke model
in the framework of the Keldysh path integral. We have shown that both the
Holstein-Primakoff approach in terms of complex bosonic variables as well as the
Ising approach in terms of real Ising variables are equivalent on the level of
the quadratic, large-$N$ limit of the theory. The critical point of the
transition, as well as the mode structure and the critical scaling behavior have
been directly derived from the quadratic Keldysh action, which illustrates the
strength of the present field theory approach. In the following section, we
discuss the extension of the Dicke model to multiple cavity modes, which
contains the possibility of frustrated atom-atom interactions and the formation
of a spin glass in the cavity system.

\subsection{Ising spins in a random multi-mode cavity}\label{sec:multi}
In the previous section, we saw that dissipation renders the superradiant transition essentially a classical phase transition with critical exponent corresponding to a $Z_2$ symmetry breaking, finite temperature transition.  However, no signature of the non-equilibrium nature of the steady state could be found in the long-wavelength dynamics governing the transition since the phonon distribution function shows a classical Rayleigh-Jeans divergence $F\sim 1/\omega$ at small frequencies.
 This changes drastically when the setup allows for multiple dissipative cavity modes, which couple atoms and photons via a random interaction potential. The random interaction potential is realized by freezing the atoms on random positions inside the cavity and by excluding atomic self-organization with a large set of modes.

The effective, cavity mediated interactions in the case of multiple cavity modes are able to induce frustration in the effective atom-atom interactions. At a critical frustration level, these will drive the system into an Ising spin glass phase. Due to the presence of drive an dissipation in the quantum optical realization of this spin glass phase, it features no analogue in condensed matter physics and the corresponding spin glass transition does not correspond to the classical finite temperature transition~\cite{Buchhold2013}. The corresponding model is motivated by recent works on ultracold atoms in cavities~\cite{Gopalakrishnan2011,Strack2010G,Strack2011G,PhysRevLett.110.075304,Buchhold2013} and is described by the master equation \eqref{Spin2}, where now the Hamiltonian 
\eq{Glass3}{
H=\frac{\omega_z}{2}\sum_{s=1}^N\sigma^z_s+\sum_{l=1}^M\anno{\omega}{l}\creo{a}{l}\anno{a}{l}+\sum_{s,l}g_{sl}\ (\creo{a}{l}+\anno{a}{l})\sigma^x_s
}
describes the energy of $M$ photon modes with frequency $\omega_l$ and $N$ Ising spins with level spacing $\omega_z$, interacting via coupling constants $g_{sl}=g_0\cos(k_lx_s)$, which depend on the atomic position $x_s$ and the photon mode function $k_l$, such that $-g_0\le g_{sl}\le g_0$. The dissipator describes the decay of each of the $M$ photon modes with a uniform decay rate $\kappa$,
\eq{Glass2}{
\mathcal{L}\rho=\kappa\sum_{l=1}^{M}\left(a_l\rho\creo{a}{l} - \frac{1}{2}
  \{a_l^{\dagger} a_l, \rho \} \right).
}

\subsubsection{Keldysh action and saddle point equations}
The corresponding Keldysh action is similar to Eq.~\eqref{Spin17} and reads
\eq{Glass4}{
S\hspace{-0.25cm}&=&\frac{1}{\omega_z}\int_{t}\sum_{s}(\phi_{cs},\phi_{qs})\left(\hspace{-0.15cm}\begin{array}{cc}0& \lambda_{c}+\partial_t^2\\ \lambda_{c}+\partial_t^2&0\end{array}\hspace{-0.15cm}\right)\left(\hspace{-0.15cm}\begin{array}{c}\phi_{cs}\\ \phi_{qs}\end{array}\hspace{-0.15cm}\right)\nonumber \\
&&+\int_t \sum_{sl}g_{sl}\left(\phi_{cs}(a^*_{ql}+a_{ql})+\phi_{qs}(a^*_{cl}+a_{cl})\right)\\
&&+\int_t\sum_l (a^*_{cl}, a^*_{ql})\left(\hspace{-0.15cm}\begin{array}{cc}0 & i\partial_t-\omega_l-i\kappa\\ i\partial_t-\omega_l+i\kappa & 2i\kappa\end{array}\hspace{-0.15cm}\right)\left(\hspace{-0.15cm}\begin{array}{c}a_{cl}\\ a_{ql}\end{array}\hspace{-0.15cm}\right).
\nonumber}
Here, the soft spin approximation has been performed and the scaling of the coupling constants $g_{sl}$ in the thermodynamic limit is implicit, i.e. $g_{sl}\sim N^{-\frac{1}{2}}$.  The thermodynamic limit is reached by taking
an extensive number of atoms $N\rightarrow\infty$ but leaving the number of photon modes $M<\infty$ finite since the consideration of an extensive number of photon modes is physically unrealistic. In order to obtain a large-$N$ effective action, the photon degrees of freedom are integrated out, leading to the atomic action in frequency space
\eq{Glass5}{
S\hspace{-0.25cm}&=&\frac{1}{\omega_z}\int_{\omega}\sum_{s}(\phi_{cs},\phi_{qs})\left(\hspace{-0.15cm}\begin{array}{cc}0& \lambda_{c}-\omega^2\\ \lambda_{c}-\omega^2&0\end{array}\hspace{-0.15cm}\right)\left(\hspace{-0.15cm}\begin{array}{c}\phi_{cs}\\ \phi_{qs}\end{array}\hspace{-0.15cm}\right)\nonumber \\
&&+ \sum_{si} J_{si}\int_{\omega}(\phi_{cs},\phi_{qs})\left(\hspace{-0.15cm}\begin{array}{cc}0& \Lambda^A(\omega)\\ \Lambda^R(\omega)&\Lambda^K(\omega)\end{array}\hspace{-0.15cm}\right)\left(\hspace{-0.15cm}\begin{array}{c}\phi_{ci}\\ \phi_{qi}\end{array}\hspace{-0.15cm}\right).
}
The frequency dependent terms are the symmetrized photon Green's functions
\begin{equation}
  \label{Glass6}
  \begin{split}
    \Lambda^R(\omega)& = \frac{-\omega_p}{(\omega+i\kappa)^2-\omega_p^2}, \\
    \Lambda^K(\omega) & =
    \frac{2i\kappa(\omega^2+\kappa^2+\omega_p^2)}{\left|(\omega+i\kappa)^2 -
        \omega_p^2\right|^2}
  \end{split}  
\end{equation}
with an average photon frequency of the cavity $\omega_p$. The effective coupling
\eq{Glass7}{
J_{si}=\sum_{l=1}^M\frac{g_{sl}g_{il}}{4}
}
fluctuates between the values $-\frac{Mg_0}{4}\le J_{si}\le \frac{Mg_0^2}{4}$ and can be either ferromagnetic $J_{si}<0$ or antiferromagnetic $J_{si}>0$ depending on the $s,i$. Assuming $g_{sl}$ to be independently and equally distributed for each atom, the couplings $J_{si}$ are for sufficiently large $M$ distributed according to a Gaussian distribution function. We set their average value $\bar{J}\equiv \langle J_{si}\rangle=0$ as it lifts the frustration in the system caused by fluctuating couplings but does not modify the universal behavior of the system at the glass transition~\cite{Buchhold2013,Strack2010G}. 

The averaged partition function of the system, including the probability distribution for the couplings $J_{si}$ is 
\eq{Glass8}{
Z=\int\mathscr{D}[\phi_s]\mathscr{D}[J_{si}] e^{iS+iS_P},
}
where 
\eq{Glass9}{
iS_P=-\frac{N}{2}\sum_{s,i} \frac{J_{si}^2}{K}}
is the action of the random couplings $J_{si}$, with correlations 
\eq{Glass10a}{
\langle J_{si}J_{lm}\rangle=(\delta_{s,l}\delta_{i,m}+\delta_{s,m}\delta_{i,l})\frac{K}{N}.
}
In this sense, the coupling of the atomic degrees of freedom to the random variables $J_{si}$ has the same structure as the coupling of the atoms to an external bath. However, the significant difference to a Markovian bath, which would be $\delta$-correlated in space and time is that the quenched disorder, while correlated locally in space, has infinite correlation time. The latter is expressed by the fact that the variables $J_{si}$ are time-independent.

Averaging over all realizations of the couplings is done by a Gaussian integration (cf. App.~\ref{sec:gauss-funct-integr}), and transforms the disorder part, i.e., the second line of Eq.~\eqref{Glass5}, to a time non-local quartic contribution
\eq{Glass10}{
S_{\mbox{\tiny at-at}}=\frac{iK}{N}\int_{\omega,\nu}\sum_{l,m}\Phi^{\Lambda}_{l,m}(\omega)\Phi^{\Lambda}_{m,l}(\nu),
}
with 
\eq{Glass11}{
\Phi^{\Lambda}_{l,m}(\omega)=(\phi_{cl},\phi_{ql})\left(\hspace{-0.15cm}\begin{array}{cc}0& \Lambda^A(\omega)\\ \Lambda^R(\omega)&\Lambda^K(\omega)\end{array}\hspace{-0.15cm}\right)\left(\hspace{-0.15cm}\begin{array}{c}\phi_{cm}\\ \phi_{qm}\end{array}\hspace{-0.15cm}\right).
}
The double sum in~\eqref{Glass10} is decoupled via a Hubbard-Stratonovich transformation, which introduces the macroscopic fields $Q_{\alpha\alpha'}$, with $\alpha,\alpha'=c,q$ being Keldysh indices~\cite{Strack2010G,Buchhold2013}. This transformation is in general not unique but can be made unique by requiring the equivalence 
\eq{Glass12}{
\frac{\delta}{\delta Q_{\alpha\alpha'}}Z=0 \Leftrightarrow Q_{\alpha\alpha'}(\omega,\nu)=\tfrac{1}{N}\sum_l \langle \phi_{\alpha l}(\omega)\phi_{\alpha'l}(\nu)\rangle.
}
This identifies the Hubbard-Stratonovich field with the average atomic correlation function.
Since one is interested in the stationary, time translational invariant state, 
\eq{Glass13}{
Q_{\alpha\alpha'}(\omega,\nu)=2\pi\delta(\omega+\nu)Q_{\alpha\alpha'}(\omega).
}
After the Hubbard-Stratonovich decoupling, the resulting action is quadratic in the atomic fields and they are integrated out, leading to the macroscopic action
\eq{Glass14}{
\mathcal{S}=iN\mbox{Tr}\left[K\Lambda Q\Lambda Q-\frac{1}{2}\ln \tilde{G}\right](\omega).
}
Here $\Lambda$ and $Q$ are matrices in Keldysh space and $\tilde{G}$ is defined as
\eq{Glass15}{
\tilde{G}(\omega)=\left(G_0^{-1}(\omega)-2K\Lambda(\omega)Q(\omega)\Lambda(\omega)\right)^{-1}.
}
In the thermodynamic limit, the partition function is determined by the saddle point value of the action, i.e. by the condition 
\eq{Glass16}{
\frac{\delta\mathcal{S}}{\delta Q_{\alpha\alpha'}}\overset{!}{=}0
}
for all $\alpha, \alpha'$. This yields the values of the fields $Q_{\alpha\alpha'}(\omega)$ at the saddle-point, which can be identified with the atomic response and correlation functions.
The saddle-point equations for the atomic response and correlation functions are
\eq{Glass20}{
Q^R(\omega)=\left(\frac{2(\lambda-\omega^2)}{\omega_z}-4K\left(\Lambda^R(\omega)\right)^2Q^R(\omega)\right)^{-1}
}
and
\eq{Glass21}{
Q^K(\omega)=\tfrac{4K|Q^R|^2\Lambda^K(Q^A\Lambda^A+Q^R\Lambda^R)}{1-4K|Q^R\Lambda^R|^2}.}
Additionally, due to Eq.~\eqref{Glass12}, the soft-spin constraint maps to the $Q$-fields according to
\eq{Glass17}{
i\int_{\omega}Q^K(\omega)=\frac{2}{N}\sum_l\langle\phi_{cl}(-\omega)\phi_{cl}(\omega)\rangle=2.
}
In the glass phase, the spins attain temporally frozen configurations, expressed by an infinite correlation time of the atomic correlator, which is expressed by a non-zero Edwards-Anderson parameter
\eq{Glass18}{
q_{\mbox{\tiny EA}}\equiv \lim_{t\rightarrow\infty} \tfrac{1}{N}\sum_{l}\langle \sigma^x_l(t)\sigma^x_l(0)\rangle.
}
Consequently, the correlation function $Q^K(\omega)$ consists of a regular part, describing the short time dynamics and a non-vanishing contribution at $\omega=0$. It can be expressed via the modified fluctuation dissipation relation~\cite{Cugliandolo1999}
\eq{Glass19}{
Q^K(\omega)=4i\pi q_{\mbox{\tiny EA}}\delta(\omega)+Q^K_{r}(\omega)
}
in terms of the order parameter and  a regular contribution $Q^K_r(\omega)$. The soft-spin condition~\eqref{Glass17} together with Eq.~\eqref{Glass21} fixes the value of the Lagrange parameter $\lambda$ and therefore fully determines the spectrum of the system. Similar to the superradiant transition, the Edwards-Anderson parameter becomes non-zero when the poles of the system become critical (approach the real axis) and its emergence is a mechanism to stabilize the critical modes in the ordered phase.

\subsubsection{Non-equilibrium glass transition}
The variance of the effective, long-range atom-atom interaction $K$ is a measure of the frustration in the system and for sufficiently strong frustration, the system enters a glass phase, described by an infinite autocorrelation time of the spins and the emergence of a non-zero Edwards-Anderson parameter $q_{\mbox{\tiny EA}}>0$. This goes hand in hand with a critical continuum of modes reaching zero, which is the characteristic feature of a critical phase of matter. The phase diagram for the fully coherent model has been analyzed in~\cite{Strack2010G} and in~\cite{Buchhold2013} it was shown that the glass phase persists in the presence of dissipation. However, the dissipative model shows new universal features of the glass phase, which correspond neither to a zero nor to a finite temperature equilibrium transition. This is attributed to the fact that the critical modes are described by poles in the complex plane which are neither purely real as in the zero temperature (or quantum) case nor purely imaginary as for the finite temperature transition.

\begin{figure}[t!]
  \centering
\includegraphics[width=\linewidth]{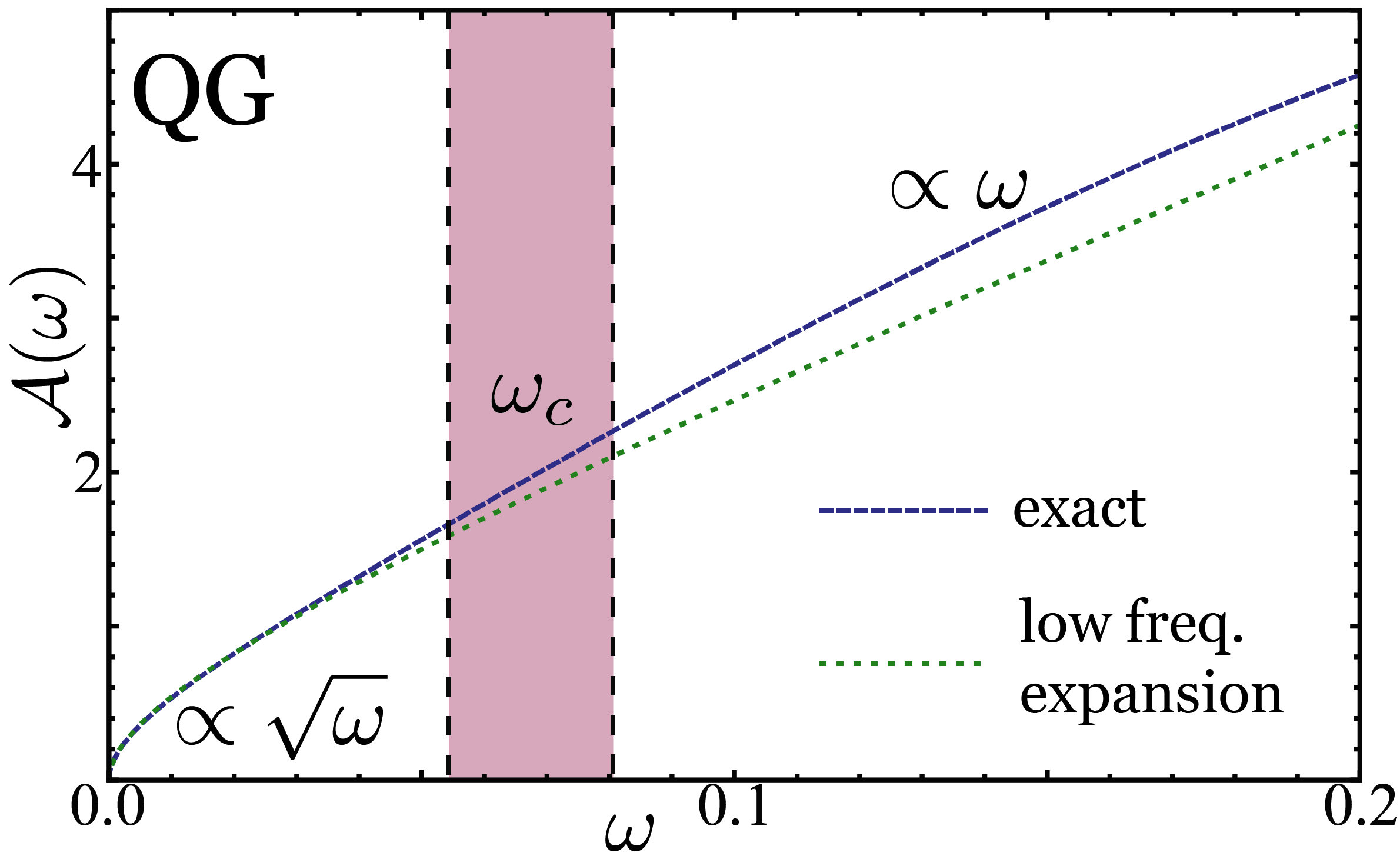}
    \caption
    {\label{fig:QGSpec}
    Spectral density $\mathcal{A}(\omega)=i\left(Q^R(\omega)-Q^A(\omega)\right)$ of the atoms in the glass phase for parameters
     ${K=0.01, \omega_z=2}$ and varying photon parameters $\omega_p, \kappa$. 
For frequencies below the crossover ${\omega<\omega_c}$, $\mathcal{A}\sim\sqrt{\omega}$ is overdamped, corresponding to Eq.~\eqref{KeyEqEx1}, while it recovers the typical thermal glass behavior $\mathcal{A}\sim\omega$ at intermediate frequencies ${\omega>\omega_c}$. Figure copied from Ref.~\cite{Buchhold2013}. (Copyright (2013) by The American Physical Society)}
\end{figure}

In the presence of dissipation, a photon that is emitted from an atom can either be absorbed by another atom, leading to an effective atom-atom interaction, which is infinite range in space, or decay from the cavity. The latter happens on a characteristic time scale $\tau_p=1/\kappa$, which is as well the time scale for which atoms can interact with each other by exchanging a photon before the photon decays. As a consequence, the photon decay reduces the effective range of the atom-atom interaction in time and reduces the strength of frustration in the system. Close to the glass transition and in the glass phase,
this leads to the emergence of a crossover scale $\omega_c=\tau_c^{-1}$, above
which the spectral properties of the system correspond exactly to a zero
temperature spin glass. Below this frequency, the spectrum reveals the breaking
of time reversal symmetry by the dissipation and formally corresponds to the
dynamical universality class of dissipative quantum glasses, other examples of which
are spin glasses coupled to an ohmic bath
~\cite{Cugliandolo1999,Cugliandolo2002,Cugliandolo2004} or a bath of metallic
electrons~\cite{Read1995,Strack2011G}.  The crossover frequency is
\begin{eqnarray}
\omega_\text{c} = \kappa\left(\frac{\omega_p^2+\tfrac{1}{2}\kappa^2}{\omega_p^2+\kappa^2}+\frac{\left(\omega_p^2+\kappa^2\right)^2}{2\sqrt{K}\omega_z^2}\right)^{-1}.\label{KeyEq1}
\end{eqnarray}
and the modifications of the spectrum below this scale are due to the damping introduced by the Markovian bath.
On the normal or disordered side of the phase diagram, the corresponding low frequency propagator is
\eq{KeyEq3}{
Q^R_n(\omega)=Z\left((\omega+ i\gamma)^2-\alpha^2\right)^{-1},}
describing the Ising spins as damped harmonic oscillators with decay rate $\gamma>0$ and frequency $\alpha$, with the physical meaning of an inverse lifetime and energy gap of the atomic excitations. When the glass transition is approached, the gap, the inverse lifetime and the residue $Z$ scale to zero simultaneously and the atomic response in the glass phase
\eq{KeyEqEx1}{
Q^R_g=\bar{Z}\left(\omega^2+\bar{\gamma}|\omega|\right)^{-\frac{1}{2}}} is non-analytic and can no longer be interpreted as a harmonic oscillator. The broken time reversal invariance manifests itself in both parameters $\gamma,\bar{\gamma}\neq0$ being non-zero, which modifies the low frequency dynamics in the entire glass phase towards a dissipative quantum glass. This is for instance expressed by a spectral density, which features an anomalous square root behavior $\mathcal{A}(\omega)=-2\mbox{Im}(Q^R(\omega))\sim \sqrt{|\omega|}$ for small frequencies and therefore has a non-analytic response at zero frequency, see Fig.~\ref{fig:QGSpec}. The latter leads to a characteristic algebraic decay of the photon correlation function, as discussed below.
 The universality class of the present dissipative quantum glass transition is determined by the critical exponents at the transition, which describe the scaling of the parameters $\alpha, \gamma, Z$ as a function of $\delta=|K-K_c|$. It is summarized in the equations
\eq{Extra3}{
 \begin{array}{ccrl}\alpha_{\delta}&=&\frac{\sqrt{2}(\omega_p^2+\kappa^2)}{8\sqrt{K^3}\kappa} & \left|\frac{\delta}{\log(\delta)}\right|^{\frac{3}{2}} \\\gamma_{\delta} &=&\tfrac{\omega_p^2+\kappa^2}{16K^2\kappa} &\left|\frac{\delta}{\log(\delta)}\right|^2 \\Z_{\delta}&=& \tfrac{\omega_p(\omega_p^2+\kappa^2)}{8\sqrt{K^5}\kappa^2} & \left|\frac{\delta}{\log(\delta)}\right|^3 \end{array},
}
which show the typical logarithmic correction to scaling at a quantum glass transition and identify the critical exponents. Indicated by the scaling behavior, dissipative and coherent dynamics rival each other when approaching the transition, separating this glass transition from equilibrium transitions and the previously discussed Dicke transition, which are either fully coherent (quantum) with $\gamma=0$ or fully dissipative (classical) with $\alpha=0$ sufficiently close to the transition. The present glass transition has therefore no counterpart in static equilibrium physics and is termed dissipative quantum glass transition. Similar behavior is present in system bath settings in equilibrium, where the bath however not only imprints a finite temperature to the system but as well modifies its spectral properties ~\cite{Cugliandolo2002, Cugliandolo1999, Strack2011G}. What these situations share in common is that the effective theory, after elimination of the bath variables, obeys no time reversal symmetry, which ensures the same asymptotic universal behavior as in the case of the Markovian photon loss in the present system.

\subsubsection{Thermalization of photons and atoms in the glass phase}
As typical for many open quantum systems (for exceptions, see the discussion in Sec.~\ref{sec:bosons}), the statistical properties of the excitations are described by a low energy effective temperature (LET) and a corresponding thermal distribution of the excitations. However, as has been found for the single mode Dicke model in the normal as well as in the superradiant phase, the LET for the photonic and the atomic subsystem did not coincide, i.e., the subsystems have not thermalized but are held at different temperatures corresponding to their individual coupling to a bath \cite{DallaTorre2013}.  This dramatically changes in the present system with multiple photon modes. As frustration is increased by driving the variance $K$ towards the critical value $K_c$, atoms and photons begin to thermalize towards the same, shared LET. 
The distribution function $F$ of photonic or atomic modes is obtained by solving the fluctuation dissipation relation 
\eq{Glasso1}{
Q^K(\omega)=Q^R(\omega)F_{\mbox{\tiny at}}(\omega)-F_{\mbox{\tiny at}}(\omega)Q^A(\omega)
}
for the atomic Green's functions. It leads to an atomic LET 
\begin{eqnarray}
T_\text{eff} = \frac{\omega_p^2+\kappa^2}{4\omega_p}\label{KeyEq2}
\end{eqnarray}
in the glass phase, which coincides with the photonic LET~\cite{Buchhold2013}.
The thermalization of the two subsystems is a consequence of the disorder induced effective long range interactions, which redistribute energy between the different modes and enable equilibration even in the presence of the Markovian dissipation in the photon sector. In the paramagnetic phase, the distribution functions of atoms and photons are identical for frequencies $\omega>\alpha$ larger than the gap, but deviate from each other for lower frequencies. The elementary excitations above this frequency are strongly correlated and can not be seen as weakly dressed photons or atoms. As a consequence, the observables in the critical glass phase are dominated by the universal low energy behavior of the glass propagator~\eqref{KeyEqEx1} and thermal statistics of the excitations.

\begin{figure}[t]
  \centering
\includegraphics[width=0.95\linewidth]{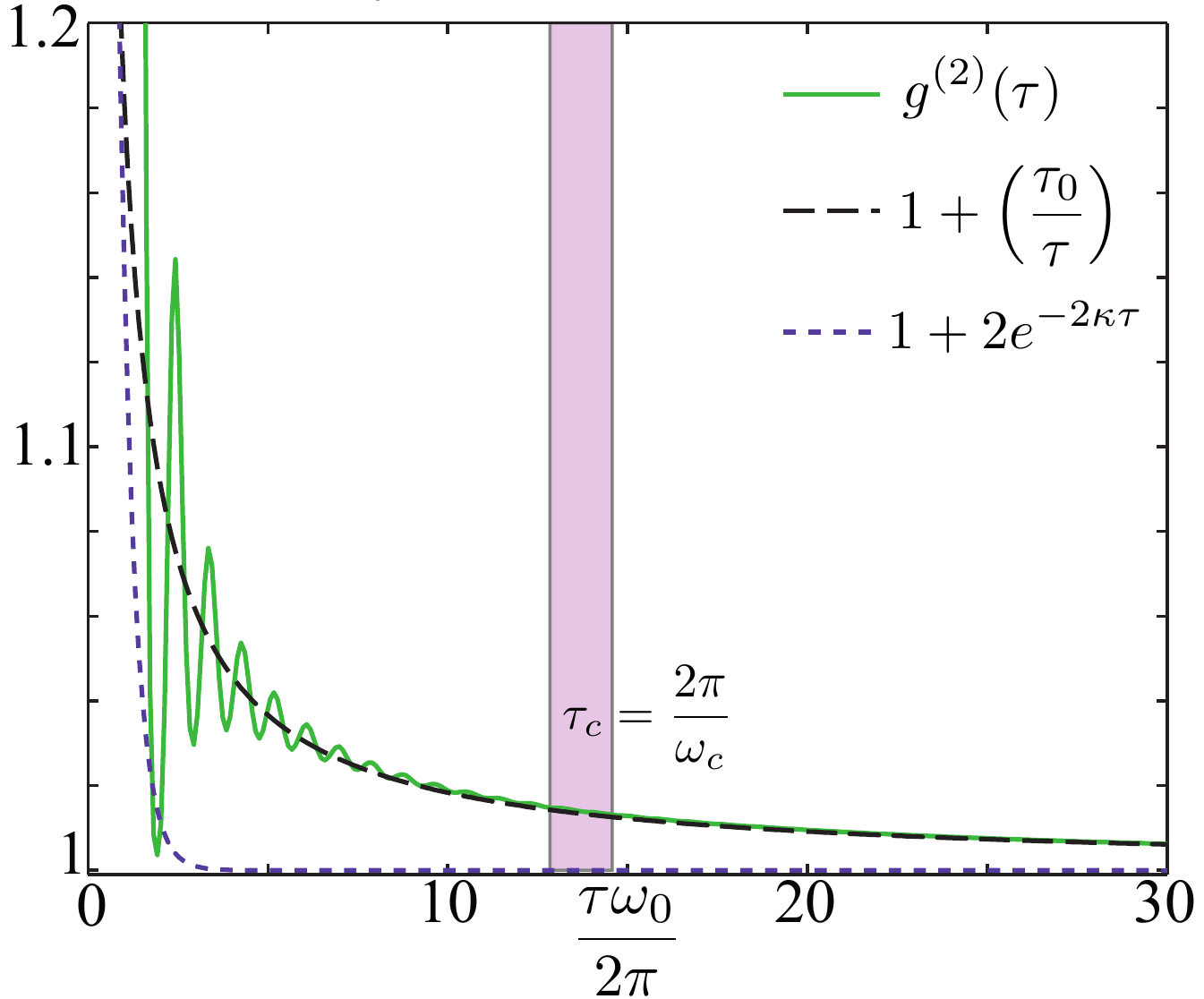}
    \caption
    {\label{fig:g2}
    Algebraic decay of the photon four point correlation function 
 $g^{(2)}(\tau)=\langle a^{\dagger}(0)a^{\dagger}(\tau)a(\tau)a(0)\rangle/n^2$ at long times, for parameters $\omega_p=1, \kappa=0.4, \omega_z=6, K=0.16$, see also Ref.~\cite{Buchhold2013}. 
The algebraic decay sets in at the inverse crossover frequency  $\omega_c$, given by Eq.~\eqref{KeyEq1}. The corresponding exponential decay of $g^{(2)}$ in the normal phase is plotted for comparison. Figure copied from Ref.~\cite{Buchhold2013}. (Copyright (2013) by The American Physical Society)}
\end{figure}

The strong light-matter interactions not only lead to thermalization of the atomic and photonic subsystems, they also feature a complete imprint of the glass dynamics of the atoms onto the cavity photons. This results in universal spectral properties of the photonic response, as has been shown in Ref.~\cite{Buchhold2013}, and concerns a complete locking of both subsystem's low energy response properties. The photons form a photon glass state, which features a large, incoherent population of photon modes, signaled by a finite Edwards-Anderson parameter in the photon correlation function
\eq{GlassEA}{
G^K_{\mbox{\tiny ph}}(\omega)=4\pi\delta(\omega)\tilde{q}_{\mbox{\tiny EA}}+G^K_{\mbox{\tiny reg}}(\omega),}
which shows the same scaling behavior as the atomic Edwards-Anderson parameter close to the glass transition $\tilde{q}_{\mbox{\tiny EA}}\sim q_{\mbox{\tiny EA}}$. A finite $\tilde{q}_{\mbox{\tiny EA}}$ implies long temporal memory in photon autocorrelation functions and an extensive number of photons permanently occupying a continuum of high energy modes. The latter is revealed by an algebraic decay of the system's correlation functions, as for instance for the four point (or $g^{(2)}$) correlation function, which is shown in Fig.~\ref{fig:g2}. 

The photon response and correlation functions are easily accessible via cavity photon output measurements, and therefore represent the natural observables for the detection of the glass dynamics in cavity QED experiments. 
In fact, in Ref.~\cite{Buchhold2013}, it has been demonstrated that a complete characterization of the glass phase can be performed by different but standard photon output measurements, such as homodyne detection and intensity correlation measurements.
Formally, the photon Green's functions can be obtained from the Keldysh action formalism by introducing source fields $\mu_{\alpha\alpha'}$ in the microscopic action, which couple to the photon variables. After integrating out the photon variables in the action, the photon Green's functions are then obtained via functional derivatives with respect to these source fields. This standard field theory technique relates the photon Green's functions to the solution for the atomic response and correlation functions~\cite{Buchhold2013}.

\subsubsection{Quenched and Markovian bath coupling in the Keldysh formalism}

The Keldysh path integral formalism represents a theoretical approach which
incorporates in a very straightforward way the coupling of the system to a
non-thermal bath. Here we compare quenched disorder and Markovian baths in more detail. As shown above in the present section, the coupling of the system
variables to quenched disorder, realized by the variables $J_{lm}$, is equivalent
to the coupling to a bath with infinite correlation time, \eq{Glassin}{ \langle
  J_{lm}(t)J_{l'm'}(t')\rangle=\frac{K}{N}(\delta_{ll'}\delta_{mm'}+\delta_{lm'}\delta_{ml'}).
} This represents the opposite limit of a Markovian bath, which has a typical
correlation time that is much shorter than the system correlation time. As a
consequence, at each single system-bath interaction event, the Markovian bath
immediately equilibrates and looses all its memory on the interaction. On the
other hand, the quenched bath equilibrates infinitely slowly and keeps memory on
each single system-bath interaction. As a consequence, it introduces effective
temporally long range interactions between the system variables. These tend to
slow down the system, pronouncing its $\omega=0$ correlations as expressed by
the Edwards-Anderson parameter.

In an equilibrium formalism, disordered systems have to be dealt with a
computationally demanding replica method~\cite{Read1993,Read1995}, which is
furthermore physically far less transparent than the Keldysh formalism, which
thus represents a convenient approach to disordered systems. In open quantum
systems, where disorder, i.e., the coupling to a quenched bath, comes together
with the dissipation introduced by a Markovian bath, both types of bath couplings
compete with each other, leading in the present glassy system to strongly
modified spectral properties, which mirror the effect of the Markovian bath by a
crossover from non-equilibrium to equilibrium scaling behavior.

%%% Local Variables:
%%% mode: latex
%%% TeX-master: "dds_review"
%%% End:

\section{Non-equilibrium stationary states: bosonic models}
\label{sec:bosons}

Recent years have seen tremendous progress in experiments on exciton-polaritons
in semiconductor microcavities (for reviews see, e.g.,
Refs.~\cite{Carusotto2013,Byrnes2014}), making these systems the prime
candidates for studying condensation phenomena under non-equilibrium
conditions. As we have already mentioned in Sec.~\ref{sec:excit-polar-syst}, the
fundamental difference from conventional condensates is due to the finite
life-time of exciton-polaritons, which makes continuous driving necessary to
maintain a steady population.

What makes these systems genuinely \emph{driven-dissipative}, is that they are
coupled to \emph{several} baths, or to time-dependent driving fields, with which
they exchange particles and energy. In the case of exciton-polaritons, which are
hybrid quasiparticles composed of a Wannier-Mott exciton and a photon, the
photonic component can leak out through the mirrors forming the cavity. Thus,
the electromagnetic vacuum outside the cavity serves as a reservoir into which
particles are lost. These losses have to be compensated by laser driving. In
many experiments, the driving laser is far blue-detuned from the bottom of the
lower polariton band, and thus coherently creates high-energy
excitations. During the relaxation of the latter, which is caused by
phonon-polariton and stimulated polariton-polariton scattering, coherence is
quickly lost. In other words, lower polaritons can be regarded as being pumped
not directly by the laser but rather by a reservoir of high energy
excitons. Several approaches have been used to model this effectively incoherent
pumping mechanism~\cite{Wouters2007a,Szymanska2006,Szymanska2007}. In this
context, the description of a driven-dissipative condensate introduced in
Sec.~\ref{sec:driv-diss-cond} might be considered as a toy model, which hides
all the microscopic details associated with coherent excitation and subsequent
relaxation of high energy excitons in the coupling of the system to several
independent baths.

In the following, we review recent investigations of the universal
long-wavelength scaling properties of driven-dissipative
condensates~\cite{Sieberer2013,Sieberer2014,Altman2015,He2014,Gladilin2014,Ji2015}.
Then, the precise details of the chosen model cease to matter: according to the
power-counting arguments given in Sec.~\ref{sec:semicl-limit-keldysh}, for any
choice of a microscopic model the effective long-wavelength description is given
by the semiclassical Langevin equation for the condensate field~\eqref{eq:17}.
In fact, the latter can be derived in the spirit of Ginzburg-Landau theory for
continuous phase transitions~\cite{Lifshitz1980} by writing down the most
general equation that is compatible with the symmetries of the problem. The
reasoning behind such an approach is that the universal properties are fully
determined by the spatial dimensionality and symmetries of a physical
system~\cite{Amit/Martin-Mayor,Zinn-Justin2002,Goldenfeld1992}.

A comparison in terms of symmetries of systems exhibiting driven-dissipative
Bose-Einstein condensation with systems in thermal equilibrium can be given most
conveniently on the basis of the semiclassical limit of
Sec.~\ref{sec:semicl-limit-keldysh}: indeed, the equation of
motion~\eqref{eq:17} for the classical field takes the form of the Langevin
equations that are used to model universal dynamics in thermal equilibrium
phenomenologically~\cite{Hohenberg1977}. Structurally, equation~\eqref{eq:17} is
similar to model A for a non-conserved order parameter, with the additional
inclusion of reversible mode couplings~\cite{DeDominicis1975}. However, the
equilibrium symmetry discussed in Sec.~\ref{sec:therm-equil-sym}, which is
present in all models of Ref.~\cite{Hohenberg1977}, is violated in
driven-dissipative systems. This violation is due to pumping and loss terms,
which moreover lead to the absence of particle number conservation. The latter typically is associated
with the symmetry under quantum phase rotations (see
Sec.~\ref{sec:energy-momentum-particle-numb-cons}). This is another crucial
difference to Bose-Einstein condensation in equilibrium: there, particle number
conservation entails the existence of a slow dynamical mode that modifies the
universal properties, and is taken into account in model F of
Ref.~\cite{Hohenberg1977}. To summarize, driven-dissipative condensation differs
from Bose-Einstein condensation in equilibrium by the absence of the equilibrium
symmetry and the symmetry under quantum phase rotations. This difference lies at
the heart of the novel universal behavior out of equilibrium discussed in the
following.

In Sec.~\ref{sec:driv-diss-cond}, we analyzed driven-dissipative condensates
within mean-field theory, disregarding fluctuations around the homogeneous
condensate mode. However, in order to access universal aspects such as the
behavior of correlations of the condensate field at large distances or dynamical
critical phenomena at the condensation transition, one has to carefully include
the effect of fluctuations. This can be done gradually --- first integrating out
short-scale fluctuations and moving on to account for fluctuations on larger and
larger scales --- by means of RG methods, such as the FRG discussed in
Sec.~\ref{sec:open-sys-FRG}. Then, an intriguing question is, whether the
non-equilibrium nature of driven-dissipative condensates at the microscopic
scale becomes more or less pronounced under renormalization. In the former case,
effective equilibrium is established at large scales, while in the latter case,
the universal physics is expected to be profoundly different from its
equilibrium counterpart. Which of these scenarios is realized depends crucially
on the spatial dimensionality of the system: in 3D the long-wavelength regime is
effectively thermal, whereas in one- and two-dimensional systems the deviation
from equilibrium is relevant in the RG sense, and these systems are governed by
strongly non-equilibrium RG fixed points.

To see how this comes about, we require a quantitative measure for the deviation
from equilibrium conditions in a driven-dissipative condensate. In
Sec.~\ref{sec:therm-equil-sym} we discussed that thermal equilibrium requires
the ratios of coherent to corresponding dissipative couplings to take a common
value. This is illustrated in Fig.~\ref{fig:noneq} and formalized in
Eq.~\eqref{eq:168067}. In turn, it implies that any deviation
$K_c/K_d \neq u_c/u_d$ in the values of these ratios directly indicates a
deviation from equilibrium conditions. Hence, the quantity $\lambda$ defined
by~\cite{Altman2015}
\begin{equation}
  \label{eq:37}
  \lambda = - 2 K_c \left( 1 - \frac{K_d u_c}{K_c u_d} \right)
\end{equation}
can serve as a quantitative measure for the departure from thermal
equilibrium. Note that $\lambda = 0$ in equilibrium, and that $\lambda$ is
well-defined also for $K_d = 0$, which is the microscopic value of the diffusion
constant $K_d$ in the model for driven-dissipative condensates introduced in
Sec.~\ref{sec:driv-diss-cond}. It turns out to be most convenient to combine
$\lambda$ with the quantities~\cite{Altman2015}
\begin{equation}
  \label{eq:47}
  D = K_c \left( \frac{K_d}{K_c} + \frac{u_c}{u_d} \right), \quad \Delta 
  = \frac{\gamma}{2 \rho_0} \left( 1 + \frac{u_c^2}{u_d^2} \right)
\end{equation}
(see Sec.~\ref{sec:driv-diss-cond} for the definition of the noise strength
$\gamma$ and the mean-field condensate density $\rho_0$) to define the
dimensionless non-equilibrium strength as~\cite{Tauber2014a}:
\begin{equation}
  \label{eq:64}
  g = \Lambda_0^{d - 2} \frac{\lambda^2 \Delta}{D^3},
\end{equation}
where $\Lambda_0$ is the UV momentum cutoff. Thus, the answer to the question of
whether equilibrium vs. non-equilibrium universal behavior is realized in
driven-dissipative condensates, is encoded in the RG flow of $g$.

\begin{figure}
  \centering
  \includegraphics[width=.7\linewidth]{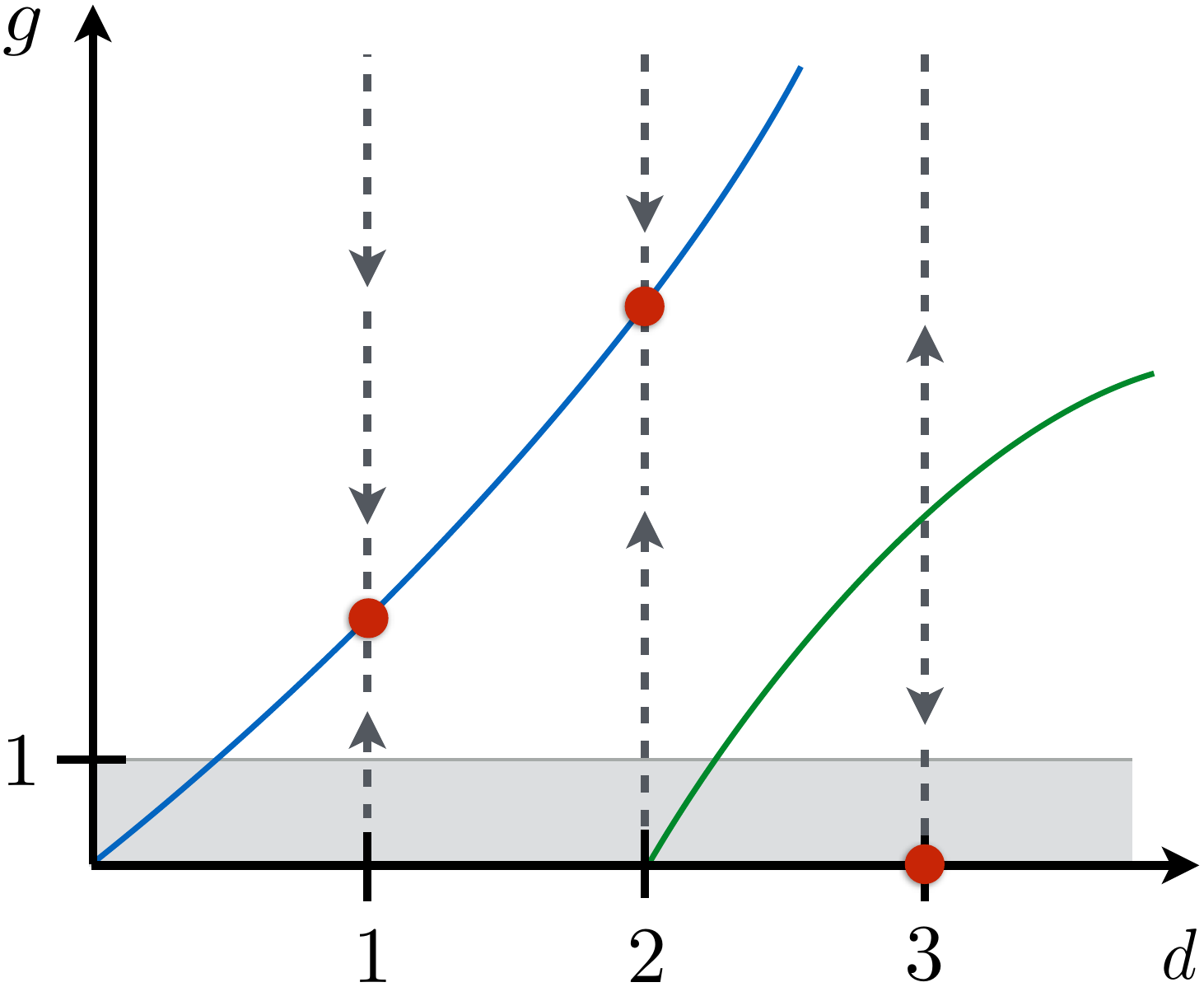}
  \caption{Equilibrium vs.\ non-equilibrium phase diagram for driven-dissipative
    condensates (cf.\ Ref.~\cite{Canet2010}). The line $g = 1$, where $g$ is
    defined in Eq.~\eqref{eq:64} and measures the deviation from equilibrium
    conditions, separates the close-to-equilibrium regime for $g < 1$ from the
    strong-coupling, far-from-equilibrium regime at $g > 1$.  Red dots indicate
    the fixed-point values of $g$ that are reached if the RG flow is initialized
    in the close-to-equilibrium regime. In dimensions one and two, the
    equilibrium fixed point at $g = 0$ is unstable, and the RG flow along the
    dashed lines is directed towards the blue line of strong-coupling fixed
    points. Thus, a system that is microscopically close to equilibrium will
    exhibit strongly non-equilibrium behavior at large scales. On the other
    hand, in three spatial dimensions, an initially small value of $g$ is
    diminished under renormalization, and the universal large-scale behavior is
    governed by the effective equilibrium fixed point at $g = 0$. The green line
    indicates the existence of a critical value $g_c$ in $d > 2$, corresponding
    to a transition between the effective equilibrium phase and a true
    non-equilibrium phase that is realized for large microscopic values
    $g > g_c$.}
  \label{fig:KPZ_phase_diagram}
\end{figure}

In the condensed phase, the RG flow of $g$ is driven dominantly by fluctuations
of the gapless Goldstone mode discussed in Sec.~\ref{sec:goldstone}, i.e., by
fluctuations of the phase of the condensate field. The latter were
shown~\cite{Sivashinsky1977,Kuramoto1984,Grinstein1993,Grinstein1996} to be
governed by the Kardar-Parisi-Zhang (KPZ) equation~\cite{Kardar1986}, in which
$\lambda$ defined in Eq.~\eqref{eq:37} appears as the coefficient of the
characteristic non-linear term, see Eq.~\eqref{eq:243456}. Below in
Sec.~\ref{sec:dens-phase-repr}, we present an alternative mapping of the
long-wavelength condensate dynamics to the KPZ equation, starting from the
Keldysh action in Eq.~\eqref{eq:112} and integrating out the gapped density mode
within the Keldysh functional integral. As a consequence of the mapping to the
KPZ equation, the RG flow of $g$ in $d$ spatial dimensions is at the one-loop
level given by~\cite{Tauber2014a}
\begin{equation}
  \label{eq:65}
  \partial_{\ell} g = - \left( d - 2 \right) g + \frac{\left( 2 d - 3 \right)
    C_d}{2 d} g^2,
\end{equation}
where $\ell = \ln(\Lambda/\Lambda_0),$ $\Lambda$ is the running momentum cutoff,
and $C_d = 2^{1-d} \pi^{-d/2} \Gamma(2 - d/2)$ is a geometric factor. The key
role that is played by spatial dimensionality becomes manifest in the canonical
scaling of $g$, which is encoded in the first term on the RHS of the flow
equation: to wit, $g$ is relevant in 1D where $d - 2 < 0$, marginal in 2D, and
irrelevant in 3D since then $d - 2 > 0$. In 2D, the loop correction --- the
second term on the RHS of Eq.~\eqref{eq:65} --- is positive, making $g$
marginally-relevant. This has far-reaching consequences for a driven-dissipative
condensate in which the microscopic value of $g$ is small, i.e., which is close
to equilibrium: upon increasing the scale at which the system is observed, the
non-equilibrium nature is more pronounced in one- and two-dimensional systems,
whereas effective equilibrium is established on large scales in
three-dimensional systems. In 1D the canonical scaling towards strong coupling
is balanced at an attractive strong-coupling fixed point (SCFP) $g_{*}$ by the
loop correction. This term vanishes at $d = 3/2,$ and for $d > 3/2$ the one-loop
equation does not have a stable SCFP, which, however, is recovered in a
non-perturbative FRG approach~\cite{Canet2010,Canet2011,Canet2012}. The RG flow
of $g$ that is found within this approach is illustrated qualitatively in
Fig.~\ref{fig:KPZ_phase_diagram}, which shows that also in 2D the flow is out of
the shaded close-to-equilibrium regime with $g < 1$, and towards a
strong-coupling, non-equilibrium fixed point. The situation is quite different
in 3D: in this case, if the microscopic value of $g$ is small, at large scales
an effective equilibrium with a renormalized value $g \to 0$ is
reached. However, for $d > 2$, there exists a critical line of unstable fixed
points $g_c$, separating the basins of attraction of the equilibrium and
non-equilibrium fixed points, for $g < g_c$ and $g > g_c$ respectively. Thus, in
addition to the effective equilibrium phase, a true non-equilibrium phase may be
reached in systems that are far from equilibrium even at the microscopic level
also in 3D~\cite{Fisher1992}. The properties of this phase have not been
explored so far.

The rest of this section is organized as follows: in
Sec.~\ref{sec:critical-dynamics-3d}, we review the dynamical critical behavior
at the driven-dissipative condensation transition in
3D~\cite{Sieberer2013,Sieberer2014,Tauber2014b}, which, according to the above
discussion, is governed by an effective equilibrium fixed point. Signatures of
the non-equilibrium nature of the microscopic model are present in the
asymptotic fade-out of the deviation from equilibrium at large scales. In
contrast, the universal scaling behavior of driven-dissipative condensates in
both 2D~\cite{Altman2015} and 1D~\cite{Gladilin2014,Ji2015,He2014} is quite
distinct from the equilibrium case and governed by the SCFP of the KPZ
equation. We review the resultant physical picture in
Secs.~\ref{sec:absence-algebr-order} and~\ref{sec:kpz-scaling-1d}, respectively.

\subsection{Density-phase representation of the Keldysh action}
\label{sec:dens-phase-repr}

Especially in reduced dimensions, at large scales the properties of condensates
are vitally influenced by fluctuations of the Goldstone mode. To name an
example, in 2D, both in~\cite{Mermin1966} and out of
equilibrium~\cite{Altman2015} these fluctuations lead to a suppression of long
range correlations. In Ref.~\cite{Altman2015}, an effective long-wavelength
description of a driven-dissipative condensate with the condensate phase as the
single dominant gapless degree of freedom (cf.\ Sec.~\ref{sec:goldstone}) has
been formulated starting from the Langevin equation~\eqref{eq:17} for the
complex order parameter.  Here we present an alternative and direct derivation
within the Keldysh functional integral
formalism~\cite{Kamenev2011,Altland2010}. It is based on the Keldysh action in
terms of which the microscopic theory of a bosonic many-body system with
particle loss and gain is formulated in Sec.~\ref{sec:driv-diss-cond}. This
differs from the derivation presented in~\cite{Altman2015}, which was based on
the detour over the Langevin equation~\eqref{eq:17} that effectively captures
the physics on a mesoscopic scale (cf.\ Fig.~\ref{fig:quantum_to_classical} and
the discussion in Sec.~\ref{sec:semicl-limit-keldysh}). On this level, amplitude
fluctuations can be eliminated leading to the KPZ equation for the phase.  In
this sense, we establish here a closer link between microscopic and mesoscopic
theories, which is both appealing from a theoretical point of view and brings
about a number of technical advantages. For example, physical observables are
usually represented by quantum mechanical operators or equivalently in terms of
fields in a Keldysh functional integral description of the microscopic theory;
here our approach comes in handy as it yields the effective long-wavelength form
of generating functionals for expectation values and correlation functions of
these observables which can then be evaluated further utilizing established
approximation strategies.

Apart from specific applications, the derivation of the action for the Goldstone
mode presented here deepens our understanding of general properties of the
Keldysh formalism with regard to phase rotation symmetries. The crucial point
is, that classical phase rotations introduced in
Sec.~\ref{sec:energy-momentum-particle-numb-cons} are a symmetry of the Keldysh
action both in a closed system and in the presence of terms that describe
incoherent pumping and losses, and as we have seen in Sec.~\ref{sec:goldstone},
the spontaneous breaking of this symmetry is sufficient to ensure the appearance
of a Goldstone mode that corresponds to fluctuations of the phase of the order
parameter~\cite{Wouters2006,Szymanska2006,Wouters2007a,Wouters2007b}. In the
basis of classical and quantum fields $\phi_{c, q}$ such phase rotations become
$\phi_{c, q} \mapsto \phi_{c, q} e^{i \alpha}$, showing that the Goldstone mode
corresponds to joint phase fluctuations of both the classical and the quantum
fields. Therefore, in order to derive the action for the Goldstone boson we
represent the fields in the form
\begin{equation}
  \label{eq:52}
  \phi_c = \sqrt{\rho} e^{i \theta}, \quad \phi_q = \zeta e^{i \theta},  
\end{equation}
where the density $\rho$ is real whereas $\zeta$ is a complex variable.

The low-energy effective action for the Goldstone boson $\theta$ in a
driven-dissipative system can be derived by integrating out fluctuations of the
density $\rho$ in Eq.~\eqref{eq:52} in the Keldysh partition function with
action $S$ given by Eq.~\eqref{eq:112},
\begin{equation}
  \label{eq:114}
  Z = \int \mathscr{D}[\phi_c, \phi_c^{*}, \phi_q, \phi_q^{*}] e^{i S} = \int
  \mathscr{D}[\rho, \theta, \zeta, \zeta^{*}] e^{i S}. 
\end{equation}
The equality of the integrals over complex classical and quantum fields and the
variables introduced in the transformation Eq.~\eqref{eq:52} follows from the
fact that this transformation leaves the integration measure invariant, i.e., we
have
$\mathscr{D}[\phi_c,\phi_c^{*},\phi_q,\phi_q^{*}] =
\mathscr{D}[\rho,\theta,\zeta,\zeta^{*}]$.
Note that this would not be the case if instead of the density we introduced the
amplitude of $\phi_c$ as a degree of freedom. Our goal is then to treat the
integrals over $\rho$ and $\zeta$ in Eq.~\eqref{eq:114} in a saddle-point
approximation, which, as we show below, is justified since fluctuations of the
density are gapped in the ordered phase with $r_d < 0$ and hence expected to be
small. The first step is thus to find the saddle point, i.e., the solutions to
the classical field equations (cf.\ Eqs.~\eqref{eq:59})
\begin{equation}
  \label{eq:115}
  \frac{\delta S}{\delta \rho} = 0, \quad \frac{\delta S}{\delta \zeta} = 0.
\end{equation}
For $r_d < 0$, we recover the mean-field solution of
Sec.~\ref{sec:driv-diss-cond}, given by $\rho = \rho_0 = - r_d/u_d = - r_c/u_c$
(note that the last equality can always be satisfied by performing a gauge
transformation to adjust the value of $r_c$ as described in
Sec.~\ref{sec:driv-diss-cond}) and $\zeta = 0$. We proceed to expand the action
Eq.~\eqref{eq:112} to second order in fluctuations of $\rho$ and $\zeta$ around
the saddle point. Note that the quantum vertex in the action Eq.~\eqref{eq:112}
that is cubic in the quantum fields does not contribute at this order. Denoting
the density fluctuations as $\pi = \rho - \rho_0$ we find
\begin{multline}
  \label{eq:67}
  S = 2 \int_{t, \mathbf{x}} \left( \sqrt{\rho_0} \left\{ - \zeta_1 \left[ \partial_t \theta +
        K_c \left( \nabla \theta \right)^2 \right] + K_c \zeta_2 \nabla^2 \theta
    \right. \right. \\ \left. \left. \vphantom{\left( \nabla \theta \right)^2} -
      \left( u_c \zeta_1 - u_d \zeta_2 \right) \pi \right\} + i \left( \gamma +
      2 u_d \rho_0 \right) \abs{\zeta}^2 \right),
\end{multline}
where $\zeta_1$ and $\zeta_2$ are, respectively, real and imaginary parts of
$\zeta$. Here, of all terms involving the products of fluctuations $\zeta_1 \pi$
and $\zeta_2 \pi$ we only keep the dominant ones in the long-wavelength limit,
i.e., we neglect contributions containing temporal derivatives
$\zeta_2 \partial_t \pi, \zeta_1 \pi \partial_t \theta$, and spatial derivatives
$\zeta_1 \nabla^2 \pi, \zeta_2 \nabla \pi \cdot \nabla \theta,$ and
$\zeta_1 \pi \left( \nabla \theta \right)^2$ which are both small as compared to
the mass-like contributions $u_c \zeta_1 \pi$ and $u_d \zeta_2 \pi$ in
Eq.~\eqref{eq:67} for the Goldstone mode $\theta$, in the regime $\omega \to 0$ for
$\mathbf{q} \to 0$. Note that terms of higher order in $\pi$ and
$\zeta$ are contained in the original action Eq.~\eqref{eq:112} both in
contributions involving derivatives and in the coherent and dissipative
vertices. The validity of the saddle-point approximation, therefore, is
restricted to the low-frequency and low-momentum sector in a weakly interacting
system.

The action~\eqref{eq:67} resulting from this expansion is linear in $\pi$ and
hence integration over this variable is trivial and yields a
$\delta$-functional, which in turn facilitates integration over the imaginary
part $\zeta_2$ of $\zeta$,
\begin{equation}
  \label{eq:68}  
  Z = \int
  \mathscr{D}[\theta,\zeta_1,\zeta_2] \, \delta[u_c \zeta_1 - u_d \zeta_2] e^{i
    S'} = \int \mathscr{D}[\theta, \tilde{\theta}] \, e^{i S_{\mathrm{KPZ}}},
\end{equation}
where at each step a normalization factor is implicitly included in the
integration measure, ensuring $Z = 1$~\cite{Kamenev2011,Altland2010}. In the
last equality we replaced $\zeta_1$ by the KPZ response field
$\tilde{\theta} = i 2 \sqrt{\rho_0} \zeta_1$, and the KPZ action
$S_{\mathrm{KPZ}}$ is given by~\cite{Tauber2014a,Kamenev2011}
\begin{equation}
  \label{eq:38}
  S_{\mathrm{KPZ}} = \int_{t, \mathbf{x}} \tilde{\theta} \left[ \partial_t \theta - D \nabla^2 \theta -
    \frac{\lambda}{2} \left( \nabla \theta \right)^2 - \Delta \tilde{\theta} \right],
\end{equation}
where the diffusion constant, non-linear coupling, and noise strength,
respectively, are expressed in terms of the microscopic parameters in the
original action Eq.~\eqref{eq:112} as
\begin{equation}
  \label{eq:32}  
  D = K_c \frac{u_c}{u_d}, \quad \lambda = - 2 K_c, \quad \Delta =
  \frac{\gamma + 2 u_d \rho_0}{2 \rho_0} \left( 1 + \frac{u_c^2}{u_d^2} \right).
\end{equation}
Along the lines of the derivation of the Langevin equation~\eqref{eq:17} from
the action in the semiclassical limit in Eq.~\eqref{eq:140}, the KPZ action can
be seen to be equivalent to the KPZ equation, which reads
\begin{equation}
  \label{eq:243456}
  \partial_t \theta = D \nabla^2 \theta + \frac{\lambda}{2} \left( \nabla
    \theta \right)^2 + \eta,
\end{equation}
where the stochastic noise $\eta$ has zero mean,
$\langle \eta(t, \mathbf{x}) \rangle = 0$, and is Gaussian with second moment
$\langle \eta(t, \mathbf{x}) \eta(t', \mathbf{x}') \rangle = 2 \Delta \delta(t -
t') \delta(\mathbf{x} - \mathbf{x}')$.
Originally~\cite{Halpin-Healy1995,Kardar1986}, Eq.~\eqref{eq:243456} was
suggested by Kardar, Parisi, and Zhang as a model to describe the growth of a
surface, e.g., due to the random deposition of atoms. In this context,
$h = \theta$ is the height of the surface, and the origin of the non-linear term
is purely geometric~\cite{Kardar1986}: the growth is assumed to occur in a
direction that is locally normal to the surface and at a rate $ds/dt = \lambda$;
If an increment $ds = \lambda dt$ is added along the normal, the corresponding
change of the surface height is
\begin{equation}
  \label{eq:125}
  dh = \sqrt{\left( \lambda dt \right)^2 + \left( \lambda dt \nabla h \right)^2}
  \approx \left[ \lambda + \frac{\lambda}{2} \left( \nabla h \right)^2 \right] dt.
\end{equation}
Removing from $dh/dt$ the average deposition rate $\lambda$ by a transformation
to a co-moving frame, $h(t, \mathbf{x}) \mapsto h(t, \mathbf{x}) + \lambda t$,
and adding a term $D \nabla^2 h$ that describes surface tension, we obtain the
(deterministic part of the) growth equation~\eqref{eq:243456}. Intuitively, it
seems clear that the growth of a surface represents a genuine non-equilibrium
process. More formally, this can be seen by noting that the non-linear term in
the KPZ equation does in general not satisfy a potential
condition~\cite{Tauber2014a}.\footnote{The most general Langevin equation
  describing (near-) equilibrium dynamics contains both (i) relaxational and
  (ii) reversible contributions to the deterministic
  dynamics~\cite{Tauber2014a}. The linear diffusion term in
  Eq.~\eqref{eq:243456} is of type (i): it can be written as
  $- \delta \mathcal{H}/\delta \theta,$ where
  $\mathcal{H} = \frac{D}{2} \int_{\mathbf{x}} \left( \nabla \theta \right)^2$,
  and this term alone would correspond to relaxation to an equilibrium
  stationary distribution $\propto e^{- \mathcal{H}/\Delta}$. The non-linear
  term, on the other hand, can not be represented as the derivative of a
  Hamiltonian functional and is hence of type (ii). However, for reversible
  contributions to be compatible with a thermal stationary state, they have to
  satisfy a potential condition. This is not the case for the non-linear term in
  the KPZ equation in dimensions $d > 1$.}

As pointed out above, an alternative derivation of the KPZ
equation~\eqref{eq:243456} as the effective long-wavelength description of
driven-dissipative condensates starts from the Langevin
equation~\eqref{eq:17}~\cite{Altman2015}. Then, the coefficients in the KPZ
equation are slightly different, and they are given by Eqs.~\eqref{eq:37}
and~\eqref{eq:47} instead of Eq.~\eqref{eq:32}. To be specific, the differences
are (i) the absence in Eq.~\eqref{eq:47} of the tree-level shift
$\propto u_d \rho_0$ of the noise strength $\Delta$ in Eq.~\eqref{eq:32}, and
(ii) the absence in Eq.~\eqref{eq:32} of the terms proportional to $K_d$ in
Eqs.~\eqref{eq:37} and~\eqref{eq:47}. (i) is due to the fact that in the
Langevin equation~\eqref{eq:17}, which is valid in the semiclassical limit (see
Sec.~\ref{sec:semicl-limit-keldysh}), the quantum vertex that is proportional to
$\phi_c^{*} \phi_c \phi_q^{*} \phi_q$ in the action in Eq.~\eqref{eq:112} is
neglected. (ii) results from the absence of the diffusion term $K_d$ in the
microscopic model~\eqref{eq:112}; on the other hand, in the Langevin
equation~\eqref{eq:17} this term is included, as it is generated by integrating
out fluctuations with wavelengths below the mesoscopic scale on which the
Langevin equation is valid (cf.\ Fig.~\ref{fig:quantum_to_classical}).

Starting from the effective long-wavelength description of the condensate
dynamics derived in this section, the universal scaling properties can be
obtained from an RG analysis. For the KPZ equation, this procedure, which leads
at lowest order in a perturbative expansion in $\lambda$ to the one-loop flow
equation~\eqref{eq:65},\footnote{Due to symmetries of the KPZ equation (for a
  comprehensive discussion see Ref.~\cite{Kloss2012}), the RG flow is described
  by the single parameter $g$ defined in Eq.~\eqref{eq:64}.} is described, e.g.,
in Refs.~\cite{Tauber2014a,Kamenev2011}.

For completeness, we note that in the absence of drive and dissipation, i.e.,
for $K_d = r_d = u_d = 0$, an analogous derivation for a weakly interacting Bose
gas in thermal equilibrium leads to an effective action for the phase alone that
is given by
\begin{equation}
  \label{eq:sfd_102}
  S =  \int_{t, \mathbf{x}} \tilde{\theta} \left( \partial_t^2 - c^2 \nabla^2 \right) \theta,
\end{equation}
where $c = \sqrt{2 K_c u_c \rho_0}$ is the speed of sound, and describes the
dissipationless propagation of sound waves with linear dispersion
$\omega = c q.$ To actually describe thermal equilibrium, this action has to be
supplemented by infinitesimal regularization terms, discussed at the end of
Sec.~\ref{sec:single-mode-cavity}.

Finally, let us comment on the relation of the approach presented here to a
Bogoliubov expansion in fluctuations of the complex fields around the mean field
average values, $\delta \phi_c = \phi_c - \sqrt{\rho_0}$ and
$\delta \phi_q = \phi_q$. The latter can be recovered formally by expanding the
transformation Eq.~\eqref{eq:52} around the arbitrarily chosen value
$\theta = 0$, which yields
\begin{equation}
  \label{eq:99}
    \phi_c  = \sqrt{\rho_0} + \frac{\pi}{2 \sqrt{\rho_0}} + i \sqrt{\rho_0}
    \theta, \quad \phi_q  = \zeta,
\end{equation}
and interpreting $\pi$ and $\theta$ as the real and imaginary parts of
$\delta \phi_c$.  In the KPZ action in Eq.~\eqref{eq:38}, such an expansion in
$\theta$ would amount to neglecting the non-linearity. Without the
non-linearity, however, the KPZ action describes a free field coupled to a
thermal bath. In other words, by performing a Bogoliubov approximation the
non-equilibrium character that is intrinsic to the microscopic model with
action~\eqref{eq:112} is lost in the long-wavelength limit. Formally, this
failure is due to the fact that the saddle-point approximation
is not valid for low-momentum modes in the Goldstone direction since they are
not gapped.

\subsection{Critical dynamics in 3D}
\label{sec:critical-dynamics-3d}

In three spatial dimensions, the deviation from thermodynamic equilibrium
conditions, which is quantified by the value of $g$ defined in
Eq.~\eqref{eq:64}, is irrelevant in the RG sense (cf.\
Eq.~\eqref{eq:65}). Hence, at large scales, effective equilibrium is
established. This immediately implies that the overall picture of Bose-Einstein
condensation in thermal equilibrium in 3D remains valid also in the
driven-dissipative context: in particular, the mean-field analysis of
Sec.~\ref{sec:driv-diss-cond}, which predicts a continuous phase transition
beyond which the classical phase rotation symmetry (cf.\
Sec.~\ref{sec:energy-momentum-particle-numb-cons}) is spontaneously broken and
off-diagonal long-range order is established, is modified \emph{quantitatively}
but not \emph{qualitatively} once fluctuations around the mean-field condensate
are taken into account. (Note, however, that in equilibrium the condensation
transition is induced by lowering the temperature to a critical value, whereas
driven-dissipative condensation is established by tuning the single-particle
pump rate.) Yet, the non-equilibrium nature of the driven-dissipative system
leaves its mark on the approach to the long-wavelength thermalized regime in a
fully universal way.

An equilibrium system, fine-tuned to a critical point, beyond which spontaneous
symmetry breaking occurs via a second order phase transition, exhibits universal
behavior. This is witnessed in non-analyticities in the free energy, and in the
long-range decay properties of correlation and response functions. The decay
properties are then governed by power laws, freed from the generic exponential
cutoff $\sim e^{-r/\xi}$ at large distances $r$ due to the divergent correlation
length $\xi \to \infty$, defining the critical point. While the concept of a
free energy is not meaningful in a non-equilibrium context, correlations and
responses can be considered also in the latter case. Then, universality entails
that the algebraic long-wavelength, long-time decay of any correlation or
response function can be characterized in terms of a set of critical exponents,
which do not depend on the microscopic details of the problem but rather on
symmetries and dimensionality. The critical point is fully characterized by this
set of critical exponents.

We emphasize a crucial difference between the critical behavior in strictly
non-interacting theories described by quadratic Hamiltonians or quantum master
equations, and the more generic --- but also much more complex --- case of
interacting problems (for an example where a Gaussian fixed point is physically
important, cf.\ Sec.~\ref{sec:ising}). The non-interacting critical theories are
described by a so-called Gaussian fixed point of the RG flow, while the
interacting ones are associated to a so-called Wilson-Fisher fixed point. The
structural difference between these fixed points is reflected by the fact that
the values of the critical exponents differ; in particular, at a Gaussian fixed
point, all exponents are rational numbers, reflecting the validity of canonical
power counting (cf. Sec.~\ref{sec:semicl-limit-keldysh}), while one obtains
non-trivial rational or non-rational numbers mirroring the importance of strong
long-wavelength fluctuation corrections at a Wilson-Fisher fixed point. The
latter is more stable than the Gaussian one (where, in fact, all couplings are
relevant in the sense of the RG), and thus the physically relevant one even if
interactions are small on the microscopic scale. This is intuitive, taking into
account that critical behavior deals with the longest distances and timescales
in a physical system.

The impact of fluctuations on the values of the critical exponents can be
described quantitatively in the framework of the renormalization group, which
provides a systematic way to deal with the intricate long-wavelength,
low-momentum divergences caused by the divergent correlation length. The
critical behavior of driven-dissipative systems in 3D was investigated
in~\cite{Sieberer2013,Sieberer2014} based on the functional renormalization
group approach discussed in Sec.~\ref{sec:open-sys-FRG}, and
in~\cite{Tauber2014b} using the field theoretic perturbative renormalization
group~\cite{Tauber2014a}. These studies gave rise to the following picture of
driven criticality:

The RG fixed point is purely dissipative (see Fig.~\ref{fig:noneq} (c)), i.e.,
all complex couplings rotate to the imaginary axis. However, it is the approach
to this fixed point that contains universal information, and this makes it
possible to distinguish equilibrium from non-equilibrium critical behavior. The
following key properties are identified:

\textit{(i) Asymptotic thermalization of correlation functions.} --- Both the
static and dynamical critical exponents, governing, e.g., the asymptotic decay
of the spatial and temporal first order coherence functions, are found to take
the same values as in the corresponding equilibrium problem. This can be made
plausible by the fact that the fixed point couplings indeed lie on a single ray
in the complex plane --- the imaginary axis. More formally, it is understood in
terms of an emergent symmetry implying asymptotic emergence of detailed balance,
or thermalization, at large spatial or temporal distances.

\textit{(ii) Universal decoherence.} --- The approach to the purely dissipative
fixed point still hosts information on the underlying quantum dynamics. The
fadeout of this coherent dynamics is described by a new critical exponent, which
can be shown to be independent of the static and dynamical exponents, i.e., it
does not relate to the latter by scaling relations. Moreover, it can be seen
that no more independent exponents could possibly be
found~\cite{Sieberer2013,Tauber2014b}. This is because the maximum number of
independent exponents is determined by the maxium number of independent
microscopic mass scales~\cite{Goldenfeld1992}, which are, in the quadratic part
of the action~\eqref{eq:140}, given by the real parameters $r_c, r_d,$ and
$\gamma$. In addition, a coupling $f$ to an external source field, corresponding
to a term
$\int_{t, \mathbf{x}} f \left( j_c^{*} \phi_q + j_q^{*} \phi_c + \cc \right)$ in
the action, has to be taken into account, leading in total to four independent
parameters. No more independent scales can be added to the quadratic part of the
action without violating the conditions of (Anti-)Hermiticity and conservation
of probability as explained in
Sec.~\ref{sec:single-mode-cavity}. Correspondingly, the set of four independent
critical exponents (the correlation length exponent $\nu$, the anomalous
dimension $\eta$, the dynamical critical exponent $z$, and new exponent) is
maximal. Finally, the value of this exponent distinguishes equilibrium from
non-equilibrium conditions. This can be understood from Fig.~\ref{fig:noneq}:
the exponent describes the fadeout of coherent couplings, i.e., the approach to
the imaginary axis, governed by a power law $\sim \Lambda^{-\eta_r}$, where
$\eta_r$ is the new exponent. In the equilibrium case, all couplings rotate
uniformly, giving rise to $\eta_r \approx -0.143$ within the FRG approach of
Refs.~\cite{Sieberer2013,Sieberer2014}. In contrast, in the non-equilibrium
case, the slowest approach to the real axis is given by $\eta_r \approx -0.101$.
The value of the non-equilibrium exponent can also be determined analytically
from the field theoretic perturbative renormalization group in a dimensional
expansion, yielding the result (to two-loop
order)~\cite{Tauber2014b}\footnote{The exponent $\eta_c$ calculated in
  Ref.~\cite{Tauber2014b} is related $\eta_r$ via $\eta_r = \eta_c - \eta$.}
\begin{equation}
  \label{eq:79}
  \eta_r = - \frac{2 \left( 4 - d \right)^2}{25} \ln \frac{4}{3}.
\end{equation}
Specifying to $d = 3$ dimensions, we obtain $\eta_r \approx -0.023$. The
discrepancy between this value and the one obtained from the FRG stated above
($\eta_r \approx -0.101$) indicates that the two-loop computation underestimates
fluctuation corrections in three dimensions.\footnote{We note that in
  Ref.~\cite{Sieberer2014} erroneously the value of $\eta_c$ from the field
  theoretic RG was directly compared to $\eta_r$ obtained from the
  FRG. Similarly, in Ref.~\cite{Tauber2014b} the value of $\eta_c$ was by
  mistake compared to $\eta_A$ of Ref.~\cite{Sieberer2014}.}  Moreover,
(possibly significant) quantitative corrections to the values of critical
exponents should also be expected from FRG calculations that go beyond the
truncation used in Refs.~\cite{Sieberer2013,Sieberer2014}.

\textit{(iii) Observability.} --- The drive exponent manifests itself, for
example, in the frequency and momentum resolved dynamical single particle
response as probed in homodyne detection, see Ref.~\cite{Sieberer2014} for
details. It thus corresponds to a direct experimental observable, though its
small value poses a challenge for experimental observation.

In summary, it is found that the \emph{correlation functions} (both static and
dynamic) thermalize, and show identical universal behavior to an equilibrium
critical system with the same symmetries. However, the dynamical \emph{response}
functions contain universal information distinguishing equilibrium from
non-equilibrium systems. The microscopic drive conditions are thus witnessed
even at the largest macroscopic distances in a fully universal way.

In the following, we review how the results summarized above can be obtained
from an open-system functional RG approach~\cite{Sieberer2013,Sieberer2014}. The
basic ingredients of this method are discussed in Sec.~\ref{sec:open-sys-FRG}.

\subsubsection{Effective action for driven-dissipative condensation}
\label{sec:effect-acti-driv}

The Wetterich equation~\eqref{eq:FRG} describes how the scale-dependent
effective action $\Gamma_{\Lambda}$ evolves from the microscopic action $S$ at
$\Lambda = \Lambda_0$ to the full effective action $\Gamma$ at $\Lambda \to 0$,
as fluctuations with momenta larger than the cutoff $\Lambda$ are integrated
out, and the latter is gradually lowered. It is an exact non-linear differential
equation for the \emph{functional} $\Gamma_{\Lambda}$. In the absence of an
exact solution, suitable schemes to approximate the functional differential
equation have to be found. For the analysis of critical behavior, progress can
be made by choosing an ansatz for $\Gamma_{\Lambda}$ that has a similar
structure as the microscopic action $S$ in Eq.~\eqref{eq:112}. In this way, the
effective action $\Gamma_{\Lambda}$ is parameterized in terms of the coupling
constants appearing in the ansatz, and consequently, the functional differential
equation for $\Gamma_{\Lambda}$ can be cast in the form of a set of ordinary
differential equations for the couplings, as we describe in detail below.

Any such ansatz will contain only a \emph{finite} number of couplings, and will
hence truncate the most general structure of $\Gamma_{\Lambda}$. In other words,
by choosing an ansatz of a particular form, one makes an approximation, and the
question arises, which couplings should be included in the ansatz or truncation
in order to obtain meaningful results. For the study of critical phenomena at a
second order phase transition, the power counting scheme of
Sec.~\ref{sec:semicl-limit-keldysh} provides a guideline: following the
arguments given there, we choose a truncation which includes all couplings that
are not irrelevant, and which therefore takes the form of the semiclassical
action in Eq.~\eqref{eq:140}:
\begin{equation}
  \label{eq:Sieberer2014-34}
  \Gamma_{\Lambda} = \int_{t, \mathbf{x}} \left\{ \bar{\phi}_q^{*} \left[ \left( i
        Z^{*} \partial_t + \bar{K}^{*} \nabla^2 \right) \bar{\phi}_c -
      \frac{\partial \bar{U}^{*}}{\partial \bar{\phi}_c^{*}} \right] + \cc + i
    \bar{\gamma} \bar{\phi}_q^{*} \bar{\phi}_q \right\}.
\end{equation}
% \begin{equation}  
%   \Gamma_{\Lambda} = \int_{t, \mathbf{x}} \left[ \left(
%       \bar{\phi}_c^{*},\bar{\phi}_q^{*} \right) 
%     \begin{pmatrix}
%       0 & \bar{D}^A \\
%       \bar{D}^R & i \bar{\gamma}
%     \end{pmatrix}
%     \begin{pmatrix}
%       \bar{\phi}_c \\ \bar{\phi}_q
%     \end{pmatrix}
%     - \left( \frac{\partial \bar{U}}{\partial \bar{\phi}_c} \bar{\phi}_q +
%       \frac{\partial \bar{U}^{*}}{\partial \bar{\phi}_c^{*}} \bar{\phi}_q^{*}
%     \right) \right].
% \end{equation}
(Recall that the variables of the effective action are the field expectation
values $\bar{\Phi}_{\nu}, \nu = c, q$ defined in Eq.~\eqref{eq:wphi}.) Here, in
addition to the complex prefactor $\bar{K} = \bar{A} + i \bar{D}$ of the
Laplacian, we are including a complex \textit{wave-function renormalization}
$Z = Z_R + i Z_I$. In the semiclassical limit, the homogeneous (i.e., not
containing derivatives with respect to time or space) part of the action can be
written in terms of an effective potential $\bar{U}$, which is a function of
$\bar{\rho}_c = \bar{\phi}_c^{*} \bar{\phi}_c$. Due to the invariance of the
microscopic action under classical phase rotations (cf.\
Sec.~\ref{sec:energy-momentum-particle-numb-cons}), which is inherited by the
effective action, only this combination of fields is allowed in the
potential. The latter is given by
\begin{equation}
  \label{eq:Sieberer2014-35}
  \bar{U}(\bar{\rho}_c) = \frac{1}{2} \bar{u}_2 \left( \bar{\rho}_c - \bar{\rho}_0
  \right)^2 + \frac{1}{6} \bar{u}_3 \left( \bar{\rho}_c - \bar{\rho}_0
  \right)^3.
\end{equation}
Here, both the two-body and three-body couplings,
$\bar{u}_2 = \bar{\lambda} + i \bar{\kappa}$ and
$\bar{u}_3 = \bar{\lambda}_3 + i \bar{\kappa}_3$, respectively, are complex. The
three-body term is marginal according to power counting, and therefore included
in the truncation. In the FRG, it is advantageous to approach the transition
from the ordered phase. Then, the form of the effective potential in
Eq.~\eqref{eq:Sieberer2014-35} corresponds to an expansion around the stationary
condensate density $\bar{\rho}_0$. Indeed, this choice implies, that the field
equations
$\delta \Gamma_{\Lambda}/\delta \bar{\phi}_c^{*} = 0, \delta
\Gamma_{\Lambda}/\delta \bar{\phi}_q^{*} = 0$
(Eqs.~\eqref{eq:var} in the absence of sources and evaluated with the
scale-dependent effective action $\Gamma_{\Lambda}$) are solved by
$\bar{\rho}_c = \bar{\rho}_0$ and $\bar{\phi}_q = 0$ on all scales $\Lambda$.

In the truncation Eq.~\eqref{eq:Sieberer2014-34}, all couplings --- including
the condensate density $\bar{\rho}_0$ --- are scale dependent. As indicated
above, by means of such an ansatz for the effective action $\Gamma_{\Lambda}$,
the functional differential equation~\eqref{eq:FRG} can be rewritten as a set of
ordinary differential equations for these \textit{running} couplings, with
initial conditions given by the microscopic action Eq.~\eqref{eq:140}. This is
achieved by applying \textit{projection prescriptions}. In the following, we
summarize this method for the problem at hand. For details we refer the reader
to Ref.~\cite{Sieberer2014}.

\subsubsection{Non-equilibrium FRG flow equations}
\label{sec:non-eq-frg-flow}

The main idea of a projection prescription on a specific coupling is to extract
this coupling from the effective action $\Gamma_{\Lambda}$ by taking appropriate
derivatives of the latter with respect to the fields and coordinates, and
subsequently setting the fields to their stationary values
$\bar{\phi}_c = \bar{\phi}_0 = \sqrt{\bar{\rho}_0}$ and $\bar{\phi}_q = 0$ (as
noted in Sec.~\ref{sec:goldstone}, choosing $\bar{\phi}_c$ to be real does not
lead to a loss of generality). Then, applying the very same projection
description to the Wetterich equation~\eqref{eq:FRG} yields the flow equation
for the corresponding coupling. For the actual evaluation of the resulting flow
equations, it is convenient to introduce rescaled fields, which are related to
the bare ones by absorbing the wave-function renormalization $Z$ in the quantum
field:
\begin{equation}
  \label{eq:23}
  \phi_c = \bar{\phi}_c, \quad \phi_q = Z \bar{\phi}_q.
\end{equation}
As a consequence of this transformation, and by rescaling all couplings
appropriately, it is possible to obtain a reduced set of flow equations, from
which $Z$ is eliminated. In a second step, the number of flow equations can be
diminished further by introducing dimensionless renormalized variables, as
detailed in Sec.~\ref{sec:scal-solut-crit} below.

We start by deriving flow equations for the non-linear couplings in the
effective potential defined in Eq.~\eqref{eq:Sieberer2014-35}. To see how one
can project the Wetterich equation onto flow equations for these couplings,
consider the effective action, evaluated for homogeneous, i.e., space- and
time-independent ``background fields:''
\begin{equation}
  \label{eq:Sieberer2014-63}
  \Gamma_{\Lambda, cq} = - \Omega \left( \bar{U}' \bar{\rho}_{cq} + \bar{U}^{\prime *}
    \bar{\rho}_{qc} - i \bar{\gamma} \bar{\rho}_q \right).
\end{equation}
Here, the subscript $cq$ in $\Gamma_{\Lambda, cq}$ indicates that both the
classical and the quantum fields are set to constant but non-zero values;
$\Omega$ denotes the quantization volume, and we introduced the following
products of fields, which are invariant under classical phase rotations (see
Sec.~\ref{sec:energy-momentum-particle-numb-cons}):
$\bar{\rho}_{cq} = \bar{\phi}_c^{*} \bar{\phi}_q = \bar{\rho}_{qc}^{*}$ and
$\bar{\rho}_q = \bar{\phi}_q^{*} \bar{\phi}_q$. From the representation
Eq.~\eqref{eq:Sieberer2014-63} it becomes immediately clear how to project the
flow equation~\eqref{eq:FRG} for $\Gamma_{\Lambda}$ onto a flow equation for the
derivative of the potential $\bar{U}$ in Eq.~\eqref{eq:Sieberer2014-35} with
respect to $\bar{\rho}_c$: one has to (i) evaluate Eq.~\eqref{eq:FRG} for
homogeneous fields, (ii) take the derivative with respect to $\bar{\rho}_{cq}$,
and finally (iii) set the quantum background fields to their stationary state
value,
\begin{equation}
  \label{eq:Sieberer2014-64}
  \partial_{\ell} \bar{U}' = - \frac{1}{\Omega}
  \left[ \partial_{\bar{\rho}_{cq}} \partial_{\ell} \Gamma_{\Lambda, cq}
  \right]_{\qnb}.
\end{equation}
Here and in the following, we specify flow equations in terms of the logarithmic
cutoff scale $\ell = \ln(\Lambda/\Lambda_0)$. From the potential $\bar{U}$ in
Eq.~\eqref{eq:Sieberer2014-35}, the couplings $\bar{u}_{2, 3}$ can be obtained
by taking further derivatives with respect to $\bar{\rho}_c$. However, instead
of projecting the flow equation~\eqref{eq:Sieberer2014-64} in this way onto
equations for $\bar{u}_{2,3}$, it is more convenient to introduce rescaled
quantities as outlined above. Inserting the representation Eq.~\eqref{eq:23} of
the bare quantum field in the effective action~\eqref{eq:Sieberer2014-63} leads
to appearance of a factor $1/Z^{*}$ in front of the term involving the effective
potential. This factor can be absorbed by introducing a rescaled potential via
$\bar{U} = Z U$. The flow equations of the bare and rescaled effective potential
are related via
\begin{equation}
  \partial_{\ell} \bar{U}' = Z \left( - \eta_Z U' + \partial_{\ell} U' \right),
\end{equation}
where $\eta_Z$ denotes the \textit{anomalous dimension} associated with the
wave-function renormalization ($\partial_{\ell} Z$ is specified below in
Eq.~\eqref{eq:86}),
\begin{equation}
  \label{eq:Sieberer2014-65}
  \eta_Z = -\partial_{\ell} Z/Z.
\end{equation}
Then, with $\partial_{\bar{\rho}_{cq}} = Z \partial_{\rho_{cq}}$ and inserting
Eq.~\eqref{eq:Sieberer2014-64} on the RHS of Eq.~\eqref{eq:Sieberer2014-66}, the
flow equation for the renormalized potential becomes (here, primes denote
derivatives with respect to $\rho_c = \phi_c^{*} \phi_c$)
\begin{equation}
  \label{eq:Sieberer2014-66}
  \partial_t U' = \eta_Z U' + \zeta', \quad \zeta' = - \frac{1}{\Omega}
  \left[ \partial_{\rho_{cq}} \partial_t \Gamma_{k,cq} \right]_{\qn}.
\end{equation}
In analogy to Eq.~\eqref{eq:Sieberer2014-35}, the renormalized effective
potential can be written as
\begin{equation}
  U(\rho_c) = \frac{1}{2} u_2 \left(
    \rho_c - \rho_0 \right)^2 + \frac{1}{6} u_3 \left( \rho_c - \rho_0 \right)^3,
\end{equation}
with renormalized couplings defined as $u_2 = \bar{u}_2/Z = \lambda + i \kappa$
and $u_3 = \bar{u}/Z = \lambda_3 + i \kappa_3$. To obtain flow equations for
$u_n, n = 2,3$, one simply has to take derivatives of the relation
$u_n = U^{(n)}(\rho_0)$ with respect to the cutoff scale $\ell$, taking into
account that also $\rho_0$ is a running coupling:
\begin{equation}
  \label{eq:Sieberer2014-67}
  \partial_{\ell} u_n = \left( \partial_{\ell} U^{(n)} \right)(\rho_0) + U^{(n
    + 1)}(\rho_0) \partial_{\ell} \rho_0.
\end{equation}
Inserting Eq.~\eqref{eq:Sieberer2014-66} on the RHS of this relation leads us to
\begin{align}
  \label{eq:Sieberer2014-68}
  \partial_{\ell} u_2 & = \beta_{u_2} = \eta_Z u_2 + u_3 \partial_{\ell} \rho_0
  + \partial_{\rho_c} \zeta' \bigr\rvert_{\mathrm{ss}},
  \\ \label{eq:Sieberer2014-69} \partial_{\ell} u_3 & = \beta_{u_3} = \eta_Z u_3
  + \partial_{\rho_c}^2 \zeta' \bigr\rvert_{\mathrm{ss}},
\end{align}
where we evaluate $\zeta'$ with $\rho_c$ set to its stationary value
$\rho_c \rvert_{\mathrm{ss}} = \rho_0$. Finally, flow equations for the real and
imaginary parts of $u_2$ and $u_3$ can be obtained by taking the real and
imaginary parts of Eq.~\eqref{eq:Sieberer2014-68}
and~\eqref{eq:Sieberer2014-69}, respectively.

The flow equation for the stationary density $\rho_0$ cannot be specified
without formal ambiguity: as we have already seen in Sec.~\ref{sec:driv-diss-cond},
both real and imaginary parts of the field equation~\eqref{eq:60} yield
conditions on $\rho_0$. However, we have seen as well that the condition
stemming from the real part can always be satisfied by choosing the proper
rotating frame. Therefore, the physically correct choice is to assume that $\rho_0$ is
actually determined by the imaginary part of the field equation, i.e., by the
condition $\Im U'(\rho_0) = 0$. Taking the derivative of this condition with
respect to the cutoff $\ell$, we find
\begin{equation}
  \label{eq:Sieberer2014-70}  
  \partial_{\ell} \rho_0 = - \left( \Im \partial_{\ell} U' \right)(\rho_0)/\Im
  U''(\rho_0) = - \Im \zeta' \bigr\rvert_{\mathrm{ss}}/\kappa,
\end{equation}
where $\zeta'$ is the same as in Eq.~\eqref{eq:Sieberer2014-66}.

Having illustrated the main idea, the flow equation for the rescaled noise
strength $\gamma = \bar{\gamma}/\abs{Z}^2$ can easily be obtained along the
lines of the derivation of Eqs.~\eqref{eq:Sieberer2014-68}
and~\eqref{eq:Sieberer2014-69}, with the result (for details of the derivation
see Ref.~\cite{Sieberer2014}; $\ezr$ is the real part of $\eta_Z$)
\begin{equation}
  \label{eq:Sieberer2014-72}
  \partial_{\ell} \gamma = \beta_{\gamma} = 2 \ezr \gamma - \frac{i}{\Omega}
  \left[ \partial_{\rho_q} \partial_{\ell} \Gamma_{\Lambda, cq}
  \right]_{\mathrm{ss}}.
\end{equation}

Thus far we have specified how to project the Wetterich equation onto flow
equations for the couplings that parameterize the homogeneous part of the
effective action given in Eq.~\eqref{eq:Sieberer2014-63}, where the classical
and quantum fields are set to constant values. In the following we review the
derivation of flow equations for the frequency- and momentum-dependent
couplings, i.e., the wave-function renormalization $Z$ and the coefficient
$\bar{K}$ of the Laplacian in Eq.~\eqref{eq:Sieberer2014-34}. This requires us
to consider non-constant values of the fields. Moreover, we work in a basis of
real fields which we introduced already in Sec.~\ref{sec:goldstone},
$\bar{\phi}_{\nu} = \frac{1}{\sqrt{2}} \left( \bar{\chi}_{\nu, 1} + i
  \bar{\chi}_{\nu, 2} \right)$
for $\nu = c, q$. Hence, the inverse propagator in this basis is given by the
second variational derivative of the effective action with respect to the fields
$\bar{\chi}_i$ (cf.\ Eq.~\eqref{eq:33}; to ease the notation, we collect these
fields in a vector
$\bar{\chi} = \left( \bar{\chi}_{c, 1}, \bar{\chi}_{c, 2}, \bar{\chi}_{q, 1},
  \bar{\chi}_{q, 2} \right)$;
the components of this vector are labeled by $i = 1, \dotsc, 4$), and the flow
equation of the inverse propagator reads accordingly:
\begin{equation}
  \label{eq:Sieberer2014-73}
  \partial_{\ell} \bar{P}_{ij}(\omega, \mathbf{q}) \delta(\omega - \omega')
  \delta(\mathbf{q} - \mathbf{q}') = \left[ \frac{\delta^2 \partial_{\ell}
      \Gamma_{\Lambda}}{\delta \bar{\chi}_i(-\omega, - \mathbf{q})
      \delta \bar{\chi}_j(\omega', \mathbf{q}')} \right]_{\mathrm{ss}}.
\end{equation}
In particular, the inverse retarded propagator is given by
\begin{equation}
  \label{eq:Sieberer2014-74}
  \bar{P}^R(\omega, \mathbf{q}) =
  \begin{pmatrix}
    - i Z_I \omega - \bar{A} q^2 - 2 \bar{\lambda} \bar{\rho}_0 & i Z_R \omega
    - \bar{D} q^2 \\
    - i Z_R \omega + \bar{D} q^2 + 2 \bar{\kappa} \bar{\rho}_0 & - i Z_I \omega
    - \bar{A} q^2
  \end{pmatrix}.
\end{equation}
Note that the Goldstone theorem (i.e., the existence of a zero eigenvalue of
$\bar{P}^R(\omega = 0, \mathbf{q} = 0)$, see Sec.~\ref{sec:goldstone}) is
preserved during the flow. At the transition, where $\bar{\rho}_0 \to 0$, both
branches of the excitation spectrum that is encoded in the zeros of
$\det \bar{P}^R(\omega, \mathbf{q})$ (cf.\ Eq.~\eqref{eq:61}) become gapless. As
pointed out in Sec.~\ref{sec:open-sys-FRG}, in the FRG, the resulting infrared
divergences are regularized by introducing an additional contribution
$\Delta S_{\Lambda}$, given in Eq.~\eqref{eq:500}, in the functional
integral. In fact, by choosing the following
optimized form of the cutoff function~\cite{Litim2000,Litim2001} (which
obviously satisfies the requirements stated in Eqs.~\eqref{eq:83}
and~\eqref{eq:2666666}),
\begin{equation}
  \label{eq:Sieberer2014-43}
  R_{\Lambda, \bar{K}}(q^2) = - \bar{K} \left( \Lambda^2 - q^2 \right)
  \theta(\Lambda^2 - q^2),
\end{equation}
in the regularized inverse propagator $\Gamma_{\Lambda}^{(2)} + R_{\Lambda}$
appearing in the Wetterich equation~\eqref{eq:FRG}, the terms $\bar{A} q^2$ in
Eq.~\eqref{eq:Sieberer2014-74} are replaced by
\begin{equation}
  \label{eq:34}  
  \bar{A} \left[ q^2 + \left( \Lambda^2 - q^2 \right)
    \theta(\Lambda^2 - q^2) \right] =
  \begin{cases}
    \bar{A} \Lambda^2 & \text{for} \quad q^2 < \Lambda^2, \\
    \bar{A} q^2 & \text{for} \quad q^2 \geq \Lambda^2,
  \end{cases}
\end{equation}
(and there is an analogous replacement for the terms $\bar{D} q^2$). Hence,
fluctuations with momenta below the cutoff scale $\Lambda$ acquire a mass
$\sim \Lambda^2$, and the infrared divergences are lifted.

It remains to specify the flow equations for $Z$ and $\bar{K}$. They can be
obtained by choosing specific values of the indices $i$ and $j$ in the flow
equation~\eqref{eq:Sieberer2014-73} for the inverse propagator, and by taking
derivatives with respect to the frequency $\omega$ and the squared momentum
$q^2$, respectively:
\begin{align}
  \label{eq:86}
    \partial_{\ell} Z & = - \frac{1}{2} \partial_{\omega} \tr \left[ \left( \id
        + \sigma_y \right) \partial_{\ell} \bar{P}^R(\omega, \mathbf{q}) \right]
    \Bigr\rvert_{\omega = 0,
      \mathbf{q} = 0}, \\
    \partial_{\ell} \bar{K} & = - \partial_{q^2} \left( \partial_{\ell}
      \bar{P}^R_{22}(\omega, \mathbf{q}) + i \partial_{\ell}
      \bar{P}^R_{12}(\omega, \mathbf{q}) \right)
    \Bigr\rvert_{\omega = 0, \mathbf{q} = 0}.   
\end{align}
Comparison with Eq.~\eqref{eq:Sieberer2014-74} shows, that these are indeed
correct projection prescriptions. Note, however, that there is some ambiguity in
choosing these projection prescriptions: for example, $\bar{A}$ appears both in
$\bar{P}^R_{11}$ and $\bar{P}^R_{22}$. Our choice extracts $\bar{K}$
corresponding to the Goldstone direction, and mixes Goldstone and gapped
directions symmetrically in the projection on $Z$ (see Ref.~\cite{Sieberer2014}
for details). Finally, the flow equation for the renormalized coefficient
$K = \bar{K}/Z$ is given by
\begin{equation}
  \label{eq:Sieberer2014-77}
  \partial_{\ell} K = \beta_K = \eta_Z K + \partial_{\ell} \bar{K}/Z.
\end{equation}

While Eqs.~\eqref{eq:Sieberer2014-68}, \eqref{eq:Sieberer2014-69},
\eqref{eq:Sieberer2014-70}, \eqref{eq:Sieberer2014-72},
and~\eqref{eq:Sieberer2014-77} define a closed system of flow equations for the
couplings $u_2, u_3, \gamma, \rho_0,$ and $K$ (as indicated above, the
wave-function renormalization $Z$ drops out of these equations), the explicit
evaluation of the various projections is rather tedious. For details of this
calculation we refer the reader to Ref.~\cite{Sieberer2014}.

\subsubsection{Scaling solutions and critical behavior}
\label{sec:scal-solut-crit}

At the critical point of a continuous phase transition, the correlation length
diverges, $\xi \to \infty$. Then, instead of exponential decay according to
$\sim e^{- r/\xi},$ correlation and response functions depend on distance as
power laws. This algebraic scaling behavior is reflected in the RG flow: indeed,
the critical point corresponds to a scaling solution to the RG flow
equations. In practice, finding a scaling solution is facilitated by absorbing
the scaling factors $\sim \Lambda^{\theta}$ (with some exponent $\theta$ for
each coupling) in new variables, which thus take constant values at the critical
point. Hence, the latter corresponds to a \emph{fixed point} of the flow
equations for the rescaled couplings. Moreover, by means of a suitable choice of
rescaled variables it is often possible to further reduce the number of flow
equations, ending up with a minimal set of independent equations. For the
present case, the flow can be specified in terms of just six real couplings.

At the beginning of this section, we introduced the quantity $\lambda$ in
Eq.~\eqref{eq:37} as a quantitative measure of the deviations from thermal
equilibrium conditions. In a similar spirit, the strength of coherent relative
to dissipative dynamics, which is encoded in the real and imaginary parts of the
couplings in the microscopic Keldysh action~\eqref{eq:112}, is measured by the
ratios
\begin{equation}
  \label{eq:Sieberer2014-78}
  \mathbf{r} = \left( r_K, r_{u_2}, r_{u_3} \right) = 
  \left( A/D, \lambda/\kappa, \lambda_3/\kappa_3 \right).
\end{equation}
The flow equations for these ratios can be obtained straightforwardly by taking
the RG scale derivatives, e.g., $\partial_{\ell} r_K = \partial_{\ell} A/D -
A \partial_{\ell} D/D^2$, and expressing $\partial_{\ell} A$ as the real part of
Eq.~\eqref{eq:Sieberer2014-77} etc. In addition to $\mathbf{r}$, we define
another three scaling variables as
\begin{equation}
  \label{eq:90}
  \mathbf{s} = \left( w, \tilde{\kappa}, \tilde{\kappa}_3 \right) = \left( \frac{2
      \kappa \rho_0}{\Lambda^2 D}, \frac{\gamma \kappa}{2 \Lambda D^2},
    \frac{\gamma^2 \kappa_3}{4 D^3} \right).
\end{equation}
The flow equations for the six dimensionless running couplings collected in
$\mathbf{r}$ and $\mathbf{s}$ form a closed set. Besides the Gaussian fixed
point at which the non-linear couplings vanish, these equations have a
non-trivial fixed point corresponding to the driven-dissipative condensation
transition at
\begin{equation}
  \label{eq:Sieberer2014-90}  
  \mathbf{r}_{*} = 0, \quad
  \mathbf{s}_{*} \approx \left( 0.475,5.308,51.383 \right).
\end{equation}
This fixed point is reached in the RG flow, when the parameters in the
microscopic action are chosen such that the system is tuned precisely to the
transition point. (Note that this point corresponds to the \emph{renormalized}
value of $w \propto \rho_0$ going to zero, and not the bare one.) What are the
physical implications of this fixed point? First, the value $\mathbf{r}_{*}$
indicates, that the effective action at the fixed point is purely
dissipative. As we have already mentioned at the beginning of
Sec.~\ref{sec:bosons}, for vanishing coherent dynamics (or, in the terminology
of equilibrium dynamical models~\cite{Hohenberg1977}: in the absence of
reversible mode couplings), the driven-dissipative model reduces to the
equilibrium model A. Thus, the values $\mathbf{s}_{*}$ are the same as in model
A, and therefore the fixed point itself does not allow to distinguish whether
the microscopic starting point of the RG flow was in or out of equilibrium.
However, the non-equilibrium nature of the driven-dissipative condensate is
witnessed in the RG flow towards this effective equilibrium fixed point. In the
following we consider the \emph{universal regime} of the RG flow, which is
reached in the deep IR (i.e., for $\Lambda/\Lambda_0 \ll 1$). In this regime,
when the couplings are close to their values at the fixed point, the RG flow can
be obtained from a linearization of the flow equations in
$\delta \mathbf{s} = \mathbf{s} - \mathbf{s}_{*}, \delta \mathbf{r}
=\mathbf{r}$.
The stability matrix governing the linearized flow takes block diagonal form,
\begin{equation}
  \label{eq:Sieberer2014-91}
  \partial_{\ell}
  \begin{pmatrix}
    \delta \mathbf{r} \\ \delta \mathbf{s}
  \end{pmatrix}
  =
  \begin{pmatrix}    
    N & 0 \\
    0 & S
  \end{pmatrix}
  \begin{pmatrix}
    \delta \mathbf{r} \\ \delta \mathbf{s}
  \end{pmatrix},
\end{equation}
with $3 \times 3$ submatrices $N$ and $S$. This block-diagonal structure
indicates, that the flow of $\mathbf{r}$ and $\mathbf{s}$ decouples in the IR.
Therefore, the flow of $\mathbf{s}$ close to the fixed point is the same as if
we would have set $\mathbf{r} = 0$ from the very beginning. In other words, not
only the values of the couplings $\mathbf{s}$ at the fixed point, but also the
critical exponents encoded in the flow of $\mathbf{s}$, which are the
correlation length exponent $\nu$, the anomalous dimension $\eta$, and the
dynamical exponent $z$, see Ref.~\cite{Sieberer2014}, are the same for both
model A and driven-dissipative condensates. (Note, however, that in model
F~\cite{Hohenberg1977}, which describes condensation with particle number
conservation in equilibrium, the dynamical exponent takes a different value than
in model A, where particle number conservation is absent.) This confirms the
asymptotic thermalization of correlation functions mentioned in
Sec.~\ref{sec:critical-dynamics-3d}~\textit{(i)}. The values of the critical
exponents we obtain from the truncation~\eqref{eq:Sieberer2014-34} are
\begin{equation}
  \label{eq:93}  
  \nu \approx 0.716, \quad \eta \approx 0.039, \quad z \approx 2.121,
\end{equation}
and agree reasonably well with results from more sophisticated calculations of
$\nu$ to and $\eta$ in the context of the static equilibrium
problem~\cite{Guida1998}.

All the information on the universal properties of the driven-dissipative
transition, which are the same as in the equilibrium model A, are encoded in
$\mathbf{s}_{*}$ and the block $S$ of the stability matrix in
Eq.~\eqref{eq:Sieberer2014-91}. The non-equilibrium nature of the microscopic
model, on the other hand, is betrayed by the block $N$, which describes the flow
of $\delta \mathbf{r}$. This block has three positive eigenvalues,
\begin{equation}
  \label{eq:Sieberer2014-98}
  n_1 \approx 0.101, \quad  n_2 \approx 0.143, \quad n_3 \approx 1.728,
\end{equation}
indicating that the ratios $\mathbf{r}$ are attracted to the fixed point value
$\mathbf{r}_{*} = 0$. The general solution to the linearized flow equation for
$\mathbf{r}$ reads
\begin{equation}
  \label{eq:Sieberer2014-100}
  \mathbf{r} = \sum_{i = 1}^3 \mathbf{u}_i c_i,
\end{equation}
where $\mathbf{u}_i$ are the eigenvectors of $N$ associated with the eigenvalues
$n_i$ in Eq.~\eqref{eq:Sieberer2014-98}. The coefficients $c_i$, which are
referred to as \textit{scaling fields}~\cite{Altland2010}, take the scaling form
$c_i \sim e^{n_i \ell} \sim \Lambda^{n_i}$. Hence, for $\Lambda \to 0$, the
dominant contribution to $\mathbf{r}$ is given by
$\mathbf{r} \sim \mathbf{u}_1 \Lambda^{n_1} = \mathbf{u}_1 \Lambda^{- \eta_r}$,
where we identified the drive exponent
\begin{equation}
  \label{eq:Sieberer2014-102}
  \eta_r = - n_1 \approx - 0.101.
\end{equation}
As anticipated in Sec.~\ref{sec:critical-dynamics-3d}~\textit{(ii)}, this
exponent governs the universal fade-out of coherent dynamics
$\propto \mathbf{r}$ in the driven-dissipative system. Note that the existence
of three distinct eigenvalues~\eqref{eq:Sieberer2014-98} is due to the
non-equilibrium character of the microscopic model. Indeed, in model A with
reversible mode couplings, the equilibrium symmetry discussed in
Sec.~\ref{sec:therm-equil-sym} allows of only one ratio
$r = r_K = r_{u_2} = r_{u_3}$ (cf.\ Eq.~\eqref{eq:168067} and
Fig.~\ref{fig:noneq}). Then, the block $N$ of the stability matrix in
Eq.~\eqref{eq:Sieberer2014-91} has only one single entry, which is given by the
``middle'' eigenvalue $n_2$ in Eq.~\eqref{eq:Sieberer2014-98}. This shows that
also in the equilibrium setting the dynamics becomes purely dissipative at the
largest scales~\cite{DeDominicis1975}, however, the value of the critical
exponent that governs universal decoherence is different. As pointed out above,
this is due to the absence of the equilibrium symmetry in the driven-dissipative
case. The fact that the different values can be traced back to a difference in
symmetry supports the strength of the result, as different symmetries are known
to give rise to quantitatively different critical
behavior~\cite{Zinn-Justin2002}.

Decoherence at large scales has clear physical signatures which facilitate
probing the drive exponent in experiments; to wit, it implies that low-momentum
excitations are diffusing rather than propagating (note that this is also
predicted by mean-field theory, cf.\ the discussion below Eq.~\eqref{eq:61}). A
careful analysis in the scaling regime reveals that the effective dispersion
relation of single-particle excitations close to criticality takes the
form~\cite{Sieberer2014}
\begin{equation}
  \label{eq:122}
  \omega \sim A_0 q^{z-\eta_r} - i D_0 q^z \sim  A_0 q^{2.223}-i D_0 q^{2.121},
\end{equation}
where the diffusive contribution is supplemented by a subdominant (by the small
difference of $\eta_r$ in the exponent) coherent part. By definition,
Eq.~\eqref{eq:122} is the location of the pole of the retarded Green's function,
and hence the coherent and diffusive parts encode respectively the position and
width of the peak of the spectral function defined in Eq.~\eqref{eq:49}. The
latter is probed, e.g., in angle-resolved spectroscopy in exciton-polariton
systems~\cite{Utsunomiya2008} or radio-frequency spectroscopy in ultracold
atoms~\cite{Stewart2008}. However, the small difference in the scaling of the
position and width of the peak predicted by Eq.~\eqref{eq:122} poses a challenge
to its experimental observation.

\subsection{Absence of algebraic order in 2D}
\label{sec:absence-algebr-order}

Semiconductor microcavities hosting exciton-polaritons are effectively
two-dimensional, and therefore this case has the greatest significance for
current experiments. Even in thermal equilibrium, two-dimensional condensates
are markedly different from their three-dimensional counterparts: first,
according to the Mermin-Wagner theorem~\cite{Mermin1966}, in a two-dimensional
condensate there cannot be true off-diagonal long-range order at any finite
temperature. Instead, at low temperatures, spatial correlations decay
algebraically with distance. Nevertheless, the system remains
superfluid. Second, the algebraic or quasi-long-range order is established in an
unusual transition, in which vortices, which proliferate at high temperatures,
form bound pairs as the temperature is tuned below the critical value. How is
this scenario modified under non-equilibrium conditions? As a first step to
answer this question, the issue of spatial correlations in two-dimensional
driven-dissipative condensates is addressed in Ref.~\cite{Altman2015}. In this
work, the results of which we describe in the following, the influence of
non-topological phase fluctuations (spin waves) on the behavior of spatial
correlations is analyzed, which leads to the conclusion, that in
driven-dissipative condensates algebraic decay is possible only on intermediate
scales, and crosses over to stretched-exponential decay on the largest
scales. The decay of correlations might be found to be even faster (i.e.,
exponential), once topological excitations (vortices) are taken into account.

In a weakly interacting Bose gas in thermal equilibrium, the absence
of true long-range order is caused by the vanishing energy cost of
long-wavelength phase fluctuations. These are governed by the quadratic
effective low-energy action~\eqref{eq:sfd_102}, from which the behavior of
spatial correlations at long distances can be obtained straightforwardly, e.g.,
by introducing sources as described in Sec.~\ref{sec:keldysh-path-integr} and
using the formulas for Gaussian functional integration collected in
Appendix~\ref{sec:gauss-funct-integr}. The result is
\begin{equation}
  \label{eq:96}
  \begin{split}
    \left\langle \psi(\mathbf{x}) \psi^{*}(\mathbf{x}') \right\rangle & \approx
    \rho_0 \langle e^{i \left( \theta(\mathbf{x}) - \theta(\mathbf{x}') \right)}
    \rangle \\ & = \rho_0 e^{- \frac{1}{2} \langle \left( \theta(\mathbf{x}) -
        \theta(\mathbf{x}') \right)^2 \rangle} \\ & \sim \abs{\mathbf{x} -
      \mathbf{x}'}^{-\alpha},
  \end{split}
\end{equation}
where $\alpha = m^2 T/(2 \pi \rho_0)$. One the other hand, the derivation in
Sec.~\ref{sec:dens-phase-repr} shows that in the case of a driven-dissipative
condensate the phase-only action is non-linear and given by the KPZ
action~\eqref{eq:38}. Hence, while the second equality in Eq.~\eqref{eq:96}
still applies to leading order in a cumulant expansion, the expectation value
$\langle \left( \theta(\mathbf{x}) - \theta(\mathbf{x}') \right)^2 \rangle$
cannot be calculated directly.\footnote{The non-linearity $\lambda$ in the KPZ
  action~\eqref{eq:38} vanishes when the equilibrium condition~\eqref{eq:168067}
  is met, leading to algebraically decaying correlations also in this case. This
  shows, that merely adding dissipation by coupling the system to a bath in
  thermal equilibrium does not have an adverse effect on correlations. (In the
  genuine case in which the condition~\eqref{eq:168067} is met, the system is
  indeed coupled to a single bath. Otherwise, realizing this condition when the
  system is coupled to several baths would require a pathological fine-tuning of
  the coupling parameters.) It is indeed the combination of independent drive
  and dissipation, which leads to the loss of algebraic coherence out of
  equilibrium.} In the original context of the KPZ equation, where $\theta$
takes the role of the height of a randomly growing
surface~\cite{Kardar1986,Halpin-Healy1995}, the behavior of this correlation
function is parameterized in terms of the \textit{roughness exponent} $\chi$ as
\begin{equation}
  \label{eq:98}
  \langle \left( \theta(\mathbf{x}) - \theta(\mathbf{x}') \right)^2 \rangle \sim
  \abs{\mathbf{x} - \mathbf{x}'}^{2 \chi}.
\end{equation}
The term roughness exponent is due to the fact that its value distinguishes
\textit{smooth} from \textit{rough} phases: for $\chi < 0$, fluctuations of the
surface height die out on large scales, and the surface is smooth; on the other
hand, if $\chi > 0$, the interface is called rough. For any finite value of
$\chi$, the scaling behavior of the correlation function in Eq.~\eqref{eq:98}
leads to stretched exponential decay of the correlations of $\psi$,
\begin{equation}
  \label{eq:97}
  \left\langle \psi(\mathbf{x}) \psi^{*}(\mathbf{x}') \right\rangle \sim e^{- c
    \abs{\mathbf{x} - \mathbf{x}'}^{2 \chi}},
\end{equation}
where $c$ is a non-universal constant. In the case that $\chi = 0$ one usually
expects logarithmic growth of
$\langle \left( \theta(\mathbf{x}) - \theta(\mathbf{x}') \right)^2\rangle$,
which would lead to the equilibrium result in Eq.~\eqref{eq:96}. However, as
discussed at the beginning of Sec.~\ref{sec:bosons}, in a 2D driven-dissipative
condensate we should expect universal behavior that is quite different from the
equilibrium case. Indeed, the FRG analysis reported in
Refs.~\cite{Canet2010,Canet2011,Canet2012,Kloss2012} and numerical
simulations~\cite{Kim1989,Miranda2008,Marinari2000,Ghaisas2006,Chin1999,Tang1992,Ala-Nissila1993,Castellano1999,AaraoReis2004,Kelling2011,Halpin-Healy2012,Pagnani2015,Halpin-Healy2013a,Halpin-Healy2013,Halpin-Healy2014}
find $\chi \approx 0.4$ for the value of the roughness exponent, which implies
that for $\abs{\mathbf{x} - \mathbf{x}'} \to \infty$ correlations in 2D
driven-dissipative condensates obey Eq.~\eqref{eq:97} and not
Eq.~\eqref{eq:96}\footnote{\label{fn:1} Note that as pointed out at the end of
  Sec.~\ref{sec:dens-phase-repr}, the KPZ non-linearity is neglected in
  Bogoliubov theory. Therefore, in 2D, this approach yields power-law decay of
  spatial correlations~\cite{Szymanska2006,Chiocchetta2013}.}, and decay
stretched-exponentially.

A notable difference between the KPZ equation for randomly growing interfaces,
and the present context of the effective long-wavelength description of a
driven-dissipative condensate is that the analogue of the interface height in
the latter case is a phase, $\theta$, and as such it is compact, i.e., defined
up to multiples of $2 \pi$. This means that topological defects --- vortices ---
in this field are possible.\footnote{This difference with the conventional KPZ
  equation also arises in ``Active Smectics''~\cite{Chen2013}.} Proliferation of
vortices would lead to an even faster, simple exponential decay of spatial
correlations. The present analysis does not take the possible presence of
vortices into account.

How do these findings compare to experimental results? Both in experiments on
incoherently pumped polariton condensates~\cite{Roumpos2012,Nitsche2014} and
simulations of parametrically pumped systems~\cite{Dagvadorj2014}, spatial
correlations have been found to decay algebraically within the confines of the
system. This, however, is not in contradiction to the present analysis based on
the KPZ equation: indeed, if the microscopic value $g_0$ of the rescaled
non-linearity~\eqref{eq:64} is small, which is actually the case in current
experiments~\cite{Altman2015}, a renormalized value of $g = 1$ is reached in the
RG flow only at the exponentially large scale
\begin{equation}
  \label{eq:74}
  L_{*} = \xi_0 e^{8 \pi/g_0},
\end{equation}
where $\xi_0$ is a microscopic scale where the RG flow is initialized, e.g., the
healing length of the condensate. Indeed, to obtain this result, we have solved
Eq.~\eqref{eq:65} with initial condition $g_0$ at the scale $\xi_0$. Then, in
systems of a size $L$ well below $L_{*}$, an effective equilibrium description
is applicable, leading to the observed algebraic decay of correlations (below
the equilibrium KT transition). In other words, even in an infinite system we
should expect a smooth crossover from algebraic to exponential decay at the
scale $L_{*}$.

So far, our analysis has been based on the semiclassical Langevin
equation~\eqref{eq:17} for the condensate dynamics, which according to the
arguments given at the beginning of Sec.~\ref{sec:bosons} correctly captures the
universal scaling properties of driven-dissipative condensates. However, in
order to obtain an estimate of $L_{*}$ for specific experimental parameters, a more
microscopic model of exciton-polariton condensates is required. Starting from a
widely used model, which has been introduced in Ref.~\cite{Wouters2007a} and
consists of a coupled system of equations for the lower polariton field and the
excitonic reservoir, the bare value $g_0$ and hence the scale $L_{*}$ can be
seen to depend on the rate at which the reservoir is
replenished~\cite{Altman2015}. For high pump rates the KPZ scale $L_{*}$ grows
rapidly, so that by pumping the system strongly enough, algebraic correlations
can be made to extend over the entire system for \emph{any} finite system size
$L$. When the pump rate is reduced, the system can loose its algebraic order in
two ways: either through the effect of the KPZ non-linearity if $L_{*}$ drops
below the system size, or --- if $L_{*}$ is still much larger than the system
size at the critical value of the pump strength for the equilibrium KT
transition to occur --- through the proliferation of vortices. Note that as
pointed out above, even when KPZ physics becomes relevant, vortices might still
modify Eq.~\eqref{eq:97}. These considerations lead to the finite-size phase
diagram reported in Ref.~\cite{Altman2015}.

The pump strengths at which the KT and KPZ crossovers occur can be estimated
based on the parameters given in Ref.~\cite{Lagoudakis2008}. It is convenient to
introduce dimensionless pumping and loss rates as follows~\cite{Altman2015}:
\begin{equation}
  \label{eq:123}  
  x  = \frac{P R}{\gamma_R \gamma_l} - 1, \quad
  \bar{\gamma} = \frac{R \gamma_l}{\gamma_R u_c}.  
\end{equation}
Here, $P$ is the rate at which the excitonic reservoir is replenished, while $R$
is the amplification rate of the condensate due to stimulated scattering of
polaritons from the reservoir; $\gamma_l$ and $\gamma_R$ are, respectively, the
decay rates of lower polaritons and reservoir excitons, and $u_c$ is the
coherent polariton-polariton interaction~\cite{Wouters2007a}. For the parameters
in~\cite{Lagoudakis2008}, the KT transition should be expected
at~\cite{Altman2015} $x_{\mathrm{KT}} \approx 0.02$. Denoting by $x_{*}$ the
pumping strength at which the KPZ scale $L_{*}$ in Eq.~\eqref{eq:74} drops below
the system size, we have~\cite{Altman2015}
\begin{equation}
  \label{eq:124}
  x_{*}/x_{\mathrm{KT}} = \bar{\gamma}^2 \ln(L/\xi_0) \approx 0.04,
\end{equation}
where we took $\bar{\gamma} \approx 0.1, \xi_0 \approx 2 \, \mu \mathrm{m}$, and
assumed a pump spot size of $L \approx 100 \, \mu \mathrm{m}$. Thus, approaching
the transition from above by lowering the pump power, the critical value
$x_{\mathrm{KT}}$ is reached first, and the system loses algebraic order through
unbinding of vortices~\cite{Roumpos2012,Nitsche2014,Dagvadorj2014}. On the other
hand, the crossover to the disordered regime will be controlled by KPZ physics
once $x_{*} \geq x_{\mathrm{KT}}$, which can be achieved by increasing the loss
rate (i.e., reducing the cavity $Q$) to $\bar{\gamma} \approx 0.5$.

While the above analysis shows that algebraic order in 2D driven-dissipative
condensates prevails only on intermediate scales, remarkably it can be restored
on all scales in strongly anisotropic systems~\cite{Altman2015}: consider a
generalization of Eq.~\eqref{eq:17}, where the gradient terms are replaced by
$\sum_{i = x,y} K_{\alpha}^i \partial_i^2 \phi_c$ for $\alpha = c,
d$. Correspondingly, Eqs.~\eqref{eq:37} and~\eqref{eq:47} are replaced by
\begin{equation}
  \label{keeling_eq:15}
  D_i  = K_c^i \left( \frac{K_d^i}{K_c^i} + \frac{u_c}{u_d} \right), \quad
  \lambda_i  = -2 K_c^i \left( 1 - \frac{K_d^i u_c}{K_c^i u_d} \right),
\end{equation}
which for $i = x,y$ are the diffusion constants and non-linearities appearing in
the anisotropic KPZ equation. The RG flow of this equation has been analyzed in
Refs.~\cite{Wolf1991,Chen2013}. It can be described in terms of the anisotropy
parameter $\Gamma = \lambda_y D_x/\lambda_x D_y$ (the system is anisotropic for
$\Gamma \neq 1$), and the non-linearity $g = \lambda_x^2 \Delta/(D_x^2 \sqrt{D_x
  D_y})$. The flow equations are given by
\begin{equation}
  \label{keeling_eq:17}
  \begin{split}    
    \frac{d g}{d \ell} & = - \frac{g^2}{32 \pi} \left( \Gamma^2 + 4 \Gamma - 1
    \right), \\ \frac{d \Gamma}{d \ell} & = -\frac{\Gamma g}{32 \pi} \left( 1 -
      \Gamma^2 \right).
  \end{split}
\end{equation}
(Note that these equations differ from the RG equations in
Ref.~\cite{Altman2015} by the sign on the RHS, which is due to the fact that
here we define the logarithmic scale as $\ell = \ln(\Lambda/\Lambda_0)$ instead
of $\ell = \ln(L/\xi_0)$.) The line $\Gamma = 0$ divides the flow into two
regions with distinct fixed-point structure: for $\Gamma > 0$, which we denote
as the regime of \textit{weak anisotropy}, at large scales isotropy is restored,
i.e., $\Gamma \to 1$ for $\ell \to \infty$, and all the results discussed above
apply.\footnote{Current experiments with exciton-polaritons are in fact slightly
  anisotropic due to the interplay between polarization pinning to the crystal
  structure, and the splitting of transverse electric and transverse magnetic
  cavity modes~\cite{Carusotto2013,Shelykh2010}. On the other hand, in
  experiments using the optical parametric oscillator regime pumping
  scheme~\cite{Stevenson2000,Baumberg2000} (see also
  \cite{Carusotto2005,Dagvadorj2014}), strong anisotropy is imprinted by the
  pump wavevector.} In the regime of \textit{strong anisotropy} corresponding to
$\Gamma < 0$, on the other hand, the flow is attracted to an effective
\emph{equilibrium} fixed point with $g = 0$ and $\Gamma = -1$. Then, algebraic
correlations of the condensate field can survive if the effective temperature at
the fixed point, which is given by the \emph{renormalized} value of the
dimensionless noise strength $\kappa = \Delta/\sqrt{D_x D_y}$, is below the
critical value $\kappa_c = \pi$ for the equilibrium KT transition. Generalizing
the microscopic model for exciton-polariton condensates mentioned above to
account for spatial anisotropy, the dependence of the effective temperature on
the strength of laser pumping can be obtained~\cite{Altman2015}. Remarkably, the
transition to the algebraically ordered phase is found to be \emph{reentrant:}
upon increasing the pump rate the ordered phase is first entered and then left
again at even higher values of the pump rate.

\subsection{KPZ scaling in 1D}
\label{sec:kpz-scaling-1d}

The marginality of $g$ in two spatial dimensions is reflected in the emergence
of the exponentially large scale $L_{*}$ in Eq.~\eqref{eq:74} beyond which KPZ
scaling can be observed. In 1D, on the contrary, the KPZ non-linearity $g$ is
relevant (cf.\ the discussion at the beginning of Sec.~\ref{sec:bosons}), which
makes one-dimensional driven-dissipative condensates even more promising candidates
to observe KPZ universality in finite-size systems. This possibility was
explored numerically in Refs.~\cite{Gladilin2014,Ji2015,He2014}, where the
scaling properties of 1D driven-dissipative condensates were studied by
simulations of the Langevin equation~\eqref{eq:17} for the condensate field.

Experimentally, the most directly accessible signatures of KPZ universality are
contained in the correlation function of the condensate field (i.e., the Keldysh
Green's function defined in Eq.~\eqref{eq:greend}),
\begin{equation}
  \label{eq:73}
  C(t - t', x - x') = \langle \psi(t, x) \psi^{*}(t', x') \rangle.
\end{equation}
Indeed, in experiments with exciton-polaritons, both spatial correlations
$C(0, x)$ and the autocorrelation function $C(t, 0)$ can be obtained by
performing interferometric measurements on the photoluminescence emitted from
the semiconductor microcavity~\cite{Kasprzak2006,Lagoudakis2008,Love2008}. Based
on the mapping of the long-wavelength condensate dynamics to the KPZ equation,
one expects exponential decay of spatial correlations $C(0, x)$ and
stretched-exponential decay of the autocorrelation function according to
$C(t, 0) \sim \exp(-c t^{2 \beta}),$ where $\beta = 1/3$ (in the original
context of the KPZ equation, which is the stochastic growth of driven
interfaces, $\beta$ is called the growth exponent~\cite{Halpin-Healy1995}) and
$c$ is a non-universal constant. This behavior was confirmed
numerically~\cite{Gladilin2014,Ji2015,He2014}. In equilibrium, i.e., for
$g = \lambda = 0$, the KPZ equation~\eqref{eq:243456} reduces to a noisy
diffusion equation, which in the surface growth context is known as the
Edwards-Wilkinson model~\cite{Edwards1982}. Then, the behavior of spatial
correlations is unchanged, whereas the exponent $\beta$ governing the decay of
temporal correlations takes the value $\beta = 1/4$. The distinction between
one-dimensional condensates in equilibrium and driven-dissipative condensates
thus becomes manifest only in the dynamical properties.\footnote{Concomitantly,
  also within Bogoliubov theory exponential decay of correlations is
  found~\cite{Wouters2006}, cf.\ the discussion at the end of
  Sec.~\ref{sec:dens-phase-repr} and Footnote~\ref{fn:1}.} Moreover, in order to
actually observe KPZ scaling in the autocorrelation function, a large value of
$g$, corresponding to a system far from equilibrium, is favorable. This can be
achieved by making drive and dissipation the dominant contributions to the
dynamics, as is the case in cavities with a reduced $Q$ factor~\cite{He2014}. To
be specific, for the parameters reported in Ref.~\cite{Wertz2012}, the $Q$
factor would have to be reduced by a factor of $\approx 30$ (corresponding to a
polariton lifetime $\approx 1 \, \mathrm{ps}$ instead of $\approx 30 \,
\mathrm{ps}$ achieved in the experiment) in order to make KPZ scaling observable
in a system of size $\approx 100 \, \mu \mathrm{m}$.

\begin{figure}
  \centering
  \includegraphics[width=3.5in]{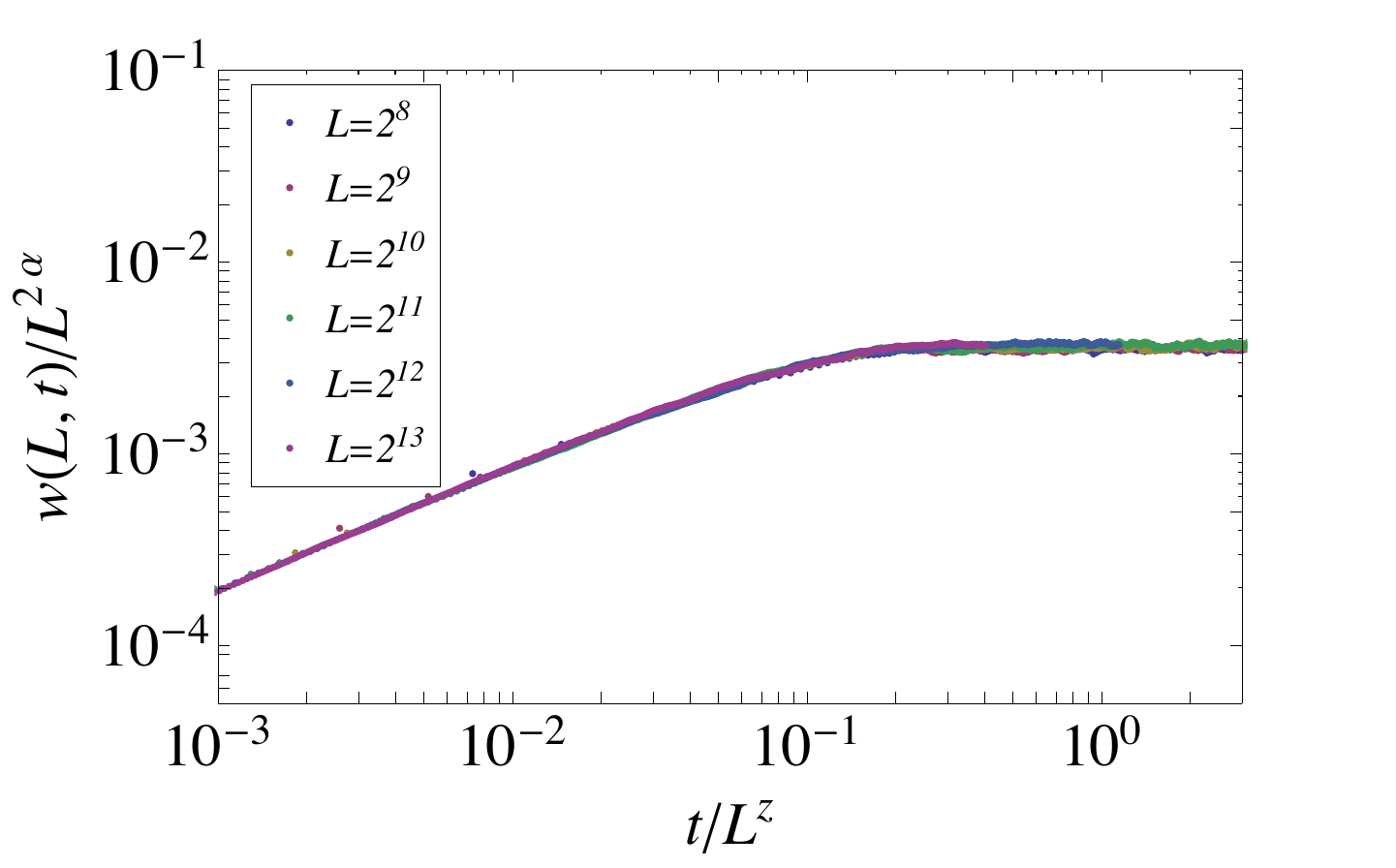}
  \caption{Finite-size scaling collapse of the roughness function $w(L, t)$
    defined in Eq.~\eqref{eq:66}. The values of the roughness exponent
    $\alpha \equiv \chi = 1/2$ and the dynamical exponent $z = 3/2$ are in the
    1D KPZ universality class. Each curve corresponding to a specific system
    size $L$ is an average over $1000$ noise realizations. The simulations were
    performed after rescaling the Langevin equation~\eqref{eq:17} to bring it to
    dimensionless form. For details of the rescaling and the values of the
    parameters used in the simulations, see Ref.~\cite{He2014}. (Copyright
    (2014) by The American Physical Society.)}
  \label{fig:wt_in_scaling_axes}
\end{figure}

Another observable, which conveniently encodes the scaling properties of the
phase $\theta$ of the condensate, is defined as
\begin{equation}
  \label{eq:66}
  w(L, t) = \left\langle
    \frac{1}{L} \int_0^L dx \, \theta(t,x)^2 - \left( \frac{1}{L} \int_0^L dx \,
      \theta(t, x) \right)^2 \right\rangle.
\end{equation}
In the context of growing interfaces, were $\theta$ takes the role of the
surface height, the quantity $w(L, t)$ is known as the \textit{roughness
  function.} It is a measure of the fluctuations of the surface height over the
linear extent of the system $L$. While the roughness function might not be
easily accessible in experiments with driven-dissipative condensates, it allows
a very compact demonstration of both static and dynamic KPZ scaling exponents if
it is obtained numerically for a range of different system
sizes~\cite{Marinari2000}. Indeed, the finite-size scaling collapse of $w(L, t)$
in Fig.~\ref{fig:wt_in_scaling_axes} shows, that after a period of growth during
which $w(L,t) \sim t^{2 \beta}$, the roughness function saturates at the time
$T_s \sim L^z$; the saturation value $w_s(L)$ scales with the system size as
$w_s(L) \sim L^{2 \chi}$, where $\chi = 1/2$ is the value of the roughness
exponent in 1D~\cite{Tauber2014a,Kamenev2011,Halpin-Healy1995}. From the growth
and roughness exponents, the usual dynamical exponent can be obtained as
$z = \chi/\beta$.

The numerical analysis reported in Ref.~\cite{He2014} was performed in the
regime of weak noise, which is characterized by the absence of phase slips in
the spatiotemporal range covered by the simulations. Indeed, the mapping of the
condensate dynamics to the KPZ equation~\eqref{eq:243456}, in which the phase is
regarded as a non-compact variable, does not take into account the possible
occurrence of such defects. However, their presence at higher noise levels is
expected to affect the scaling properties of driven-dissipative condensates.

%%% Local Variables:
%%% mode: latex
%%% TeX-master: "dds_review"
%%% End:

\section{Universal heating dynamics in 1D}
\label{sec:heating-dynamics}
The notion of universality is not restricted to systems in thermal
equilibrium. As discussed in the previous sections, non-thermal steady states of
driven-dissipative systems can show a large variety of universal features, such
as scale invariance and effective long-wavelength thermalization. However, in a
plethora of setups, aspects of universality can even be found in the time
evolution, which approaches a steady state only in the limit
$\tau\rightarrow\infty$~\cite{Gagel2014,Gagel2015,Chiocchetta2015b,StrackKPZ,MarcuzziNEU,Cai2013,Poletti2013,Schachenmayer2014}. An
example which identifies generic, universal features in the far from equilibrium
dynamics in a strongly interacting one-dimensional system is discussed in the
present section.

The setting we consider here differs from the one in the previous section not
only in its focus on time evolution, but also in terms of underlying
symmetries. Above we have studied systems that are open in the sense that both
energy and particle number were not conserved, witnessed by the absence of the
thermal symmetry (cf. Sec.~\ref{sec:therm-equil-sym}) and the quantum phase
rotation symmetry (cf. Sec.~\ref{sec:energy-momentum-particle-numb-cons}), and
leading to a breaking of detailed balance and a low momentum diffusive Goldstone
mode (cf. Sec.~\ref{sec:goldstone}), respectively. Here we consider an open
system, where only energy is not conserved, but particle number is. In fact, the
absence of energy conservation here is reflected by a permanent inflow of energy
into the system. The Lindblad operators are Hermitian in the present case, and
this leads to continuous heating and ultimately to an entirely classical,
infinite temperature stationary state, described by a density matrix
$\rho \sim \mathbf{1}$, where the latter unit matrix is understood in the entire
Fock space of the problem. This motivates us to study the time evolution of
heating, with a focus on the short time dynamics following initialization in a
pure zero temperature ground state, where quantum effects are still present. The
presence of particle number conservation leads us to take a different strategy
than in the previous sections. Here, accomodating number conservation, which is
at the heart of the strongly collective behavior of one dimensional systems, we
first map the quantum master equation in the operatorial formalism to an
effective long-wavelength description in terms of an open Luttinger liquid. In
this way, we can carefully account for the linear sound mode that is expected on
the grounds of exact particle number conservation,
cf. Sec.~\ref{sec:goldstone}. Only after this procedure, we perform the mapping
to the Keldysh functional integral, similar to our strategy for the spins in
Sec.~\ref{sec:neqspin}.

\subsection{Heating an interacting Luttinger liquid}
\label{sec:intll}
Consider a one-dimensional lattice of interacting bosons for which the dynamics
is described by the following master equation: 
\eq{Heat1}{
  \partial_t\rho=-i\left[H,\rho\right]+\gamma_e\sum_i\left[2n_i\rho n_i-\left\{n^2_i,\rho\right\}\right].
} Here, $H$ is a bosonic lattice Hamiltonian, whose long-wavelength physics is
described by an interacting Luttinger liquid. For concreteness, one can consider
a Bose-Hubbard model in one dimension away from integer filling, with Hamiltonian 
\begin{equation}
  \label{eq:113}
  H=-J\sum_{i}\left(\creo{b}{i}\ann{b}{i+1}+\creo{b}{i+1}\ann{b}{i}\right)+\tfrac{U}{2}\sum_i  n_i(n_i-1), 
\end{equation}
which describes nearest neighbor hopping of particles with
hopping amplitude $J$ and local interactions with interaction energies $U$. The dissipative contributions
which drive the system out of equilibrium, are the Hermitian jump
operators $n_i=\creo{b}{i}\ann{b}{i}$, measuring the local particle
number in terms of bosonic creation and annihilation operators
$\creo{b}{i}, \ann{b}{i}$. For cold bosonic atoms in optical lattices, these jump
operators represent the leading order contribution of dissipation induced by
spontaneous emission from the lattice drive laser~\cite{Pichler2010} (see also the discussion in Sec.~\ref{sec:cold-atoms-optical-lattices}), but this
kind of dissipation can
as well be realized by coupling the bosonic particles to a phonon reservoir with
a large effective temperature, which is equivalent to a locally fluctuating
chemical potential
\begin{equation}
     \label{Heat2}  
  \begin{split}
    \mu(x,t)&= \mu_0+\delta\mu(x,t),\nonumber\\
    \langle \delta\mu(x,t)\delta\mu(x',t')\rangle&=
    \gamma_e\delta(t-t')\delta(x-x').
  \end{split}
\end{equation}
It causes dephasing and leads to a linear increase of the energy in the system,
$\langle H\rangle_t\propto J\gamma_et$~\cite{Pichler2010,Pichler2013}. As a
consequence of the linear energy increase, the system will never thermalize and
is constantly driven away from equilibrium, approaching the
$T\rightarrow\infty$ state described by $\rho \propto \mathbf{1}$ at infinite time $t$. We note that, for hermitean Lindblad operators, $\rho \propto \mathbf{1}$ is always a solution to the quantum master equation. 

In order to analyze the heating dynamics for short and transient times, the
master equation is transformed to the Luttinger representation on the operatorial level, and only later on we perform the mapping to the Keldysh functional integral. The Luttinger
description is valid as long as the occupation of quasi-particle modes does not
exceed a critical value, which is determined by the Luttinger cutoff
$\Lambda$~\cite{heating}. Starting with a zero or low temperature
initial state to which this applies, there exists a cutoff time $t_{\Lambda}$,
up to which the system can be described in terms of Luttinger liquid
variables. In this regime, one can take the continuum limit $b_i\rightarrow b_x$
and express the bosonic operators in a phase and amplitude
representation:
\begin{equation}
  \label{Heat3}
  \begin{split}    
b_x&=\sqrt{\rho_x}e^{i\theta_x},\\
\rho_x&=\rho_0+\partial_x\phi_x/\pi,
  \end{split}
\end{equation}
in terms of the Luttinger variables $\partial_x\phi_x$ and $\theta_x$, which
represent smooth density and phase fluctuations and fulfill the commutation
relation $\left[\partial_x\phi_x,\theta_{x'}\right]=i\pi\delta(x-x')$.  The
long-wavelength description of the Bose-Hubbard model is expressed by the
Hamiltonian
\begin{equation}
  \label{Heat4}
  H=\frac{1}{2\pi}\int_x
  \left[uK\left(\partial_x\theta_x\right)^2-\tfrac{u}{K}\left(\partial_x\phi_x\right)^2
     +\kappa\left(\partial_x\phi_x\right)\left(\partial_x\theta_x
      \right)^2\right],
\end{equation}
which describes interacting Luttinger phonons on length scales
$x\ge \left(\rho_0 Um\right)^{-1/2}$, above which a continuum representation of
the Hamiltonian is appropriate. For weak interactions, the effective parameters
can be estimated to be $u=\left(\tfrac{\rho_0U}{m}\right)^{1/2}$,
$K=\tfrac{\pi}{2}\left(\tfrac{\rho_0}{Um}\right)^{1/2}$.  The non-linearity in
the Hamiltonian accounts for the leading order quasi-particle scattering term in
the low energy regime ($\kappa=1/m$), which is irrelevant in the sense of the
renormalization group and does not modify static, equilibrium
correlations. However, it is vital for the quantitative description of dynamic
correlation functions, and is non-negligible in a non-equilibrium setting where
the dynamics is affected by the non-linearity even on a qualitative level.  The
dissipative part of the master equation becomes quadratic in the Luttinger
representation, such that the equation of motion reads
\begin{equation}
  \label{Heat5}
  \partial_t\rho=-i\left[H,\rho\right]
  +\tfrac{\gamma}{\pi^2}\int_x\left[2(\partial_x\phi_x)\rho(\partial_x\phi_x)
    -\left\{(\partial_x \phi_x)^2,\rho\right\}\right].
\end{equation}
This decoherence term is the leading order contribution of a $U(1)$-symmetric
(i.e., particle number conserving) decoherence mechanism in one-dimensional
quantum wires, which features a linear increase of the energy in
time. Furthermore, the $U(1)$ symmetry guarantees the existence of a linear
sound mode and permits the transformation to the Luttinger framework in the
coherence dominated regime. From a microscopic perspective, it is evident that
the Luttinger description has to break down for sufficiently strong decoherence,
i.e., after the system has been heated up sufficiently. This breakdown can be
estimated by the usual Luttinger criterion $n_q<\Lambda/|q|$, and leads to a
good estimate for the relevant time scales up to which the dynamics of the
system is dominated by coherent sound modes and therefore properly expressed in
the Luttinger framework~\cite{heating}. The corresponding time regime is set by
the condition $t<u^2 (\kappa\gamma)^{-1}$.

The quadratic part of the equation of motion is diagonalized by the canonical Bogoliubov transformation
\begin{equation}
  \label{Heat6}
  \begin{split}    
    \theta_x&=\theta_0+i\int_q(\tfrac{\pi}{2|q|K})^{1/2} (\creo{a}{q}-\anno{a}{-q})e^{-iqx},\\
    \phi_x&=\phi_0-i\int_q(\tfrac{\pi
      K}{2|q|})^{1/2}\mbox{sgn}(q)(\creo{a}{q}+\anno{a}{-q})e^{-iqx},
  \end{split}
\end{equation}
which leads to the master equation
\begin{multline}
  \label{Heat7}  
\partial_t\rho= -i \int_q \left[u|q|\creo{a}{q}\anno{a}{q} , \rho\right] \\ +\int_q
\tfrac{\gamma|q|}{\pi
  K}\left[\anno{A}{q}\rho\creo{A}{q}-\tfrac{1}{2}\{\creo{A}{q}\anno{A}{q},\rho\}\right]
 -\int_q[H^{(3)}_{\mbox{\tiny ph}},\rho].
\end{multline}
In this equation, the dissipative part is expressed via the operators
$\creo{A}{q}=\creo{a}{q}+\anno{a}{-q}=\anno{A}{-q}$ in terms of bosonic phonon
operators $[\anno{a}{q},\creo{a}{p}]=\delta(q-p)$. The cubic Hamiltonian
incorporates resonant phonon scattering processes \eq{Heat8}{
  H^{(3)}_{\mbox{\tiny
      ph}}=3\kappa\sqrt{\tfrac{\pi}{2K}}\int_{q,p}'\sqrt{|qp(p+q)}(\creo{a}{p+q}\anno{a}{q}\anno{a}{p}+\mbox{h.c.}),
} which conserve momentum and the phonon energy. This is expressed by
$\int'_{qp}$, which signals to integrate only over configurations $\{p,q\}$ with
$|q|+|p|=|p+q|$. The non-resonant processes create only short-lived quantum
states which do not contribute in the long time dynamics due to dephasing, and which are
therefore not relevant for the forward dynamics of the system.

In the Luttinger representation, the dissipative contribution is quadratic and
its effect is a constant population of the individual phonon modes. This can be
seen most easily by computing the time evolution of the phonon densities in the
quadratic framework, which yields
\begin{align}
  \label{Heat9}
\partial_t\langle \creo{a}{q}\anno{a}{q}\rangle_t&=\tfrac{\gamma|q|}{2\pi K},\\
\partial_t\langle \creo{a}{q}\creo{a}{-q}\rangle_t&=-\tfrac{\gamma|q|}{2\pi K}-2iu|q|\langle \creo{a}{q}\creo{a}{-q}\rangle_t.\label{Heat10}
\end{align}
For a system prepared initially in the ground or finite temperature state, the initial phonon densities are $\langle \creo{a}{q}\anno{a}{q}\rangle_0=n_{\mbox{\tiny B}}(u|q|)$ and $\partial_t\langle \creo{a}{q}\creo{a}{-q}\rangle_0=0$, where $n_{\mbox{\tiny B}}(u|q|) = (e^{\beta  u |q| } -1)^{-1}$ is the Bose-Einstein distribution evaluated on-shell. In this case, the solution of Eqs.~\eqref{Heat9}, \eqref{Heat10} is 
\begin{align}
  \label{Heat11}
\langle \creo{a}{q}\anno{a}{q}\rangle_t&=n_{\mbox{\tiny B}}(u|q|)+\tfrac{\gamma|q|t}{2\pi K},\\
\langle \creo{a}{q}\creo{a}{-q}\rangle_t&=\tfrac{\gamma e^{-iu|q|t}}{2\pi u K}\sin(u|q|t).
\end{align}
The first term describes an increase of the phonon density linear in time and momentum and leads to a linear increase of the system energy, i.e., heats up the system according to
\eq{Heat13}{
\Delta E_t=\langle H^{(2)}\rangle_t-\langle H^{(2)}\rangle_0=\tfrac{u\gamma\Lambda^2t}{4\pi K}.
}
Equation~\eqref{Heat13} relates the effective long-wavelength Luttinger
parameters $u, K, \gamma$ of the heating setup to the macroscopic heating rate
$\partial_t\Delta E_t$ via the microscopic cutoff $\Lambda$. Since this heating
rate is not model specific but depends on the individual realization of the
heating dynamics, it is not surprising that it depends on macroscopic and
microscopic parameters. In this sense, Eq.~\eqref{Heat13} should be viewed as
the definition of the effective heating parameter $\gamma$ for a generic model
in the presence of heating~\cite{heating}.

The off-diagonal phonon density oscillates in the complex plane, thereby taking
absolute values
$|\langle \creo{a}{q}\creo{a}{-q}\rangle_t|\le\tfrac{\gamma}{2\pi u K}$, which
are negligibly small in the weak heating regime $\gamma\ll uK$, i.e., in the
coherence dominated dynamics. It is therefore sufficient to consider only the
diagonal elements in the phonon basis when extending the analysis to the
interacting model.

The jump operators in the microscopic master equation~\eqref{Heat1} are the
Hermitian, local density operators $n_i$, which preserve the $U(1)$ invariance
of the dynamics even in the presence of dissipation. This leads to a decay of
the off-diagonal elements of the density matrix, i.e., to decoherence in the
local number state representation, and an evolution of the density matrix
towards its diagonal ensemble.  In the Luttinger representation, this
decoherence expresses itself in a permanent production of photons \eqref{Heat11}, i.e. a permanent heating of the Luttinger liquid, which features no
compensation mechanism in the quadratic sector and the consequent lack of a well
defined steady state. The energy increases constantly, which will lead to an
obvious breakdown of the Luttinger description as soon as the energy stored in
the long-wavelength modes exceeds a critical value. At this point, the dynamics
is no longer dominated by the coherences of $\rho$ but by its diagonal elements,
including the breakdown of superfluidity and quasi-long range order.

The way in which energy is distributed amongst the long-wavelength modes by the
heating~\eqref{Heat9} is not typical for an interacting system at low energies,
since it deviates strongly from a Bose-Einstein distribution and the associated
detailed balance of energy. The phonon scattering terms in
$H^{(3)}_{\mbox{\tiny ph}}$ favor detailed balance and strongly modify the actual
distribution function compared to~\eqref{Heat11}, which makes them
non-negligible in the present non-equilibrium setting.

\subsection{Kinetic equation}
\label{sec:kineq}
In order to determine the time evolution of the excitation densities in
interacting systems out of equilibrium, a common and often successful strategy
is the so-called kinetic equation approach~\cite{Kamenev2011} (for an application to periodically driven Floquet systems, cf. Ref.~\cite{Genske2015}), which determines
the time evolution of the distribution function of the excitations in terms of
the system's self-energies. For the present setup, this approach has to be
modified in order to take into account the driven-dissipative nature of the
system and the resonant character of the interactions. The latter lead to a
breakdown of perturbation theory and require non-perturbative techniques beyond
one-loop corrections. A detailed derivation and discussion of the applicability
and limitations of such an approach can be found
in~\cite{Buchholdmethod,heating}. Other forms of kinetic equations for
interacting Luttinger liquids, which focus on a different set of non-equilibrium
conditions, include a perturbative treatment of phonon backscattering terms,
resulting from additional disordered or lattice
potentials~\cite{Tavora2014,Tavora2013}, as well as a treatment of cubic phonon
interactions in the presence of a smooth  background potential and a curved
phonon dispersion~\cite{Gutman2014}.

The quantity of interest in this section is the time-dependent occupation of phonon modes $n_{q,t}=\langle \cre{a}{q}\ann{a}{q}\rangle_t$ in the presence of heating and phonon scattering. For bosonic modes and in the steady state, the occupation of the modes is related to the Keldysh Green's function via 
\begin{align}\label{FDR2}
i\int_{\omega}G^K_{q,\omega}=2n_q+1.
\end{align}
For a system in thermal equilibrium, $n_q=n_{\text{B}}(\epsilon_q,T)$ is the Bose distribution function, see Sec.~\ref{sec:therm-equil-sym}, while for a general non-equilibrium steady state, $n_q$ is a positive function, which has to be determined from the specific context. One can now introduce the hermitian distribution function $F_{q,\omega}$ as in Eq.~\eqref{eq:FDR-general}, but generalized to a system with a continuum of momentum modes. In terms of the hermitian distribution function, the anti-hermitian Keldysh Green's function can be parameterized according to
\begin{align}\label{FDR3}
G^K_{q,t,t'}=\left(G^R_{q}\ \tilde{\circ}\ F_q-F_q\ \tilde{\circ}\ G^A_q\right)_{t,t'}.
\end{align}
Here, $G^R_q,\ G^A_q=\left(G^R_q\right)^{\dagger}$ and $F_q$ are two-time functions evaluated at equal momentum $q$ and $\tilde{\circ}$ represents the convolution with respect to time and matrix multiplication according to the Nambu structure of the Green's functions. For a system, which is diagonal in Nambu space, $\tilde{\circ}=\circ$ reduces to a simple multiplication. In the presence of off-diagonal occupations, i.e. for $\langle \cre{a}{-q}\cre{a}{q}\rangle\neq0$, the operation $\tilde{\circ}$ has to respect the symplectic structure of bosonic Nambu space and is promoted to $\tilde{\circ}=\sigma^z\circ$, as described in Ref.~\cite{Buchholdmethod}. For a bosonic system in steady state, all terms in \eqref{FDR3} are time-translational invariant and Fourier transformation yields
\begin{align}\label{FDR4}
G^K_{q,\omega}=G^R_{q,\omega}\sigma^zF_{q,\omega}-F_{q,\omega}\sigma^zG^A_{q,\omega}.
\end{align}
In contrast, for a system out of equilibrium undergoing a non-trivial time
evolution, the mode occupations $n_{q,t}$ remain well defined, but
Eq.~\eqref{FDR3} is not time-translational invariant anymore. One then has to
find a representation for the Green's functions and $F$, which reveals the
time-dependent occupations. This is done in the following, leading to the Wigner
representation of the bosonic distribution function.

A convenient parameterization of non-equilibrium correlation and response
functions is the so-called \emph{Wigner representation} in time, which introduces a
forward and a relative time coordinate ($t, \delta$) for the phonon Green's
functions, according to
\begin{align}
  \label{Heat14}
  G^R_{q,t,\delta}&=-i\theta(\delta)\langle [\anno{a}{q,t+\delta/2},\creo{a}{q,t-\delta/2}]\rangle,\\
  G^K_{q,t,\delta}&=-i\langle\{\anno{a}{q,t+\delta/2},\creo{a}{q,t-\delta/2}\}\rangle.
\end{align}
A  two-time function $C(t_1,t_2)$ can always be
transformed to Wigner coordinates, $C(t_1,t_2)=C(t,\delta)$, by defining
$t=\tfrac{t_1+t_2}{2}$ and $\delta=t_1-t_2$. Here, the explicit dependence on
$t$ expresses the forward time evolution of non-equilibrium systems, while
for equilibrium systems in the presence of time-translational invariance, the
forward time dependence of generic two-time functions just drops out,
$C(t,\delta)\equiv C(0,\delta)$, for all $ t$. In Wigner coordinates, the
parametrization of the Keldysh correlation function is
\eq{Heat15}{
  G^K_{q,t,\delta}=\left(G^R_{q}\circ F_q-F_q\circ G^A_q\right)_{t,\delta}. }
Here, we choose to neglect the subleading off-diagonal contributions according to the above discussion, and consider only diagonal modes in Nambu space. Eq.~\eqref{Heat15} contains the full Green's functions of the system, which can be expressed via the Dyson relation in terms of the self-energies $\Sigma^{R/A/K}$,
\eq{DysonR}{
\left(\begin{array}{cc}G^K&G^R\\ G^A&0\end{array}\right)^{-1}=\left(\begin{array}{cc}0&G_0^A-\Sigma^A\\ G_0^R-\Sigma^R&-\Sigma^K\end{array}\right).
}
The self-energies $\Sigma^{R/A/K}$ represent the correction to the bare Green's functions $G^{R/A}_0=(i\partial_t-u|q|)^{-1}$ due to phonon scattering and heating events. Inserting the Dyson representation into Eq.~\eqref{Heat15}, it can
be inverted and rearranged to read \eq{Heat16}{
  \partial_tF_{q,t,\delta}=i\Sigma^K_{q,t,\delta}-i\left(\Sigma^R_q\circ
    F_q-F_q\circ G^A_q\right)_{t,\delta}.  } Here, the notion $(...)_{t,\delta}$
expresses the fact that the whole expression in brackets should be transformed
to Wigner coordinates after performing the convolution
$\Sigma^R_q\circ F_q\equiv \int_{t'}\Sigma^R_{q,t_1,t'} F_{q,t',t_2}$ in
ordinary time representation. The functionals $\Sigma^{R/A/K}$ are the
self-energies in the retarded, advanced, and Keldysh sectors, which incorporate
the effect of interactions and the heating on the quadratic sector. The
temporal derivative on the LHS of~\eqref{Heat16} is the Wigner representation
of the bare, non-interacting Green's functions without heating. Taking the
Fourier transform of Eq.~\eqref{Heat16} with respect to the relative time
coordinate $\delta$ yields the Wigner representation of the distribution
function \eq{Heat17}{ F_{q,t,\omega}=\int_{\delta}
  e^{i\omega\delta}F_{q,t,\delta}, } for which \eq{Heat18}{
\partial_tF_{q,t,\omega}=i\Sigma^K_{q,t,\omega}-i\left(\Sigma^R\circ F-F\circ\Sigma^A\right)_{q,t,\omega}.
}
The corresponding transformation for the convolution inside the parenthesis is
\eq{Heat19}{
\left(\Sigma^R\circ F\right)_{q,t,\omega}=\Sigma^R_{q,t,\omega}e^{\frac{i}{2}\left(\overset{\leftarrow}{\partial}_t\overset{\rightarrow}{\partial}_{\omega}-\overset{\leftarrow}{\partial}_{\omega}\overset{\rightarrow}{\partial}_t\right)} F_{q,t,\omega}.
}
Its explicit evaluation is nontrivial and in most cases simply
impossible. However, it is possible to approximate the complex exponential by
the leading order expansion for many typical relaxation
dynamics~\cite{Kamenev2011}. In order to understand Eq.~\eqref{Heat19}, one should
take a closer look at its expansion up to first order in derivatives:
\begin{equation}
  \label{Heat20}
  \left(\Sigma^R\circ
    F\right)_{q,t,\omega}\hspace{-0.2cm}=\Sigma^R_{q,t,\omega}F_{q,t,\omega}\left(1-\tfrac{i}{2}\tfrac{\partial_tF_{q,t,\omega}}{F_{q,t,\omega}}\tfrac{\partial_{\omega}\Sigma^R_{q,t,\omega}}{\Sigma^R_{q,t,\omega}}+(\omega\leftrightarrow
    t)\right)+O(\partial^2).
\end{equation}
The ratio
$\kappa^f_q\equiv \left|\tfrac{\partial_tF_{q,t,\omega}}{F_{q,t,\omega}}\right|$
is the rate with which the distribution $F$ is changing in forward time, i.e.,
the forward time evolution rate, which is determined by the interplay between
heating and collective quasi-particle scattering. On the other hand,
$\left(\kappa^r_q\right)^{-1}\equiv\left|\tfrac{\partial_{\omega}\Sigma^R_{q,t,\omega}}{\Sigma^R_{q,t,\omega}}\right|$
is identified with the inverse rate of the relative time dynamics, which is
dominated by fast single phonon propagation. As a consequence
$\kappa^r_q\gg\kappa^f_q$ and the correction terms in~\eqref{Heat20} can be
safely neglected. This is a typical situation for many kinetic equation
approaches and is termed the {\it{Wigner approximation}}. For the present setup, a more
careful analysis has shown that the Wigner approximation is indeed satisfied as
long as the Luttinger representation of the problem is
valid~\cite{heating}.

In order to project Eq.~\eqref{Heat18} onto the quasi-particle densities, it is multiplied by the spectral function $\mathcal{A}_{q,t,\omega}=iG^R_{q,t,\omega}-iG^A_{q,t,\omega}$, followed by a subsequent integration over frequencies $\omega$. The spectral function fulfills the sum rules
\eq{Heat21}{
\int_{\omega}\mathcal{A}_{q,t,\omega}&=&\langle [\anno{a}{q,t,0},\creo{a}{q,t,0}]\rangle=1,\\
\int_{\omega}\hspace{-0.2cm}\mathcal{A}_{q,t,\omega}F_{q,t,\omega}&\hspace{-0.4cm}\overset{\text{Wigner approx.}}{=}\hspace{-0.4cm}&\int_{\omega}\hspace{-0.2cm}\left(\mathcal{A}\circ F\right)_{q,t,\omega}=G^K_{q,t,0}=2n_{q,t}+1,\ \ \ \ \ \ \ \ \ 
}
with $n_{q,t}=\langle \creo{a}{q,t}\anno{a}{q,t}\rangle$ being the phonon density. Applying this to Eq.~\eqref{Heat18} in the Wigner approximation yields
\eq{Heat22}{
\partial_tn_{q,t}=\tfrac{i}{2}\int_{\omega}\left(\Sigma^K_{q,t,\omega}-\Sigma^R_{q,t,\omega}F_{q,t,\omega}+\Sigma^A_{q,t,\omega}F_{q,t,\omega}\right)\mathcal{A}_{q,t,\omega}. 
}

For well defined quasi-particles, i.e., excitations with a well defined energy-momentum relation, the spectral function reflects the well-defined structure of the excitations and is sharply peaked at the quasi-particle dispersion $\omega=u|q|$, with a typical width $\sigma^R_{q,t}\ll u|q|$, which is the imaginary part of the self-energy evaluated on the mass shell $\sigma^R_{q,t}=-\mbox{Im}(\Sigma^R_{q,t,u|q|})$. If this is the case, the full self-energies in Eq.~\eqref{Heat22} multiplied with the spectral function can be approximated by their on-shell values, as they are expected to vary only smoothly in the region where $\mathcal{A}$ is non-zero. The approximation $\Sigma^{R/A/K}_{q,t,\omega}\mathcal{A}_{q,t,\omega}
\approx \Sigma^{R/A/K}_{q,t,u|q|}\mathcal{A}_{q,t,\omega}$ is called the {\it{quasi-particle approximation}} and in the present case, similar to the Wigner approximation, it is applicable in the entire Luttinger regime~\cite{heating,Buchholdmethod}. The latter is a consequence of the subleading, RG-irrelevant nature of the interactions, which lead to self-energies $\sigma^R_{q,t}\ll u|q|$.
Performing the quasi-particle approximation, Eq.~\eqref{Heat22} obtains the simple form
\eq{Heat23}{
\partial_tn_{q,t}=\tilde{\sigma}^K_{q,t}-\sigma^R_{q,t}(2n_{q,t}+1).
}
In this equation, the anti-Hermitian Keldysh self-energy $\Sigma^K_{q,t,\omega}$ has been replaced by its on-shell value $\Sigma^K_{q,t,u|q|}=-2i\tilde{\sigma}^K_{q,t}$, with the real function $\tilde{\sigma}^K_{q,t}$.

The kinetic equation~\eqref{Heat23} describes the time evolution of the phonon density $n_{q,t}$ in terms of the on-shell self-energies $\sigma^{R}, \tilde{\sigma}^{K}$, which in turn are determined by both the interactions and the heating term. In order to identify the contribution from the heating, one has to identify the impact of the dissipative contribution in Eq.~\eqref{Heat7} on the action $S$. Following the steps in Sec.~\ref{sec:deriv-keldysh-acti} carefully and setting the off-diagonal density contributions to zero, according to Eq.~\eqref{Heat11}, the dissipative contribution to the microscopic action $S$ is
\eq{Heat24}{
S_{D}=i\int_{p,t}\tfrac{\gamma|p|}{\pi K}a^*_{q,p,t}\ann{a}{q,p,t}.
}
Here, $p$ is the momentum variable and $q$ labels the quantum component of the Keldysh field variable. The dissipation thus enters the action only in the quantum-quantum sector and does not modify the spectrum in the quantum-classical sector. This is again a consequence of the Hermitian nature of the Lindblad operators, which lead to a continuously increasing occupation of phonons but do not introduce a compensating dissipative mechanism in the retarded and advanced sector of the action \footnote{The heating mechanism is operative for generic situations. For example, mean-field Mott initial states, which are the exact ground states of the Bose-Hubbard model for fine-tuned $J=0$, are pure states which are not touched by the dissipator considered in this section.  One point of view on this phenomenology is that the infinite temperature state is an attractive fixed point of dynamics, but there are other unstable ones which need additional symmetries to be physically relevant (such as a spatially local gauge symmetry in the $J=0$ example). }. This is drastically different from the situation in the models of Secs.~\ref{sec:neqspin} (cf. Eqs.~\eqref{Spin8} and~\eqref{Spin7}) and~\ref{sec:bosons} (cf. Eqs.~\eqref{eq:Sieberer2014-34} and~\eqref{eq:Sieberer2014-35}), where the interplay of dissipation in the retarded/advanced sectors and fluctuation or noise in the Keldysh sector of the action lead to non-equilibrium fluctuation-dissipation relations describing well-defined stationary states different from the trivial state $\rho \propto \mathbf{1}$.

With the form of~\eqref{Heat24}, the Keldysh self-energy is $\tilde{\sigma}^K_{q,t}=\tfrac{\gamma|p|}{2\pi K}+\sigma^K_{q,t}$, where the bare $\sigma^K_{q,t}$ in this form is determined by the interactions alone.  The resulting kinetic equation consists of three contributions
\eq{Heat25}{
  \partial_tn_{q,t}=\underbrace{\tfrac{\gamma|q|}{2\pi K}}_{\text{in-term
      heating}}+\underbrace{\sigma^K_{q,t}}_{\text{in-term
      scattering}}-\underbrace{\sigma^R_{q,t}(2n_{q,t}+1)}_{\text{out-term
      scattering}}.  } The first term represents the population of phonon modes
due to the constant heating term, while the second and the third term are effects of the elastic collisions redistributing energy. The second term, proportional to the Keldysh self-energy,
describes scattering of phonons into the mode $q$ due to the interactions, while
the third term, proportional to the retarded self-energy, describes scattering of
phonons out of the mode $q$, and is therefore directly proportional to
$n_{q,t}$. Setting the interactions to zero, both self-energies $\sigma^{R/K}=0$
vanish and only the heating term remains, rendering the time evolution of the
phonon density in the absence of interactions into Eq.~\eqref{Heat9}.

The kinetic equation~\eqref{Heat25} represents the foundation of the analysis of the non-equilibrium dynamics in the presence of heating and phonon scattering. In order to solve for the time-evolution of the phonon densities, one has to compute the self-energies $\sigma^{R/K}_{q,t}$ for each momentum mode and at each time step. The self-energies have to be determined by a non-perturbative approach, which we discuss in the following.

\subsection{Self-consistent Born approximation}
\label{sec:scb}
For an interacting model of resonantly scattering phonons, the phonon
self-energies are functionals of the phonon density, such that the RHS
of~\eqref{Heat25} contains an implicit, non-linear dependence on $n_{q,t}$. In
order to make this implicit dependence explicit, the self-energies are typically
evaluated perturbatively at one loop order and higher order corrections to the
time evolution are neglected~\cite{Kamenev2011}. However, for the present
scenario, the interactions are resonant, i.e., describe scattering events inside
a continuum of degenerate states, and therefore perturbative computations
diverge at any order. This defines the need for non-perturbative approaches to
compute the phonon self-energies, the simplest of which is the so-called
{\it{self-consistent Born approximation}}, which we discuss in the following.

\begin{figure}[t]
  \centering
  \includegraphics[width=\linewidth]{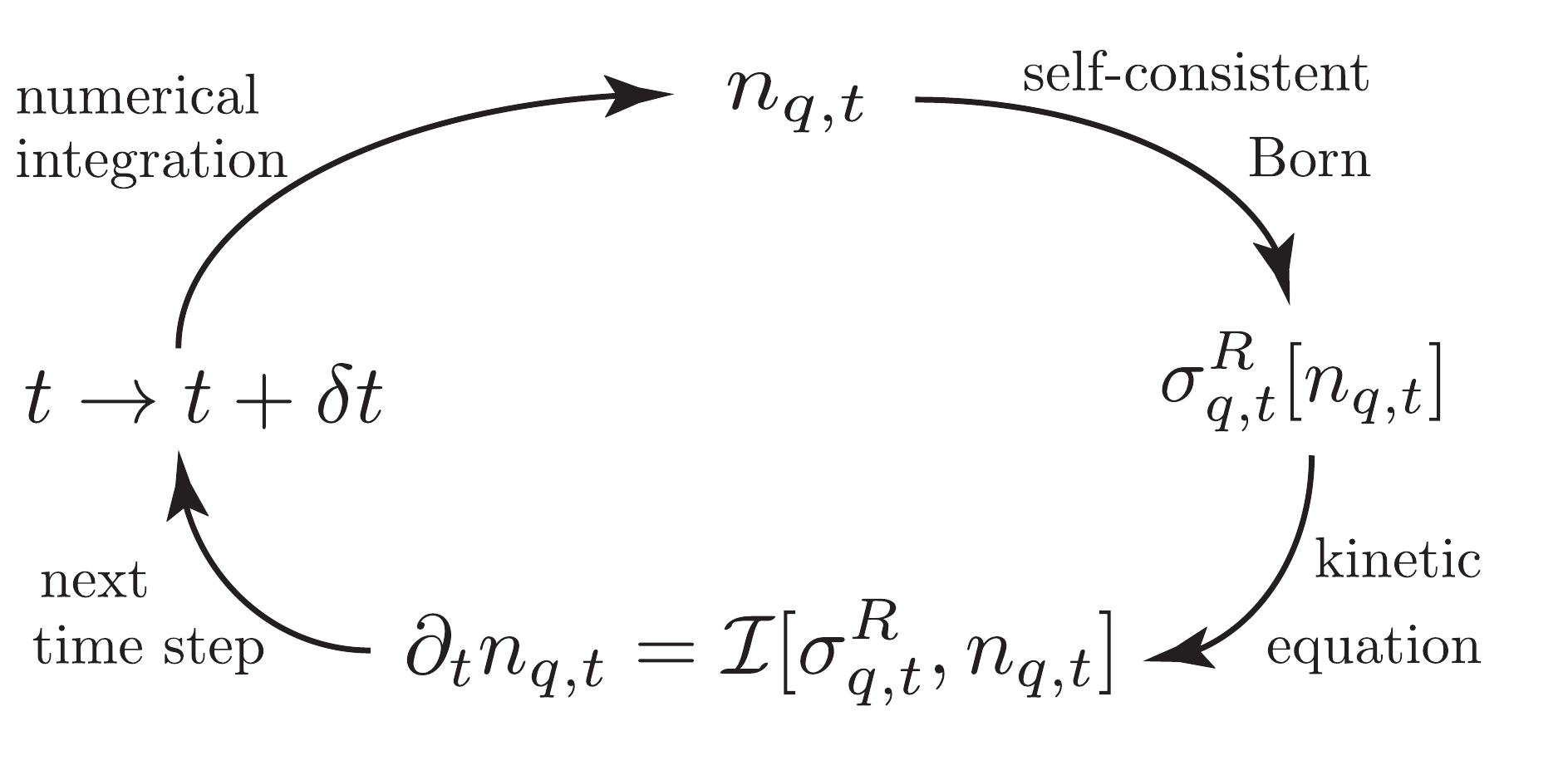}
  \caption{Illustration of the iterative process to compute the time evolution of the phonon density $n_{q,t}$. For a specific forward time $t$, the on-shell self-energy $\sigma^R_{q,t}$ is determined via Eq.~\eqref{Heat33} and subsequently the result is inserted into the kinetic equation~\eqref{Heat34}. In order to integrate the phonon density, a Runge-Kutta solver for differential equations is used, which determines $n_{q,t}$ numerically.}
  \label{fig:Scheme}
\end{figure}

The Keldysh action for interacting Luttinger liquids with heating is
composed of a dissipative part $S_D$, which has been discussed in
Eq.~\eqref{Heat24}, and a Hamiltonian part $S_H$, which results from the
Hamiltonian dynamics in Eqs.~\eqref{Heat7},~\eqref{Heat8}. In the Keldysh
representation, the action is 
\begin{multline}
  \label{Heat26}
  S =
  \int_{p,t}\hspace{-0.2cm}(a^*_{c,p,t},a^*_{q,p,t})\left(\hspace{-0.2cm}\begin{array}{cc}
                                                                                 0
                                                                                 &
                                                                                   \hspace{-0.4cm}i\partial_t-u|p|-i0^+\\ i\partial_t-u|p|+i0^+ & i\tfrac{\gamma|p|}{\pi K}\end{array}\hspace{-0.2cm}\right)\left(\hspace{-0.2cm}\begin{array}{c}\ann{a}{c,p,t}\\ \ann{a}{q,p,t}\end{array}\hspace{-0.2cm}\right) \\
  +\int'_{p,k,t}v(p,k)\Big[2a^*_{c,k+p,t}\ann{a}{c,k,t}\ann{a}{q,p,t}\\
  +a^*_{q,k+p,t}\Big(\ann{a}{c,k,t}\ann{a}{c,p,t}+\ann{a}{q,k,t}\ann{a}{q,p,t}\Big)+\text{h.c.}\Big]
\end{multline}
with the vertex function $v(p,k)=3\kappa\sqrt{\tfrac{\pi}{2K}|pk(p+k)}$. One way
of computing the one-loop self-energy is to determine the one-loop correction to
the effective action $\Gamma[a^*_{\alpha},\ann{a}{\alpha}]$ defined in
Eq.~\eqref{eq:1700} for general bosonic fields $\Phi$. The effective action in
the absence of an external source is defined as 
\begin{equation}
  \label{Heat27}
  e^{i \Gamma[a^*_{\alpha},\ann{a}{\alpha}]} = \int \mathscr{D}[\delta
  a^*_{\alpha},\delta \ann{a}{\alpha}] \, e^{i S[a^*_{\alpha} + \delta
    a^*_{\alpha},\ann{a}{\alpha} + \delta \ann{a}{\alpha}]},
\end{equation}
and fulfills the equation of motion $\frac{\delta\Gamma[a^*_{\alpha},\ann{a}{\alpha}]}{\delta a^*_{\alpha}}=\frac{\delta\Gamma[a^*_{\alpha},\ann{a}{\alpha}]}{\delta\ann{a}{\alpha}}=0$. The one-loop effective action is then obtained by expanding the action $S$ up to second order in the fluctuation fields and subsequently integration over the fluctuations 
\eq{Heat28}{
e^{i \Gamma^{(\text{1-loop})}[a^*_{\alpha},\ann{a}{\alpha}]}\hspace{-0.3cm}&=\hspace{-0.2cm}&e^{i S[a^*_{\alpha},\ann{a}{\alpha}]}\hspace{-0.15cm}\int \hspace{-0.2cm}\mathscr{D}[\delta
  a^*_{\alpha},\delta \ann{a}{\alpha}] \, e^{\frac{i}{2} (\delta a^*_{\alpha},\delta\ann{a}{\alpha})S^{(2)}[a^*_{\alpha},\ann{a}{\alpha}](\delta\ann{a}{\alpha},\delta a^*_{\alpha})^T}\nonumber\\
&=\hspace{-0.2cm}&e^{i S[a^*_{\alpha},\ann{a}{\alpha}]+\frac{1}{2}\text{Tr}\log\left(S^{(2)}[a^*_{\alpha},\ann{a}{\alpha}]\right)}.}
This identifies the one-loop effective action 
\eq{Heat29}{
\Gamma^{(\text{1-loop})}[a^*_{\alpha},\ann{a}{\alpha}]=S[a^*_{\alpha},\ann{a}{\alpha}]-\tfrac{i}{2}\text{Tr}\log\left(S^{(2)}[a^*_{\alpha},\ann{a}{\alpha}]\right)
}
in terms of the microscopic action $S$ and its second variation with respect to the fields 
\eq{Heat30}{
S^{(2)}[a^*_{\alpha},\ann{a}{\alpha}]=\left(\begin{array}{cc}\frac{\delta^2S}{\delta a^*_{\alpha}\delta\ann{a}{\alpha'}}&\frac{\delta^2S}{\delta a^*_{\alpha}\delta a^*_{\alpha'}}\\\frac{\delta^2S}{\delta\ann{a}{\alpha}\delta\ann{a}{\alpha'}} &\frac{\delta^2S}{\delta\ann{a}{\alpha}\delta a^*_{\alpha}}\end{array}\right).
}
In order to determine the correction to the bare action, the logarithm in
Eq.~\eqref{Heat29} is expanded in powers of the fields
$a^*_{\alpha},\ann{a}{\alpha}$. The quadratic self-energy is the second
order expansion of the logarithm and its matrix elements are determined by the
integrals
\begin{align}
  \label{Heat31}
  \Sigma^R_{Q}&= 2i\hspace{-0.15cm}\int_{P}'\hspace{-0.15cm}G^K_{P}\left(v^2(q,-p)G^R_{Q-P}\hspace{-0.1cm}+v^2(p,-q)G^A_{P-Q}\hspace{-0.1cm}+v^2(p,q)G^R_{P+Q}\right),\\
\Sigma^K_{Q}&=2i\int_P'\left[v^2(q-p)\left(G^K_{P}G^K_{Q-P}+G^R_PG^R_{Q-P}+G^A_PG^A_{Q-P}\right)\right.\nonumber\\
&\left.+2v^2(p,q)\left(G^K_{P+Q}G^K_{P}+G^A_{P+Q}G^R_P+G^R_{P+Q}G^R_P\right)\right],\label{Heat32}
\end{align}
where we used the collective indices $Q=(q,\omega), P=(p,\nu)$ for momentum and
relative frequency. In Wigner approximation, the Green's functions are diagonal
in forward time and therefore evaluated at equal forward time $t$. The integrals
in~\eqref{Heat31}, \eqref{Heat32} are performed only over resonant momentum
configurations, see the discussion around Eq.~\eqref{Heat8}. In perturbation theory, the Green's functions under the integral
are the bare, non-interacting Green's functions
$G^R_Q=\left(\omega-u|q|+i0^+\right)^{-1}$, which diverge on the mass-shell
$\omega=u|q|$ and lead to a summation of infinities for the self-energy. On the
other hand, in self-consistent Born approximation, the bare Green's functions in
Eqs.~\eqref{Heat31}, \eqref{Heat32} are replaced by the full Green's functions
$G^R_{Q}=\left(\omega-u|q|-\Sigma^R_Q\right)^{-1}$. As a consequence, the
on-shell Green's function is regularized by the self-energy $\Sigma^R$ and takes
the value \eq{Heat32a}{ G^R_{q,u|q|}=-\left(\Sigma^R_{q,u|q|}\right)^{-1}=-i
  \left(\sigma^R_q\right)^{-1}.  } Inserting~\eqref{Heat32a} in the definition
for the retarded self-energy and evaluating the self-energy on-shell, one
obtains
\begin{equation}
  \label{Heat33}
  \sigma^R_{q,t}=v_0^2\int_{0<p}\left(\tfrac{\partial_tn_{p,t}}{\sigma^R_{q,t}}+2n_{p,t}+1\right)\left(\frac{qp(q-p)}{\sigma^R_{p,t}+\sigma^R_{q-p,t}}+\frac{pq(p+q)}{\sigma^R_{p,t}+\sigma^R_{p+q,t}}\right)
\end{equation}
with the vertex prefactor $v_0=v(1,1)$, cf. the definition below Eq.~\ref{Heat26}. A similar equation is obtained for the Keldysh on-shell self-energy $\sigma^K_{q,t}$. Inserting Eq.~\eqref{Heat33} and the result for the Keldysh self-energy, which we do not discuss here but can be found in Ref.~\cite{Buchholdmethod}, into the kinetic equation~\eqref{Heat25}, one finds
\begin{multline}
  \label{Heat34}
\partial_tn_q =\frac{\gamma|q|}{2\pi K}\\
+2v_0^2\int_{0<p<q}\hspace{-0.6cm}\frac{pq(q-p)(n_pn_{q-p}-n_q(1+n_p+n_{q-p}))}{\sigma^R_q+\sigma^R_p+\sigma^R_{q-p}}
\\ +4v_0^2\int_{0<p}\hspace{-0.4cm}\frac{pq(q+p)(n_{q+p}(n_q+n_p+1)-n_qn_p)}{\sigma^R_q+\sigma^R_p+\sigma^R_{q+p}}.
\end{multline}
The kinetic equation~\eqref{Heat34} and the equation for the on-shell self-energy~\eqref{Heat33} determine the forward time evolution of the system's phonon density $n_{q,t}$ and self-energy $\sigma^R_{q,t}$ in a self-consistent manner. For a general phonon density, both equations have to be solved iteratively according to the scheme depicted in Fig.~\ref{fig:Scheme}.
Before the numerical results for dynamics in the presence of heating, i.e., the numerical solution of Eqs.~\eqref{Heat33}, \eqref{Heat34}, are discussed, it is useful to study certain limiting cases. This facilitates the understanding of the numerical results in the subsequent section.
\subsubsection{Kinetics for small momenta}
For sufficiently small momenta $q\ll1$, the kinetic equation simplifies considerably. In this case, the second line of Eq.~\eqref{Heat34} can be discarded completely, since the integral is performed over a very small momentum interval $0<p<q$. On the other hand, for the integral in the third line of Eq.~\eqref{Heat34}, all terms $(q+p)\approx p$ for the dominant part of the integral. Therefore
\eq{Heat35}{
  \partial_tn_{q,t}=|q|\left(\frac{\gamma}{2\pi K}+\mathcal{J}_t\right), } where
the time-dependent integral $\mathcal{J}_t$ is $q$-independent and reads
\eq{Heat36}{ \mathcal{J}_t=2v_0^2\int_{0<p}\frac{p^2n_p(n_p+1)}{\sigma^R_{p}}.
} As a consequence, the resulting change of the phonon density for small momenta
$|q|$ is linearly proportional to $|q|$, i.e.  \eq{Heat37}{
  n_{q,t}=n_{q,0}+|q|\left(\frac{\gamma t}{2\pi
      K}+\int_{0<t'<t}\mathcal{J}_{t'}\right).  } The momentum regime
$0<|q|<q_2$, defined as the regime where the non-linear contributions in $|q|$ are negligible,
depends on the actual value of the phonon density $n_{q,t}$, and has to be
determined for each specific scenario individually. However, the existence of
$q_2$ is guaranteed by the above general arguments, and the fact that
$\partial_tn_{q=0,t}=0$ for all times $t$. The latter is a consequence of the
$U(1)$ symmetry of the present setup and the global particle number
conservation, as discussed in~\cite{Buchholdmethod}.

\subsubsection{Scaling of the self-energy}\label{sec:scaling}
The fact that the present system is described by a $U(1)$-invariant, massless field theory is reflected by the absence of a scale in the self-energy equation~\eqref{Heat33}. One important consequence is that $\sigma^R_{q,t}\overset{q\rightarrow0}{\rightarrow}0$ generically, i.e., the generation of a mass gap is forbidden by symmetry. 
A further consequence of Eq.~\eqref{Heat33} is that, whenever the term in brackets obeys a scaling law, 
\eq{Heat38}{
\tfrac{\partial_tn_p}{\sigma^R_{p}}+2n_p+1\sim \gamma_n |p|^{\eta_n},
}
the solution for the self-energy will be a scaling function $\sigma^R_q=\gamma_R |p|^{\eta_R}$ as well. Inserting this scaling ansatz in Eq.~\eqref{Heat33} yields
\eq{Heat39}{
\gamma_R|p|^{\eta_R}=\frac{v_0^2\gamma_n}{\gamma_R}|p|^{4+\eta_n-\eta_R}\mathcal{I}_{\eta_n,\eta_R}
}
and identifies
\eq{Heat40}{
\eta_R=2+\frac{\eta_n}{2} \text{ and } \gamma_R=v_0\sqrt{\gamma_n\mathcal{I}_{\eta_n,\eta_R}}.
}
The dimensionless integral $\mathcal{I}_{\eta_n,\eta_R}$ is defined as
\eq{Heat41}{
\mathcal{I}_{\eta_n,\eta_R}=\int_{0<x}x^{\eta_n}\left(\frac{x(1-x)}{x^{\eta_R}+|1-x|^{\eta_R}}+\frac{x(1+x)}{x^{\eta_R}+(1+x)^{\eta_R}}\right),
}
and depends only on the exponents $\eta_R, \eta_n$. This self-energy integral is dominated by contributions around $x=1$ and converges for all physically reasonable phonon densities. In this sense, the scaling behavior is universal, i.e. robust against the influence from high energy modes~\cite{Buchholdmethod,heating}.
\subsection{Heating and universality}
\label{sec:heatison} With the preparations from the previous sections, one can simulate the heating dynamics of an interacting Luttinger liquid in terms of the time dependent phonon population $n_{q,t}$ and self-energy $\sigma^R_{q,t}$, as has been done in Refs.~\cite{Buchholdmethod,heating}. Considering the system initially to be in the ground state, the simulation is initialized with a phonon density $n_{q,t=0}=0$, which for $t>0$ is continuously increased due to heating. The central result of the analysis is a scaling solution for the self-energies $\sigma^R_{q,t}\sim |q|^{\eta_R}$ with a new non-equilibrium exponent $\eta_R=5/3$, which is observable in the long-wavelength regime, i.e., on distances $x>x_{\text{th}}(t)$. Here, $x_{\text{th}}(t)$ marks a thermal distance, below which the dynamics is dominated by thermalized short distance modes and above which the phonon density increases linearly in momentum $n_{q,t}\sim|q|$, as it was the case for the bare heating~\eqref{Heat9}. The thermal distance increases sub-ballistically $x_{\text{th}}(t)\sim t^{4/5}$ in time, with a characteristic heating exponent $\eta_h=4/5$, while at the same time, the effective temperature describing the distribution of the short distance modes increases linearly in time $T(t)\sim t$.

\subsubsection{Phonon densities}
\begin{figure}
  \centering
  \includegraphics[width=\linewidth]{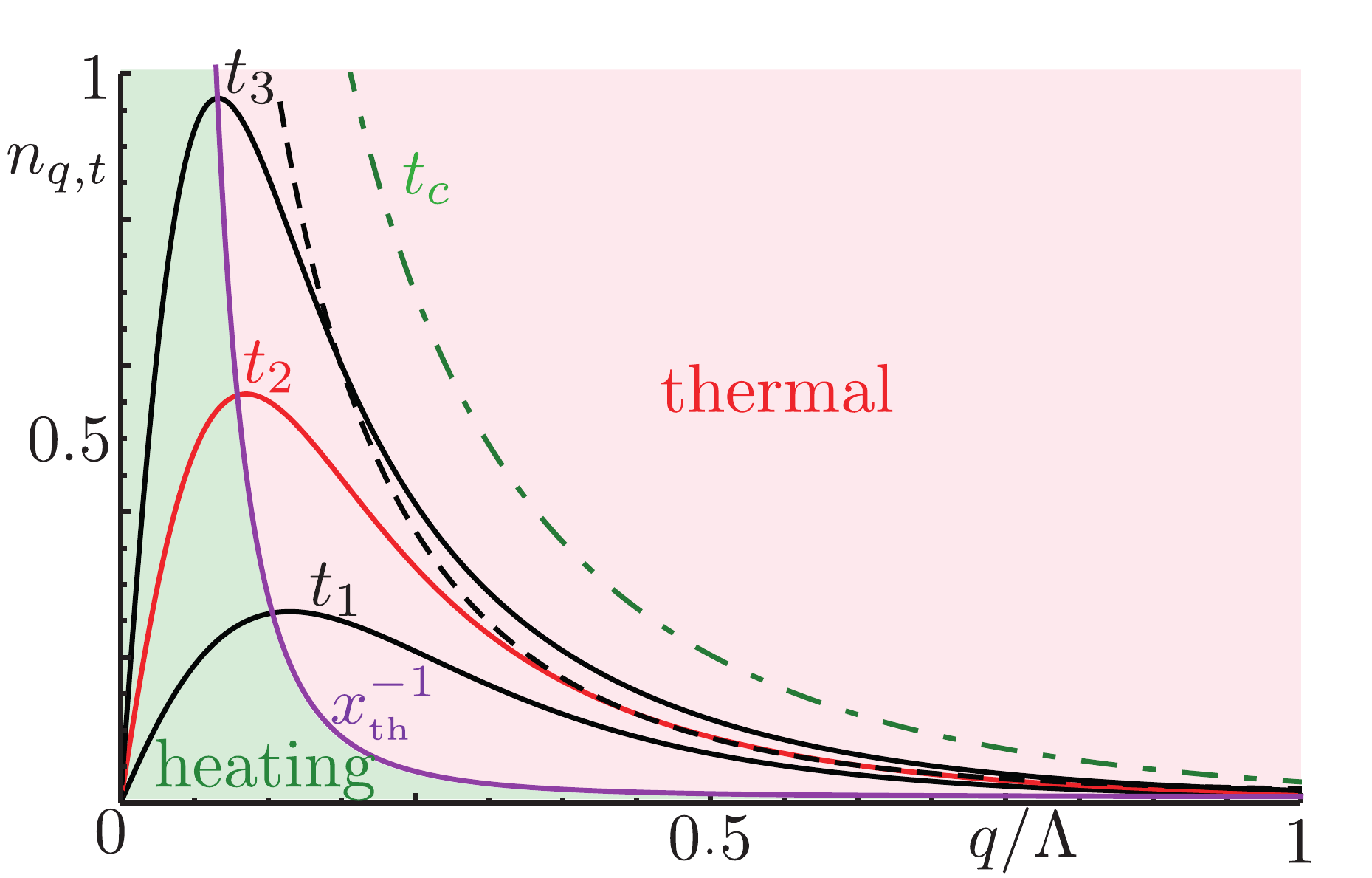}
  \caption{Time evolution of the phonon density $n_{q,t}$ in a sequence
    $(t_1,t_2,t_3)=(2,3,4)\cdot\frac{10}{v_0\Lambda^2}$ in terms of $q/\Lambda$
    and for a heating rate $\gamma=0.06 v_0\Lambda$. In the heating regime, for
    small momenta, the distribution increases linearly in momentum
    $n_{q,t}\sim|q|$, while it decreases as $n_{q,t}\sim 1/|q|$ in the
    interaction dominated thermal regime. The crossover
    $x_{\text{th}}^{-1}\sim t^{-4/5}$ approaches zero as time evolves. The
    dashed line represents a Bose-Einstein distribution corresponding to the
    phonon density at $t=t_3$.  The dash-dotted line indicates $n_B(T(t_c))$ at
    the time $t_c$, for which the Luttinger description breaks down, i.e.,
    $T(t_c)=u\Lambda$.  }
  \label{fig:DistDephTemp}
\end{figure}

The results of a numerical simulation of the phonon densities are shown in Fig.~\ref{fig:DistDephTemp}. 
One can clearly identify the crossover from an interaction dominated thermal regime with $n_{q,t}=\frac{T(t)}{u|q|}$ at large momenta to a heating dominated regime with $n_{q,t}\sim|q|$. The crossover momentum between the two regimes represents the inverse thermal length $(x_{\text{th}}(t))^{-1}\sim t^{-4/5}$.

In the large momentum regime, the dominant contribution to the kinetic equation is the collision term, which establishes an approximate detailed balance between phonon absorption and emission in the presence of the heating. This is expressed by the evolution of the phonon density towards a Bose-Einstein distribution, which is the fixed point of the collision term alone. In this regime $n_{q,t}=n_{\text{B}}(u|q|,T(t))\approx \frac{T(t)}{u|q|}$ with very good agreement and the only indicator of the permanent heating is the continuously increasing temperature $T(t)\sim t$.
This is contrasted by the evolution of the phonon density in the low momentum regime, which is dominated by strong phonon production. In this regime, the scattering of high-momentum phonons into low-momentum modes enhances the effect of the heating and leads, due to the structure of the vertex, to an increase of the linear production rate according to Eq.~\eqref{Heat35}.
The pinning of the phonon occupation of the $q=0$ mode is an exact result for the underlying $U(1)$-symmetric, i.e. particle number conserving dynamics. It can be shown that the phonon number fluctuations in the zero momentum mode are proportional to the fluctuations of the global particle number in the system, see Ref.~\cite{heating}, and consequently they are integrals of motion for a $U(1)$-symmetry preserving dynamics. The pinning effect for the low momentum distribution at $n_{q=0,t} =n_{q=0,0}$  leads to a very slow, sub-ballistic thermalization of the low momentum regime, since the formation of the typical, thermal Rayleigh-Jeans divergence $n_{q,t}\sim 1/|q|$ is only achieved by the scaling of the inverse thermal length $(x_{\text{th}}(t))^{-1}$ to zero, instead of a direct filling of these modes.
\subsubsection{Non-equilibrium scaling}
Away from the crossover scale $q \neq x_{\text{th}}^{-1}$, deep in the heating or thermal regimes, the phonon density exhibits scaling behavior and can be written as
\eq{Heat42}{
n_{q,t}=\left\{\begin{array}{cl}c(t)|q| & \text{ for }|q|\ll x_{th}^{-1} \\
\frac{T(t)}{u|q|}& \text{ for } |q|\gg x_{th}^{-1}\end{array}\right. ,
}
where the functions $c(t)$, $T(t)$ have to be determined numerically. In these regimes, according to Sec.~\ref{sec:scaling}, one finds scaling behavior of the phonon self-energy as well. The corresponding scaling exponent is determined by
\eq{Heat43}{
f_{q,t}\equiv\frac{\partial_tn_{q,t}}{\sigma^R_{q,t}}+2n_{q,t}+1\sim |q|^{\eta_n}.
}
In the heating dominated regime, $n_{q,t}=c(t)|q|$ and therefore
\eq{Heat44}{
f_{q,t}=\frac{c'(t)}{\gamma_R}|q|^{1-\eta_R}+c(t)|q|+1\overset{\eta_R>1}{\rightarrow} \frac{c'(t)}{\gamma_R}|q|^{1-\eta_R}
}
for small momenta, since $\eta_R>1$ is guaranteed by the subleading nature of the interactions. This directly implies $\eta_n=1-\eta_R$ and a super-diffusive exponent $\eta_R = \frac{5}{3}$ in this regime, which lacks any equilibrium counterpart.

On the other hand, in the large momentum regime $f_{q,t}$ is obviously dominated
by the term proportional to $n_{q,t}$ and $\eta_n=-1$, which leads to the known
thermal equilibrium exponent $\eta_R=\frac{3}{2}$~\cite{andreev80} and a thermal
scaling behavior for large momenta not only in the distribution function but
also in the self-energy.  The crossover between the two scaling regimes can be
estimated by equating the two relevant terms, which tend to dominate $f_{q,t}$
in the corresponding regimes. This yields \eq{EqNonEq}{
  \frac{\partial_{t}n_{q_{\text{th}}}}{2n_{q_{\text{th}}}+1}=\sigma^R_{q_{\text{th}}},} for
$q_{\text{th}}=x_{\text{th}}^{-1}$ and can be used to analytically estimate the scaling of
$x_{\text{th}}(t)$ in time~\cite{heating}, resulting in
$x_{\text{th}}(t)\sim t^{4/5}$, which agrees well with the numerical findings.

\subsubsection{Observability} 
The universal scaling of the phonon self-energies as well as the scaling regimes of the phonon distribution function can be observed in cold atom experiments via Bragg spectroscopy~\cite{heating}, which gives direct access to the characteristic universal features in the heating dynamics.
In Bragg experiments, the detected Bragg signal is directly proportional to the Fourier transform of the two-point density-density correlation function~\cite{Ketterle1999,Esslinger2004B,Sengstock2011,Meinert2015} or dynamical structure factor $S_{q,t,\omega} = \int d\delta dx e^{i (qx - \omega \delta)}\langle \{ n (t+\delta/2,x),n (t-\delta/2,0)\}\rangle$. In the Luttinger framework in the Keldysh formalism this translates into  $S_{q,t,\omega} = - \langle\rho_c(-q,t,-\omega)\rho_c(q,t,\omega)\rangle$.
Explicit evaluation of the structure factor yields
\begin{equation}
  S_{q,t,\omega} = \tfrac{(2n_{q,t}+1)|q|K}{\pi\sigma^R_{q,t}} \left(
    \tilde{f}\left(\tfrac{\omega-\epsilon_q}{\sigma^R_{q,t}}\right) +
    \scriptstyle{(\omega\to -\omega)} \right).
\end{equation}
Here, $\tilde f (x) = 1/(1+x^2)$ is a dimensionless scaling function, which is centered at $x=0$ and has unit width. As a consequence, the dynamic structure factor is peaked at the mass shell $\omega=u|q|$ and has a typical width $\delta\omega=\sigma^R_{q,t}$, which reveals the scaling of the self-energies. 

The scaling of the distribution function and the time dependent crossover $x_{\text{th}}(t)$ does not necessitate dynamical (frequency resolved) information, and can be obtained from the static structure factor alone. The latter represents the equal time density-density correlation function. It is determined as the frequency integrated dynamic structure factor
\begin{eqnarray}\label{eq:staticstruct}
S_{q,t} = \int_{\omega}S_{q,t,\omega}= \tfrac{|q|K}{\pi}(2n_{q,t}+1)
\end{eqnarray}
and scales quadratically in the momentum in the heating regime while approaching a constant in the thermalized regime, thereby revealing the crossover between these two regimes.

%%% Local Variables:
%%% mode: latex
%%% TeX-master: "dds_review"
%%% End:

\section{Conclusions and outlook}
\label{sec:conclusions}

We have reviewed here recent progress in the theory of driven open quantum
systems, which are at the interface of quantum optics, many-body physics, and
statistical mechanics. In particular, we have developed a quantum field
theoretical approach based on the Keldysh functional integral for open systems,
which underlies these advances. The formalism developed here paves the way for
many future applications and discoveries. We may structure these into four
groups of topics.

\emph{Semi-classical regime} --- A key challenge here is to sharpen the contrast
between equilibrium and genuine non-equilibrium physics. One way of succeeding
in this respect is to uncover new links of driven open quantum systems to
paradigmatic situations in non-equilibrium (classical) statistical mechanics. We
have discussed one such connection between the phase dynamics of driven Bose
condensates and surface growth in Sec.~\ref{sec:bosons}. Further instances have
been pointed out in the literature recently: for example, driven Rydberg gases
can be brought into regimes where they connect to the physics of
glasses~\cite{Olmos2012} or the universality class of directed
percolation~\cite{MarcuzziDP}, and the late stages of heating of atoms in
optical lattices show slow decoherence dynamics described by non-linear
diffusion, reminiscent of glasses as
well~\cite{Cai2013,Poletti2013,Sciolla2015}. Connections of such settings to
field theoretical non-linear reaction-diffusion models~\cite{Cardy1996,CardyRev}
still await their exploration. A new kind of equilibrium to non-equilibrium
phase transition may be expected in three dimensional driven bosonic systems on
the basis of the phase diagram for surface roughening, and it is intriguing to
investigate whether ultracold atom setups could provide a physical platform to
explore such physics.

Even more ambitiously, such driven open quantum systems hold the potential for
 truly new paradigms in non-equilibrium statistical mechanics. One example is presented by the fundamental open question on
the driven open analogue of the Kosterlitz-Thouless scenario in two dimensions, directly relevant for experiments with exciton-polariton systems.  Technically, this requires to analyze a KPZ equation with a \emph{compact}
variable, allowing for the presence of vortex defects. Keldysh field theory
offers the flexibility to address such questions.

\emph{Quantum regime} --- The \emph{quantum} dynamical field theory framework developed
here also allows us to address problems where the limit of \emph{classical} dynamical
field theories (see Sec.~\ref{sec:semicl-limit-keldysh}) is not applicable. Here
one challenge is to identify traces of non-equilibrium quantum effects at
macroscopic distances. An instance of such a phenomenon has been established
recently in terms of a driven analogue of quantum critical behavior in a system
with a dark state, a state which is decoupled from noise~\cite{Marino2015}. It remains
to be seen whether a full classification of driven Markovian (quantum)
criticality can be achieved, complementing the seminal analysis of equilibrium
classical criticality by Hohenberg and Halperin~\cite{Hohenberg1977}. Beyond
bosonic systems, this also includes fermionic systems, which have been shown to
exhibit critical scaling~\cite{Eisert2010,Horstmann2013,Hoening2012}, but so far were
analyzed at a Gaussian fixed point only. 

Another challenge in this direction is to identify effects, which unambiguously reveal the microscopic quantum mechanical origin of the underlying dynamics. Here an example was provided recently in the context of driven Rydberg systems, where a short distance constraint in the coherent Hamilton dynamics gives rise to an additional relevant direction in parameter space, leading to a new kind of absorbing state phase transition without immediate counterpart in models of classical origin~\cite{Marcuzzi2016}.

Certainly, progress  in this respect will necessitate a more comprehensive understanding of the structure of quantum dynamical field theories. One relevant issue is to reveal universal aspects of the low frequency dynamics tied to the presence of conservation laws. A concrete goal is the systematic construction of  dynamical slow modes on the basis of the symmetries (and their breaking patterns) of the Keldysh functional integral.

\emph{Topology in open quantum systems} -- Recently, experiments with photonic lattices have started to address quantum Hall physics in driven open quantum systems~\cite{Hafezi14,Hafezi16}. Although these systems can be idealized to some extent as closed systems, it is a fundamental challenge to explore the fate of the quantum Hall effect -- or more generally, physical phenomena related to topology -- under general non-equilibrium conditions. Another angle is provided by theoretical proposals, where drive and/or dissipation do not occur as a small perturbation, but rather as the dominant resource of many-body dynamics, guiding the system density matrix into topologically nontrivial states, which sometimes even do not have a direct equilibrium counterpart. This concerns topologically non-trivial dark states in driven open atomic fermion systems~\cite{Diehl2011,Bardyn2013} and periodically driven (Floquet) dynamics~\cite{Lindner11,Karzig15} alike. 

The density matrices describing such systems typically do not correspond to pure, but rather to mixed states. While such density matrices can still host non-trivial topological properties~\cite{Avron2012,Delgado14,Budich2015}, the extent to which this translates into physically observable consequences in the correlations and responses to (artificial) external gauge fields is at the moment a wide open issue. This calls for the development of non-equilibrium topological field theories in the framework of the Keldysh functional integral: the established equilibrium counterpart has proven to be able to efficiently describe both bulk correlations and responses to external gauge potentials, as well as to provide a proper notion of the bulk-boundary correspondence (cf. e.g.~\cite{Qi11}), giving access to the edge physics in both non-interacting and interacting systems.

\emph{Dynamics} --- Addressing the time evolution of open systems adds a new
twist to the question of
thermalization~\cite{Rigol2008,Calabrese2006,Moeckel2008,Gring2012} or, more
generally, equilibration of quantum systems. This is also a necessary step to
achieve a realistic description of broad classes of experiments, in particular,
with ultracold atoms, as well as certain solid states systems in pump-probe
setups~\cite{PumpProbe1,PumpProbe2}. While first instances of universal behavior
have been identified in low dimension in the short~\cite{heating} and
long~\cite{Poletti2013,Schachenmayer2014,Lang2016} dynamics, it is certainly
fair to say that general principles so far remain elusive.

\section{Acknowledgements}
\label{sec:ack}

The authors thank E. Altman, L. Chen, A. Chiocchetta, E. G. Dalla Torre,
A. Gambassi, L. He, S. D. Huber, S. E. Huber, M. Lukin, J. Marino, S. Sachdev,
P. Strack, U. T\"auber, and J. Toner for collaboration on projects discussed in
this review. We are also grateful for inspiring and useful discussions with many
people at the KITP workshop ``Many-Body Physics with Light.'' In particular, we
thank A. Altland, B. Altshuler, I. Carusotto, A. Daley, A. Gorshkov, M. Hafezi,
M. Hartmann, R. Fazio, M. Fleischhauer, A. Imamoglu, J. Koch, I. Lerner,
I. Lesanovsky, P. Rabl, A. Rosch, G. Shlyapnikov, M. Szymanska, J. Taylor and
H. T\"ureci for critical remarks and feedback on the manuscript. We are indebted
to J. Koch for proofreading the manuscript. We acknowledge support by the
Austrian Science Fund (FWF) through the START grant Y 581-N16 and the German
Research Foundation (DFG) via ZUK 64 and through the Institutional Strategy of
the University of Cologne within the German Excellence Initiative (ZUK
81). S. D. also acknowledges support by the European Research Council (ERC) under the European UnionÕs Horizon 2020 research and innovation programme (grant agreement No 647434) and by the Kavli Institute for Theoretical
Physics, via the National Science Foundation under Grant No. NSF PHY11-25915,
and L. M. S. support from the European Research Council through Synergy Grant
UQUAM.
%%% Local Variables:
%%% mode: latex
%%% TeX-master: "dds_review"
%%% End:

\appendix

\section{Functional differentiation}
\label{sec:funct-diff}

In this appendix, we give a brief account of functional or variational
differentiation, following the presentation in Ref.~\cite{Boettcher2012}. The
most transparent way to introduce the basic relations of functional
differentiation is by drawing an analogy to the familiar formulas of partial
differentiation. Indeed, we can consider a field $\phi(x)$ with $x \in \R^d$ to
be the continuum limit of a function $\phi_i$ defined on lattice points
$i \in \Z^d$. With this identification, relations involving partial derivatives
can be translated into corresponding ones for functional derivatives simply by
replacing discrete indices $i$ by continuous ones $x$, sums $\sum_i$ by
integrals $\int_x = \int d^d x$, and partial derivatives
$\partial/\partial \phi_i$ by functional derivatives $\delta/\delta
\phi(x)$.
Following this prescription, the basic formula
$\partial \phi_i/\partial \phi_j = \delta_{i j}$ leads to
\begin{equation}
  \label{eq:129}
  \frac{\delta \phi(x)}{\delta \phi(x')} = \delta(x - x').
\end{equation}
When we are working with complex fields, $\phi$ and $\phi^{*}$ are usually
treated as independent variables. Then, the generalization of
\begin{equation}
  \label{eq:130}
  \frac{\partial}{\partial \phi_k^*}  \left( \sum_{i,j} \phi_i^* A_{ij}
    \phi_j \right) = \sum_j A_{kj} \phi_j
\end{equation}
reads
\begin{equation}
  \label{eq:131}
  \frac{\delta}{\delta \phi^{*}(x)}  \int_{y,y'} \phi^*(y) A(y, y') \phi(y')
  = \int_{y'} A(x,y') \phi(y').
\end{equation}
The expression $A(x,x')$ is the continuum limit of a matrix $A_{ij}$. In
particular, it can be a differential operator. As an example, we calculate the
second functional derivative for the case that $A(x, x') = \nabla^2 \delta(x -
x').$ By straightforward differentiation we find
\begin{equation}
  \label{eq:132}
  \begin{split}
    \frac{\delta^2}{\delta \phi^{*}(x) \delta \phi(x')} \int_y \phi^*(y) \nabla^2
    \phi(y) & = \frac{\delta}{\delta \phi^{*}(x)} \int_y \phi^{*}(y) \nabla^2
    \delta(y - x') \\ & = \frac{\delta}{\delta \phi^{*}(x)} \nabla^2 \phi(x')
    \\ & = \nabla^2 \delta(x - x').
  \end{split}
\end{equation}
Finally, we note that the chain rule applies also in the case of functional
differentiation. This last ingredient is required to perform the second
variational derivatives in Eq.~\eqref{eq:greend} in order to obtain the Green's
functions from the generating functional.

\section{Gaussian functional integration}
\label{sec:gauss-funct-integr}

Here we summarize a number of useful formulas for Gaussian functional
integration, which can be found in any textbook on field theory (see, e.g.,
Refs.~\cite{Altland2010,Negele1998,Folland2008}). The basic formula for real
fields $\chi(x) = \left( \chi_1(x), \dotsc, \chi_n(x) \right)$, $x \in \R^d$ (in
the applications discussed in the main text, the components of $\chi(x)$ are
fields on the closed time path or --- after performing the Keldysh rotation
Eq.~\eqref{eq:41} --- classical and quantum fields, and $x$ collects temporal
and spatial coordinates, $x = \left( t, \mathbf{x} \right)$), is given by
\begin{multline}
  \label{eq:Gaussian-int-198}
  \int \mathscr{D}[\chi] \, e^{\frac{i}{2} \int_{x, x'} \chi(x)^T D(x,x')
    \chi(x') + i \int_x j(x)^T \chi(x)} \\ = \left( \det D \right)^{-1/2}
  e^{-\frac{i}{2} \int_{x, x'} j(x)^T D^{-1}(x,x') j(x')},
\end{multline}
where $\int_x = \int d^d x,$ $j(x) = \left( j_1(x), \dotsc, j_n(x) \right)$, and
the inverse of the integral kernel $D(x,x')$ is defined by means of the relation
\begin{equation}
  \label{eq:Gaussian-int-199}
  \int_{\xi} D(x,\xi) D^{-1}(\xi,x') = \delta(x - x') \id.
\end{equation}
The above formula is valid for invertible symmetric kernels,
$D(x,x') = D(x',x)^T$, with positive semi-definite imaginary part. If $D(x,x')$
is not symmetric, on the LHS of Eq.~\eqref{eq:Gaussian-int-198} it can be
replaced by the symmetrized kernel
\begin{equation}
  \label{eq:Gaussian-int-200}
  \tilde{D}(x,x') = \frac{1}{2} \left[ D(x,x') + D(x',x)^T \right],
\end{equation}
leaving the value of the exponent invariant. Then,
Eq.~\eqref{eq:Gaussian-int-198} can again be applied.

In case that the integral kernel is translaitionally invariant, i.e., it
satisfies $D(x,x') = D(x - x')$, it is advantageous to work in Fourier
space. Then, Eq.~\eqref{eq:Gaussian-int-198} can be written as
\begin{multline}
  \label{eq:Gaussian-int-201}
  \int \mathscr{D}[\chi] \, e^{\frac{i}{2} \int_q \chi(-q)^T D(q) \chi(q) + i
    \int_q j(-q)^T \chi(q)} \\ = \left( \det D \right)^{-1/2} e^{-\frac{i}{2}
    \int_q j(-q)^T D^{-1}(q) j(q)},
\end{multline}
where we are using the shorthand notation
$\int_q \equiv \int \frac{d q}{(2 \pi)^d}$. The Fourier transformation turns
the convolution in Eq.~\eqref{eq:Gaussian-int-200} into a multiplication,
showing that $D^{-1}(q)$ can be obtained by inversion of the matrix $D(q)$,
\begin{equation}
  \label{eq:Gaussian-int-202}
  D(q) D^{-1}(q) = \id.
\end{equation}
As above, in Eq.~\eqref{eq:Gaussian-int-201} we are assuming $D(q) = D(-q)^T$.
If this is not the case, $D(q)$ has to be replaced by the symmetrized version
\begin{equation}
  \label{eq:Gaussian-int-203}
  \tilde{D}(q) = \frac{1}{2} \left[ D(q) + D(-q)^T \right],
\end{equation}
before the formula can be applied.

For the case of complex fields $\psi(x) = \left( \psi_1(x), \dotsc, \psi_n(x)
\right)$ (and with corresponding definitions of $\phi(x)$ and $\chi(x)$),
Eq.~\eqref{eq:Gaussian-int-198} is replaced by
\begin{multline}
  \label{eq:88}
  \int \mathscr{D}[\psi, \psi^{*}] \, e^{i \int_{x, x'} \psi^{\dagger}(x)
    D(x,x') \psi(x') + i \int_x \left( \phi^{\dagger}(x) \psi(x) +
      \psi^{\dagger}(x) \chi(x) \right)} \\ = \left( \det D \right)^{-1} e^{-i
    \int_{x, x'} \phi^{\dagger}(x) D^{-1}(x,x') \chi(x)},
\end{multline}
where we are assuming that $-i \left( D - D^{\dagger} \right)$ is positive
semi-definite, but $D(x, x')$ does not have to be symmetric. The corresponding
formula for the case of a translationally invariant integral kernel reads (note
the different signs of $q$ in comparison to Eq.~\eqref{eq:Gaussian-int-201})
\begin{multline}
  \label{eq:89}
  \int \mathscr{D}[\psi, \psi^{*}] \, e^{i \int_q \psi^{\dagger}(q) D(q) \psi(q)
    + i \int_q \left( \phi^{\dagger}(q) \psi(q) + \psi^{\dagger}(q) \chi(q)
    \right)} \\ = \left( \det D \right)^{-1} e^{-i \int_q \phi^{\dagger}(q)
    D^{-1}(q) \chi(q)}.
\end{multline}

%%% Local Variables:
%%% mode: latex
%%% TeX-master: "dds_review"
%%% End:

% \bibliographystyle{iopart-num}
% \bibliographystyle{apsrev4-1}
% \bibliographystyle{unsrt}
\bibliography{dds_review}

\end{document}